%% file: main.tex
\documentclass[AMA,STIX1COL]{WileyNJD-v2}
\pdfoutput=1
\usepackage{moreverb}
\usepackage{NJDnatbib}

\input{sections/0-commands}

\usepackage[pagewise]{lineno}

\articletype{Research Article}%

\received{<day> <Month>, <year>}
\revised{<day> <Month>, <year>}
\accepted{<day> <Month>, <year>}

\begin{document}

\title{Comparing methods addressing multi-collinearity when developing prediction models} 

\author[1]{Artuur M. Leeuwenberg}
\author[1]{Maarten van Smeden}

\author[2]{Johannes A. Langendijk}
\author[2]{Arjen van der Schaaf}

\author[3]{Murielle E. Mauer}

\author[1]{Karel G.M. Moons}
\author[1]{Johannes B. Reitsma}
\author[1]{Ewoud Schuit}

\authormark{LEEUWENBERG \textsc{et al}}

\address[1]{\orgdiv{Julius Center for Health Sciences and Primary Care}, \orgname{University Medical Center Utrecht}, Utrecht University, \orgaddress{\state{Utrecht}, \country{The Netherlands}}}
\address[2]{\orgdiv{Department of Radiation
Oncology}, \orgname{University Medical Center Groningen}, Groningen University,  \orgaddress{\state{Groningen}, \country{The Netherlands}}}
\address[3]{\orgname{European Organisation for Research and Treatment of Cancer Headquarters}, \orgaddress{\state{Brussels}, \country{Belgium}}}



\corres{Artuur Leeuwenberg, Julius Center for Health Sciences and Primary Care, University Medical Center Utrecht, Utrecht University, 3508 GA, Utrecht, The Netherlands. \\\email{aleeuw15@umcutrecht.nl}}



\input{sections/01-ABSTRACT}

\maketitle

\blfootnote{\textbf{Abbreviations:} NTCP, normal tissue complication probability; OAR, organ at risk}

\input{sections/1-INTRO}
\input{sections/2-METHODS}
\input{sections/3-NTCP-SIM}

\input{sections/4-DISCUSSION}

\input{sections/5-ACKNOWLEDGEMENTS}

\bibliography{main}

\input{sections/6-APPENDIX}


\end{document}

%% file: sections/0-commands.tex
\usepackage{pdflscape} 

\usepackage{amsmath}
\usepackage{multicol}
\usepackage{tikz}
\usepackage{collcell}
\usepackage{colortbl}
\usepackage{pdfcomment}
\usepackage{etoolbox} 
\usepackage{chemmacros}
\usepackage{color, colortbl}
\usepackage{array}
\usepackage{hyperref}

\newcommand{\tuurfootnote}[1]{}

\newcommand\BibTeX{{\rmfamily B\kern-.05em \textsc{i\kern-.025em b}\kern-.08em
T\kern-.1667em\lower.7ex\hbox{E}\kern-.125emX}}
\DeclareMathOperator*{\argmin}{arg\,min}

\newcommand{\xerbasiclow}{A}
\newcommand{\xerbasichigh}{A$_{\vartriangle}$}
\newcommand{\xerextlow}{B}
\newcommand{\xerexthigh}{B$_{\vartriangle}$}
\newcommand{\dysbasiclow}{C}
\newcommand{\dysbasichigh}{C$_{\vartriangle}$}
\newcommand{\dysextlow}{D}
\newcommand{\dysexthigh}{D$_{\vartriangle}$}

\newcommand{\xerbasiclowreal}{A$^*$}
\newcommand{\xerexthighreal}{B$_{\vartriangle}^*$}
\newcommand{\dysbasiclowreal}{C$^*$}
\newcommand{\dysexthighreal}{D$_{\vartriangle}^*$}

\newcommand{\citl}[0]{Intercept} 
\newcommand{\cslope}[0]{Slope}
\newcommand{\auroc}[0]{AUROC}
\newcommand{\rsquared}[0]{R$^2$}

\newcommand{\expjacc}[0]{MJICS}
\newcommand{\mse}[0]{MSE}
\newcommand{\sign}[0]{\text{sgn}}

\newcommand{\LRnn}[0]{LR$_\textsc{nn}$}

\newcommand{\grayline}[0]{\arrayrulecolor[gray]{0.5}\midrule}

\newcommand\blfootnote[1]{%
  \begingroup
  \renewcommand\thefootnote{}\footnote{#1}%
  \addtocounter{footnote}{-1}%
  \endgroup
}


%% file: sections/01-ABSTRACT.tex
\abstract[Abstract]{
Clinical prediction models are developed widely across medical disciplines. When predictors in such models are highly collinear, unexpected or spurious predictor-outcome \mbox{associations} may occur, thereby potentially reducing face-validity and explainability of the prediction model. Collinearity can be dealt with by exclusion of collinear predictors, but when there is no a priori motivation (besides collinearity) to include or exclude specific predictors, such an approach is arbitrary and possibly inappropriate. We compare different methods to address collinearity, including shrinkage, dimensionality reduction, and constrained optimization. The effectiveness of these methods is illustrated via simulations.
In the conducted simulations, no effect of collinearity was observed on predictive outcomes. However, a negative effect of collinearity on the stability of predictor selection was found, affecting all compared methods, but in particular methods that perform strong predictor selection (e.g., Lasso).}

\keywords{Multi-collinearity, prediction models, normal-tissue complication probability models}

%% file: sections/1-INTRO.tex

\section{Introduction}\label{sec:intro}


Multi-collinearity between predictors 
is a common phenomenon in clinical prediction modeling, for example, in
prediction of Alzheimer's disease from MRI images \cite{teipel2015relative},
prediction of metabolic acidosis in laboring women that had a high-risk singleton pregnancy in cephalic presentation beyond 36 weeks of gestation,\cite{westerhuis2012prediction}
prediction of lung function in children,\cite{narchi2018prediction} and prediction of complications of radiotherapy in cancer patients.\cite{van2020key,van2019method} Multi-collinearity is caused by dependence between predictors.\cite{schisterman2017collinearity} 
When collinearity among predictors is high, the data in itself provides limited information on how the explained variance in the outcome should be distributed over the collinear predictor coefficients. In other words, there is not just one model, but there are multiple ways to assign coefficients that can predict the outcome in the data used to develop the model (almost) equally well. \cite{farrar1967multicollinearity} Consequently, model coefficients of collinear variables generally show large variance (large standard errors) even in large data sets. Although this is generally not considered problematic with regard to predictive performance,\cite{harrell2015regression} it 
can result in unexpected coefficients for individual predictors, reducing the face-validity and explainability of predictors included in the resulting model and thus the model in general.\cite{schuit2013unexpected,moons2009prognosis}

Two common methods to address collinearity are \textit{predictor selection}\tuurfootnote{add example?}, and \textit{predictor averaging}\tuurfootnote{add example?}. Both make strong assumptions about the predictive value of the collinear predictors. Predictor selection assumes that the excluded predictors have no added predictive value over the predictors that are retained in the model with respect to the outcome (essentially imposing coefficients of zero). Predictor averaging assumes that the averaged predictors have the same predictive relation to the outcome (imposing exact equivalence of the coefficients). In some cases it is possible to convincingly motivate such assumptions using prior clinical knowledge or by resorting to data-driven approaches (e.g., backward selection). However, finding evidence in the data for such strong assumptions can be difficult, especially when collinearity is high, and the outcome is only weakly associated with the difference between collinear predictors.
Therefore, further research into more sophisticated methods to address collinearity is needed.

This article is organized as follows: In section \ref{sec:methods}, we describe different methods for handling multi-collinearity. In section \ref{sec:ntcp}, we compare the described methods via simulations in a case study on the development of models for the prediction of complications of radiotherapy in cancer patients, in terms of predictive performance, and in terms of coefficient estimation, including the choice of predictors in the final model. In section \ref{sec:discussion}, we discuss and summarize our conclusions.















%% file: sections/2-METHODS.tex
\section{Methods for Collinearity}\label{sec:methods}

\subsection{Penalization of large coefficients}

We assume the interest is in a binary outcome ($y$) and candidate predictors $X$. The aim is to estimate the risk of $y$ conditioned on the predictor values, $P(y=1|X)$. As a base model, we assume standard logistic regression (\textbf{LR}), shown in Equation \ref{eq:lr_func} , where $\beta=\langle \beta_1, \beta_2, ..., \beta_d \rangle$ is the vector of coefficients, $\beta_0$ the intercept, and coefficient are estimated by maximizing the likelihood of the outcome in the data used for model development.
\begin{align}
\label{eq:lr_func}
    P(y| X, \beta, \beta_0) = \frac{1}{e^{-(\beta X  + \beta_0)}}
\end{align}

In addition to the maximum likelihood of the outcome in the development data, approaches like \textbf{Lasso} and \textbf{Ridge} include the size of the model's coefficients (excluding the intercept) as an extra penalty for coefficient estimation. Adding this penalty results in models with smaller coefficients that make less extreme predictions (closer to the outcome proportion). The penalty can also reduce the variance in the estimated coefficients induced by collinearity. Although Lasso and Ridge have similar structure penalizing high regression coefficients, Ridge was originally designed to address collinearity, and Lasso to perform predictor selection in high-dimensional data.
Lasso penalizes large coefficients linearly, by extending the cost function with the $\ell_1$-norm of the coefficients, which generally results in predictor selection of the most predictive features.\cite{tibshirani1996regression}
Ridge penalizes coefficient size quadratically, resulting in a grouping effect of collinear predictors, instead of selection.\cite{hoerl1970ridge} 
In practice, the desire to perform predictor selection may be independent of the degree of collinearity present in the data, and rather to enhance usability of a more parsimonious prediction model. To facilitate a balance between predictor selection and grouping, the \textbf{Elastic Net} method was developed, \cite{zou2005regularization} which combines the penalties of Lasso and Ridge.

Penalization of coefficient size is a popular method in clinical prediction, aimed to improve predictive performance over maximum likelihood. Recent simulation studies suggest these penalization approaches often improve the predictive performance on average, but can show poor performance in small and low-dimensional datasets.\cite{riley2020penalisation}


\subsection{Dropout regularization}
\textbf{Dropout} regularization is a method aimed directly at reducing co-adaptation of coefficients during model estimation, and is widely used for regularization of neural networks.\cite{hinton2012improving} Co-adaptation refers to the degree to which the value of one regression coefficient depends on that of other coefficients.
Dropout works in iterative gradient-based training procedures, like the one used in the current work (described in Appendix \ref{appendix:loss_minimization}). When using Dropout, at each (gradient-based) learning step, each predictor has a non-zero probability $\delta$ to be dropped from the model with a certain probability, effectively selecting a random sub-model at each iteration. This selected sub-model is used to make predictions as part of that learning step, and the involved coefficients are updated accordingly. The coefficients selected at each step are updated independently of the dropped-out predictors, preventing co-adaptation in the final model. 
An alternative view to Dropout is to consider it as an efficient approximation to taking the mean over the predictions of an exponentially large set of sub-models, without having to estimate all those models individually.

Alternatively, Dropout can also be expressed as a penalty, which for logistic regression models is most similar to Ridge regularization, and includes a quadratic penalty on the size of coefficients. In contrast to Ridge, Dropout does not assign the penalty uniformly across the predictors. Dropout rewards infrequent predictors that enable the model to make confident predictions (predicted risks close to 0 or 1) whenever the predictor of interest is active.\cite{wager2013dropout}


\subsection{Dimensionality reduction}

The multi-collinearity of predictors may be due to shared dependence on a smaller set of unobserved underlying variables, that could themselves be related to the outcome. Principal component analysis (PCA) can reduce the dimensionality of the original predictor space, to obtain a smaller set of variables that explain (most of) the variance in the original predictors, but is in itself uncorrelated. These uncorrelated variables, the principal components, can be used as input to a logistic regression model to relate them to the outcome. This combination of PCA with logistic regression is called (incomplete) principal component logistic regression (\textbf{PCLR}).\cite{kendall1965course,aguilera2006using,suarthana2010diagnostic} With regard to the original model, the effect of using PCLR is that predictors that correlate strongly, and are thus likely related the same principal components, obtain similar coefficients. 

In this study, we focus on linear PCA as this gives the possibility to rewrite the PCLR model to an equivalent logistic model from the original predictors to the outcome (details on this are given in Appendix \ref{appendix:models}). This enables direct comparison of the coefficients with the other methods, and reduces the importance of interpretability of the principal components, as we can always observe the coefficients of each of the predictors in the final model.

Linear autoencoders (LAE) are similar to PCA but do not find the exact same projection as PCA. However, their components span the same directions.\cite{kunin2019loss} In contrast to PCA or LAE, which determine the components based on the explained variance in the original predictors irrespective of the outcome, we extend the training criterion of LAE to find components that not only explain the variance of the original predictors but are also predictive of the outcome (from now on referred to as \textit{linear autoencoder logistic regression}; \textbf{LAELR}). The relative importance of (1) explaining the variance in the predictors, and (2) maximizing the likelihood of the outcome, is determined by an additional parameter that (like the number of used components) needs to be tuned. How to tune such parameters is discussed later in section \ref{sec:hyperparams}. To summarize, LAELR can be seen as a compromise between PCLR and logistic regression (a more detailed formulation can be found in appendix \ref{appendix:models}). 

\subsection{Constrained optimization}
Besides penalizing the absolute size of coefficients, as in Lasso or Ridge, other penalties or criteria can be incorporated, possibly using knowledge from the clinical domain or setting. For example, in some cases, it may be valid to assume a priori that it is unlikely that certain predictors have a negative relation to the outcome (e.g., in the later described case study one could assume that increasing radiation dosage to healthy tissue does not reduce the probability of complications). In a logistic regression model, not having a negative relation to the outcome means that these predictors should not get negative coefficient values in the final model. Encouraging the non-negativity (\textbf{NN}) of certain coefficients can be modeled by adding a penalty for negative coefficient values to the maximum likelihood criterion. \cite{hull1997xerox} Alternatively, if the non-negativity constraints are to be respected at all times they can be incorporated as hard constraints during the maximum likelihood estimation of the model through, for example, gradient projection.\cite{calamai1987projected}

If the additional assumptions based on domain knowledge are correct and complementary to the information already present in the training data, incorporating them can reduce the coefficients' search space. This may prevent selection of implausible models that satisfy maximum likelihood but are in fact inconsistent with clinical knowledge, and consequently reduce the coefficient variance due to multi-collinearity.



%% file: sections/3-NTCP-SIM.tex
\section{Motivating example}\label{sec:ntcp}

\subsection{Clinical background}
Cancer patients receiving radiation therapy often experience complications after the therapy due to radiation damage to healthy tissue surrounding the tumor. For example, common complications for head and neck cancer patients are xerostomia (decreased salivary flow resulting in dry mouth syndrome), or dysphagia (swallowing problems). Prediction models, called normal-tissue complication probability (NTCP) models, are used to predict the risk for individual patients of developing complications after radiation-based therapy, based on patient, disease, and treatment characteristics including the dose distributions given to the healthy tissue surrounding the tumor, the so-called organs at risk (OAR). 
Besides informing patients about the expected risks of radiation-induced complications, NTCP models are clinically used to guide treatment decisions by looking at the difference in predicted risk of complications ($\Delta$NTCP) between treatment plans: sometimes by pair-wise treatment plan comparison \citep{dritschilo1978complication,langendijk2013selection}, but also for complete treatment plan optimization \citep{wolbarst1982optimization,kierkels2016multivariable}, where the planned dosage is adjusted to minimize the risk of complications, by minimizing the model-predicted NTCP, while maintaining tumor control.

For this setting, proper handling of collinearity is crucial, as in the process of treatment optimization unexpected coefficients may result in steering dosage to OAR that due to collinearity may not seem important (e.g., if coefficients are zero or negative), but in fact are associated with increased complication risks.

\subsection{Simulation Study}
We report and planned the simulation study using the ADEMP strategy, following Morris and colleagues.\cite{morris2019using}

\subsubsection{Aims}
The aims of this simulation study are to:
\begin{enumerate}
    \item Study the \textit{effect of collinearity} on development of clinical prediction models in terms of discrimination, calibration, and coefficient estimation in low dimensional settings (the number of predictors is smaller than the number of events).
    \item Compare the \textit{effectiveness of eight methods} in handling the potentially negative effects of collinearity (logistic regression, Lasso, Ridge, ElasticNet, PCLR, LAELR, Dropout, and non-negativity-based constrained optimization).
\end{enumerate}

\subsubsection{Data‐generating mechanisms}
The simulations are based on four prediction modeling settings: mimicking two outcomes in our motivating example (xerostomia and dysphagia), and two predictor sets per outcome: a smaller predictor set with less collinearity, where the given radiation is only indicated by the mean dose per OAR, and a larger predictor set with higher collinearity, where more detailed dose-volume predictors are added as well.\footnote{More detailed descriptions of the used predictor sets are given in appendix Table \ref{tab:variable_sets}.} These four initial settings are colored gray in Table \ref{tab:generated_settings}: \xerbasiclow{} and \dysbasiclow{} for the settings with small predictor sets, and \xerexthigh{} and \dysexthigh{} for the larger predictor sets. For these four settings, predictor data were simulated from a multivariate normal distribution, using the means and covariance matrix of the observed predictors of 740 head-and-neck patients (with primary tumor locations: pharynx, larynx, or the oral cavity) that underwent radiotherapy at the University Medical Center Groningen (UMCG), and were selected for having no missing data in the predictors or outcome. The simulated ground-truth relation between predictors and outcome is constructed by fitting a logistic regression with Ridge penalization on the corresponding real data.\footnote{In a 5-fold cross validation on the real data, Ridge yielded good results in terms of calibration and discrimination, but also included the largest proportion OAR in the model (relevant for this case study).} 

To study the effect of collinearity independently of the number of predictors and the number of events-per-variable (EPV) we generated another four simulation settings: for each setting with a large predictor set that inherently exhibits high collinearity (\xerexthigh{} and \dysexthigh{}) we generate\footnote{We change the degree of collinearity by scaling the covariance matrix of the multi-variate Gaussian (without changing the diagonal). In doing so, the degree of class separation and the outcome prevalence of the ground-truth model may change. To maintain the same degree of class separation in the data, we scale the slope of the ground-truth model. Additionally, to maintain the same outcome prevalence, we adjusted the ground-truth intercept accordingly. This way we change the degree of collinearity, but maintain ground-truth area under the receiver operator curve, and outcome prevalence.} low-collinearity variants (\xerextlow{} and \dysextlow{} respectively), and for each setting with a small predictor set that inherently exhibits a lower degree of collinearity (\xerbasiclow{} and \dysbasiclow{}) we generate high-collinearity variants (\xerbasichigh{} and \dysbasichigh{} respectively). Finally, we end up with a total of eight simulation settings, for which four pair-wise comparisons can be made to assess the effect of collinearity.

\begin{table}[h!]
    \centering
    \caption{\label{tab:generated_settings}Eight simulation settings that are evaluated \textit{for each} method. The sub-scripted triangle ($\vartriangle$) is used to indicate high collinearity settings. The star (*) refers to the real-data version of a simulated setting.}
    \begin{tabular}{lcccccc|ccccccc|ccccc|}
    \toprule
    Setting & y & N & No. predictors & EPV & Median VIF \\
      \midrule
      \rowcolor{lightgray}
         \xerbasiclow{} / \xerbasiclowreal{} & Xerostomia & 592 & 7 & 23  & 5 \\ 
         \xerbasichigh{} & Xerostomia & 592 & 7 & 23  & 43 \\
               \rowcolor{lightgray}
         \xerextlow{} & Xerostomia & 592 & 19 & 8 & 5 \\
         \xerexthigh{} / \xerexthighreal{} & Xerostomia & 592 & 19 & 8 & 43 \\ 
         \midrule
               \rowcolor{lightgray}
         \dysbasiclow{} / \dysbasiclowreal{} & Dysphagia & 592 & 13 & 6  & 7 \\
         \dysbasichigh{} & Dysphagia & 592 & 13 & 6  & 43 \\
               \rowcolor{lightgray}
         \dysextlow{} & Dysphagia & 592 & 43 & 2 & 7 \\
         \dysexthigh{} / \dysexthighreal{} & Dysphagia & 592 & 43 & 2 & 43 \\
    \bottomrule
    \end{tabular}
\end{table}

To assess to what degree the simulation is accurate for the actual clinical prediction modeling problem, we compare the results of the simulation to a comparable real-data setting. These real-data experiments are indicated by a star ($^*$) in Table \ref{tab:generated_settings}, and have the same modeling characteristics as the corresponding simulations: the same predictor covariance, outcome prevalence, and sample size.

\subsubsection{Estimators/Target of analysis}
We quantify collinearity by the median variance inflation factor (VIF). The VIF of a predictor reflects the relative increase in coefficient variance for that predictor due to the presence of other predictors. A VIF of 1 indicates absence of collinearity, whereas a VIF larger than 10 is often considered to reflect a high degree of collinearity. \cite{neter1989applied}

\subsubsection{Application of the methods}
\label{sec:hyperparams}
Besides standard logistic regression (\textbf{LR}), we compare all methods discussed in section \ref{sec:methods}: Lasso, Ridge, ElasticNet, PCLR, LAELR, Dropout, and \LRnn{} (the use of non-negativity constraints for dosage coefficients through gradient projection). These are listed in Table \ref{tab:methods}.

For a fair comparison, we perform equal hyperparameter\footnote{Parameters that are not part of the model itself, but steer how the coefficients are determined (e.g., the relative importance of the shrinkage penalty in Lasso and Ridge, or the number of components for PCLR, among others).} tuning across methods. For all models, we tune hyperparameters using Bayesian optimization \cite{snoek2012practical} in a (nested) 3-fold cross-validation setting, with a log-likelihood tuning criterion. As general data preprocessing we standardize all predictors to have zero-mean and unit variance. More details about the exact training criteria for each method, hyperparameter tuning, and optimization can be found in Appendix section \ref{appendix:models}. 

\begin{table}[h!]
        \caption{\label{tab:methods}List of compared methods.}
\resizebox{\textwidth}{!}{
    \begin{tabular}{llll}
    \toprule
        Method & Abbreviation & Hyperparameters \\
        \midrule
        Logistic regression & LR & - \\
         Lasso penalization & Lasso & $c_{\ell_1}$ (inverse penalty importance)\\
         Ridge penalization & Ridge & $c_{\ell_2}$ (inverse penalty importance)\\ 
         Elastic Net penalization & ElasticNet & $c_{\ell_1}$, $c_{\ell_2}$ (inverse importance per penalty)\\ 
         Dropout regularization & Dropout & $\delta$ (dropout ratio)\\
         Principal component logistic regression & PCLR & $d_{\textsc{pca}}$ (number of components)\\
         Linear auto-encoder logistic regression & LAELR & $d_{\textsc{lae}}$ (number of components), $c_{LAE}$ (inverse importance of reconstruction loss)\\
         Non-negative logistic regression & \LRnn{} & - \\
         \bottomrule
    \end{tabular}}
    
\end{table}

\subsubsection{Performance measures}
We analyze our aims with regard to the measures stated in Table \ref{tab:sim_metrics}. A less explored measure we use is the expected proportion of included coefficients that has the same direction of effect (positive, negative, or zero) across two simulated model construction repetitions (the mean Jaccard index of the coefficient signs: \expjacc, ranging from 0 to 1). This measure is formally defined in Equation \ref{eq:mjics}, for arbitrary samples $i$ and $j$, to assess the robustness of the coefficient interpretation when developing a prediction model: we consider methods that include the same predictors in the model and assign the same directions of effect when repeating the model construction process to be more robust than methods that include different coefficients or assign different direction of effect across iterations.
\begin{multicols}{2}
\begin{align}
\label{eq:mjics}
    \text{\expjacc} = \mathbb{E}\left[\frac{|\sign(\hat{\theta}_{i}) \cap \sign(\hat{\theta}_{j})|}{|\sign(\hat{\theta}_{i}) \cup \sign(\hat{\theta}_{j})|}\right]
\end{align}\break
\begin{align}
    \sign(x) = \begin{cases} \hspace{.1cm}-1, & \text{ iff } x < -0.01  \\
                            \hspace{.3cm}1, & \text{ iff } x > 0.01 \\
                            \hspace{.3cm}0, & \text{ otherwise } \end{cases}
\end{align}
\end{multicols}
All measures are estimated by repeatedly sampling ($n_{\text{rep.}}=100$) a new dataset from the constructed Gaussian distributions, refitting the models, and evaluating them in a validation set generated from the same distributions as the development set of size N=10,000. The reported 95\% confidence intervals are based on these repetitions, and reflect variability of the entire model construction procedure: sampling training data, developing the model (including hyperparameter tuning), and sampling a new validation set.
For the real-data settings, a repeated 5-fold cross validation (N=592 per fold) on the real data is used to estimate each measure, and their respective confidence intervals ($n_{\text{rep.}}=100$). \cite{kim2009estimating}

\begin{table}[h!]
        \caption{\label{tab:sim_metrics}Overview of the measures used to compare methods on predictive performance, and coefficient estimation.}
    \begin{tabular}{llc}
    \toprule
         Measure& Abbreviation & Ideal value  \\
         \midrule
         \textbf{Predictive performance}\\
         \hspace{.3cm}Area under the receiver-operator characteristic curve & \auroc & 1\\
         \hspace{.3cm}Calibration intercept & \citl & 0\\
         \hspace{.3cm}Calibration slope & \cslope & 1\\
         \hspace{.3cm}Nagelkerke R-squared & \rsquared & 1\\
         \textbf{Coefficient estimation}\\
         \hspace{.3cm}Mean squared error between the estimated and the true coefficients & \mse & 0\\
         \hspace{.3cm}Mean proportion of coefficients with the same direction of effect after repetition & \expjacc & 1\\
         
   \bottomrule \end{tabular}
\end{table}


\subsubsection{Coding and execution} 
All experiments were implemented in Python 3.6 (primarily using Scikit-learn\cite{pedregosa2011scikit} and PyTorch\cite{paszke2019pytorch}). Predictive performance measures are calculated in R (using the \textit{val.prob.ci.2} function \cite{van2016calibration}). The code used to conduct the experiments is available at \href{https://github.com/tuur/collinearity}{https://github.com/tuur/collinearity}.


\subsection{Analysis of the results}
This section presents the simulation results with regard to predictive performance and coefficient estimation. Based on a comparison between our simulations and the real-data experiments in terms of predictive performance we concluded that the simulations are in accordance with the real-data settings. Results of the real-data experiments can be found in appendix section \ref{appendix:real_data_results}.

\input{sections/CALIB-GRID-AB}

\newpage
\subsubsection{Predictive performance}
Simulation results regarding calibration and discrimination for the xerostomia settings are reported in Figure \ref{fig:results_performance_xerostomia} (results for dysphagia can be found in appendix Figure \ref{fig:calibration_fans_dysphagia}). We observed no effects of collinearity on the predictive performance of any of the compared methods: in terms of \auroc, \rsquared, \citl, \cslope, nor the calibration plots (comparing \xerbasiclow{} with \xerbasichigh{}, and \xerextlow{} with \xerexthigh{}). Based on the calibration plots in Figure \ref{fig:results_performance_xerostomia}, we do observe a slight overall overestimation of risk for LR compared to the other methods when extending the predictor set (comparing \xerextlow{} to \xerbasiclow{}, and \xerexthigh{} to \xerbasichigh{}), probably due to the lower EPV. 

We obtained similar results for the simulated dysphagia settings, finding no effect of collinearity on predictive performance, and little to no difference between the compared methods in any of the performance measures (\auroc, \rsquared, \citl, \cslope). Again, LR yielded worse calibration compared to the other methods (irrespective of the degree of collinearity). As expected, the difference between LR and the other compared methods was largest in terms of both calibration and discrimination in the setting with the lowest EPV (setting D, with an EPV of 2), indicating that LR suffers most from overfitting.

\subsubsection{Coefficient estimation}
Observing the estimation of the regression coefficients, we find that in terms of MSE between the estimated coefficients and the true coefficients, particularly LR has a somewhat lower MSE, which is negatively affected by the increase in collinearity (shown in Figure \ref{fig:mse_results_ab} for xerostomia, and in appendix Figure \ref{fig:mse_results_cd} for dysphagia). The remaining methods obtained lower MSE, and did not show a consistent effect of collinearity.

\input{sections/BOXPLOTS-AB}
\vspace{-.5cm}

When observing the effect of collinearity on the stability of the predictor selection (to what degree the same predictors were selected with the same directions of effect when repeating the model development process across simulations), we find a negative effect of collinearity on selection stability for setting A (EPV of 23) across methods (in Figure \ref{fig:mjics_results_ab}, and in appendix Figure \ref{fig:mjics_results_cd}). For settings B and C (EPVs of 8 and 6 respectively) we only find a negative effect of collinearity for LR, Lasso, and \LRnn{}, and no clear effect for the remaining methods: Ridge, ElasticNet, PCLR, LAELR, and Dropout. For setting D (EPV of 2) we observe a clear positive effect of collinearity on the stability of predictor selection for Ridge, PCLR, and LAELR, a positive but weaker effect for ElasticNet and Dropout, and a slight negative effect for LR, Lasso and \LRnn{}.

Our results suggests that the effect of collinearity can be explained by two aspects. First, collinearity negatively affects the stability of maximum likelihood-based coefficient selection (reducing \expjacc), due to the increased variance in coefficient estimation. This explains why the negative effect remains present for LR and \LRnn{} across all settings: coefficient estimation for these methods is purely likelihood-based. 

The second aspect is that of regularization, which can have a stabilizing effect of coefficient selection.
The degree of regularization is determined by the hyperparameter tuning process, which is indirectly impacted by the EPV: low EPV settings are more likely to result in overfitting, and consequently obtain a larger degree of regularization. High EPV settings are less prone to overfitting and are consequently obtain less regularization. This can be observed from Figure \ref{fig:xer_hyperparam_results} for the xerostomia settings (and in appendix Figure \ref{fig:dys_hyperparam_results} for dysphagia).\footnote{Notice that $c_\ell{_1}$, $c_\ell{_2}$, and $c_{ENet}$ are \textit{inverse} shrinkage factors, meaning that lower values indicate a larger degree of shrinkage. A larger degree of dropout ratio $\delta$ indicates a larger degree of regularization. With regard to PCLR and LAELR it is important to notice that the number of components is only indicative of the degree of regularization \textit{within the same predictor set}: In Figure \ref{fig:xer_hyperparam_results}.3, the larger number of components in setting B in comparison to setting A does not imply less regularization, as the original dimension of B is in itself much larger than that of A (19 compared to 7).}

\input{sections/HYPPARAMPLOTS-AB}
\vspace{-.5cm}

Ridge, Dropout, and ElasticNet all quadratically penalize coefficient size, resulting in a grouping effect of collinear predictors. When regularization is strong, and collinearity is high, this constitutes a strong grouping effect, which in turn stimulates stable predictor selection. 

For PCLR and LAELR, a larger degree of regularization implies a heavier dependence on the principal components that explain the variance among predictors. As collinearity increases, a smaller number of components is required to explain the same amount of variance among predictors. This can be directly observed in Figure \ref{fig:xer_hyperparam_results}.3, where for the large predictor set (B) hyperparameter tuning resulted in a smaller number of components for PCLR and LAELR when collinearity was higher. This reliance on less components can in turn result in more stable coefficient estimation.

For Lasso, and partially ElasticNet, more regularization implies a stronger predictor selection effect, resulting in smaller models. Stronger selection in itself decreases the likelihood of (by chance) selecting the same coefficients when developing the model on a different sample. We conjecture that this is the reason why Lasso and \LRnn{} have low overall \expjacc{}, independently of collinearity compared to the other methods. Additionally, as Lasso's selection is likelihood-based, the negative impact of collinearity on predictor selection, as observed for LR and \LRnn{}, also affects Lasso. This can be observed by the reduction of \expjacc{} in the high collinearity settings in Figure \ref{fig:mjics_results_ab} (but also in appendix Figure \ref{fig:mjics_results_cd}).

%% file: sections/CALIB-GRID-AB.tex
\clearpage
\newcommand\plotwidth{0.33\textwidth}
\newcommand{\centered}[1]{\raisebox{1.8cm}{#1}}
\begin{figure}[h!]
    \centering
    \resizebox{.98\textwidth}{!}{
    \begin{tabular}{lccccc}
         & \xerbasiclow{} (EPV=23, mVIF=5) & \xerbasichigh{} (EPV=23, mVIF=43) & \xerextlow{} (EPV=8, mVIF=5) & \xerexthigh{} (EPV=8, mVIF=43)  \vspace{.5cm}\\
         \centered{LR} & \includegraphics[width=\plotwidth]{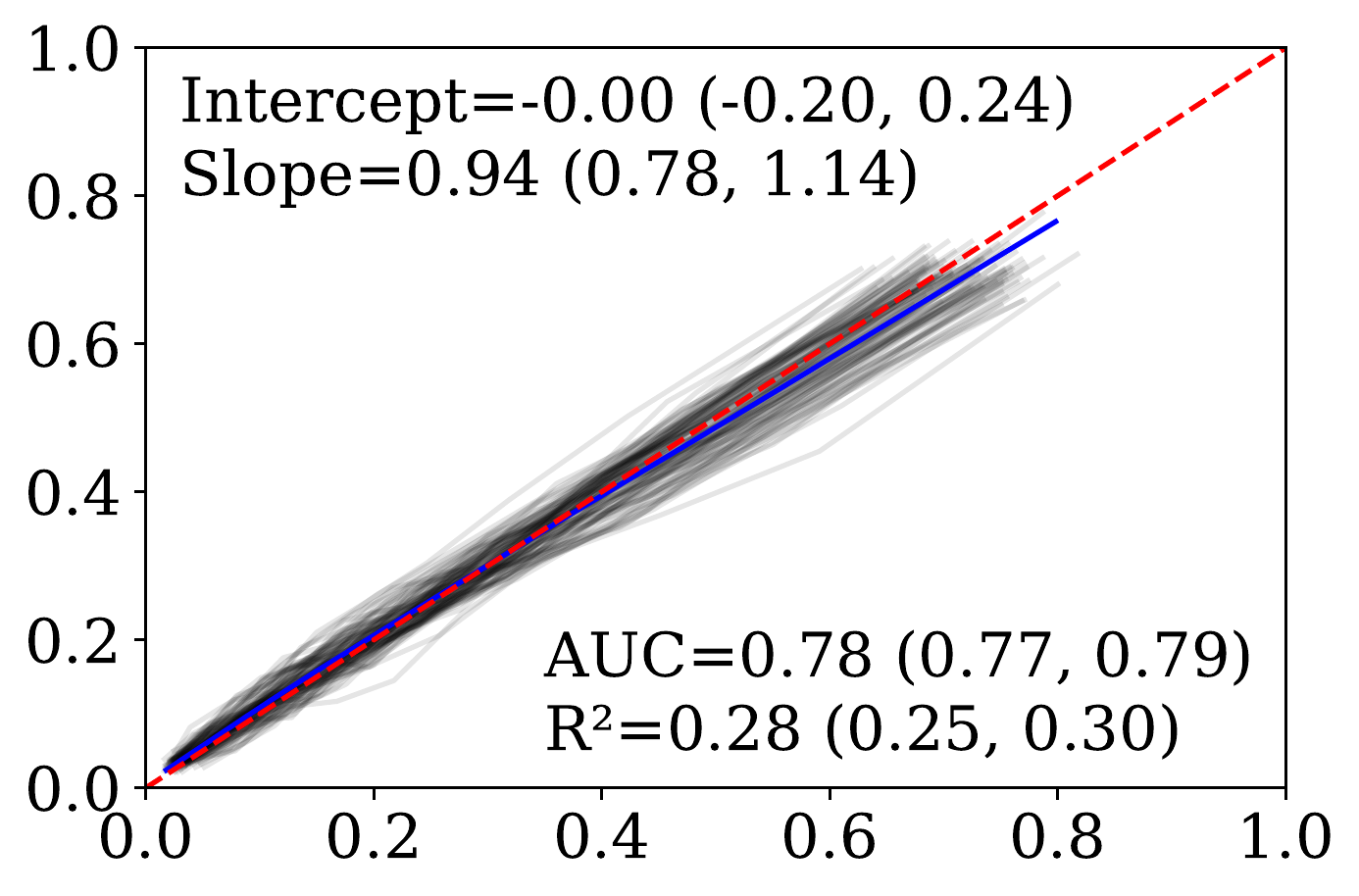} & \includegraphics[width=\plotwidth]{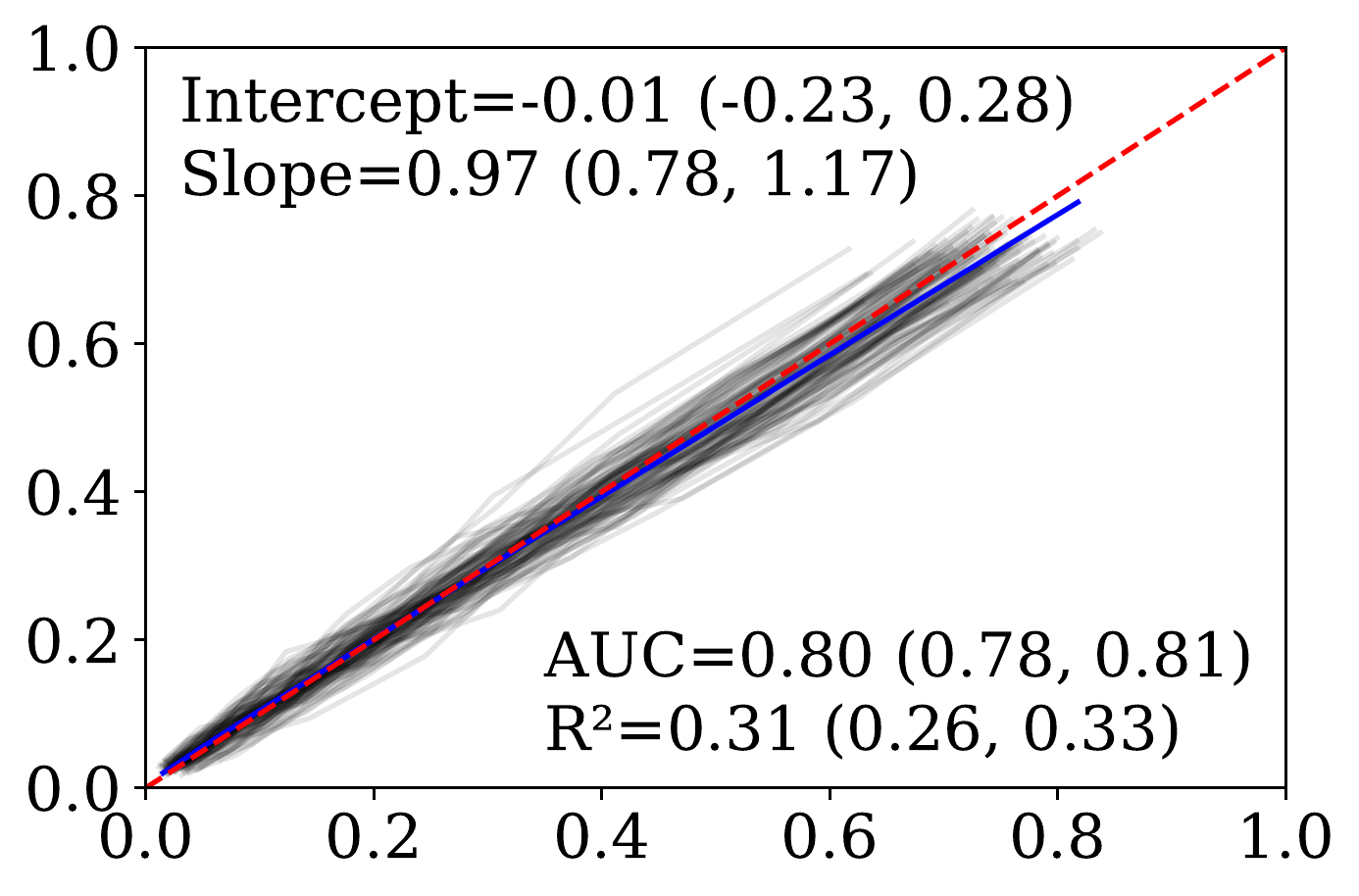} & \includegraphics[width=\plotwidth]{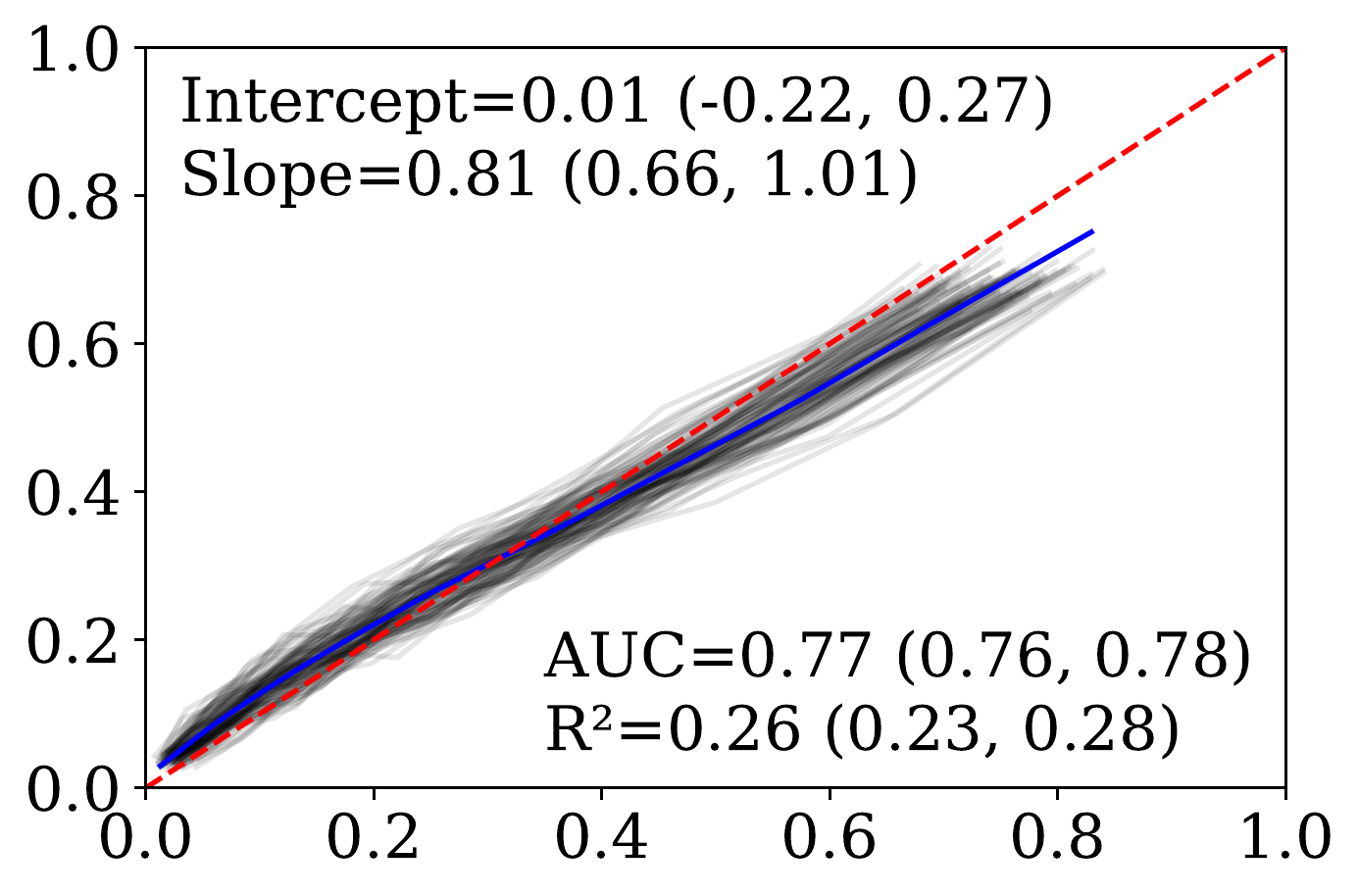} & \includegraphics[width=\plotwidth]{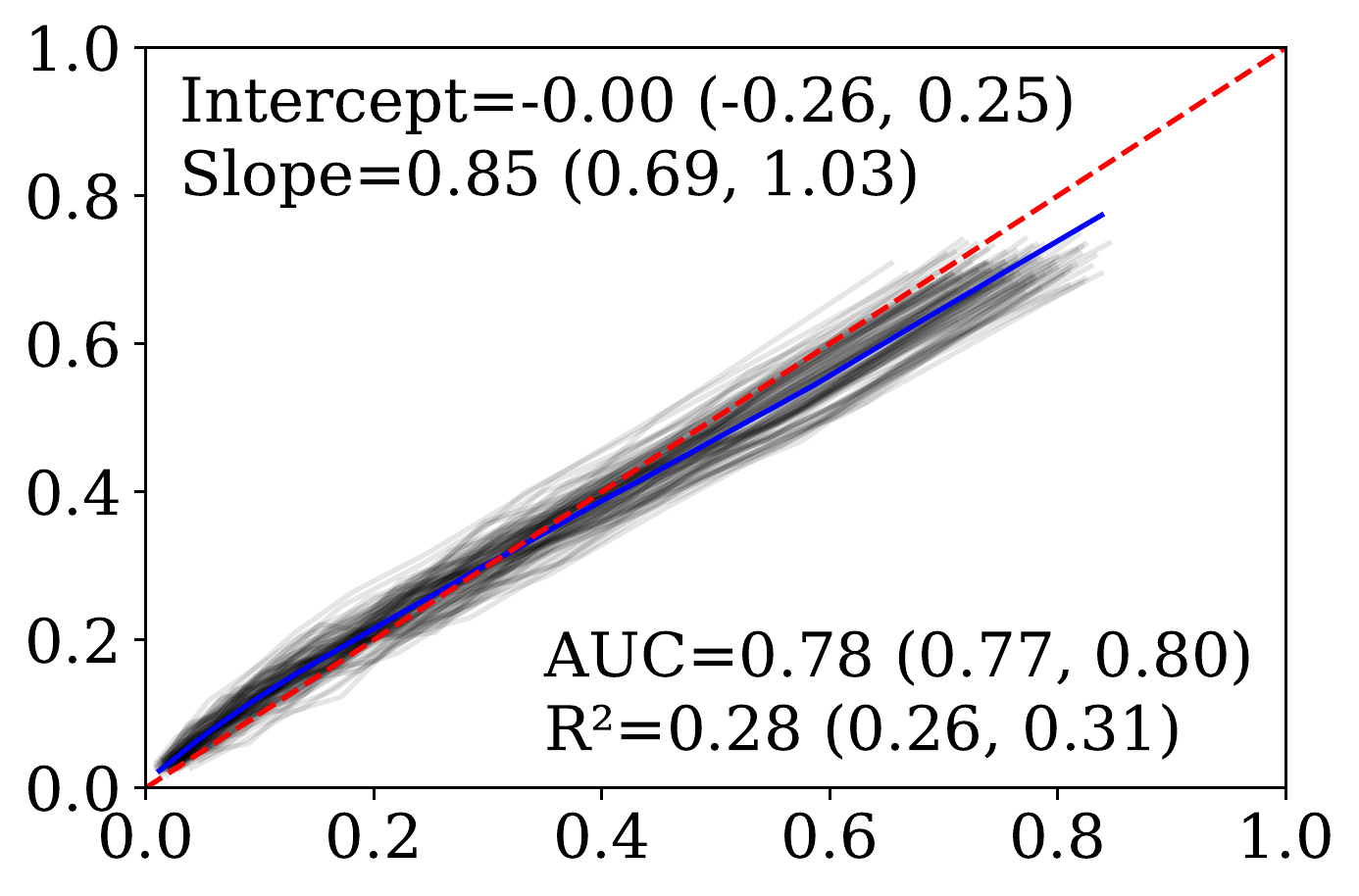}\\
         \centered{Lasso} & \includegraphics[width=\plotwidth]{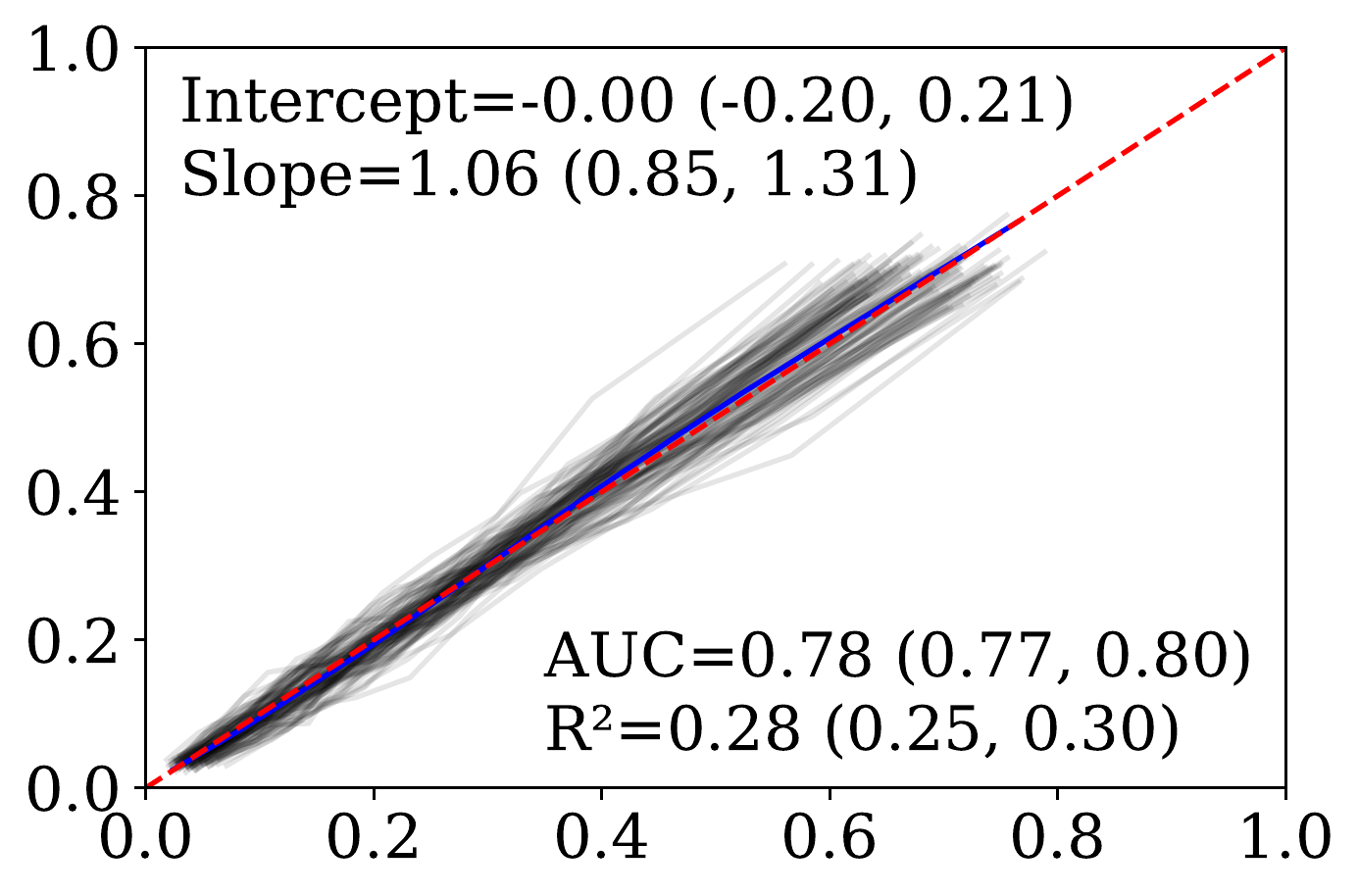} & \includegraphics[width=\plotwidth]{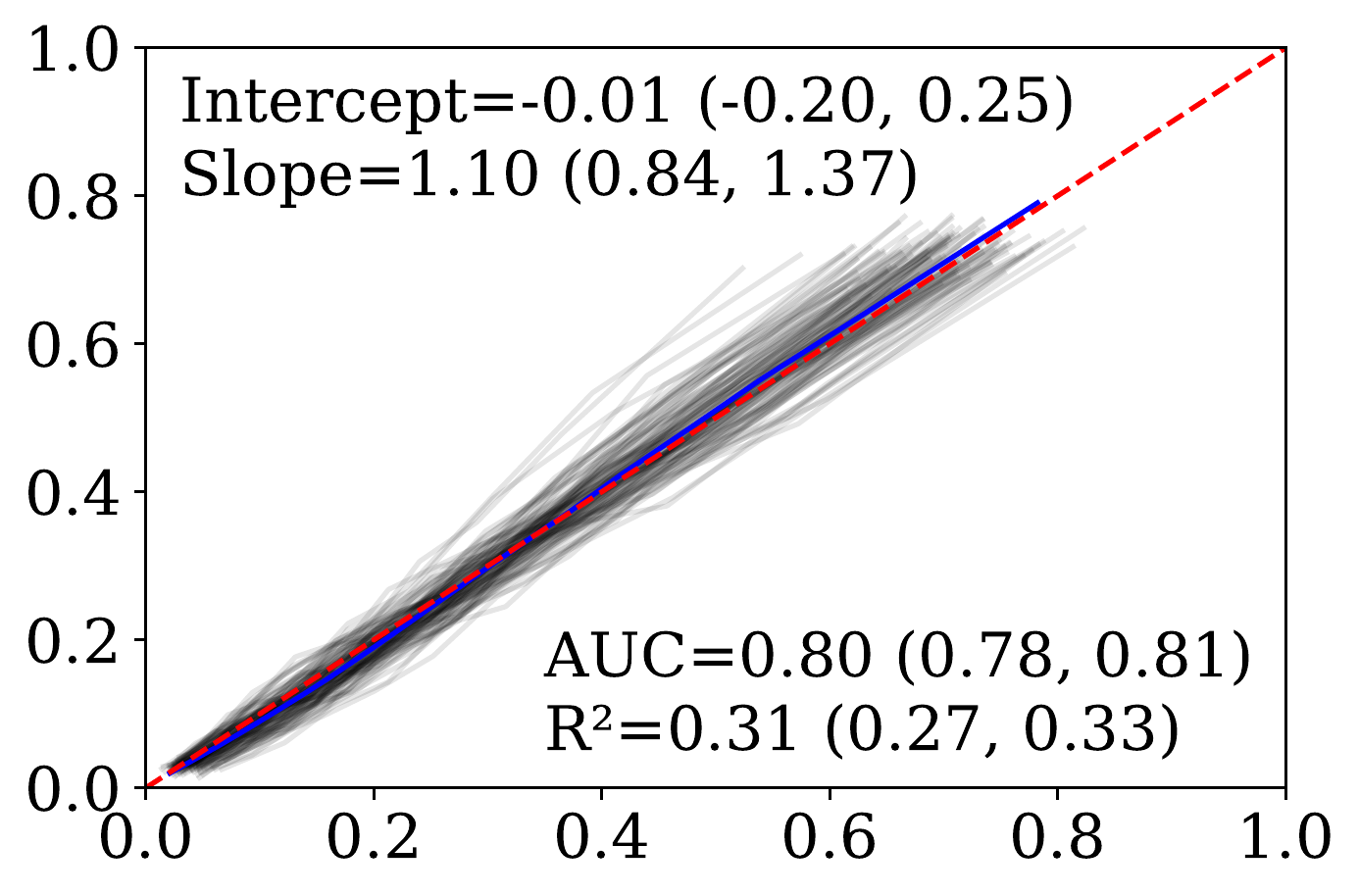} & \includegraphics[width=\plotwidth]{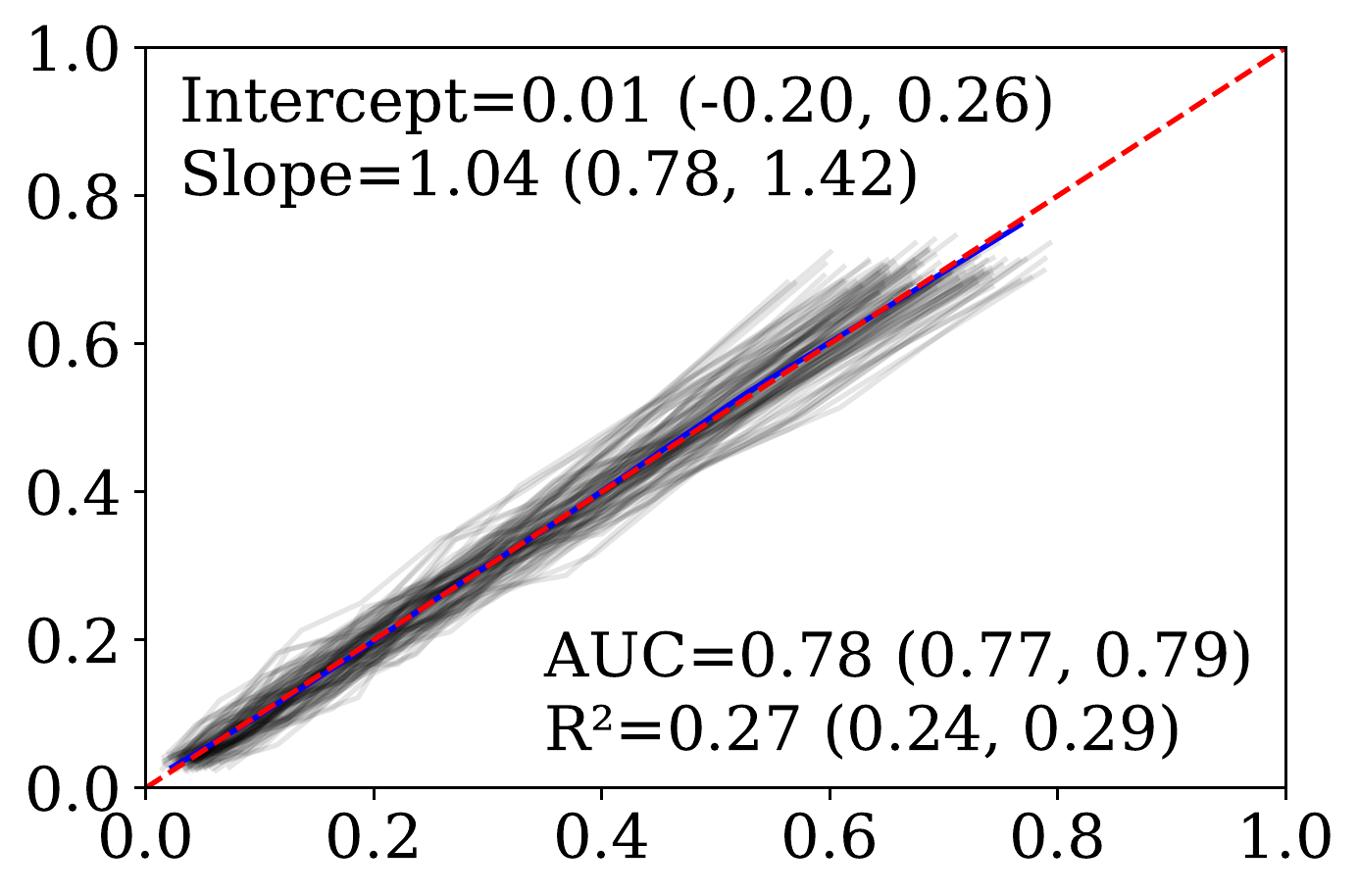} & \includegraphics[width=\plotwidth]{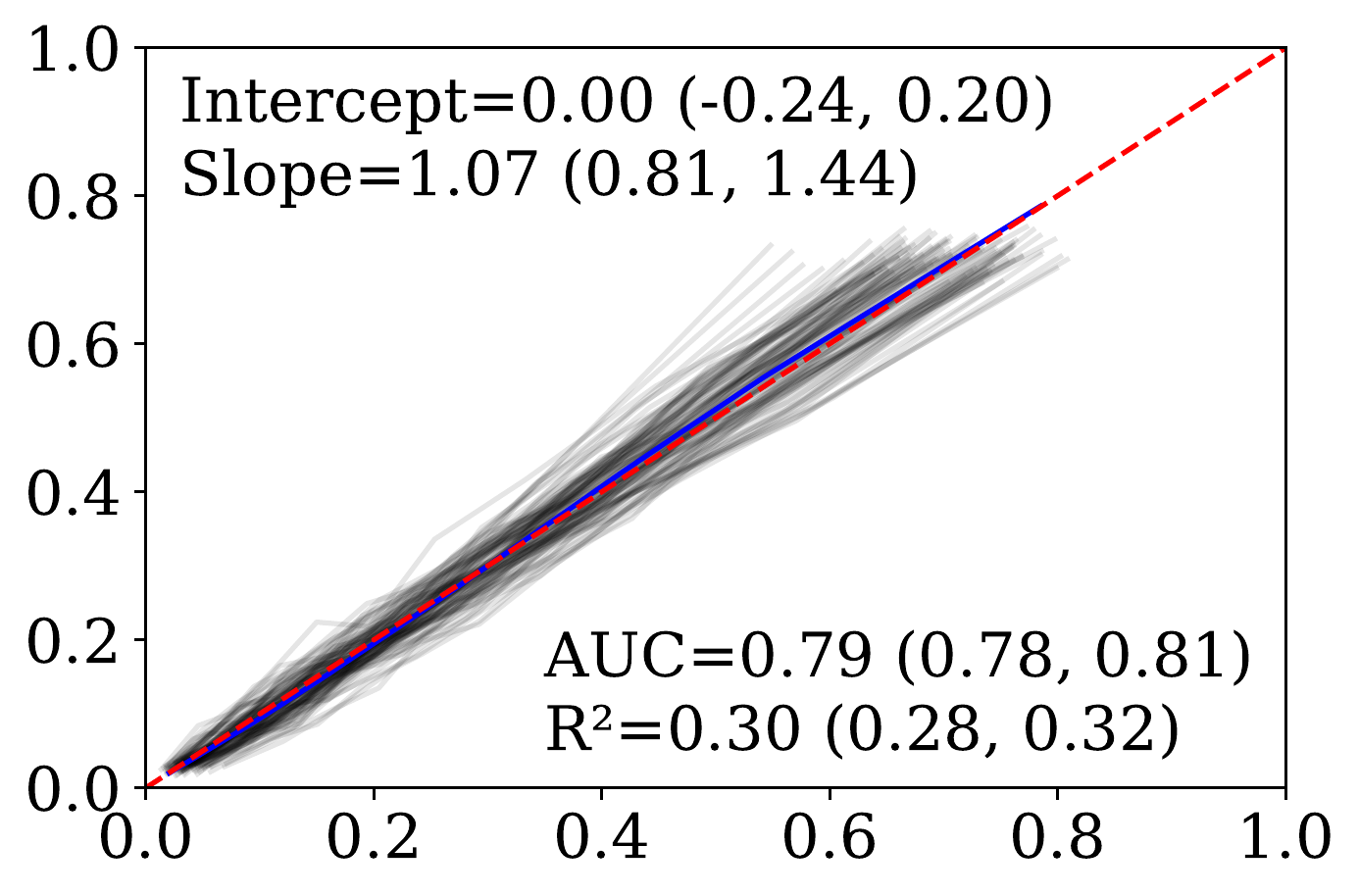}\\
         \centered{Ridge} & \includegraphics[width=\plotwidth]{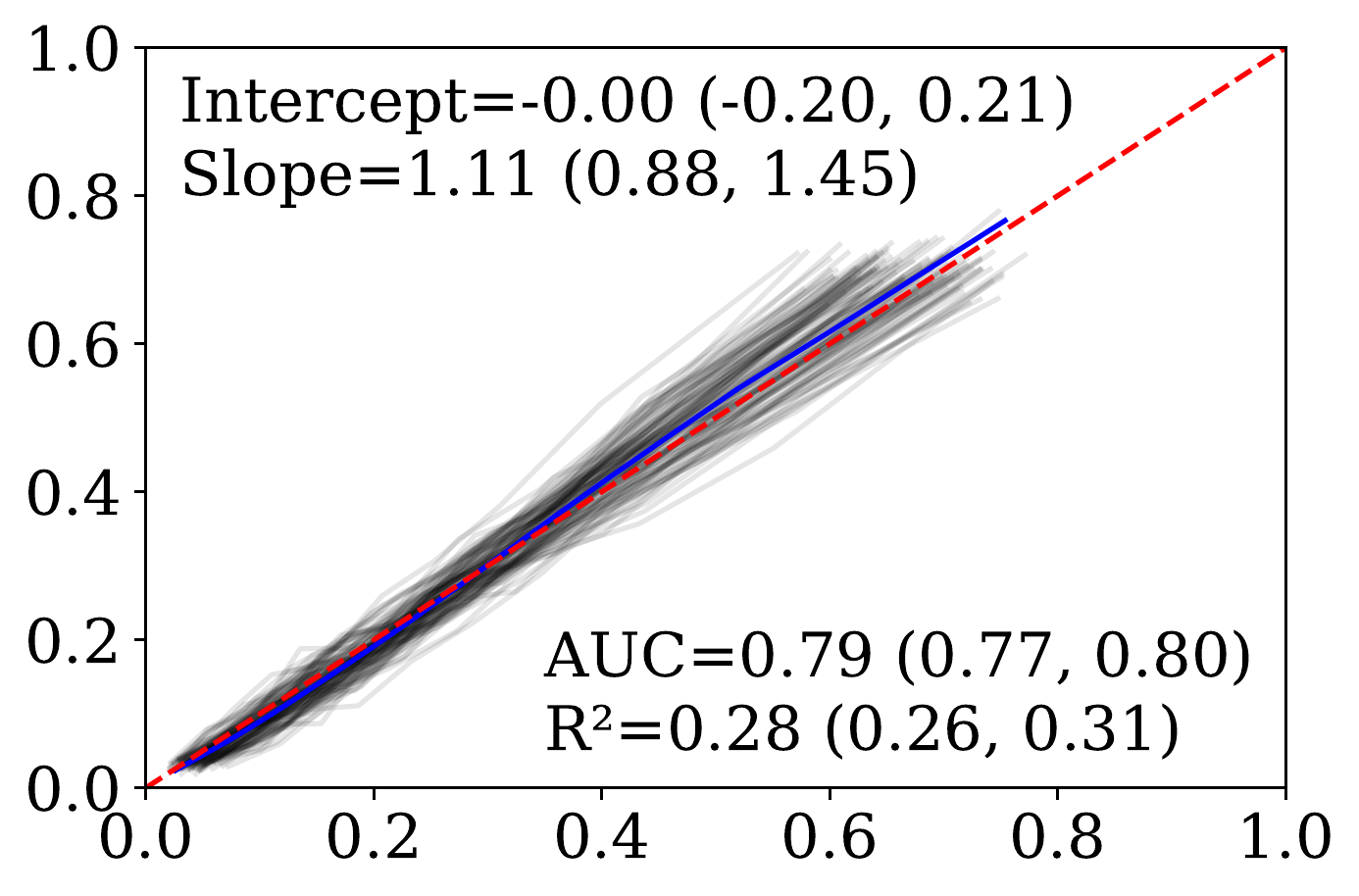} & \includegraphics[width=\plotwidth]{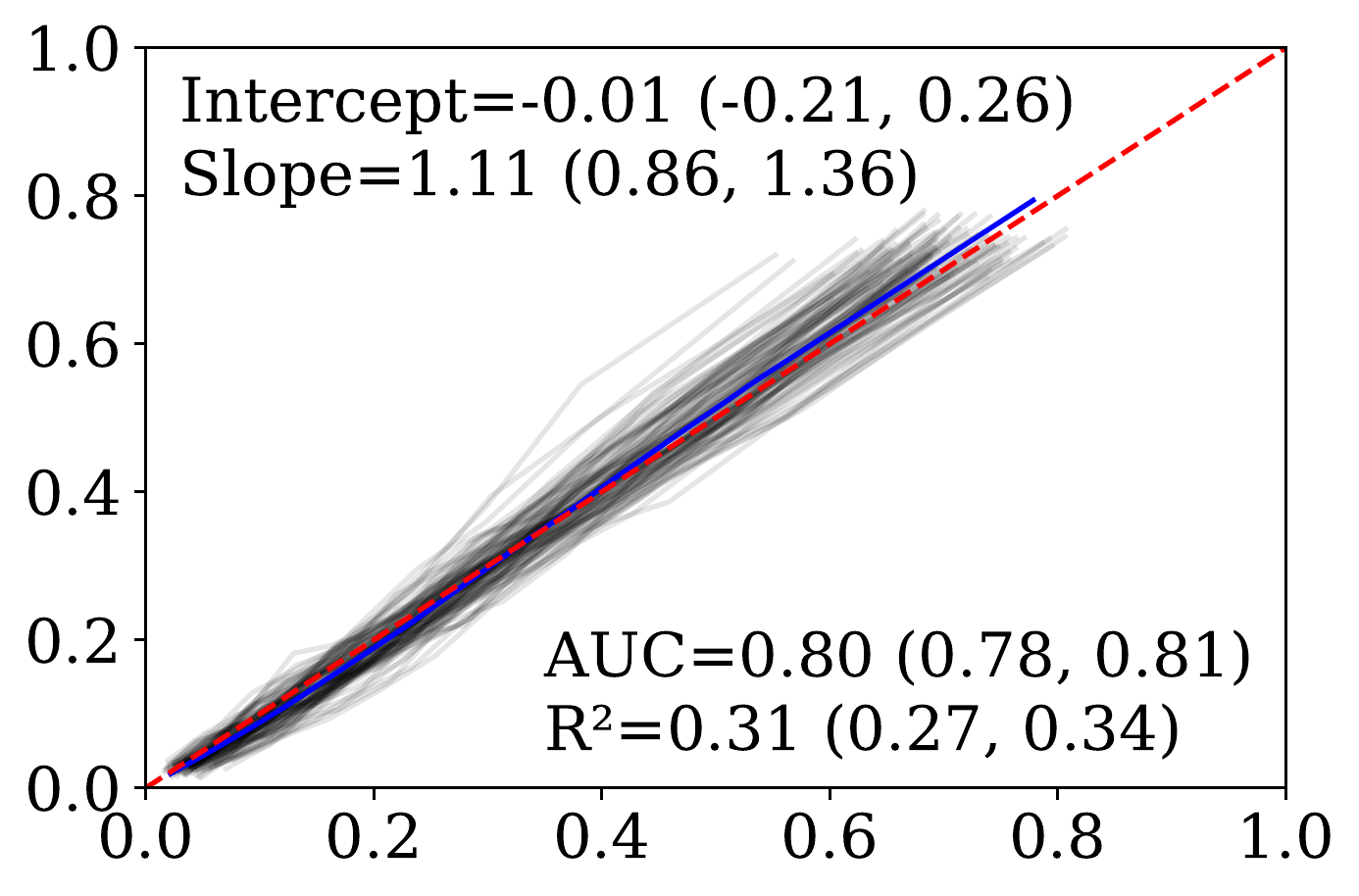} & \includegraphics[width=\plotwidth]{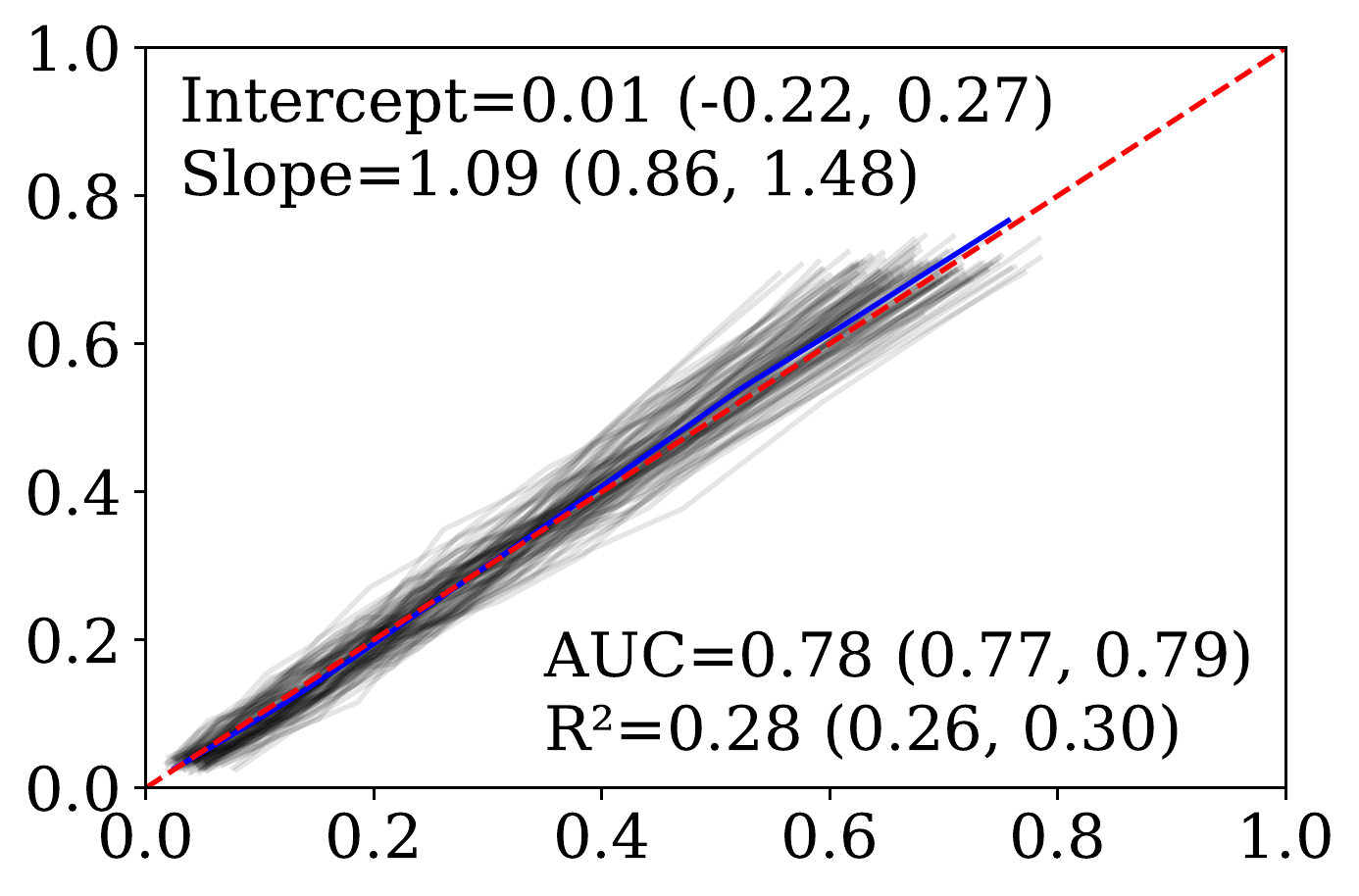} & \includegraphics[width=\plotwidth]{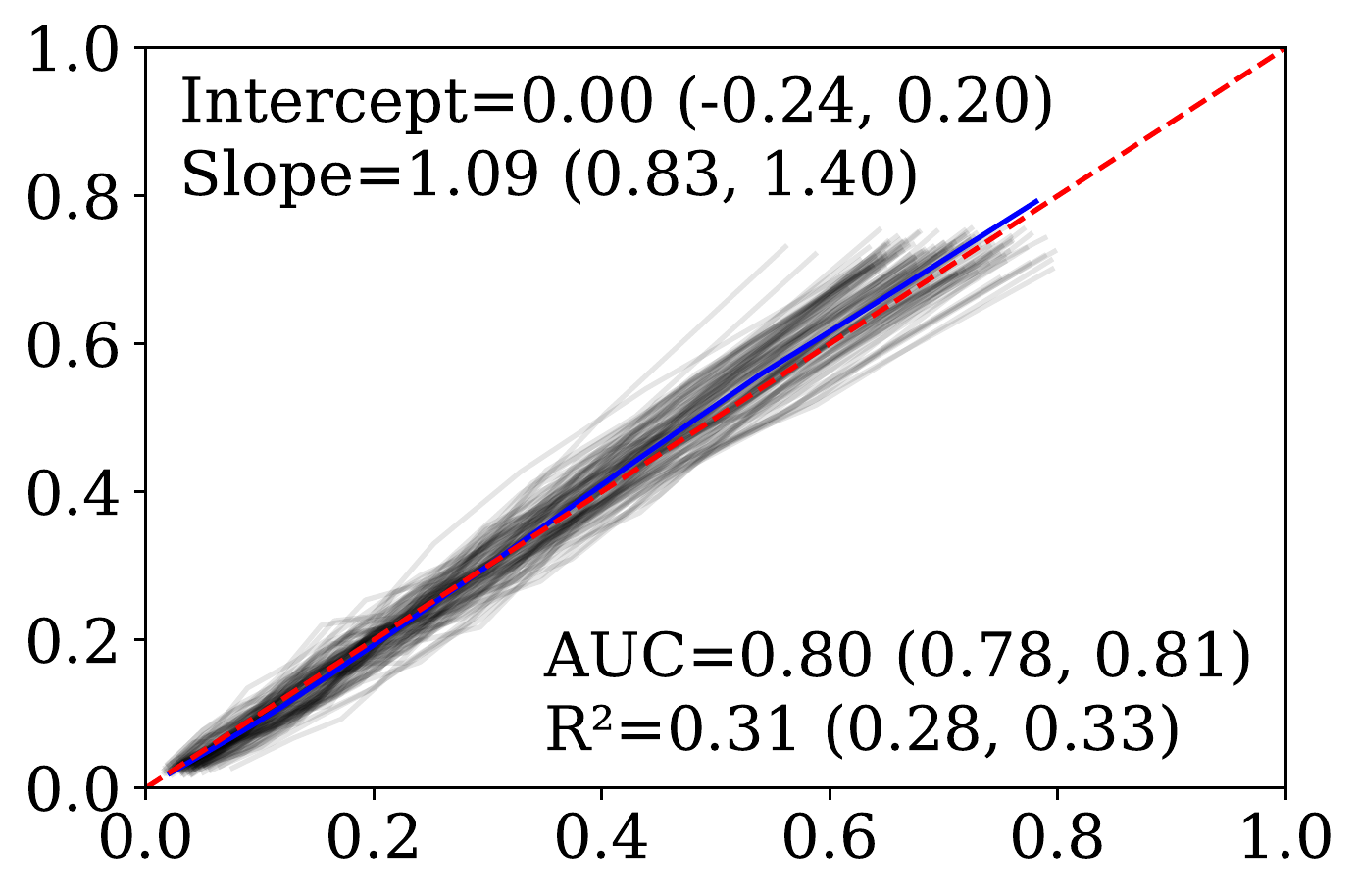}\\
         \centered{ElasticNet} & \includegraphics[width=\plotwidth]{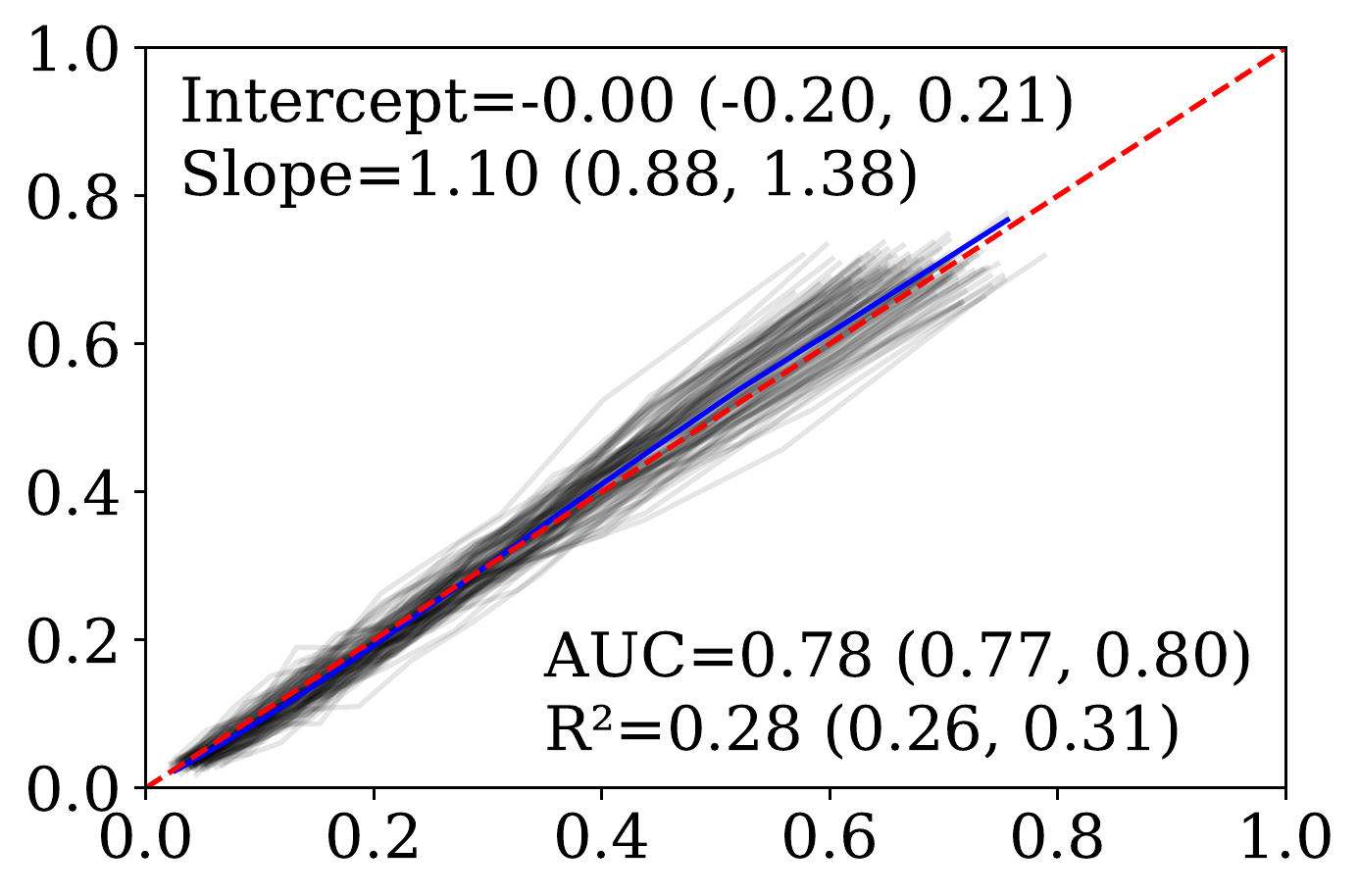} & \includegraphics[width=\plotwidth]{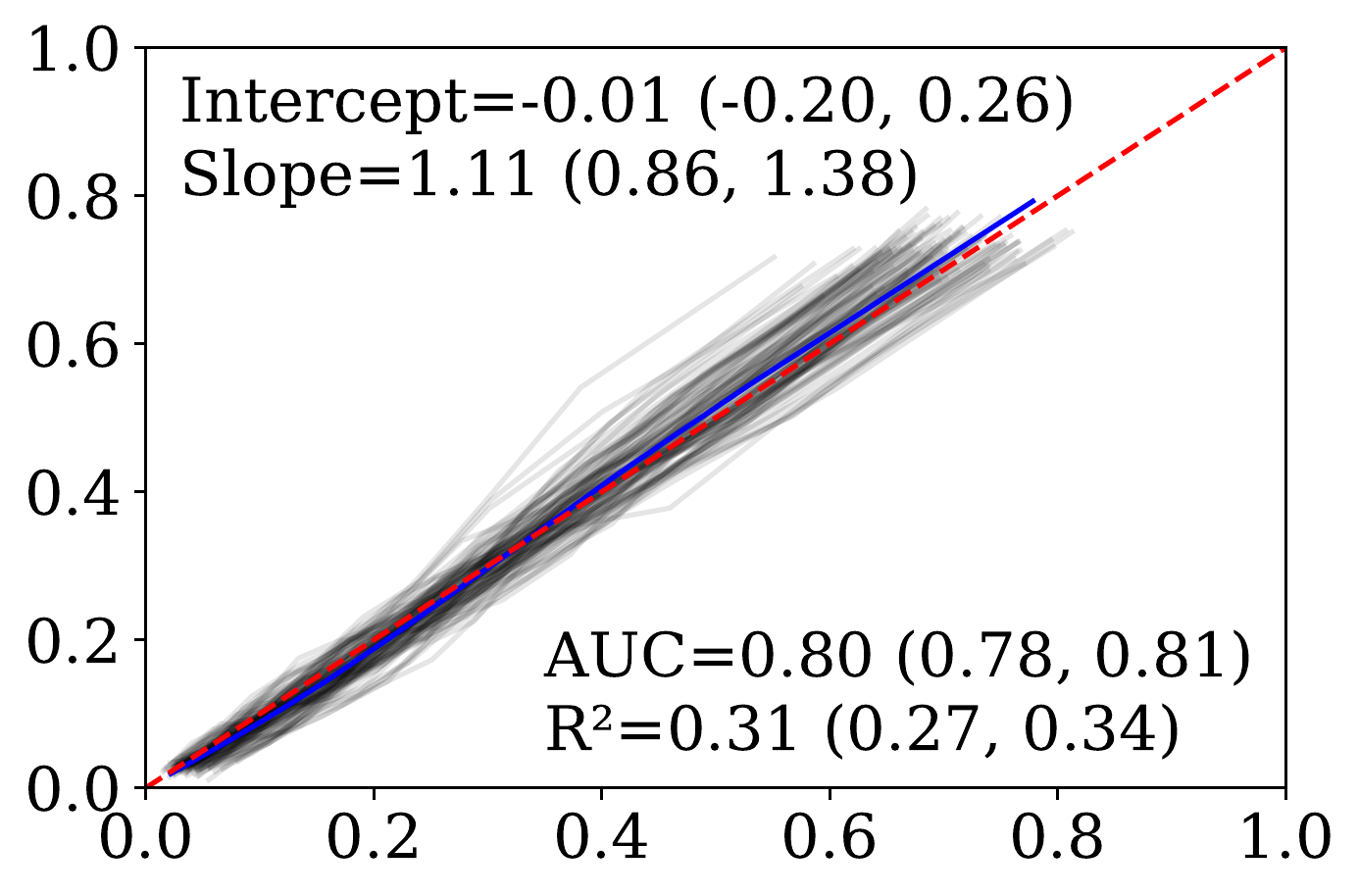} & \includegraphics[width=\plotwidth]{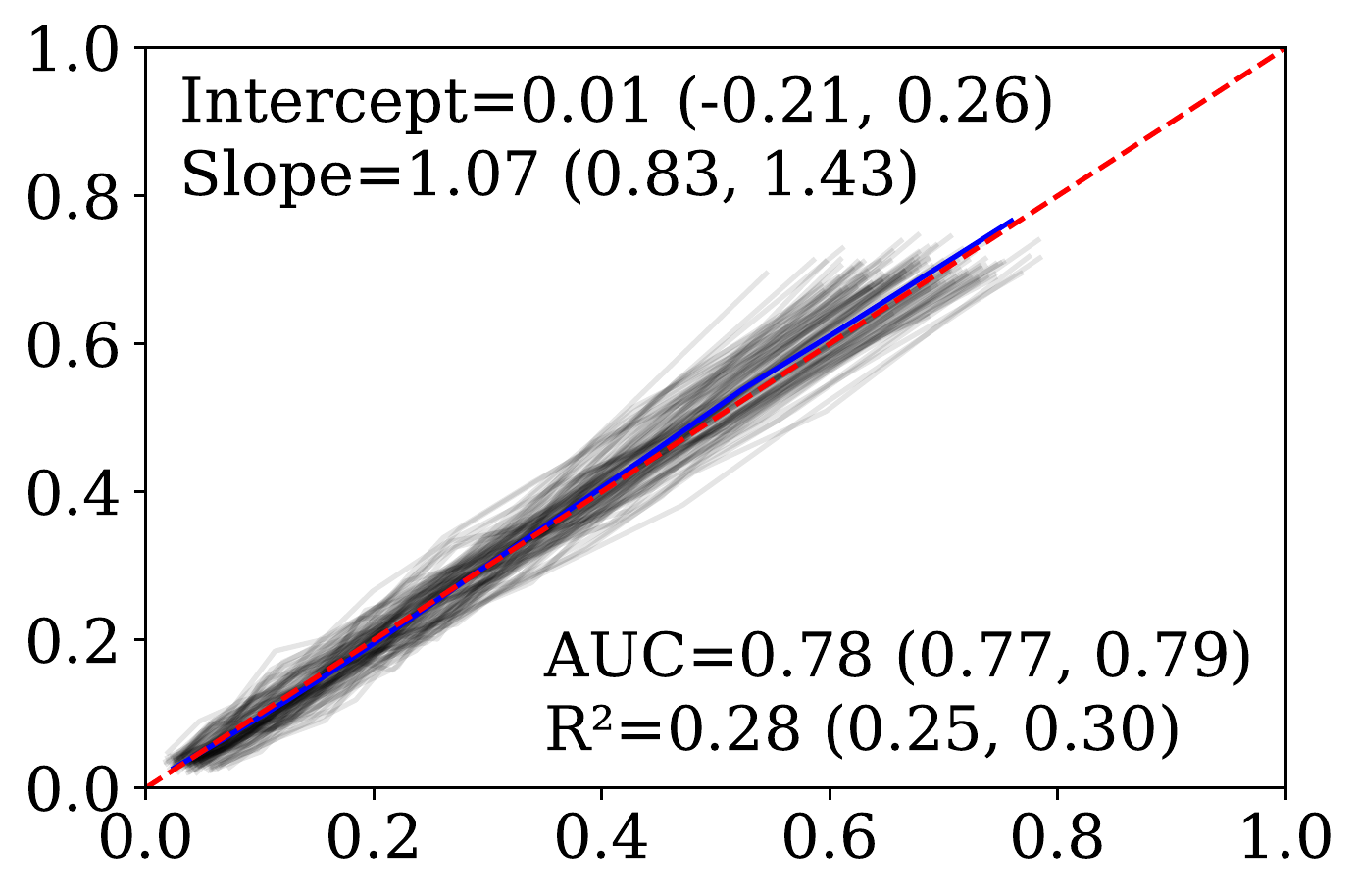} & \includegraphics[width=\plotwidth]{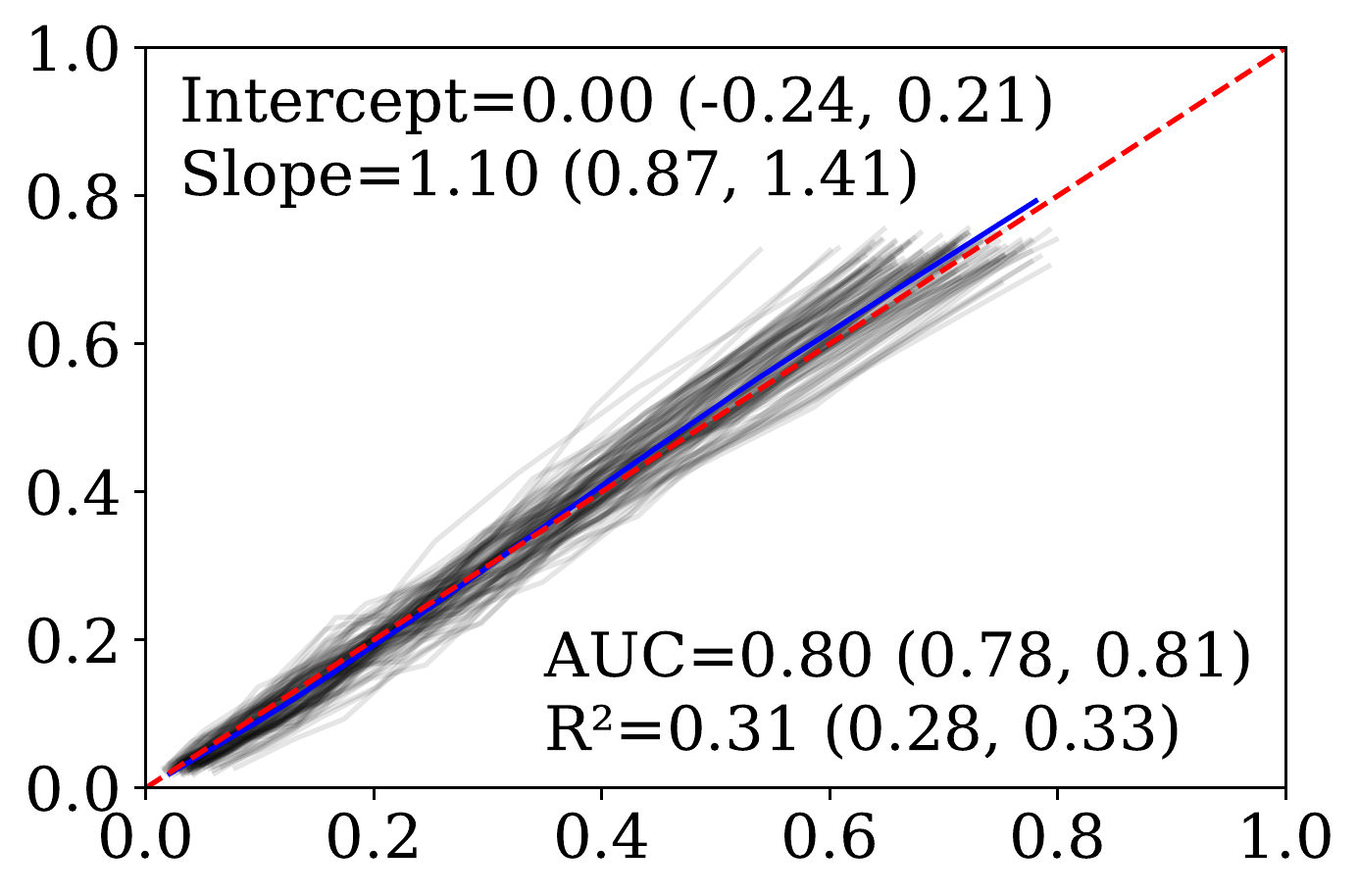}\\
         \centered{PCLR} & \includegraphics[width=\plotwidth]{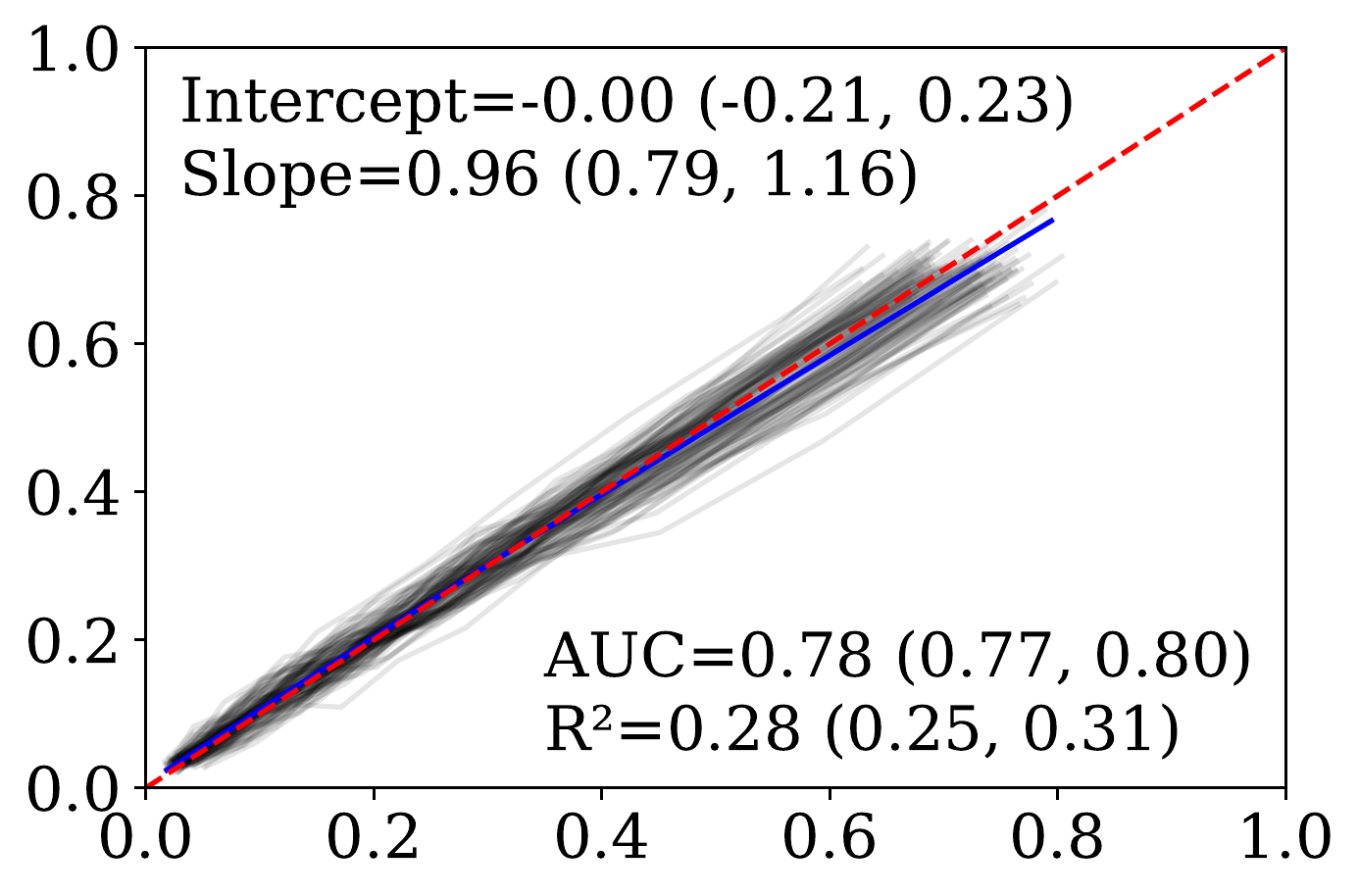} & \includegraphics[width=\plotwidth]{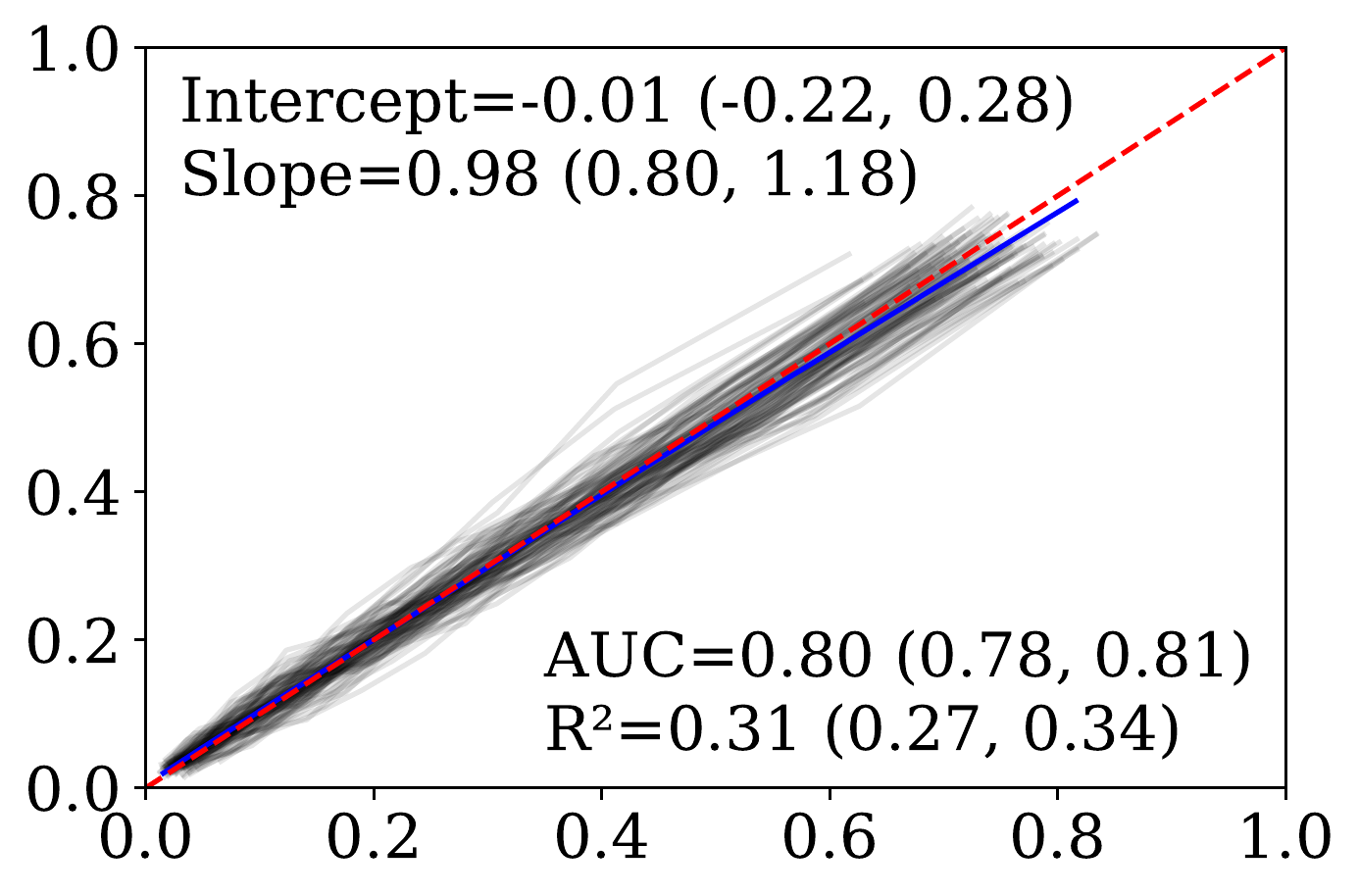} & \includegraphics[width=\plotwidth]{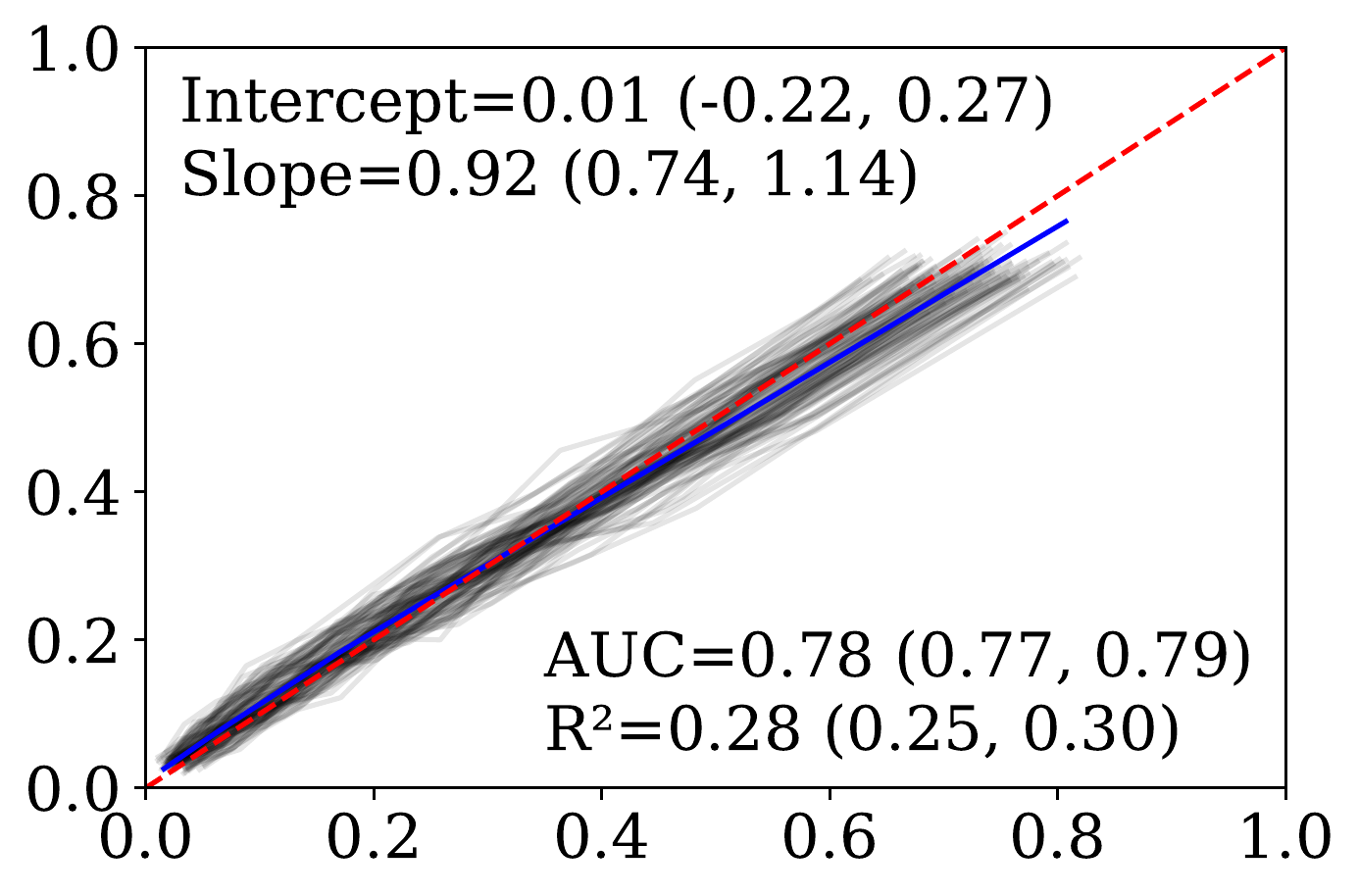} & \includegraphics[width=\plotwidth]{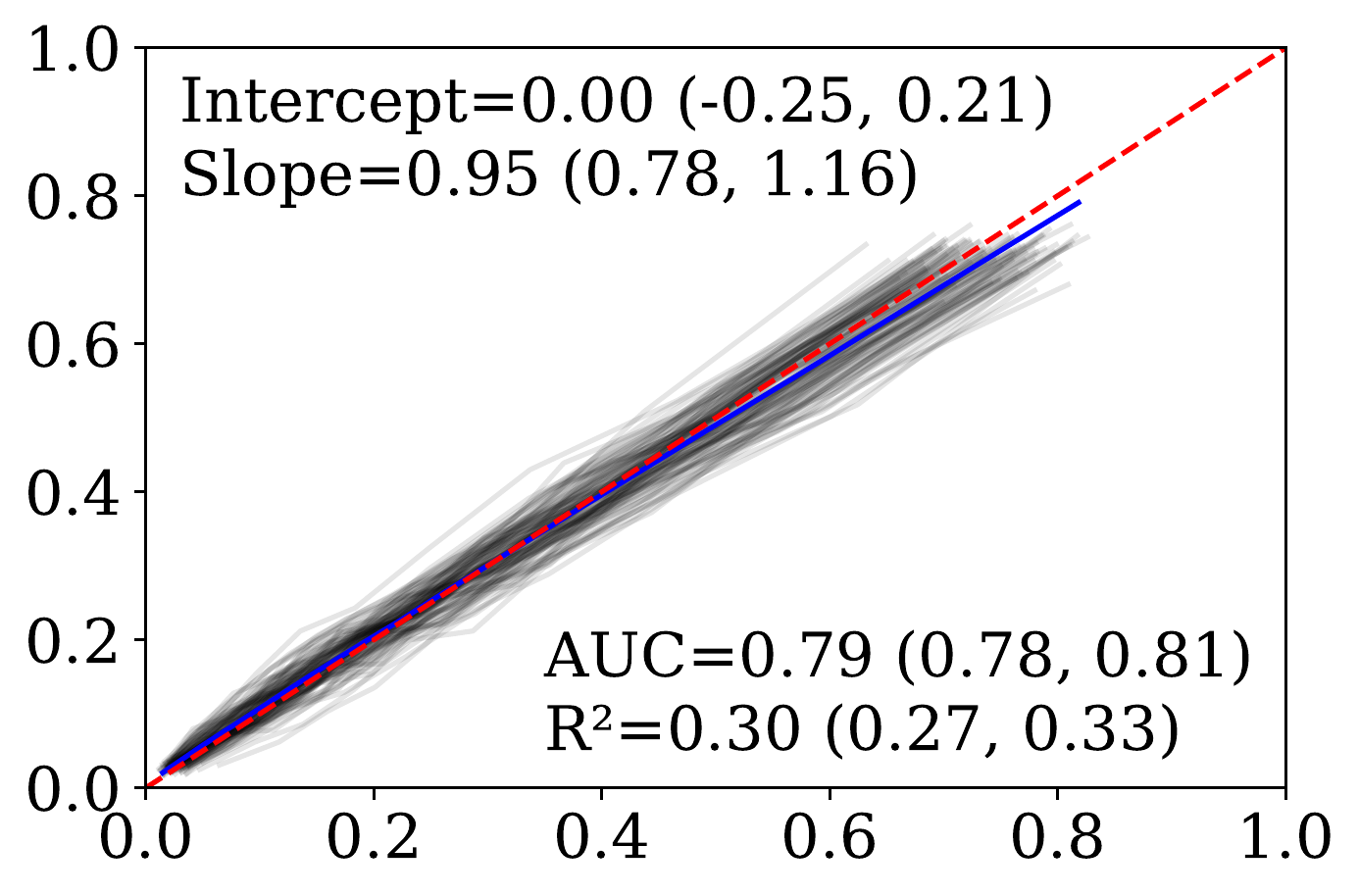}\\
         \centered{LAELR} & \includegraphics[width=\plotwidth]{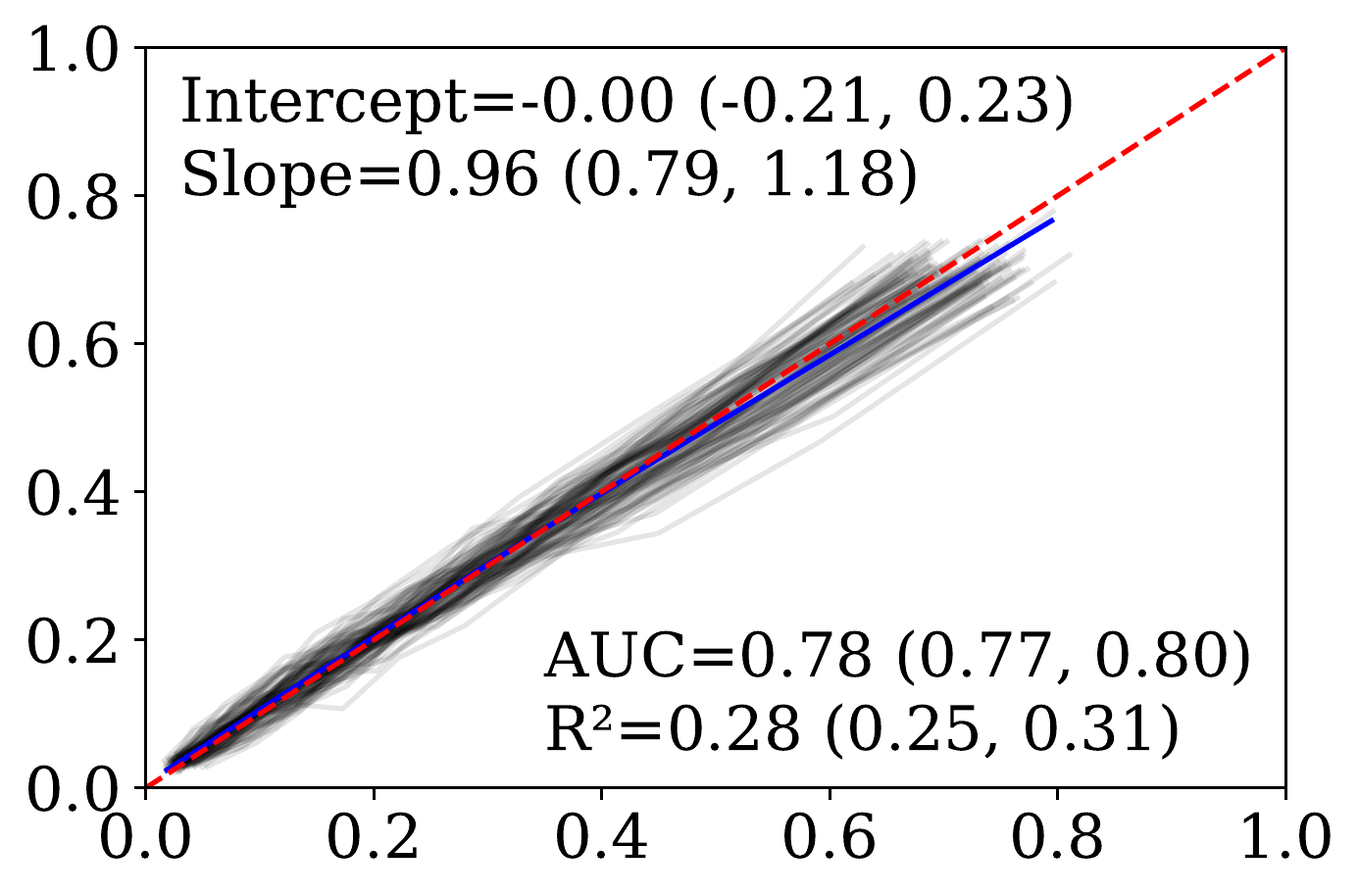} & \includegraphics[width=\plotwidth]{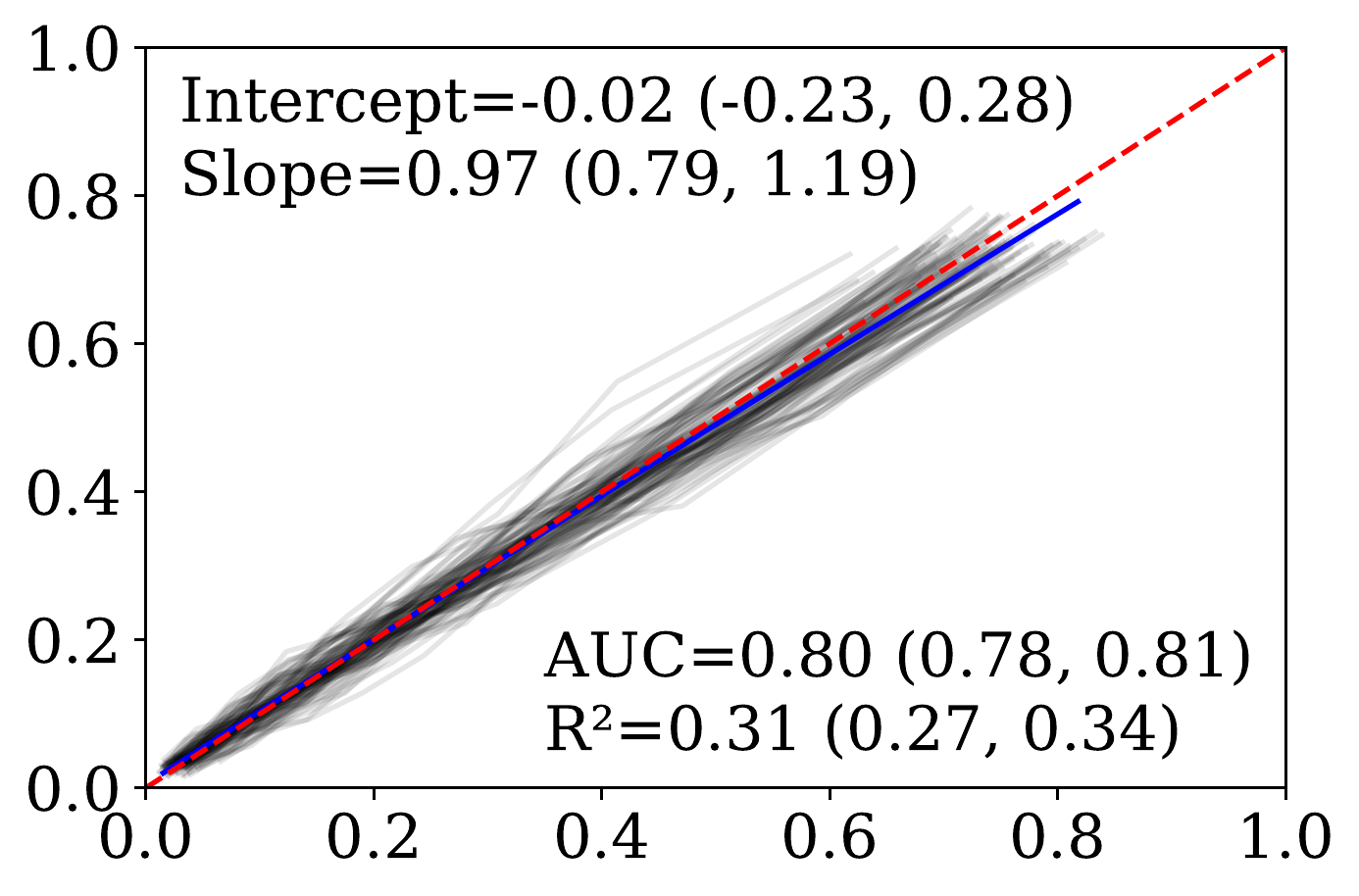} & \includegraphics[width=\plotwidth]{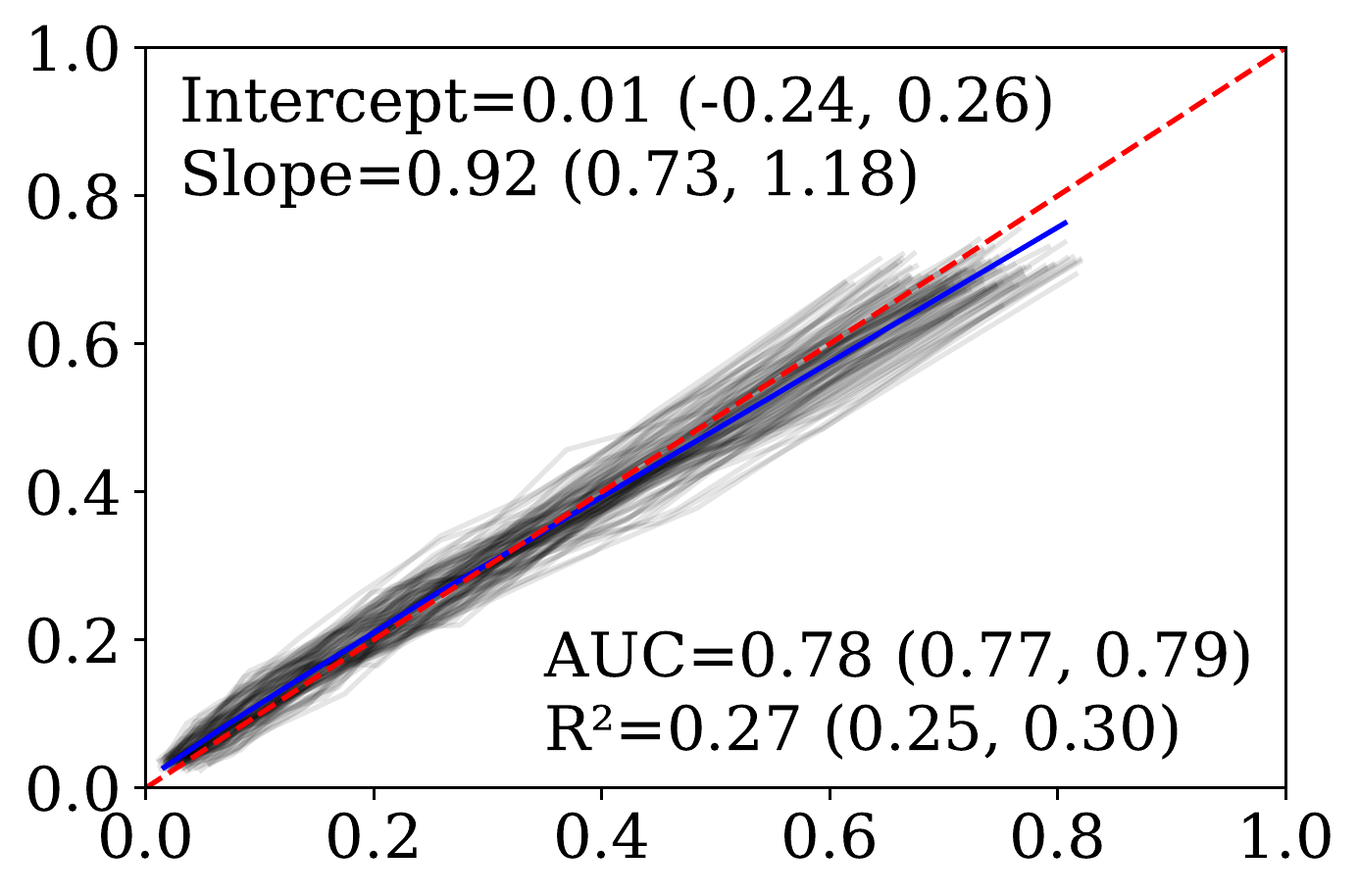} & \includegraphics[width=\plotwidth]{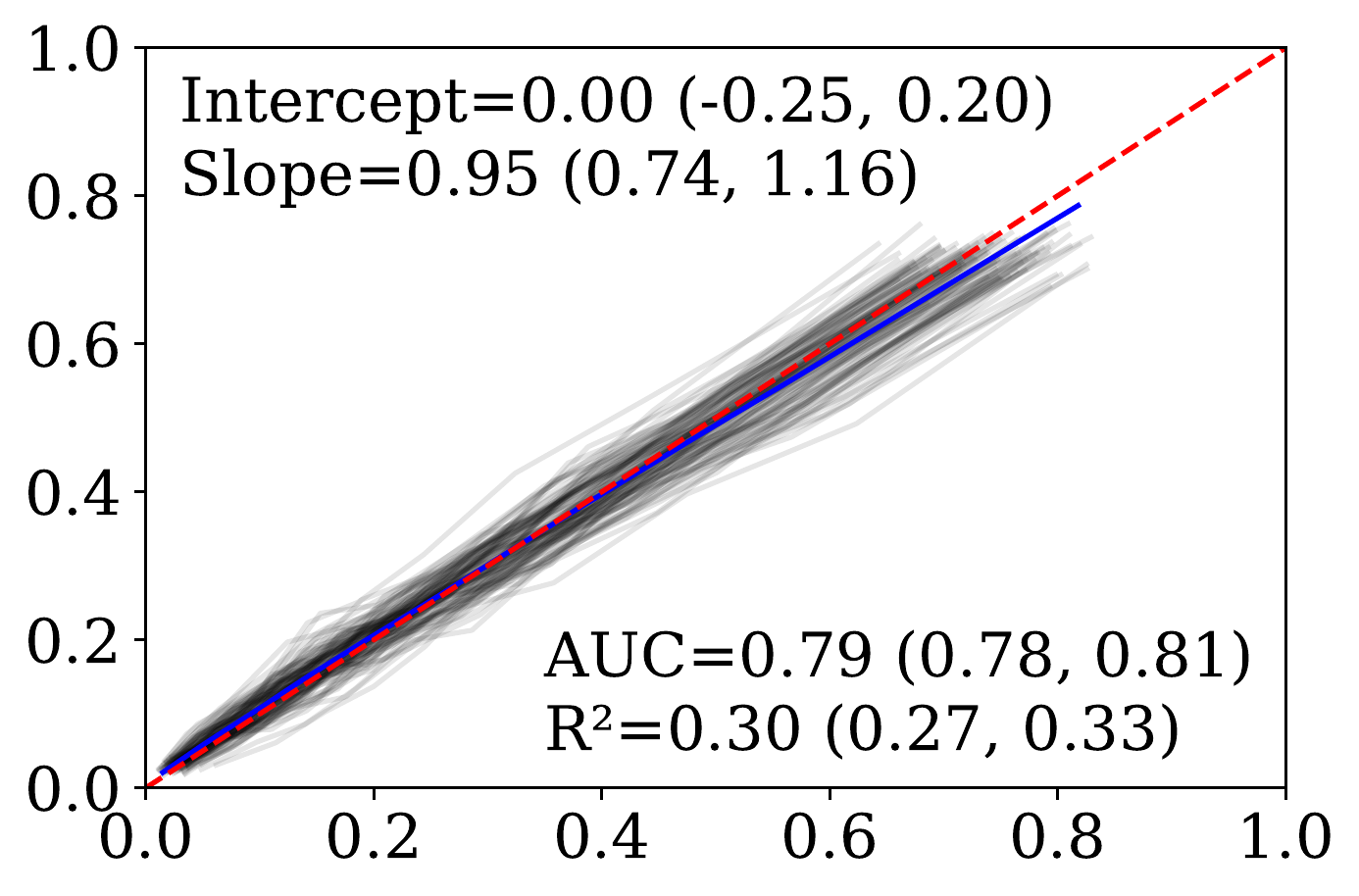}\\
         \centered{Dropout} & \includegraphics[width=\plotwidth]{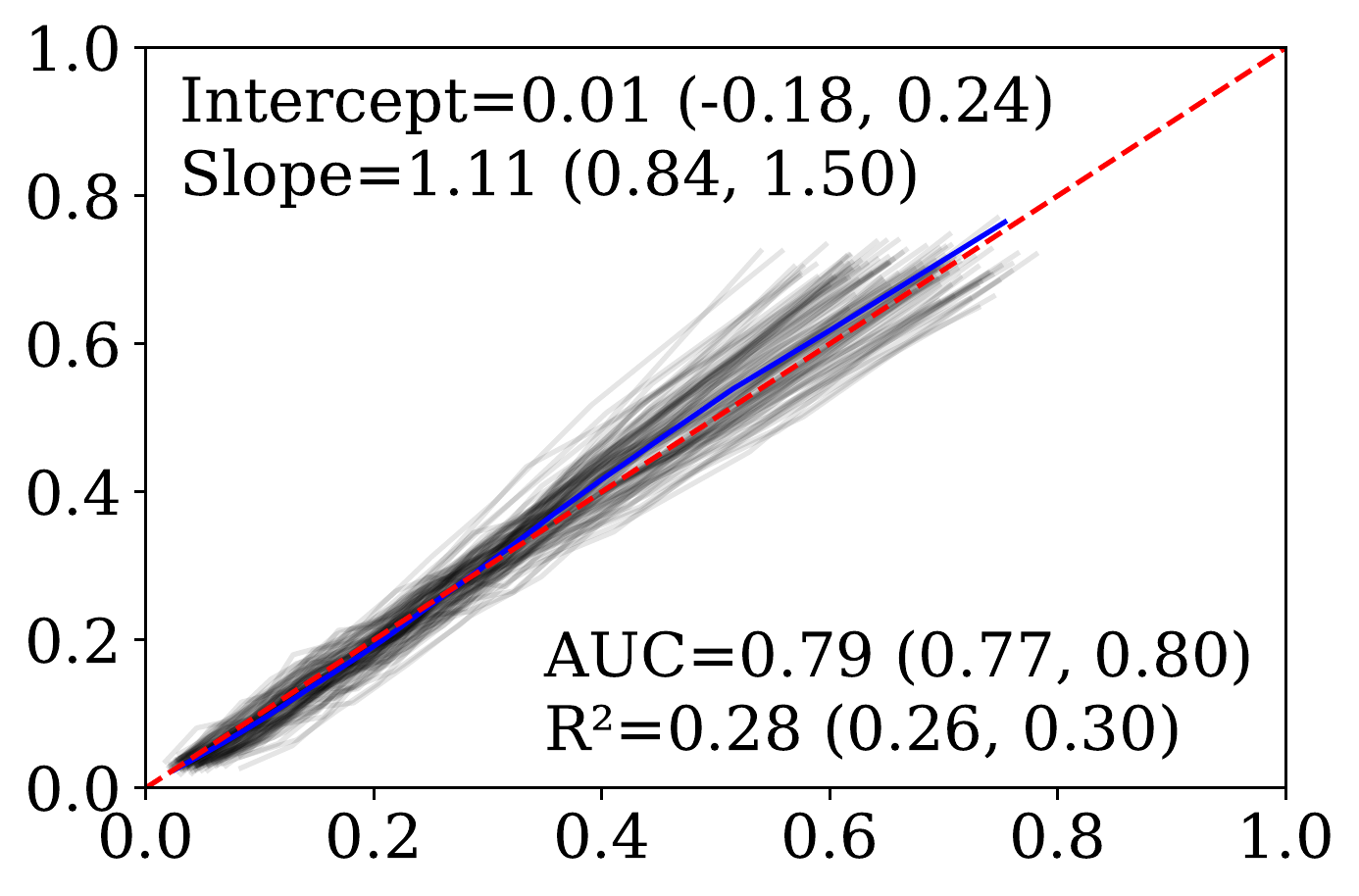} & \includegraphics[width=\plotwidth]{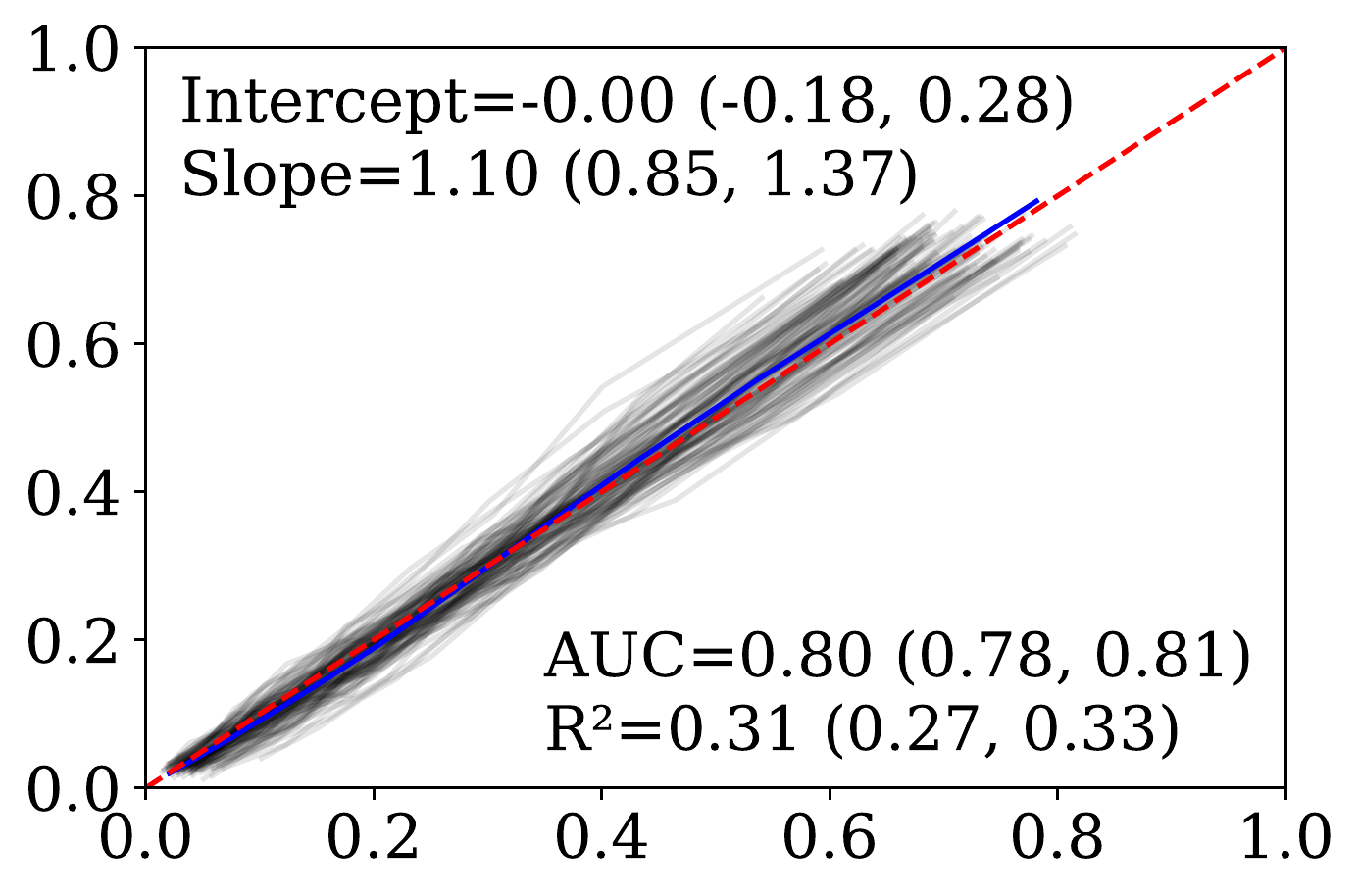} & \includegraphics[width=\plotwidth]{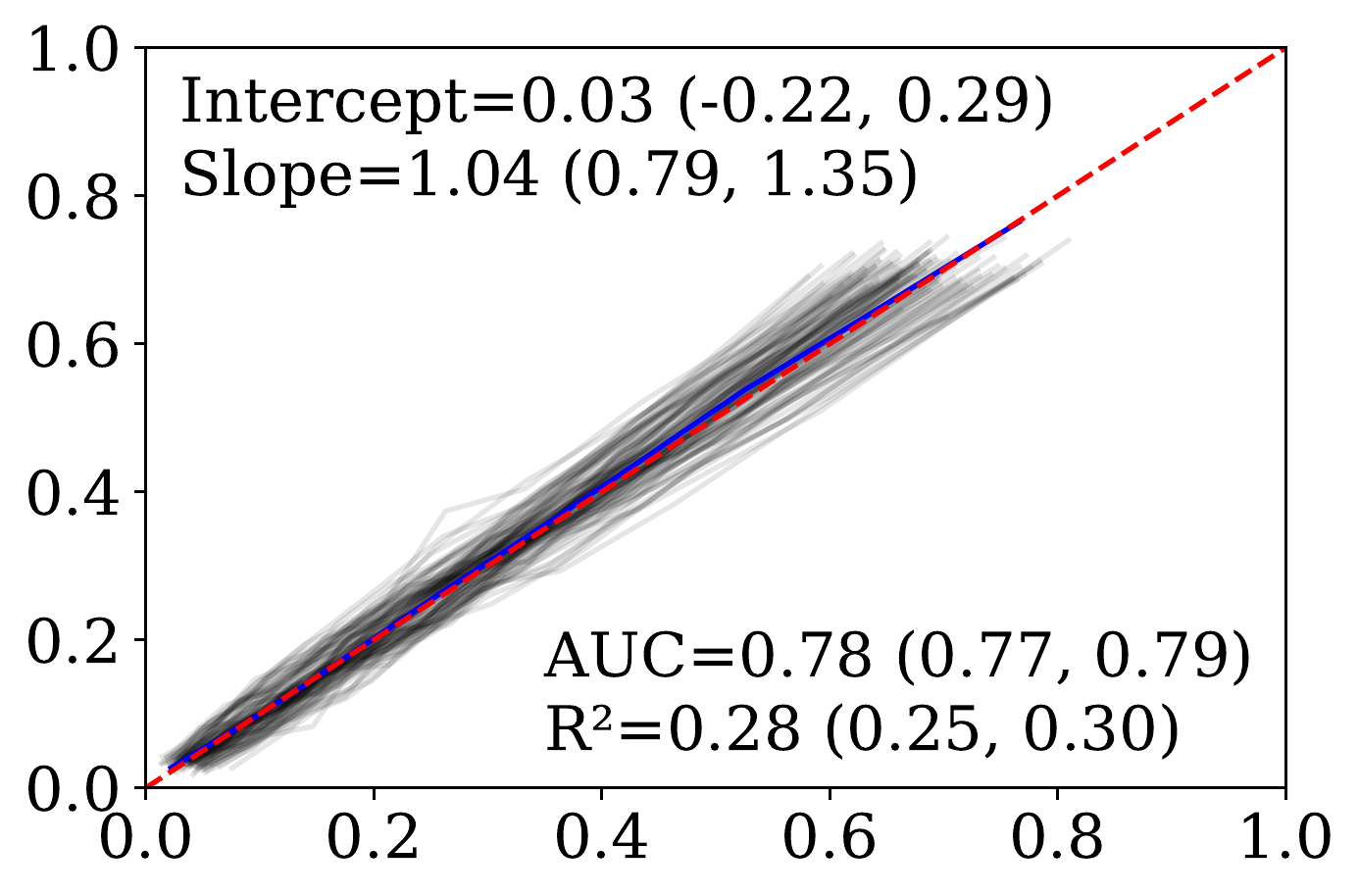} & \includegraphics[width=\plotwidth]{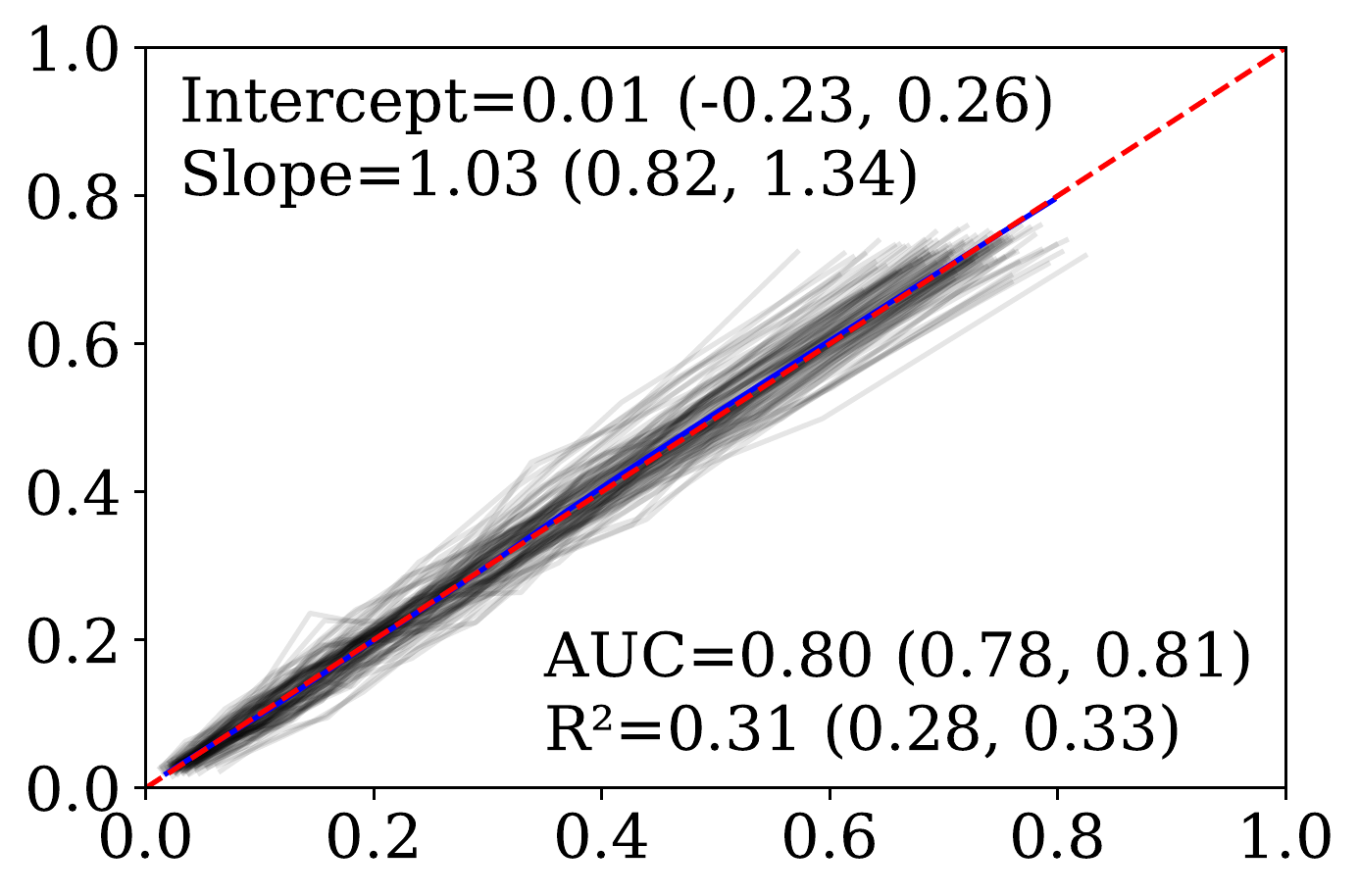}\\
         \centered{\LRnn} & \includegraphics[width=\plotwidth]{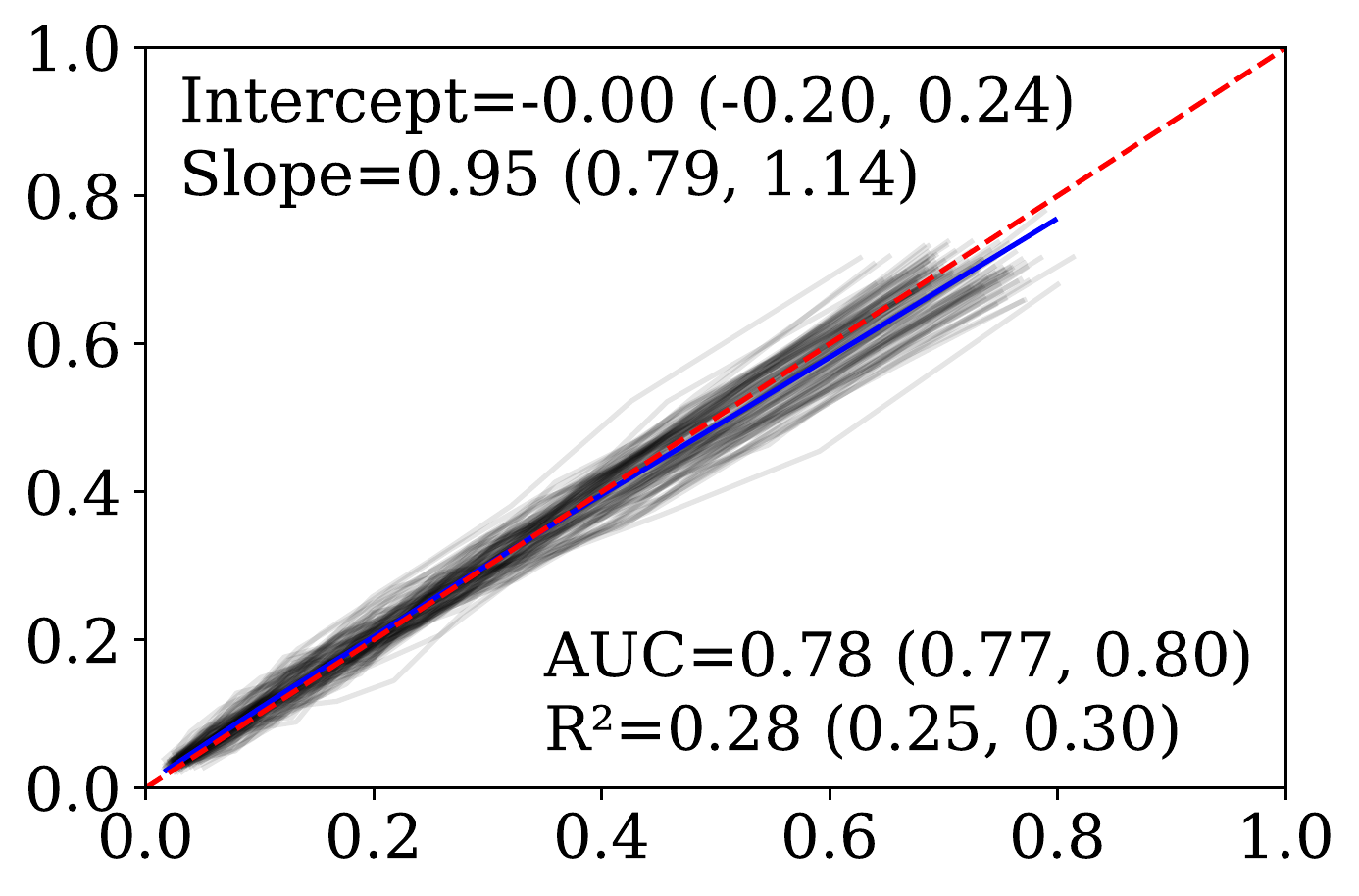} & \includegraphics[width=\plotwidth]{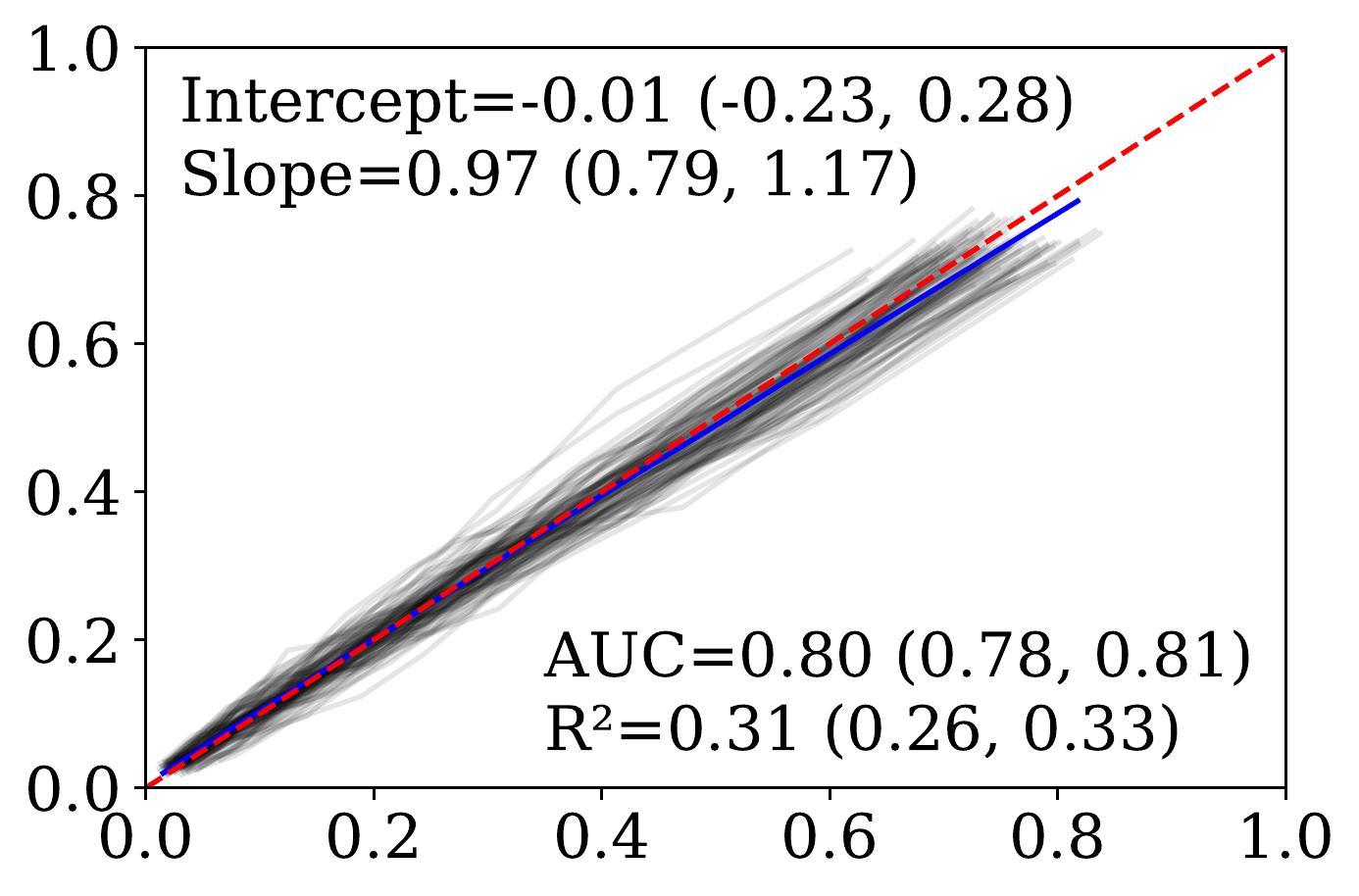} & \includegraphics[width=\plotwidth]{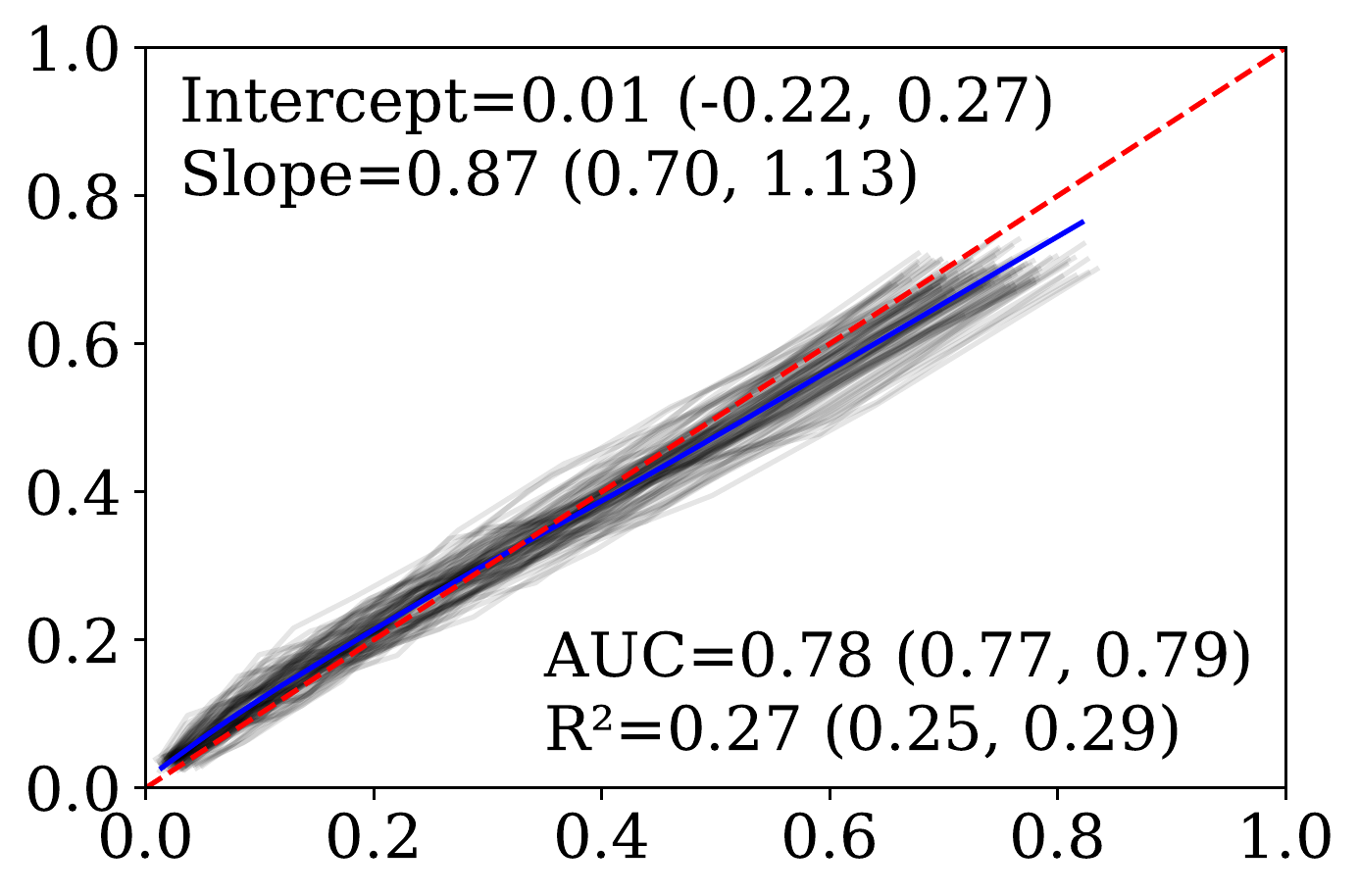} & \includegraphics[width=\plotwidth]{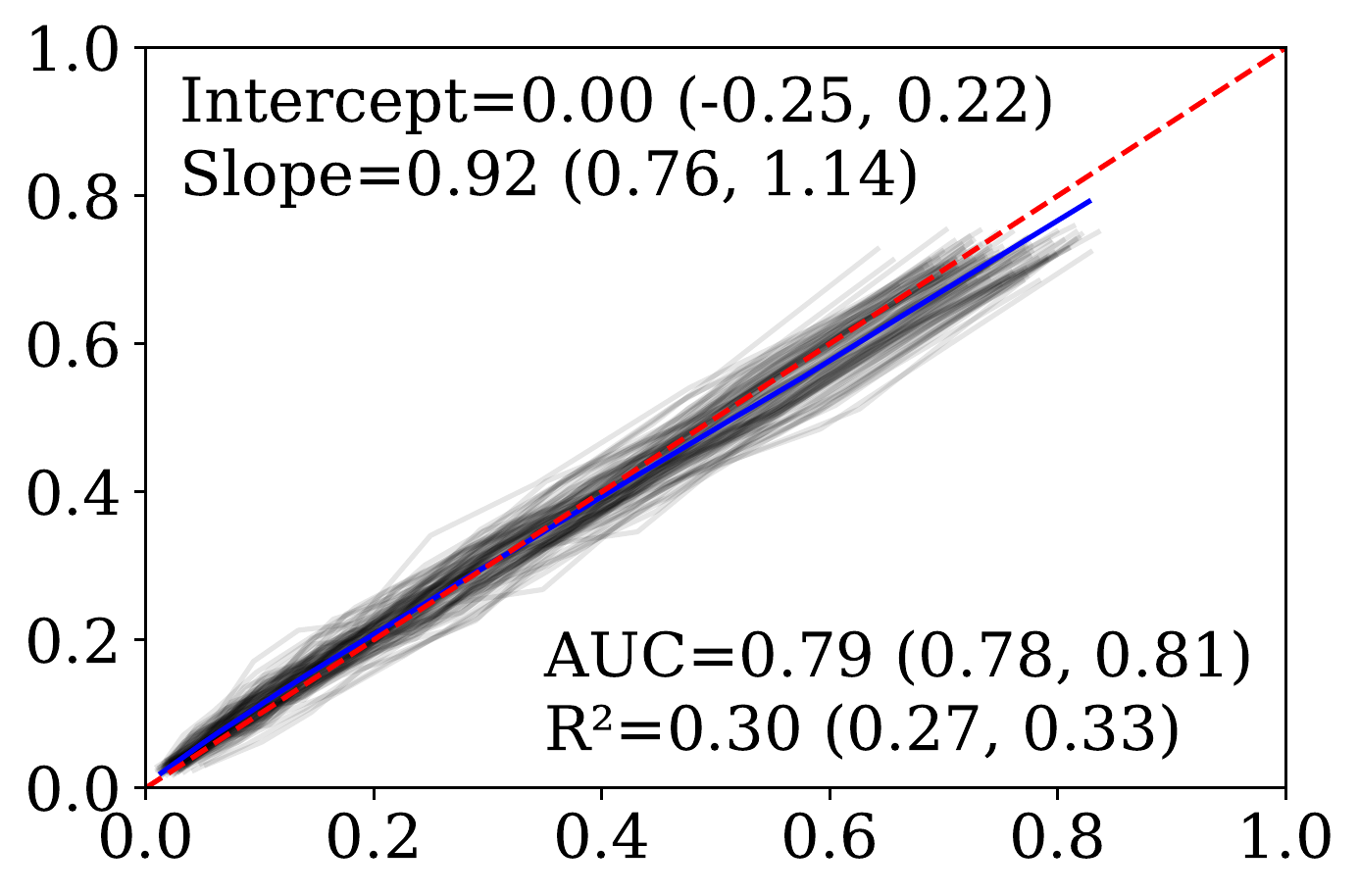}\\
    \end{tabular}}
    \caption{\label{fig:results_performance_xerostomia}Predictive performance results for the xerostomia simulations. Lowess-smoothed calibration curves per simulation are plotted in grey. The calibration curve over all repetitions is shown in blue. Perfect calibration, the diagonal, is dashed in red.}
\end{figure}

%% file: sections/BOXPLOTS-AB.tex
\newcommand\boxplotwidthab{0.48\textwidth}
\begin{figure}[h!]
    \centering
    \begin{tabular}{cc}
    \includegraphics[width=\boxplotwidthab]{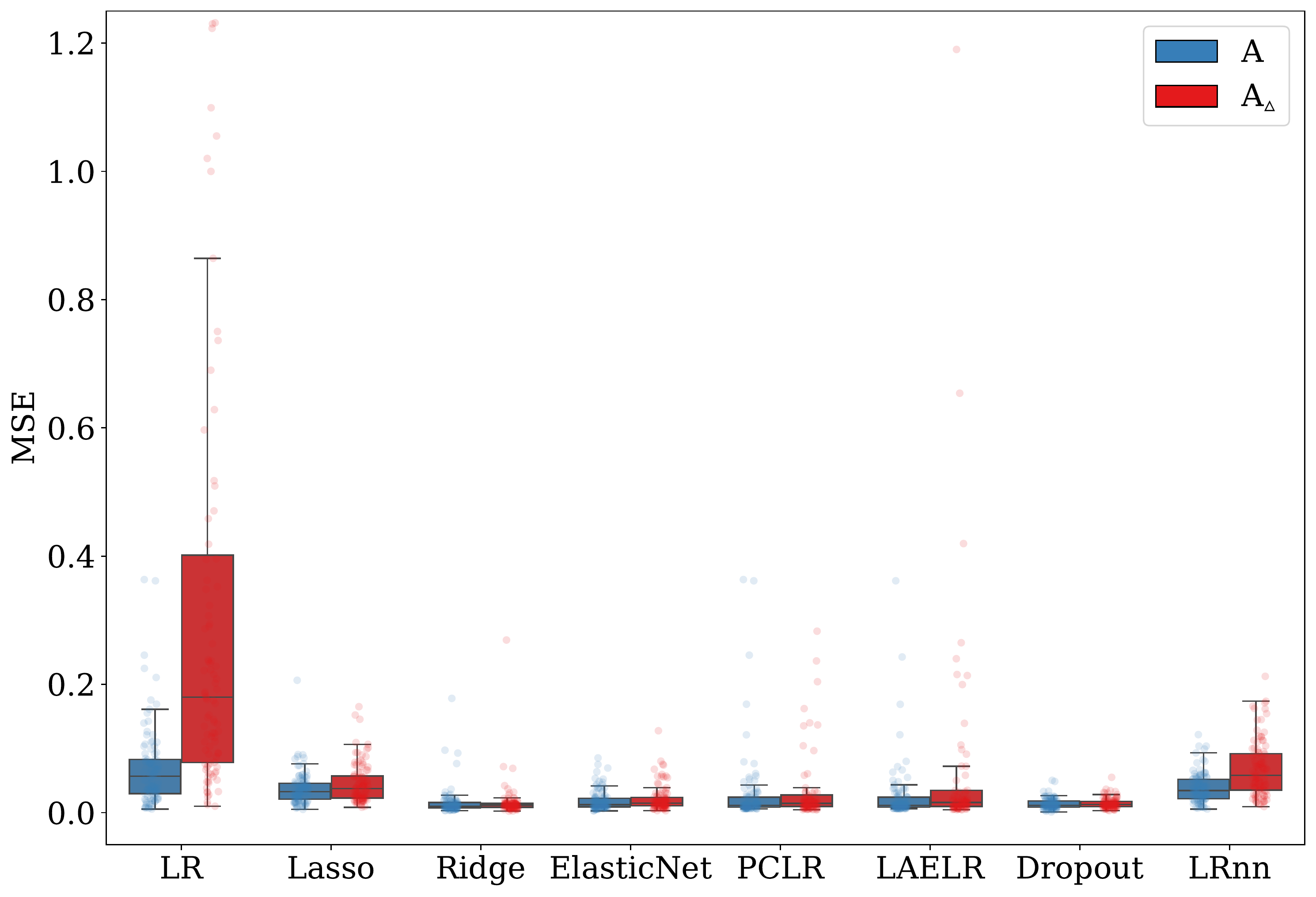}& \includegraphics[width=\boxplotwidthab]{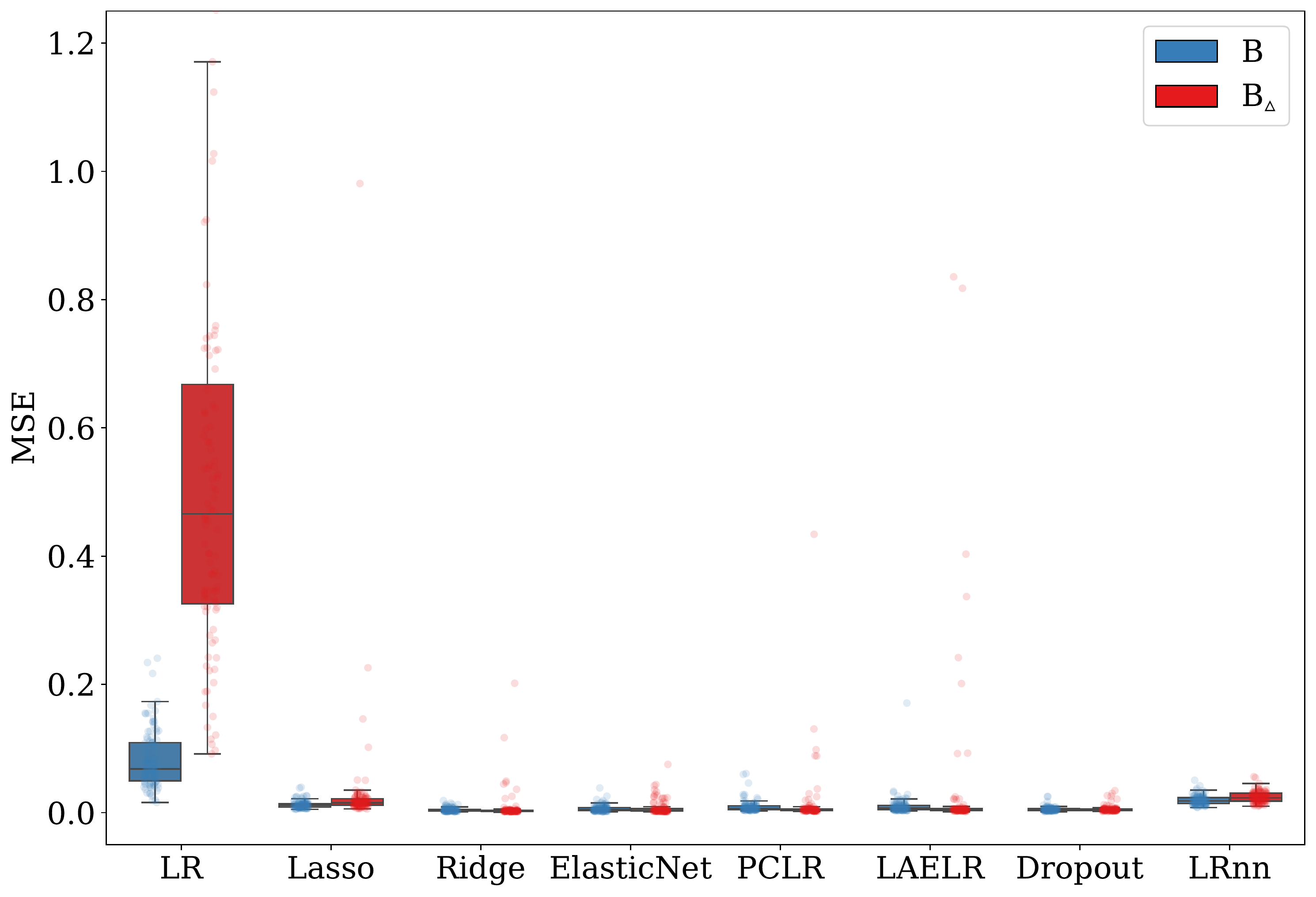}\\
    \end{tabular}
    \caption{Across models, the mean squared error between the estimated and the true coefficients for each method, for the xerostomia settings. \textbf{Red} indicates \textbf{high collinearity}, and blue low collinearity.}
    \label{fig:mse_results_ab}
\end{figure}
\vspace{-.5cm}

\begin{figure}[h!]
    \centering
    \begin{tabular}{cc}
    \includegraphics[width=\boxplotwidthab]{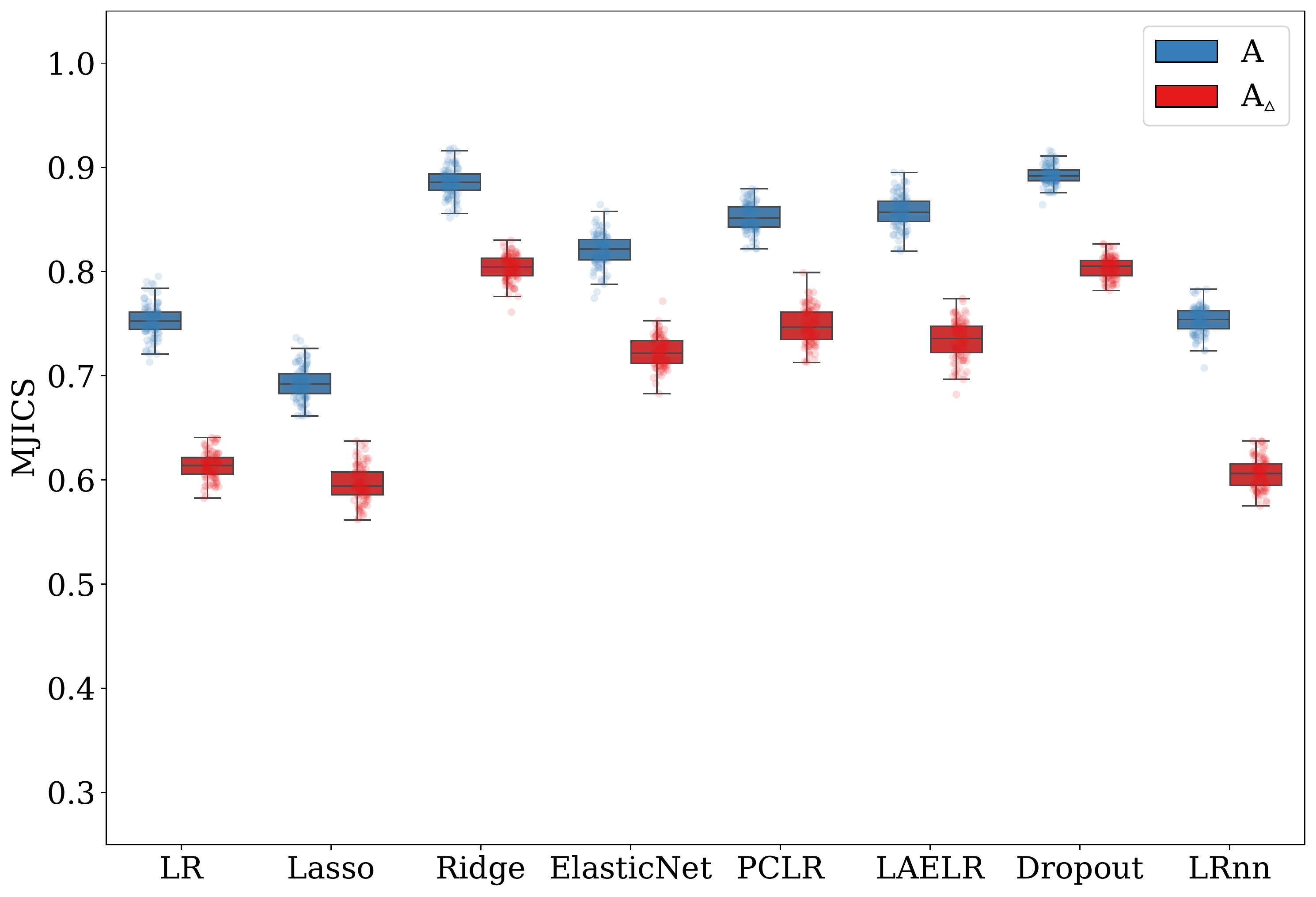}& \includegraphics[width=\boxplotwidthab]{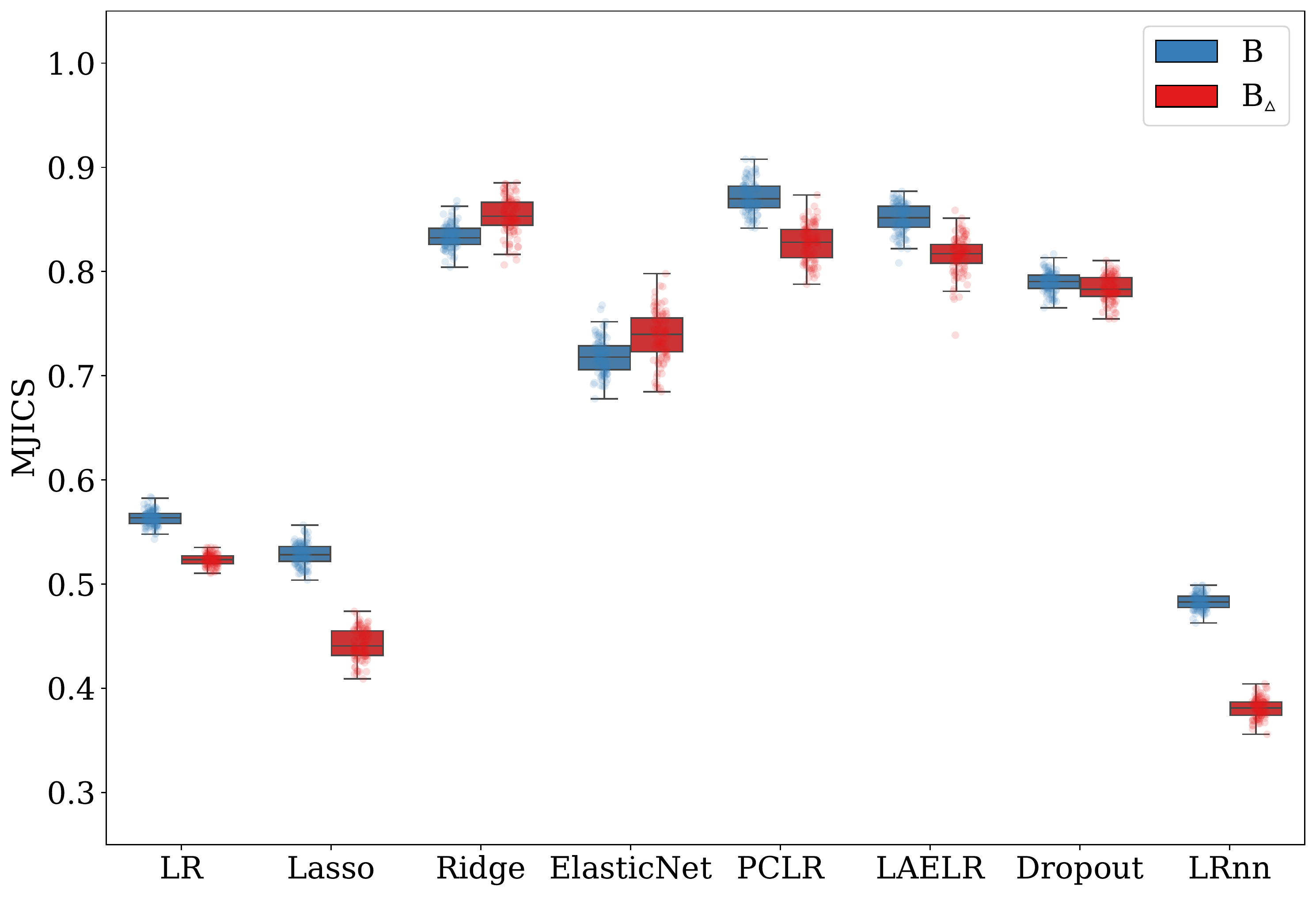}\\
    \end{tabular}
    \caption{Across models, the mean proportion of coefficients with the same direction of effect after repetition for the xerostomia settings. \textbf{Red} indicates \textbf{high collinearity}, and blue low collinearity.}
    \label{fig:mjics_results_ab}
\end{figure}

%% file: sections/HYPPARAMPLOTS-AB.tex
\newcommand{\hypplotwidthab}{0.3\textwidth}
\begin{figure}[h!]
    \centering
    \begin{tabular}{cccc}
       \rotatebox{90}{\hspace{1.3cm}\footnotesize Hyperparameter value} &  \includegraphics[width=\hypplotwidthab]{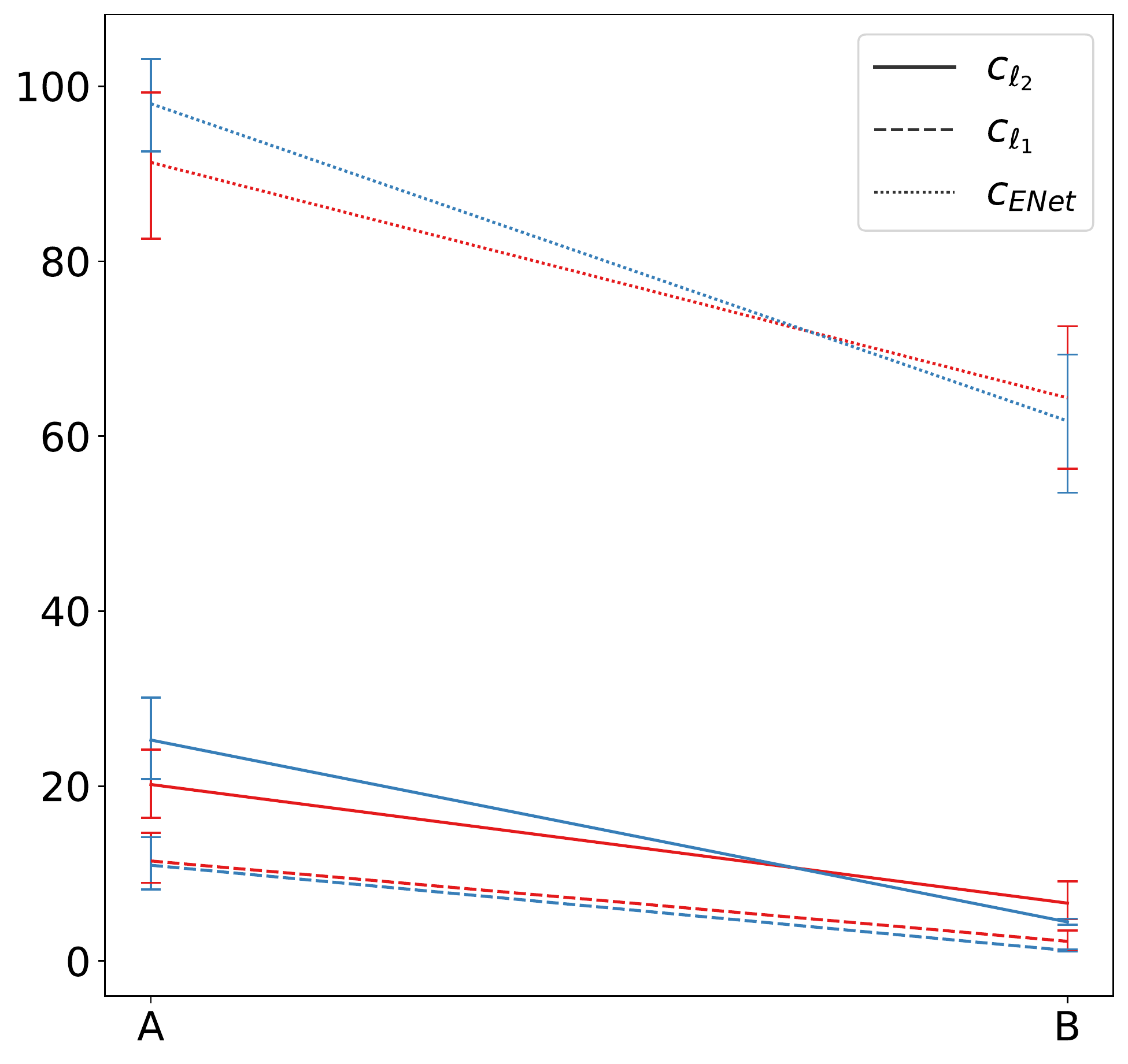} &  \includegraphics[width=\hypplotwidthab]{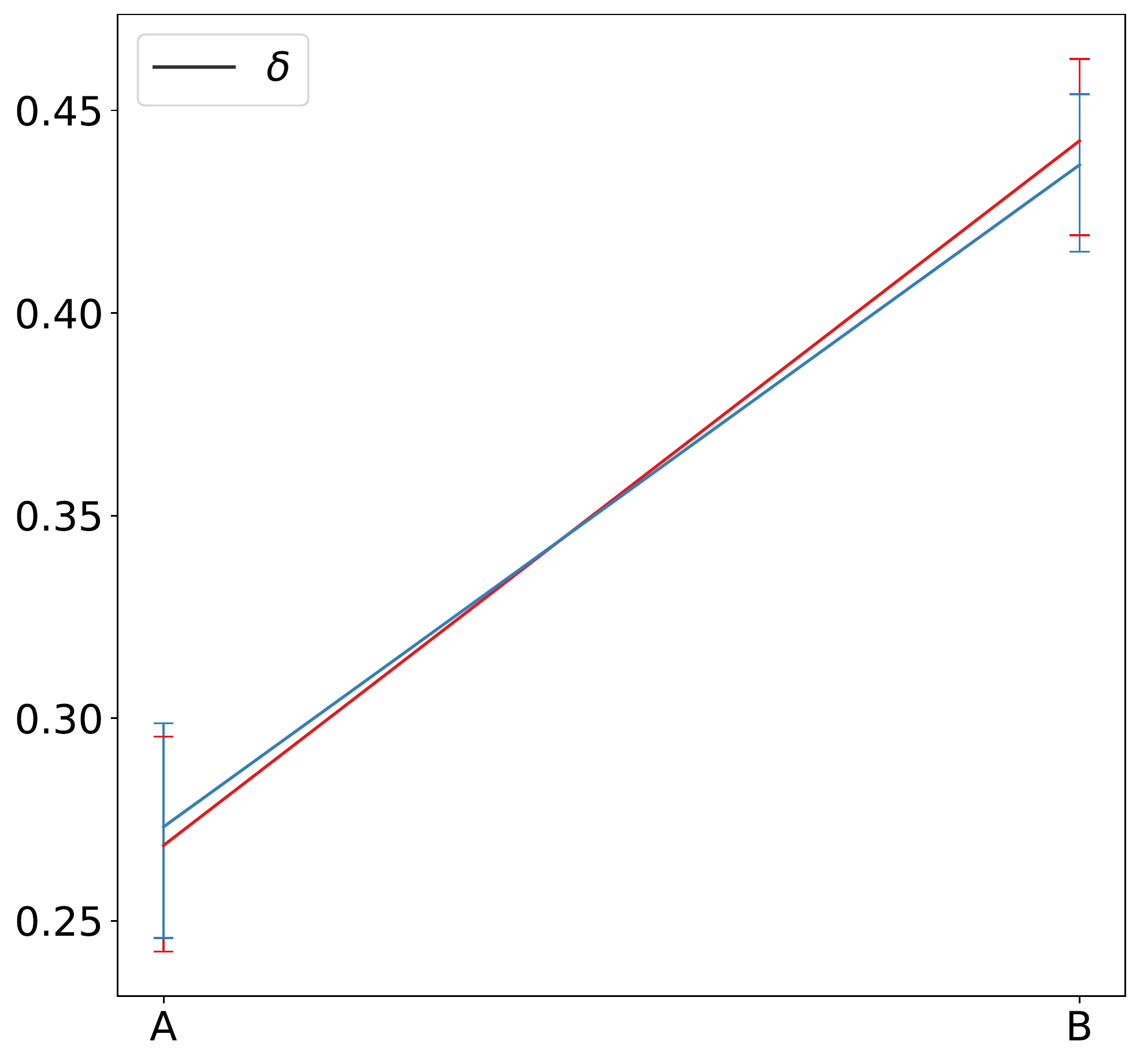} & \includegraphics[width=\hypplotwidthab]{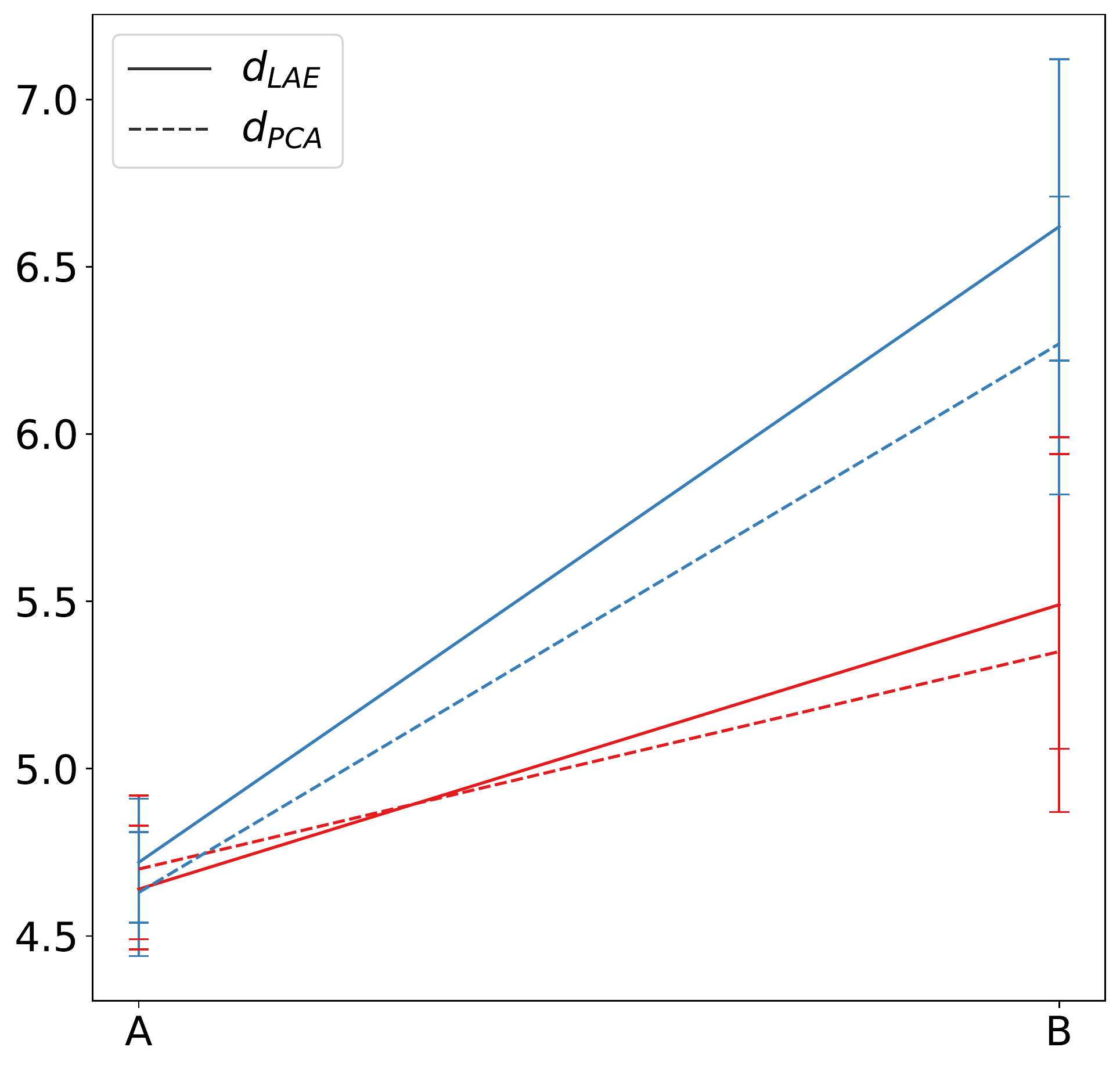}\\
        & (1) & (2) & (3) \\
    \end{tabular}
    \caption{\label{fig:xer_hyperparam_results}Hyperparameter values for xerostomia: per predictor set, setting A being the small predictor set with high EPV (EPV=23), and setting B the large predictor set with lower EPV (EPV=8). The \textbf{high collinearity} settings \textbf{in red}, and the \textbf{low collinearity} setting \textbf{in blue}. The methods are distributed across three plots due to their different scales. Hyperparameter notation follows Table \ref{tab:methods}, except for $c_{\textit{ENet}}$, which is the total shrinkage factor for ElasticNet ($c_{\ell_1} + c_{\ell_2}$).}
    
\end{figure}

%% file: sections/4-DISCUSSION.tex
\vspace{-.5cm}
\section{Discussion}
\label{sec:discussion}

The current study investigated the effect of collinearity on predictive performance and the stability of coefficient estimation, comparing eight different methods in a simulation study on the construction of prediction models that estimate the risk on complications after radiotherapy in head-and-neck cancer patients.

In this paper, we found little to no impact of collinearity on predictive performance (discrimination and calibration of the fitted models) across methods and simulation settings. As expected, we found that collinearity has a negative impact on the stability of coefficient selection in high EPV settings for all methods. For settings with a lower EPV, that consequently required a larger amount of regularization, the negative impact of collinearity on the stability of selected coefficients was lower for methods that distribute the explained outcome variance more evenly across collinear predictors: via grouping (Ridge, ElasticNet, and Dropout), or based on principal components (PCLR, and LAELR). For standard logistic regression, or methods that have a strong predictor selection effect (Lasso, and non-negative logistic regression) the stability of predictor selection was lower compared to the other methods, and was negatively influenced by collinearity across all simulations.


Harrell (2001)\cite{harrell2015regression} mentioned that when there is no difference in the degree of collinearity between development and validation data, collinearity is generally not considered a problem for predictive performance, but can be problematic for reliable variable selection when performing stepwise selection. This was also confirmed by Cohen, Cohen, West and Aiken (2003)\cite{cohen2013applied}, and later also by Dormann et al. (2013)\cite{dormann2013collinearity}, who compared 23 methods (including various dimensionality reduction techniques and shrinkage-based methods) to address collinearity in five simulated ecological predictor-response relationships. 
The current study findings are in line with these earlier works and provides additional evidence to support this. An important note to make is that in low-dimensional settings (where the number of predictors is smaller than the number of samples) with correlating predictors earlier work by Tibshirani (1996)\cite{tibshirani1996regression,zou2005regularization} and Pavlou et al., (2015)\cite{pavlou2016review} empirically found in their experiments that selection-based approaches like Lasso yielded lower predictive performance compared to for example Ridge. The current study did not find such a difference in predictive performance between Lasso and Ridge in any of the eight settings.


Nevertheless, for addressing collinearity in clinical prediction models, we would recommend refraining from data-driven predictor selection approaches (like Lasso), because of the increased instability of predictor selection in the presence of collinearity, even in relatively high EPV settings. If a model is model is interpreted at some point after its development (e.g., to perform face validity checks or to explain predictions in clinical practice), predictor selection could incorrectly give the impression that there is strong evidence in the data for the individual contribution of certain collinear predictors.


There are several limitations that should be considered when interpreting this study. Firstly, the current work has our research has focused only on low-dimensional settings and binary logistic regression models.
Future studies may evaluate the effect of collinearity, for instance in settings with multiple outcomes (e.g., multinomial regression).
Finally, we focused on evaluation of predictive performance in the same population, under no change of collinearity structure between the development and validation data. Collinearity has been shown to have a negative impact on performance under changes between development and validation data, and is considered a difficult challenge to overcome, for which a good understanding of the underlying mechanism causing the collinearity is crucial.\cite{dormann2013collinearity}

We believe that beside being able to anticipate how harmful a change in collinearity between train and test data may be for predictive performance, an interesting direction of future research is to study how background knowledge about the underlying collinearity mechanism, and about potential changes of collinearity across development and validation data can be used to adapt prediction models accordingly.

%% file: sections/5-ACKNOWLEDGEMENTS.tex
\section*{Acknowledgements}
This work is supported by the European Union's Horizon 2020 research and innovation programme under grant agreement No 825162 [HTx project]. 


%% file: sections/6-APPENDIX.tex
\clearpage

\section*{Appendix}
\setcounter{subsection}{0}

\renewcommand{\thesubsection}{\Alph{subsection}}

\subsection{Results for the simulated dysphagia settings.}
\label{appendix:dysphagia_results}
\vspace{-1cm}
\input{sections/CALIB-GRID-CD}

\input{sections/BOXPLOTS-CD}
\input{sections/HYPPARAMPLOTS-CD}

\newpage
\clearpage
\subsection{Results for the real-data settings.}
\label{appendix:real_data_results}
\vspace{-.5cm}
\input{sections/CALIB-GRID-REAL}

\newpage
\clearpage
\subsection{Baseline tables}

\input{sections/BASELINE-TABS}

\newpage
\clearpage
\subsection{Definition of methods}
\label{appendix:models}

\input{sections/APP-METHOD-DEFS}

\subsection{Performed sample size calculations}
\label{sec:appendix:samplesizecalculations}

\input{sections/SAMPLE-SIZE-CALS}

\newpage
\clearpage
 \subsection{Correlation plots}
 In all correlation plots, negative correlations are indicated in blue, and positive correlations in red.
\input{sections/3-TABS-CORR}

\newpage
\clearpage
\subsection{Coefficients}
\newcommand{\negsum}{$\sum_{\beta_{< -0.01}}$}
\newcommand{\possum}{$\sum_{\beta_{> 0.01}}$}
\newcommand{\negprop}{$P_{\beta_{ < -0.01}}$}
\newcommand{\posprop}{$P_{\beta_{ > 0.01}}$}

In this section, we report the mean coefficients of the estimated models in all simulated, and real-data settings, together with some general statistics about the coefficients: the sum, and proportion of all negative dose-coefficients (\negsum{} and \negprop{} respectively), and the sum and proportion of all positive dose-coefficients (\possum{} and \posprop{} respectively).

\input{sections/COEFFS}


%% file: sections/CALIB-GRID-CD.tex
\begin{figure}[h!]
    \centering
    \resizebox{.95\textwidth}{!}{
    \begin{tabular}{lcccc}
         Method& \dysbasiclow{} (EPV=6 ,mVIF=7) & \dysbasichigh{} (EPV=6 ,mVIF=43) & \dysextlow{} (EPV=2 ,mVIF=7) & \dysexthigh{} (EPV=2 ,mVIF=43) \\
         \centered{LR}  & \includegraphics[width=\plotwidth]{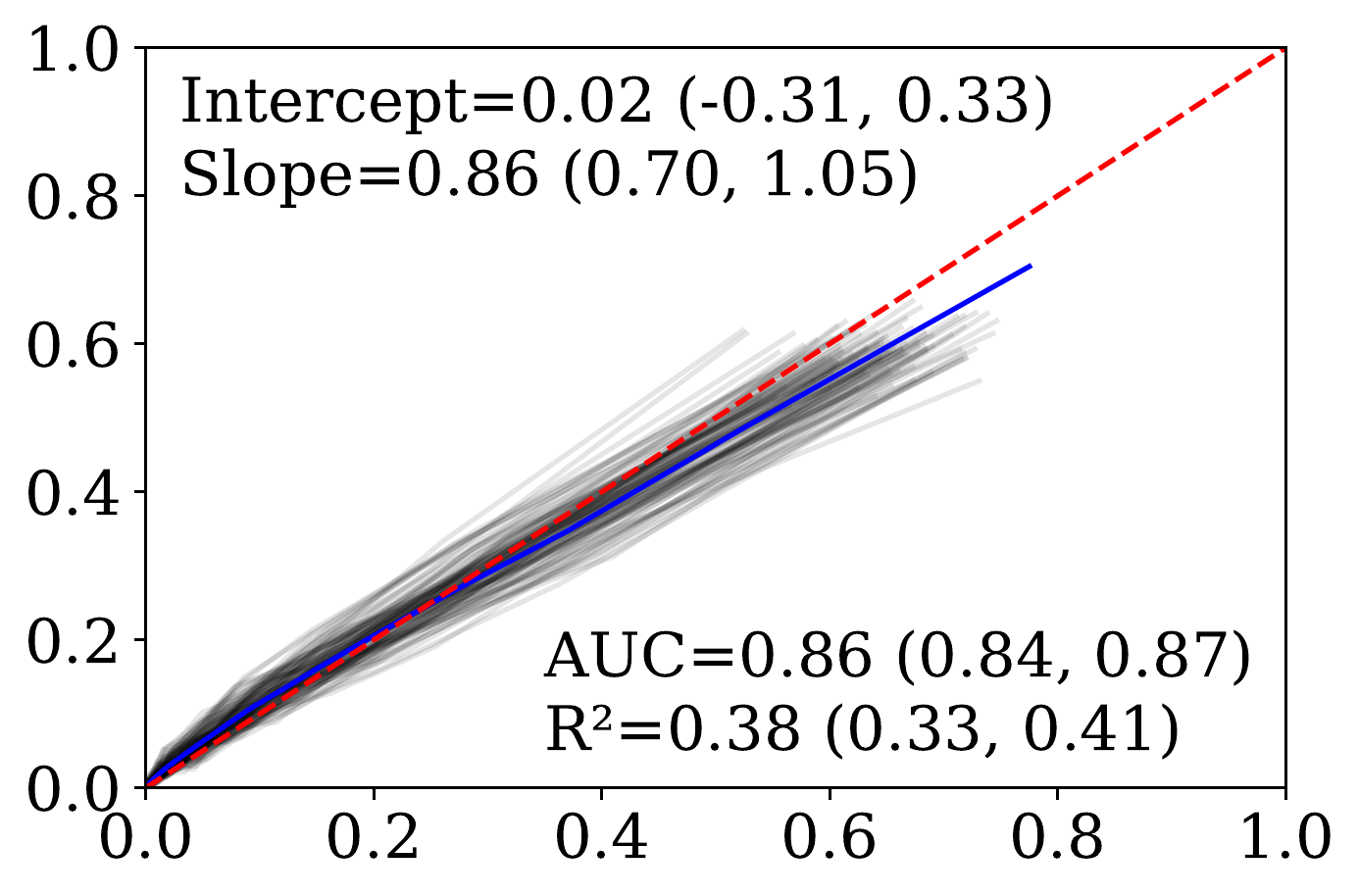} & \includegraphics[width=\plotwidth]{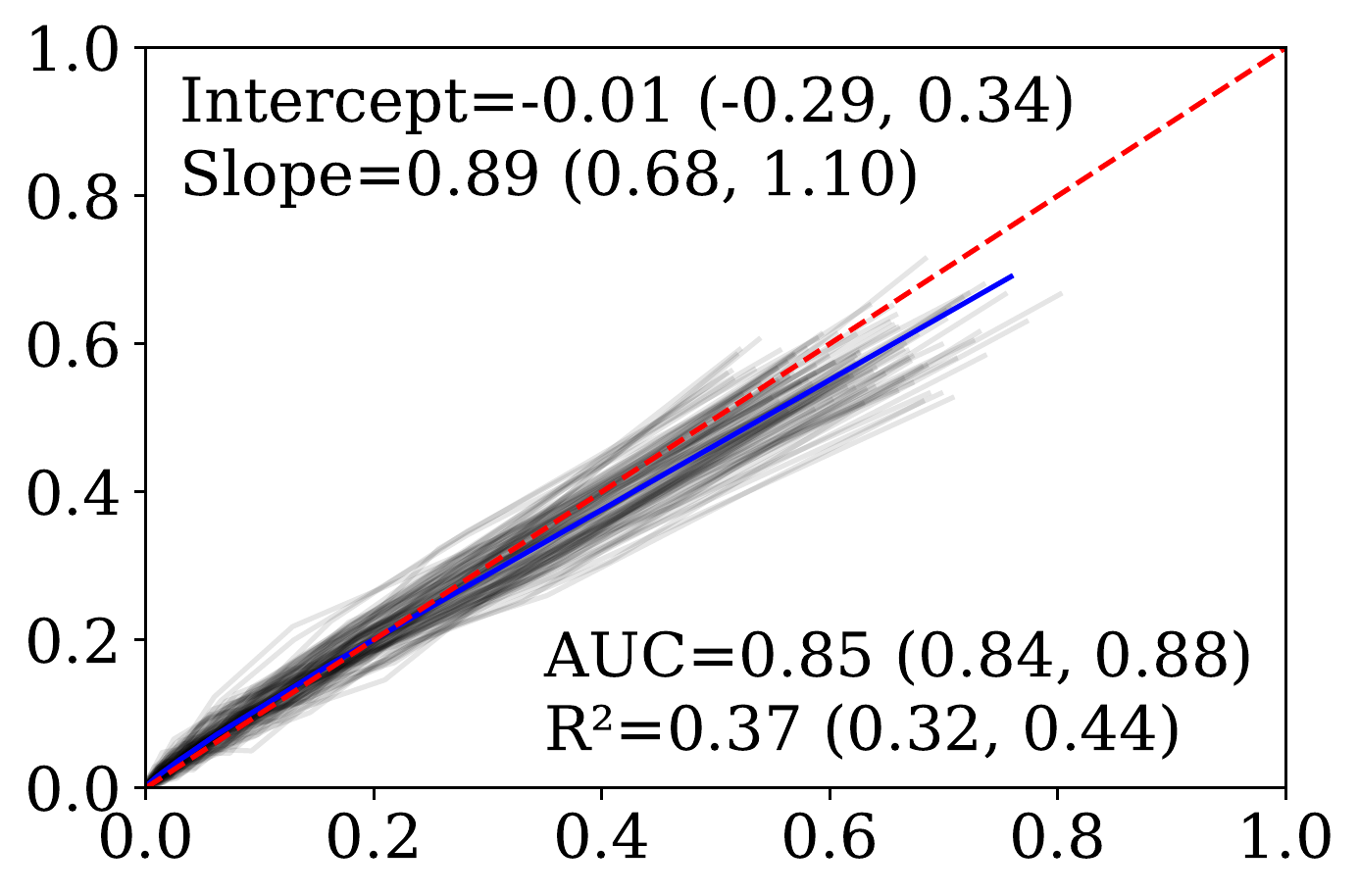} & \includegraphics[width=\plotwidth]{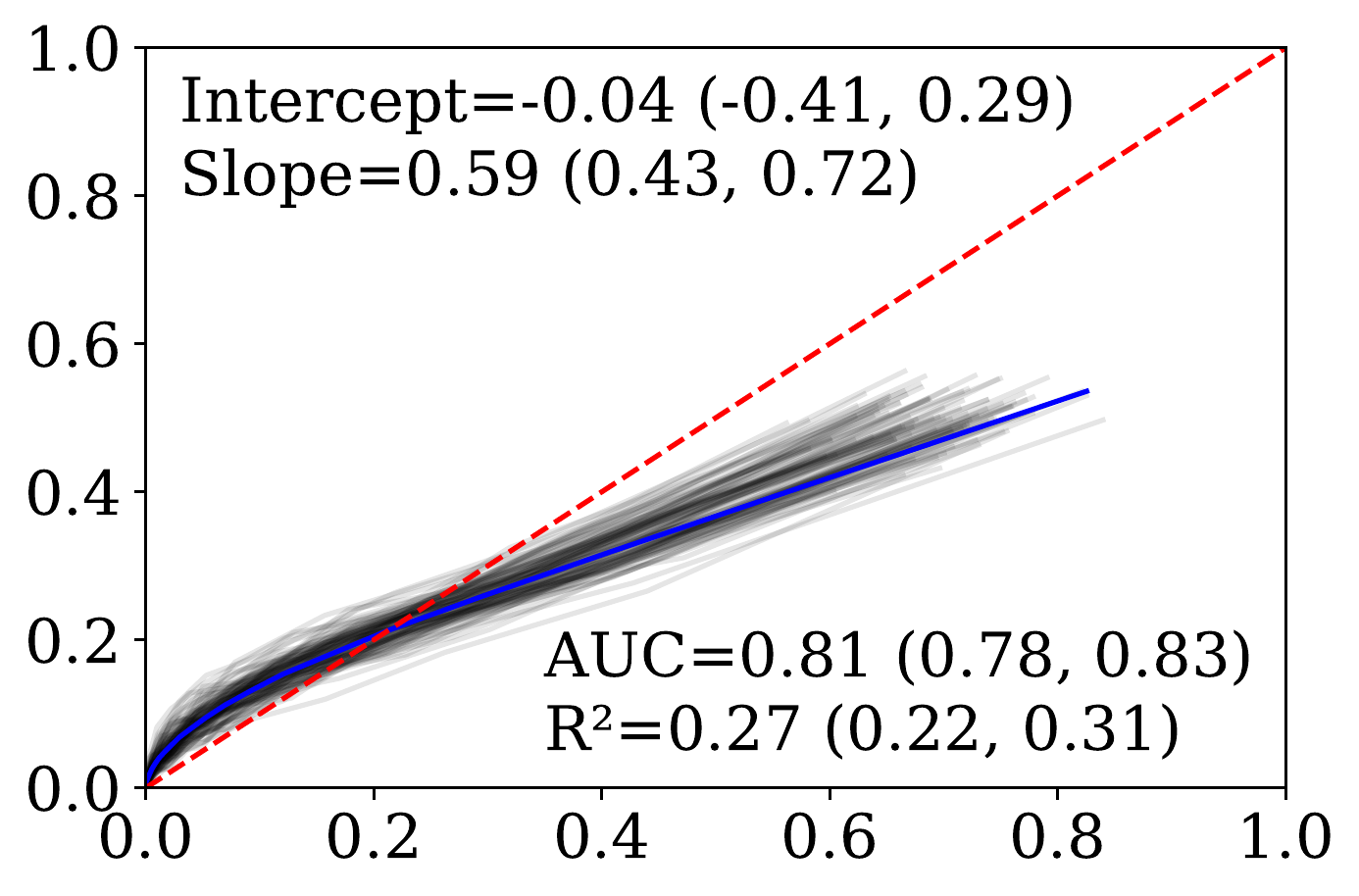} & \includegraphics[width=\plotwidth]{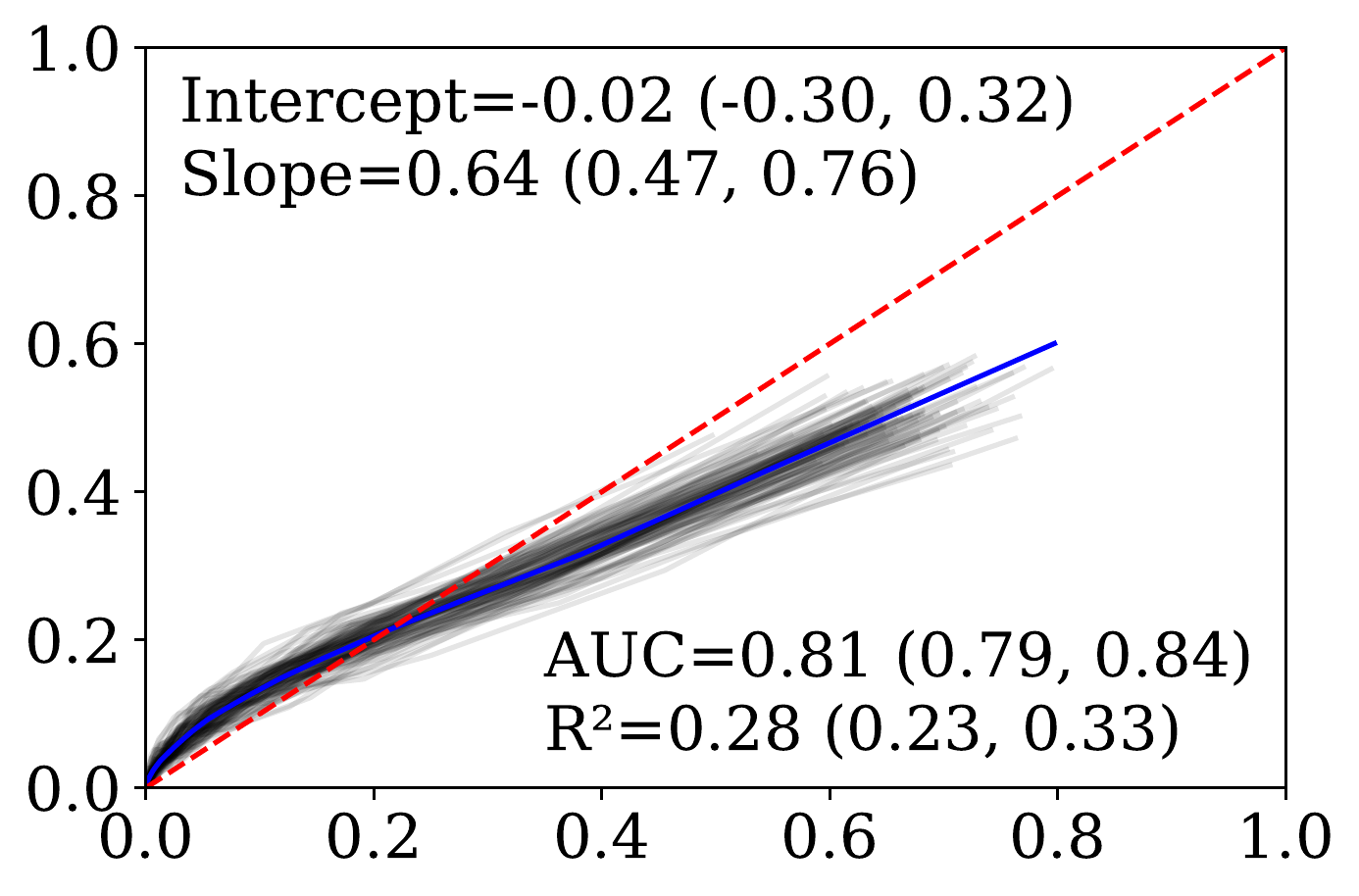}\\
         \centered{Lasso} & \includegraphics[width=\plotwidth]{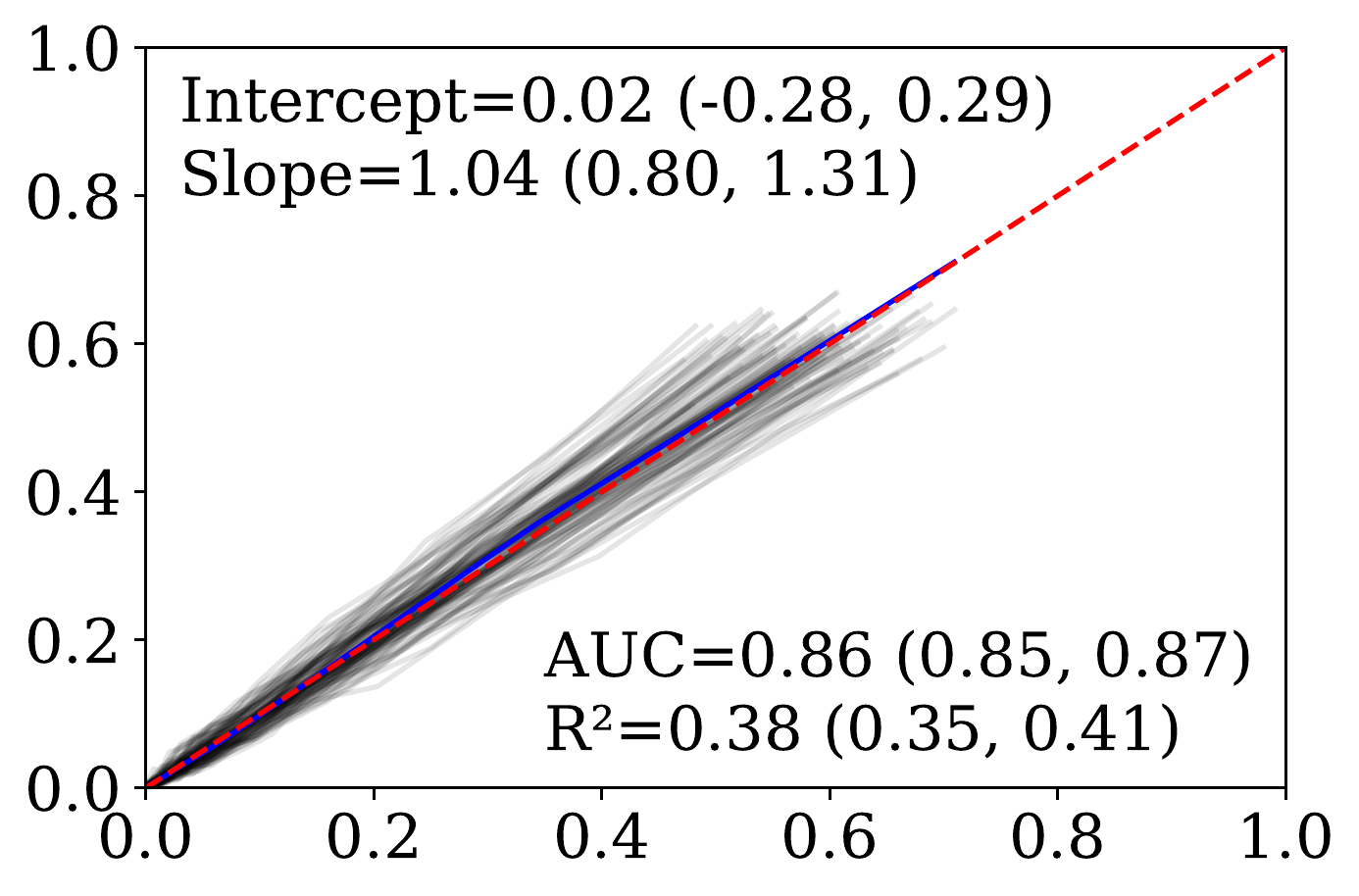} & \includegraphics[width=\plotwidth]{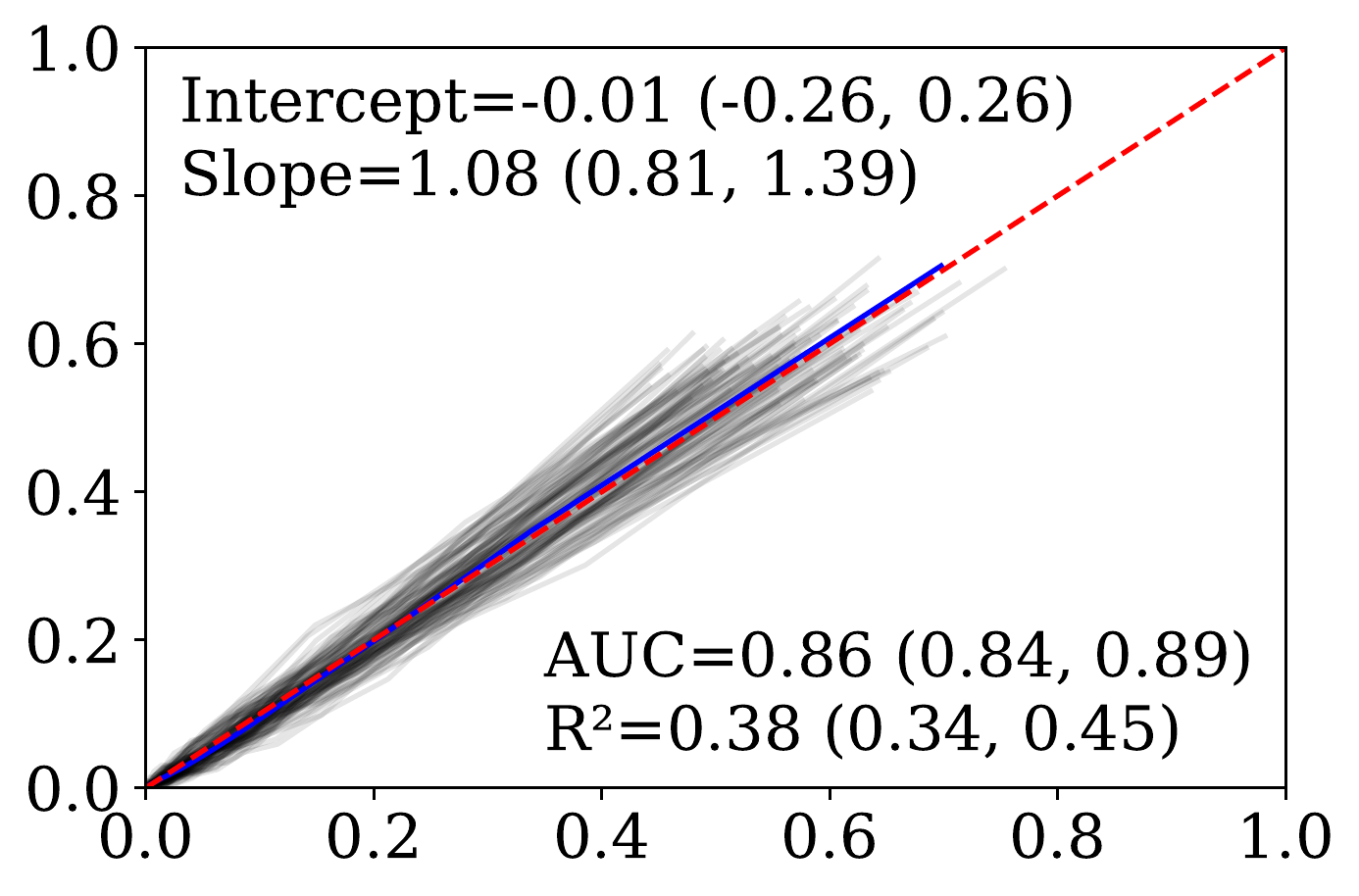} & \includegraphics[width=\plotwidth]{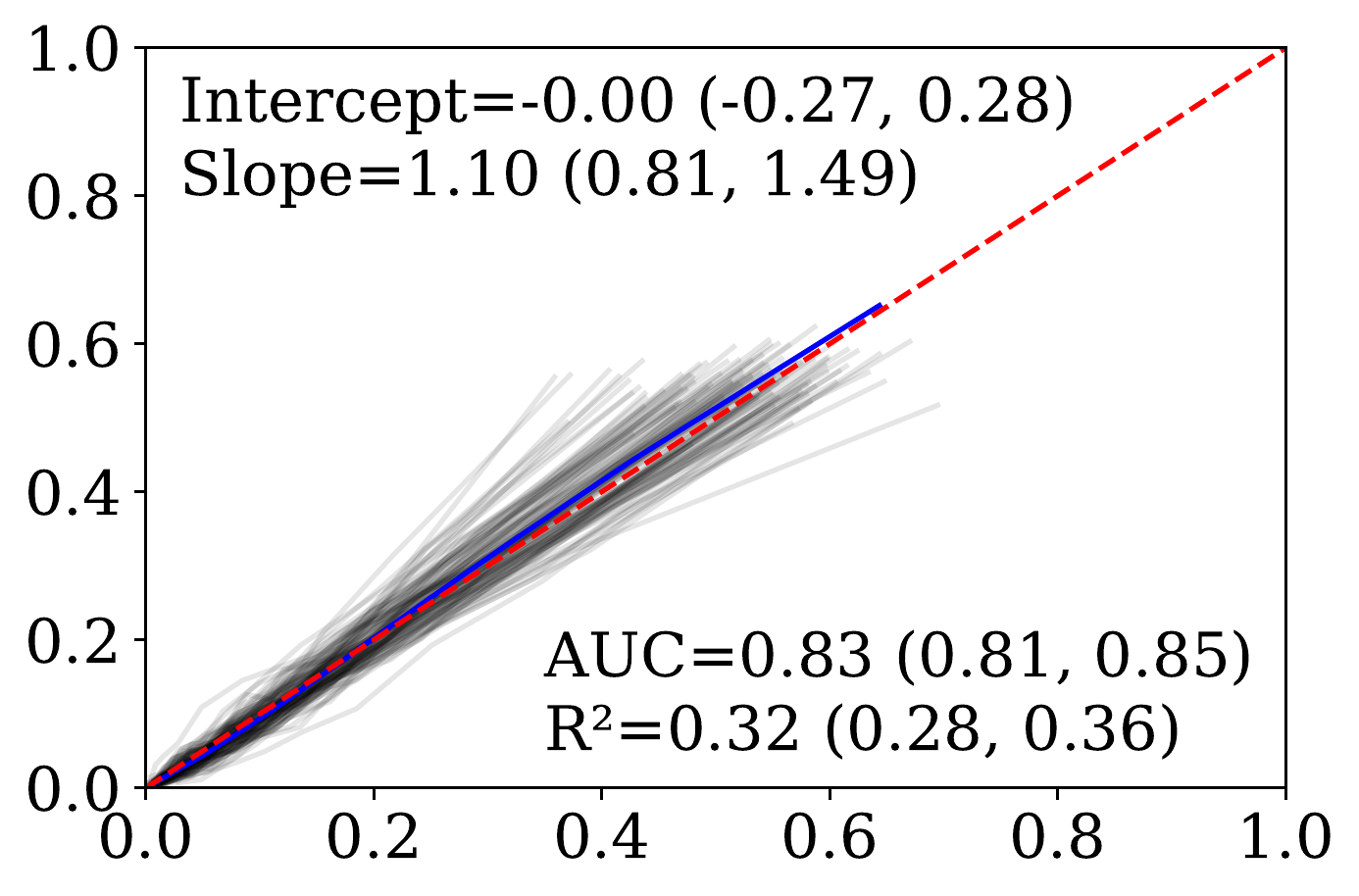} & \includegraphics[width=\plotwidth]{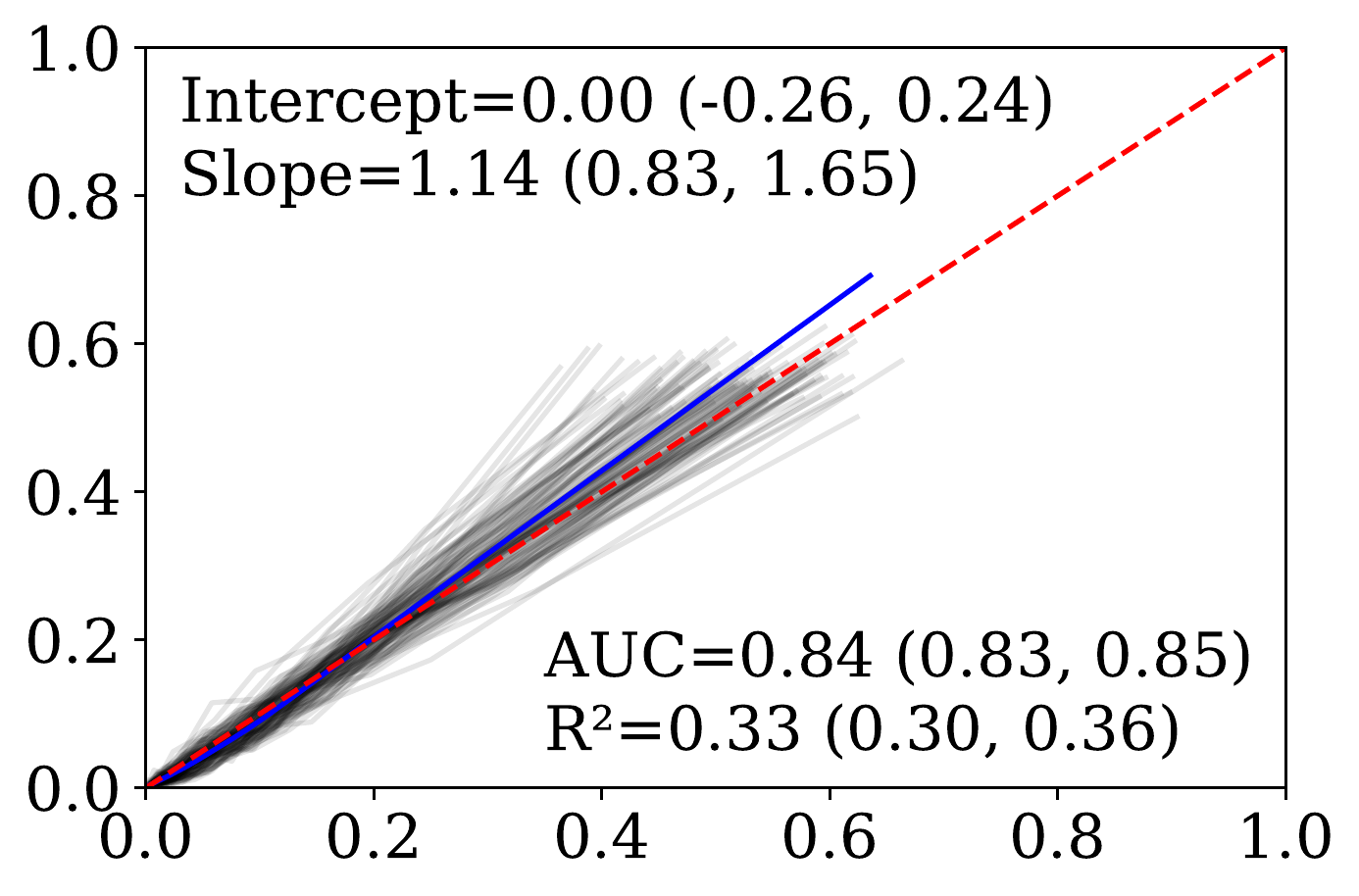}\\
         \centered{Ridge} & \includegraphics[width=\plotwidth]{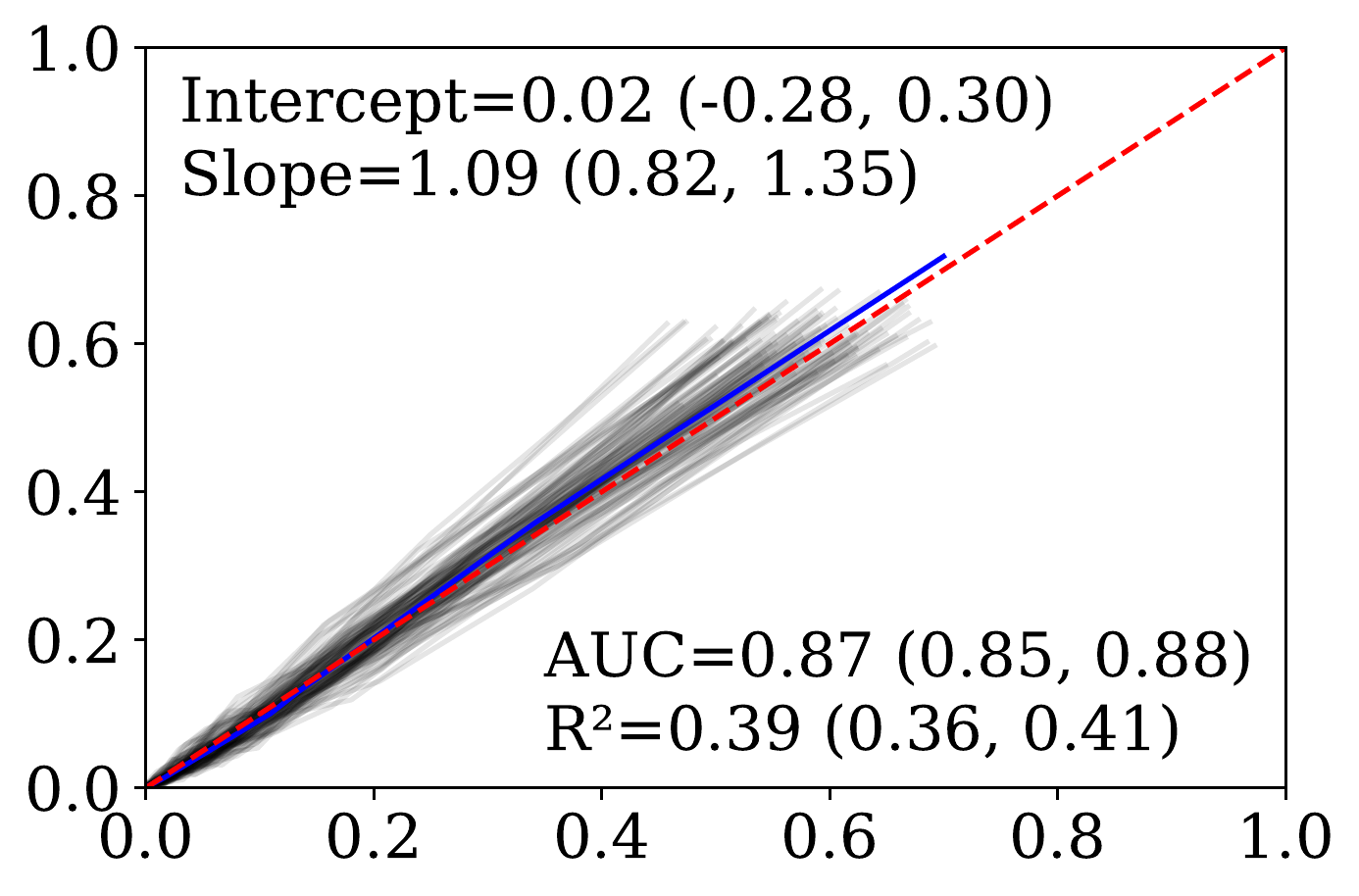} & \includegraphics[width=\plotwidth]{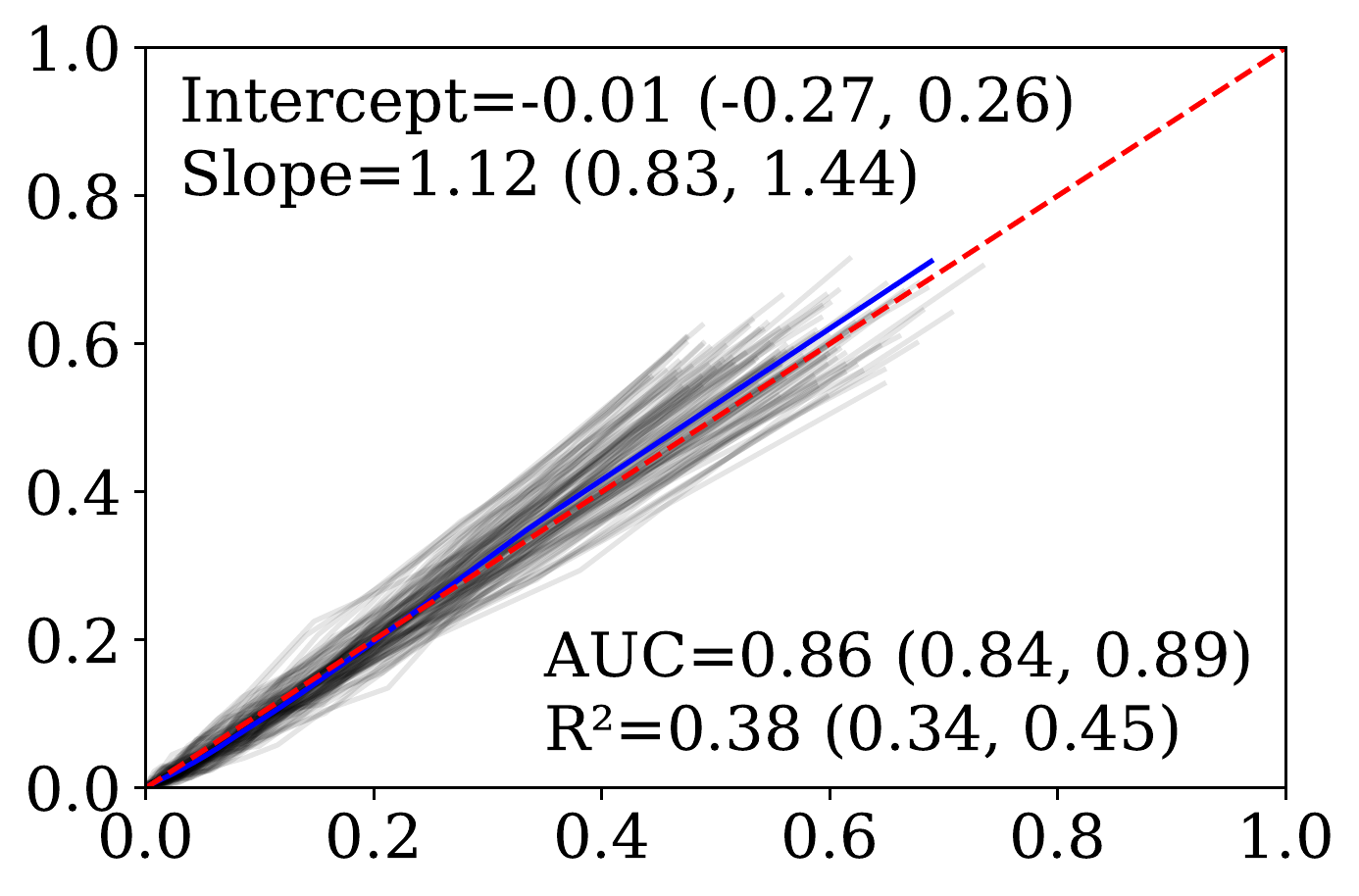} & \includegraphics[width=\plotwidth]{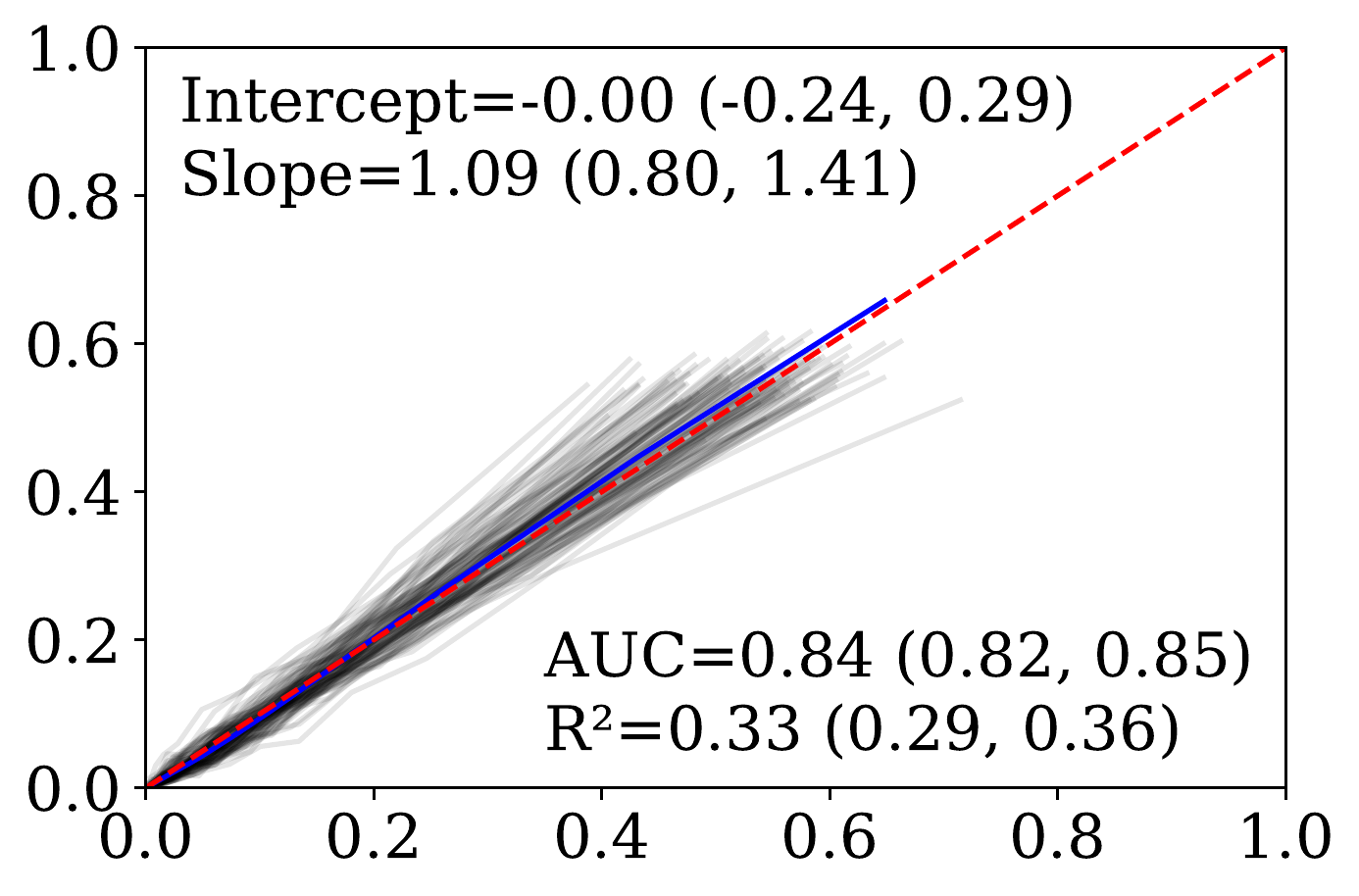} & \includegraphics[width=\plotwidth]{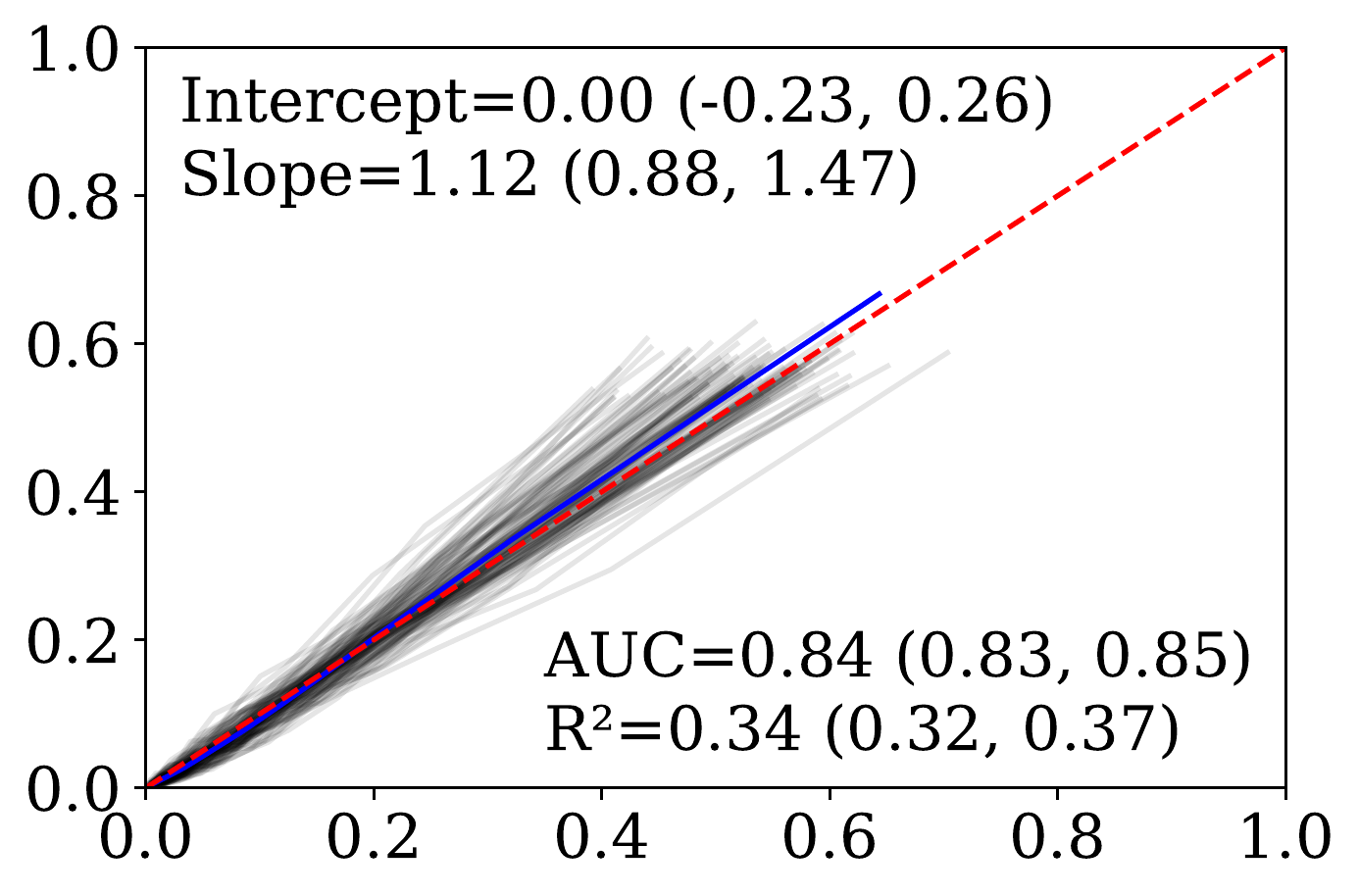}\\
         \centered{ElasticNet} & \includegraphics[width=\plotwidth]{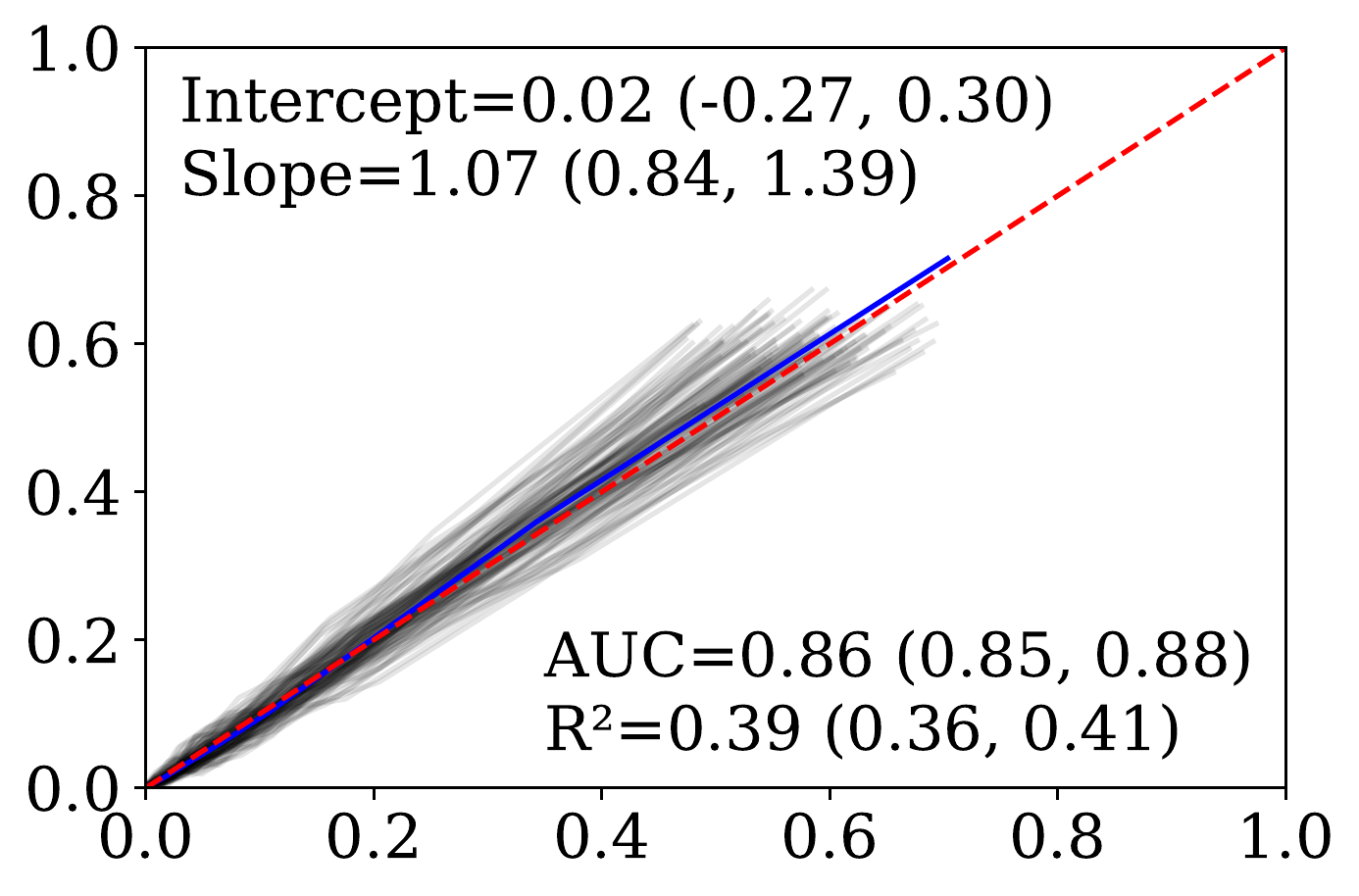} & \includegraphics[width=\plotwidth]{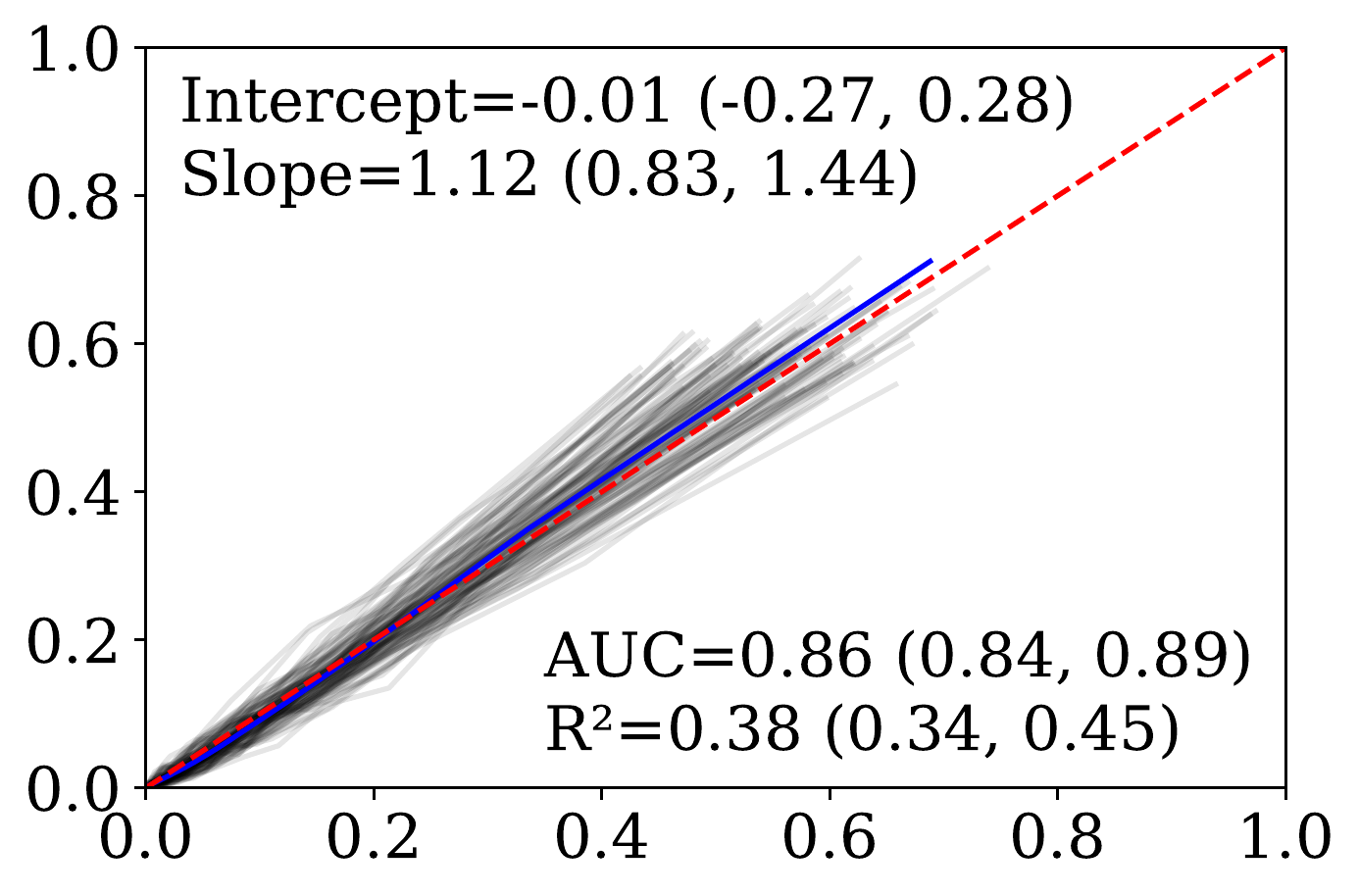} & \includegraphics[width=\plotwidth]{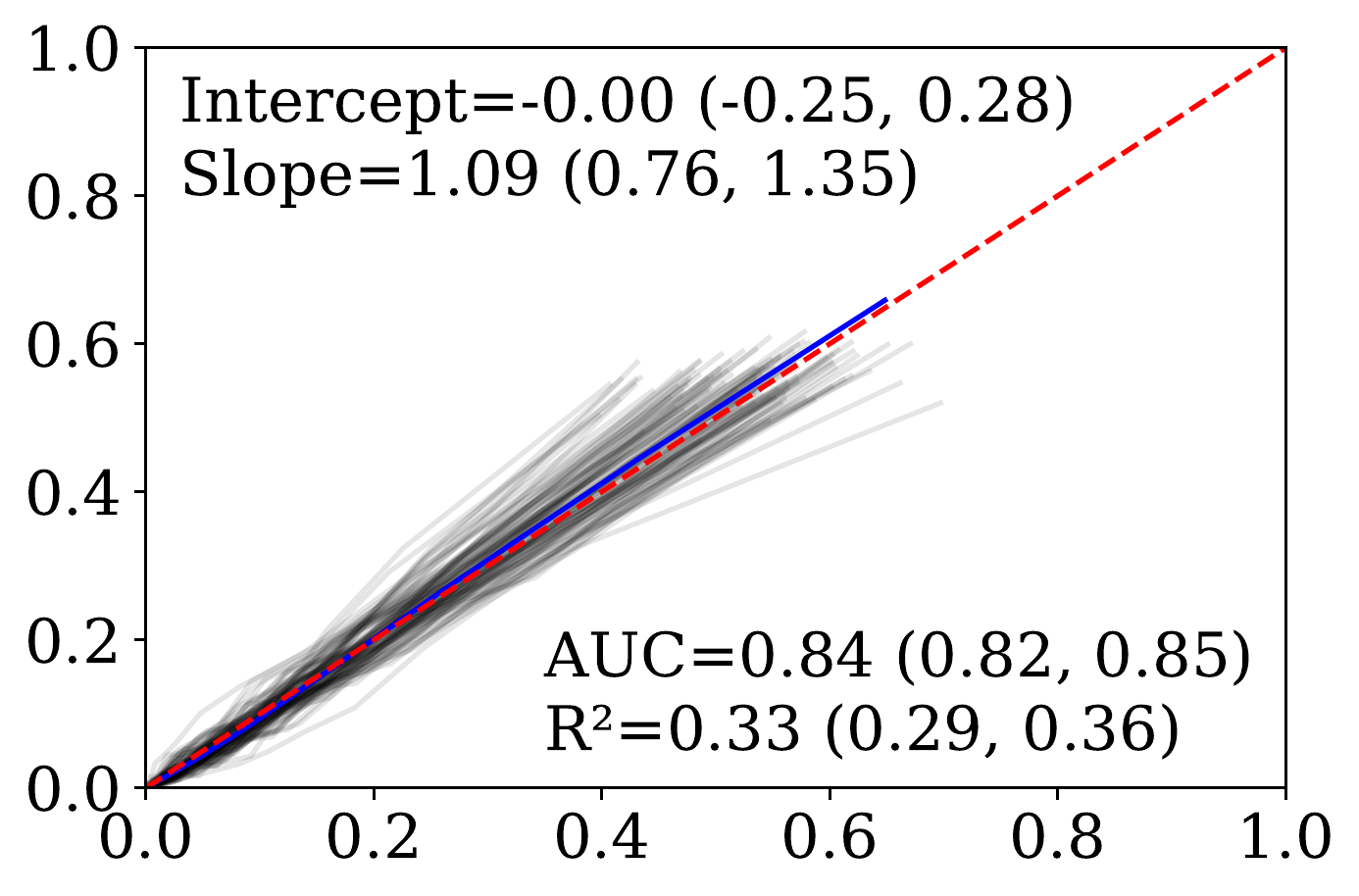} & \includegraphics[width=\plotwidth]{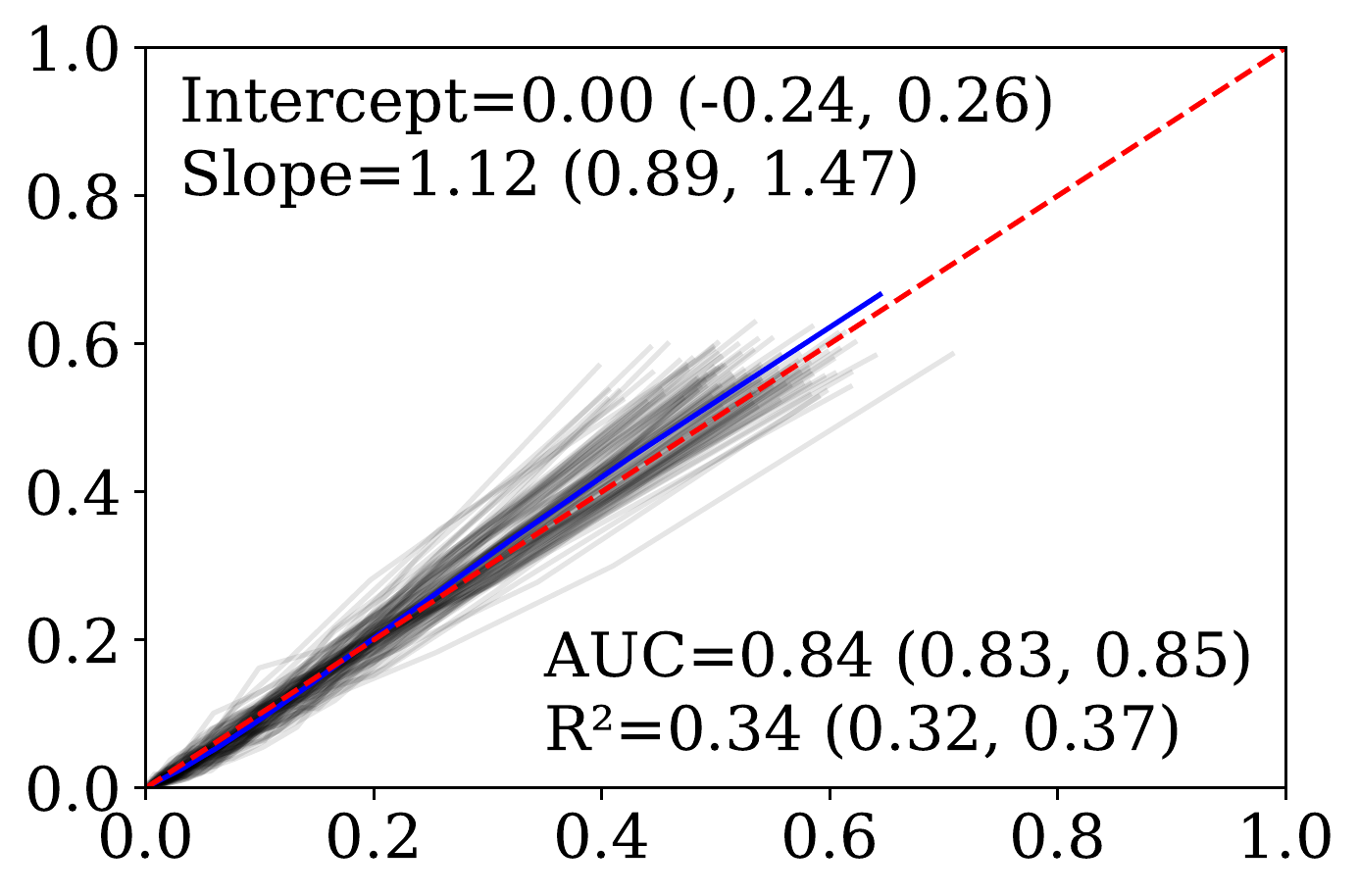}\\
         \centered{PCLR} & \includegraphics[width=\plotwidth]{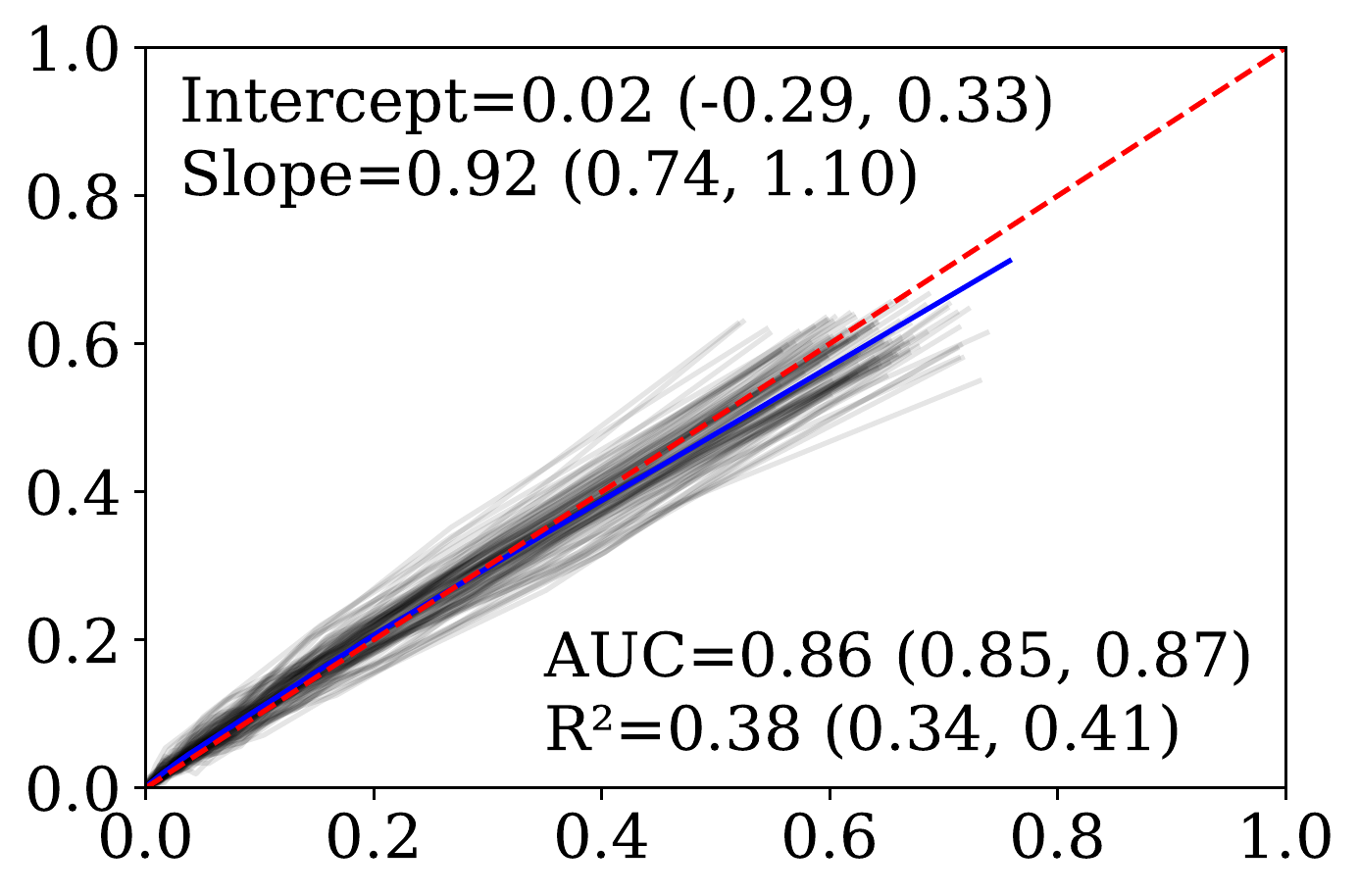} & \includegraphics[width=\plotwidth]{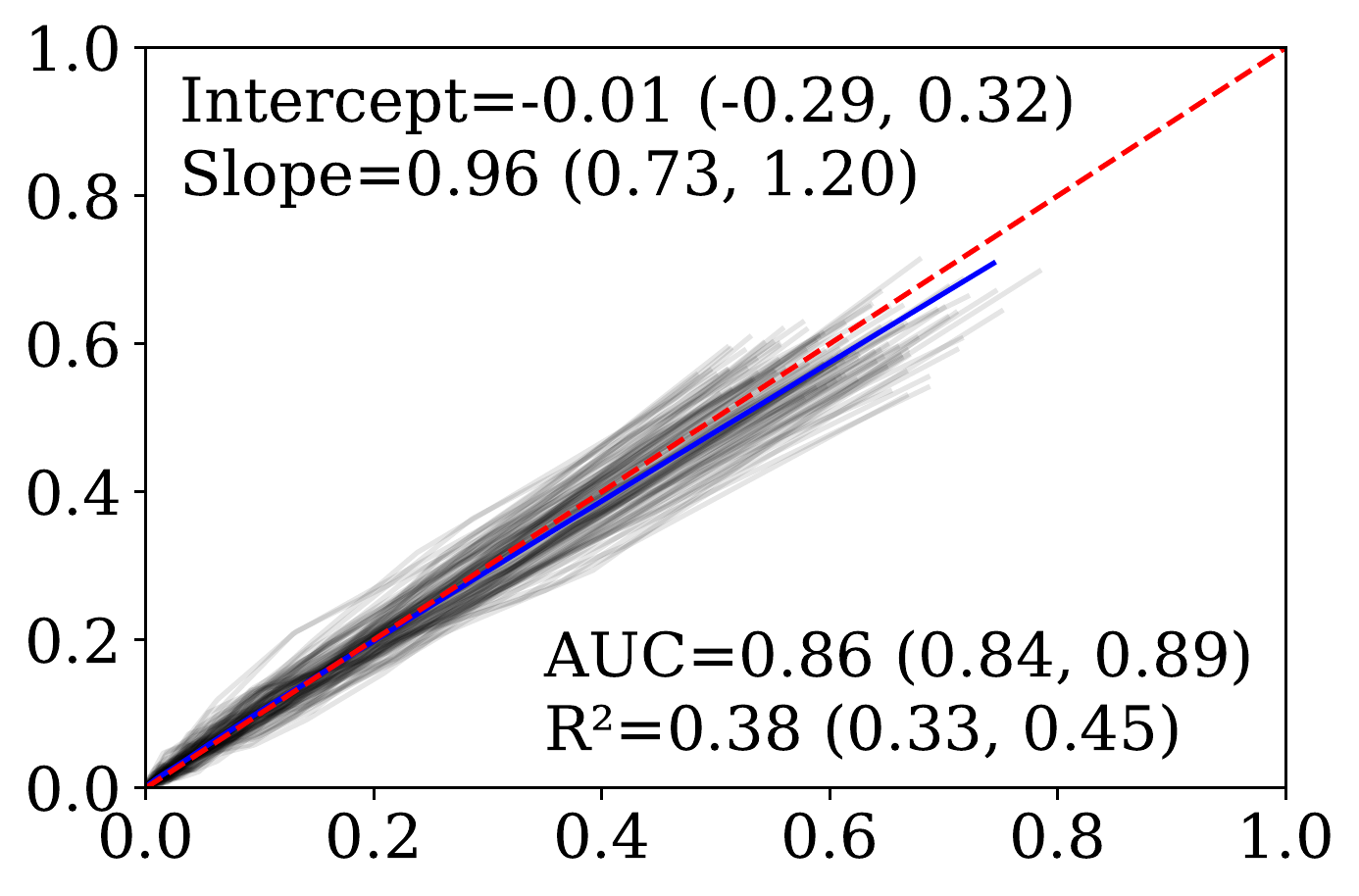} & \includegraphics[width=\plotwidth]{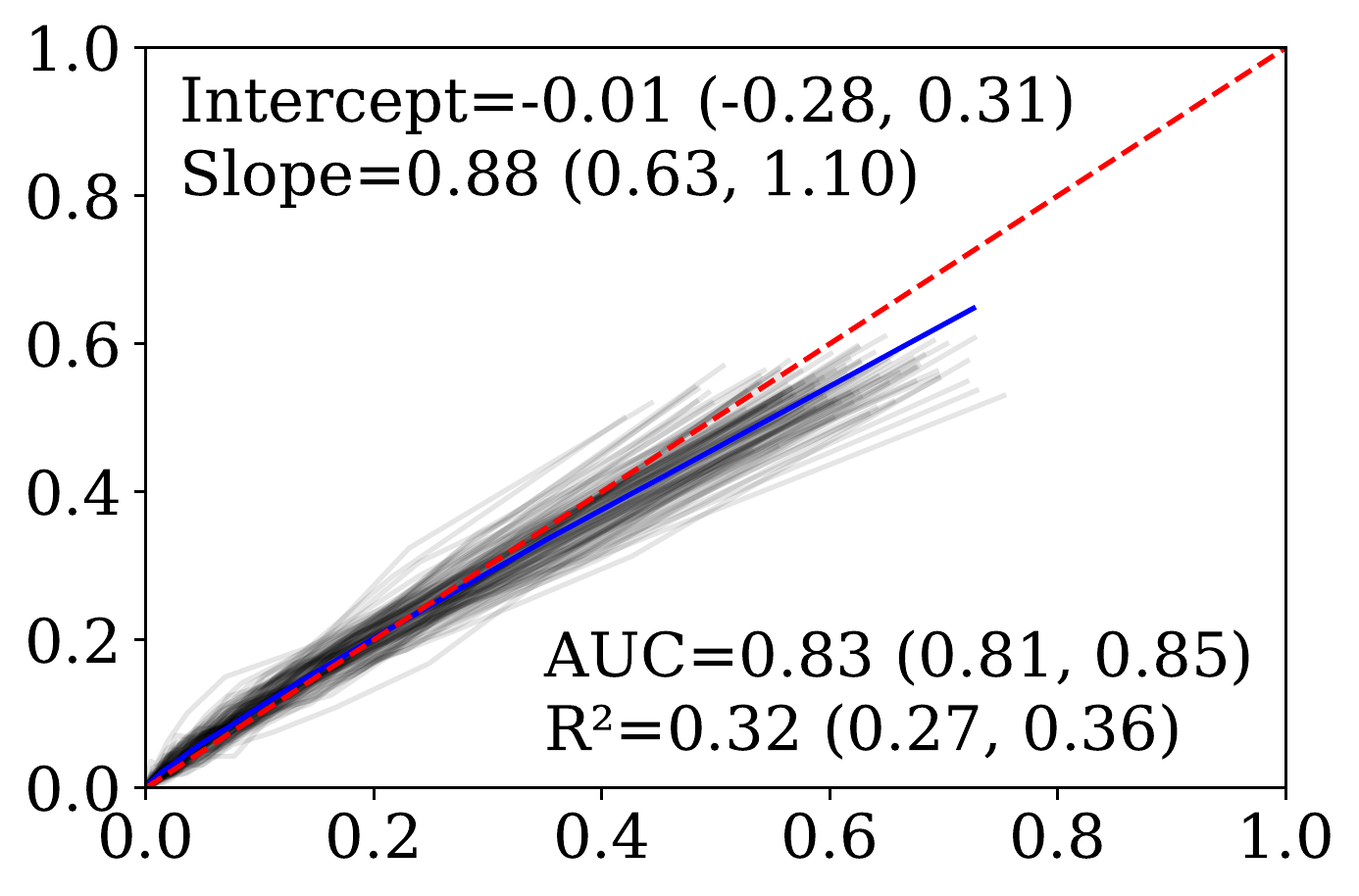} & \includegraphics[width=\plotwidth]{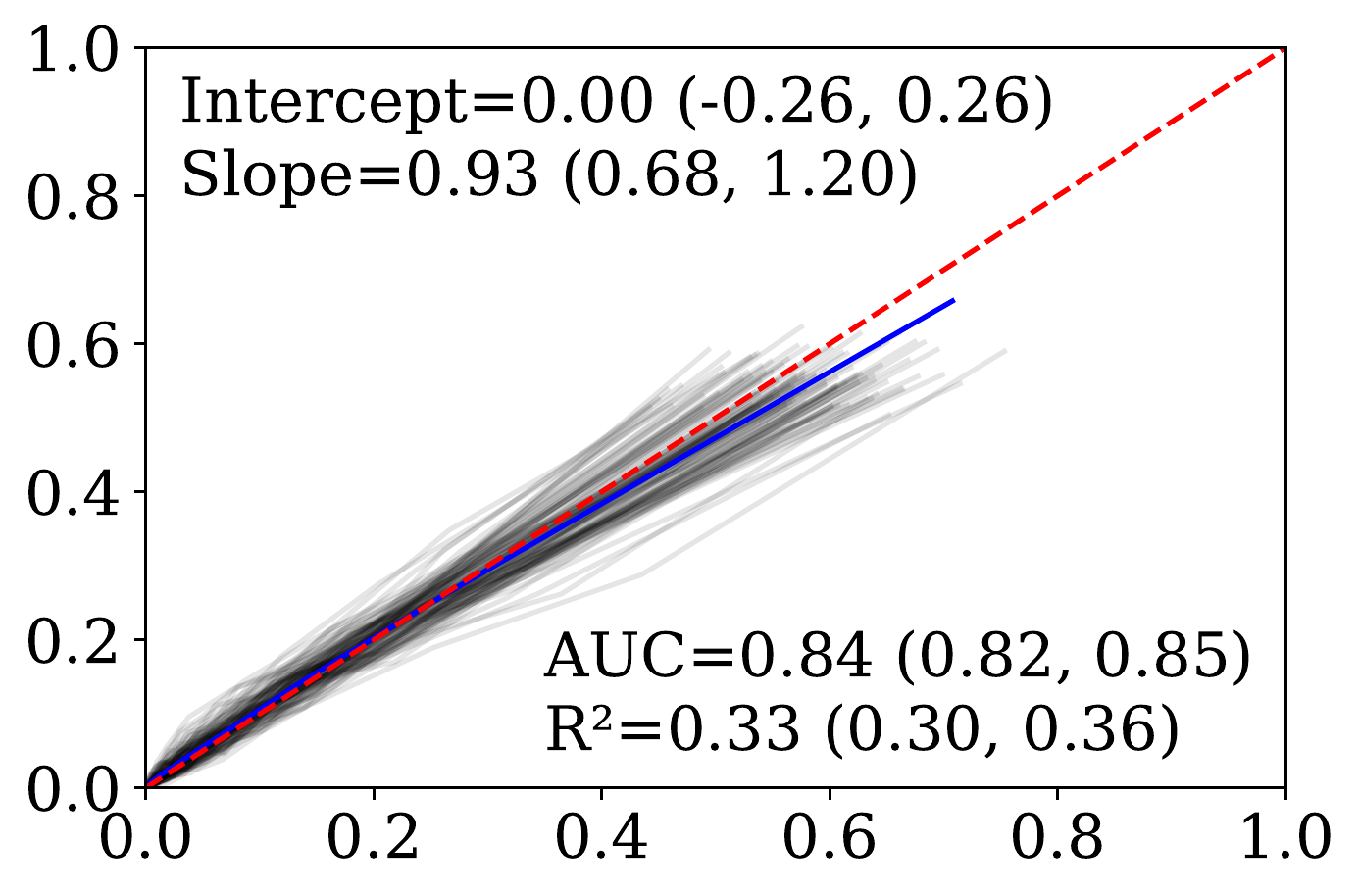}\\
         \centered{LAELR} & \includegraphics[width=\plotwidth]{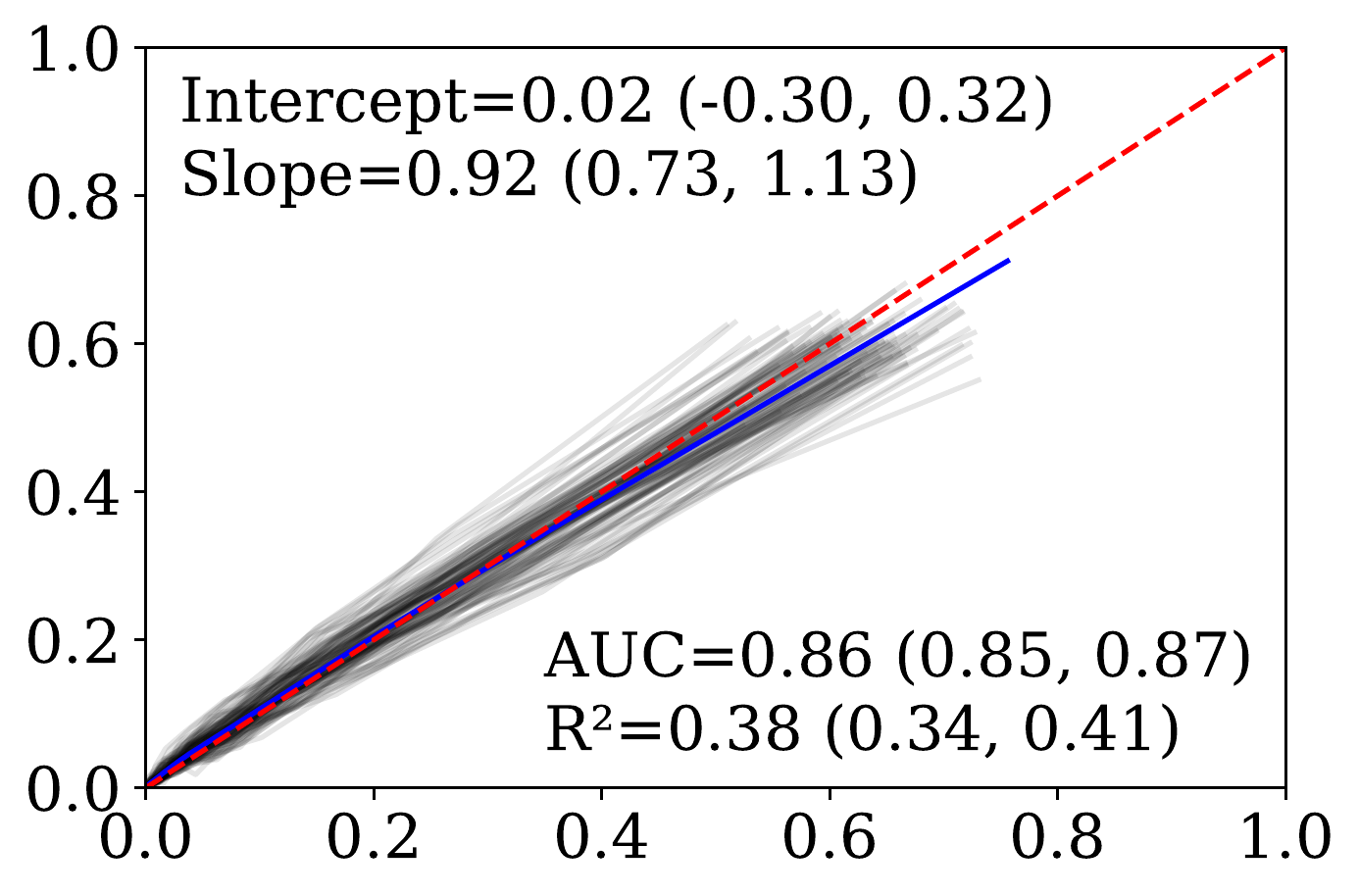} & \includegraphics[width=\plotwidth]{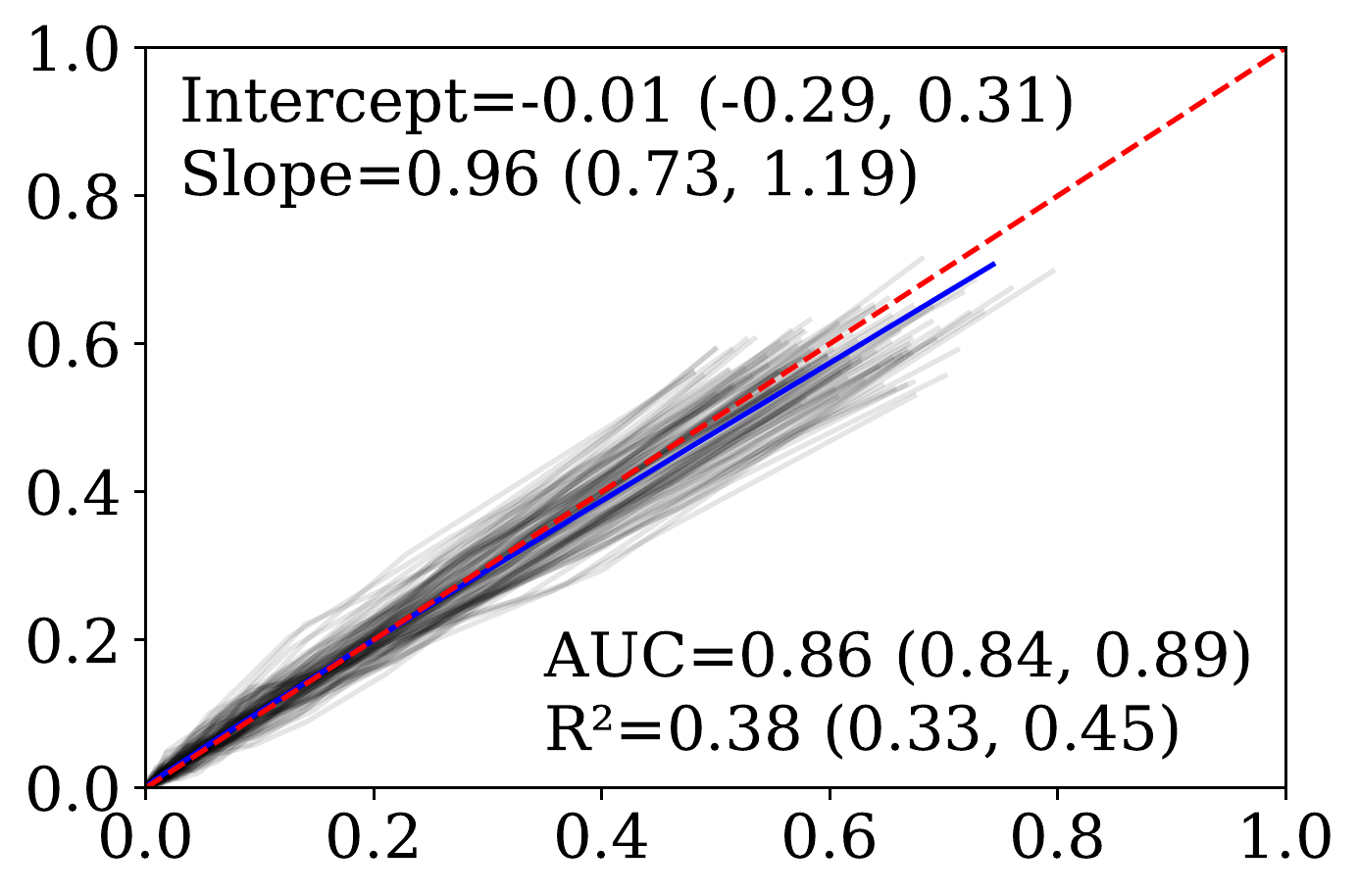} & \includegraphics[width=\plotwidth]{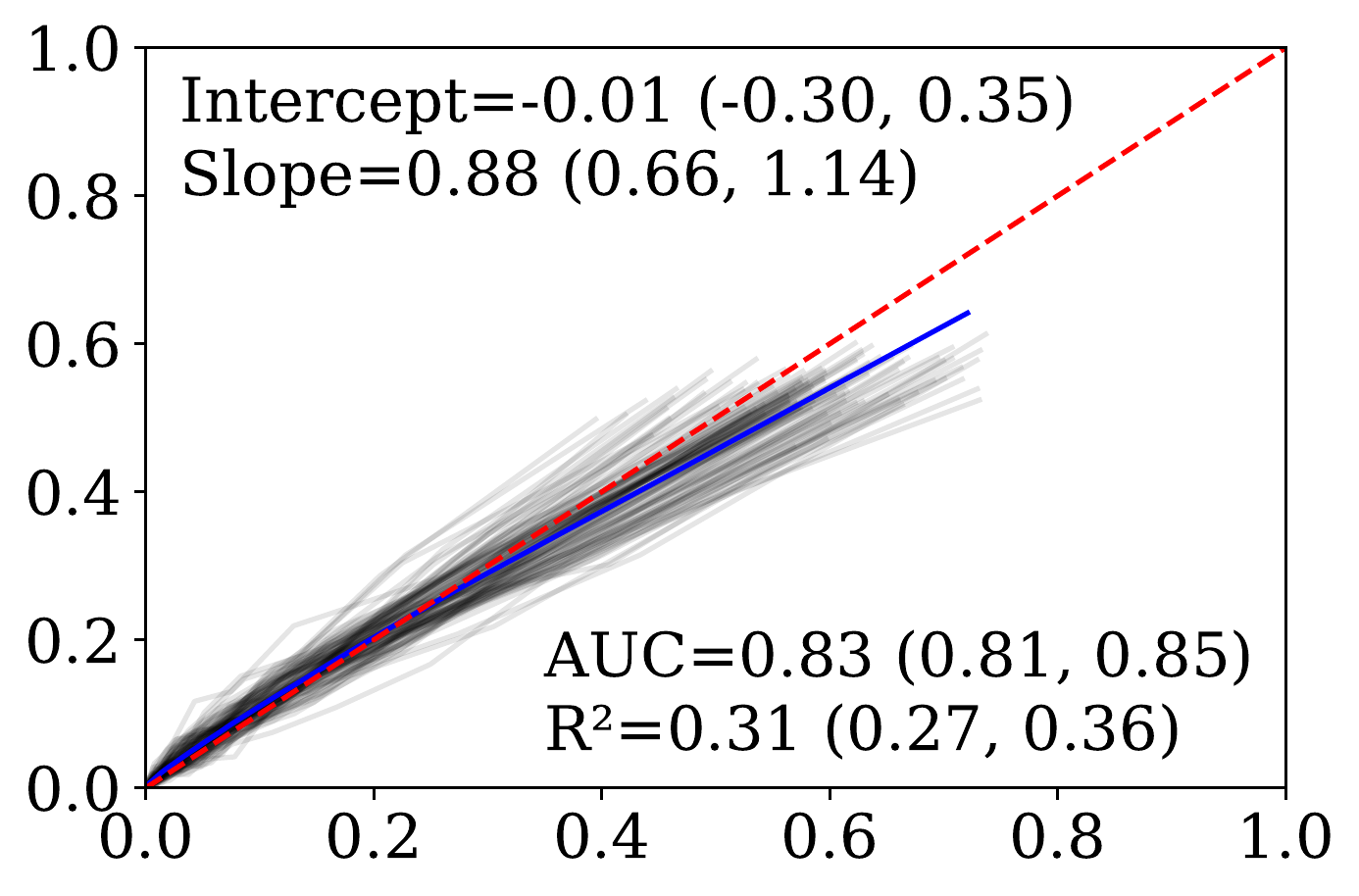} & \includegraphics[width=\plotwidth]{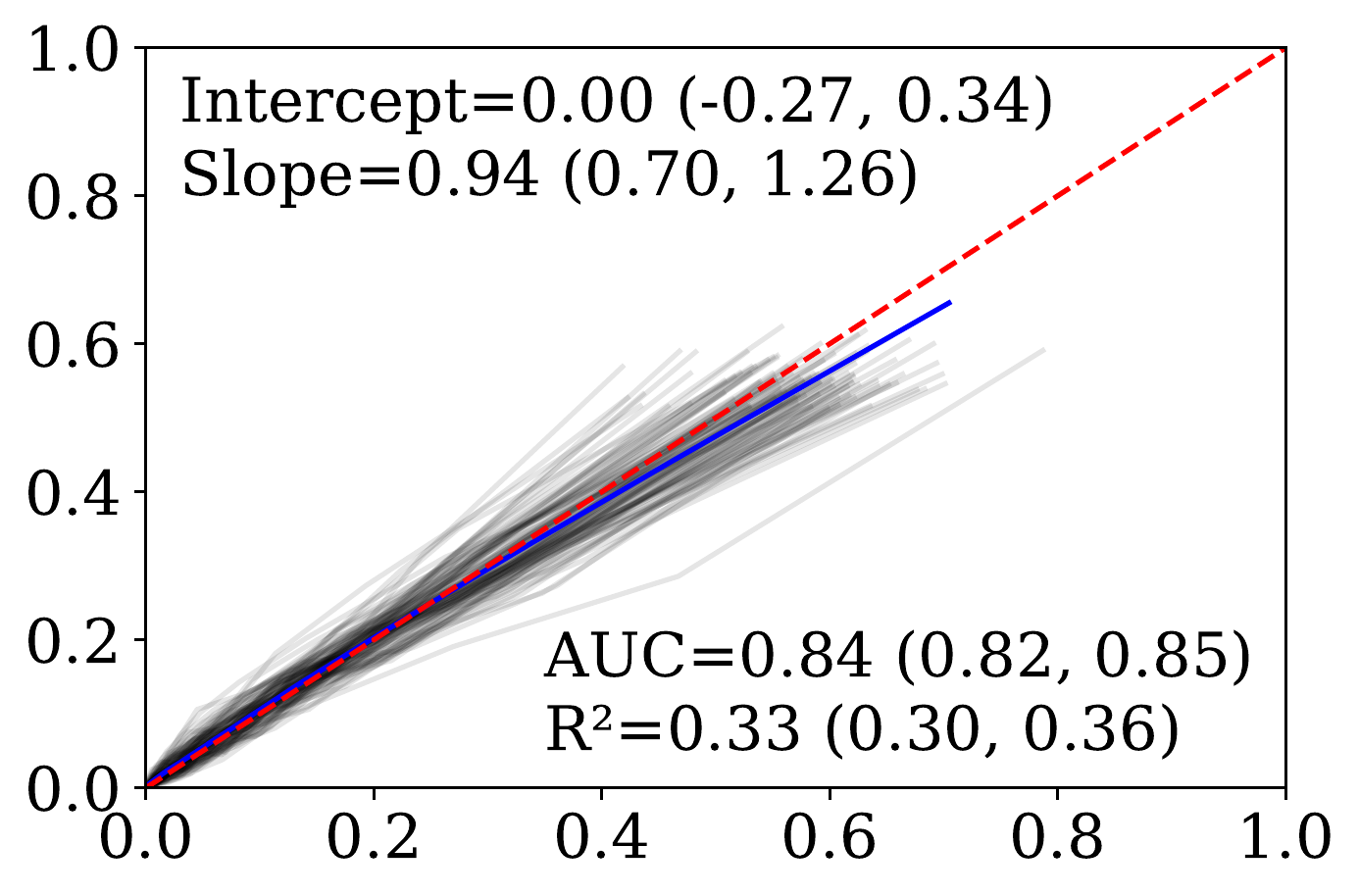}\\
         \centered{Dropout} & \includegraphics[width=\plotwidth]{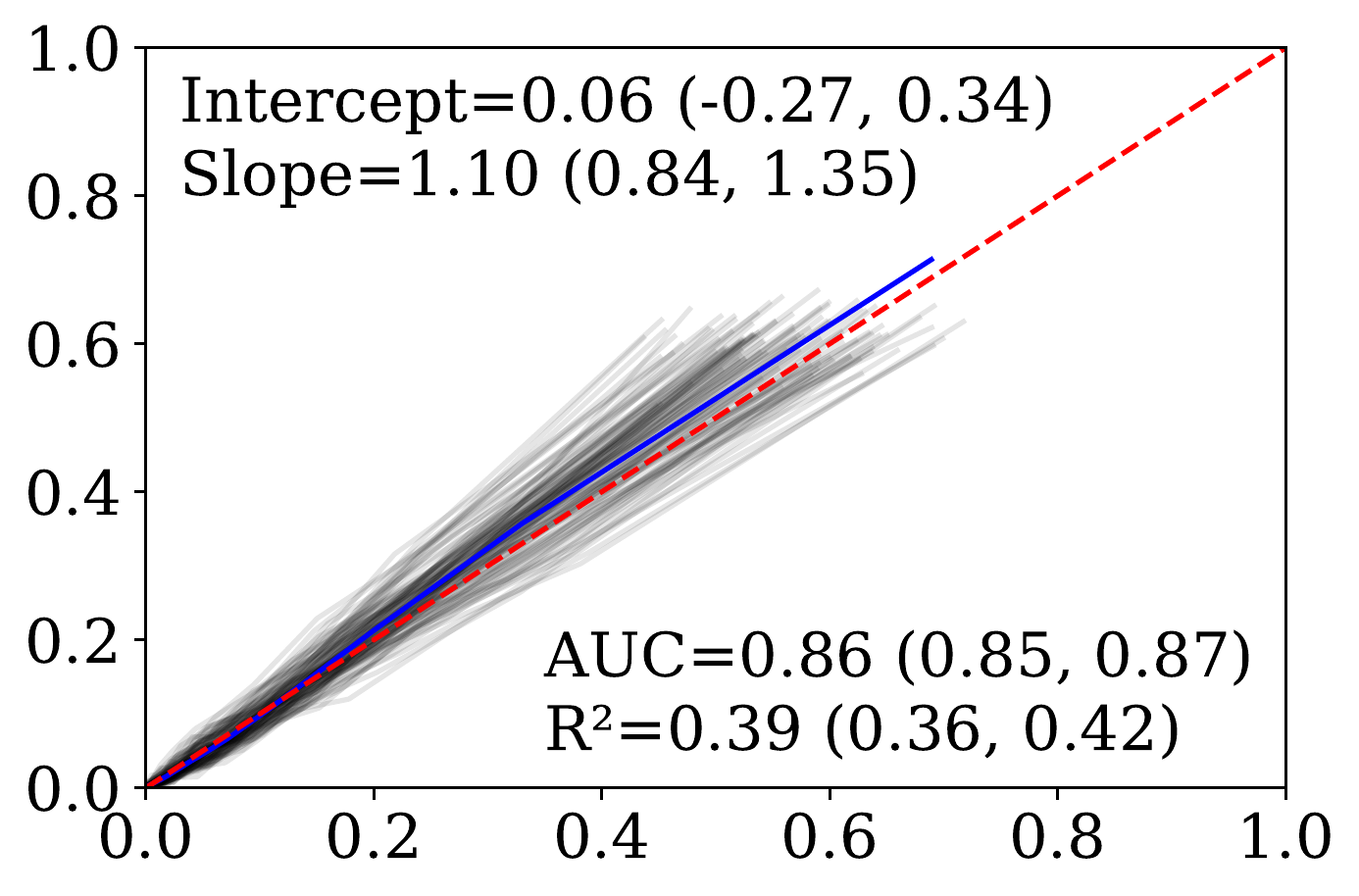} & \includegraphics[width=\plotwidth]{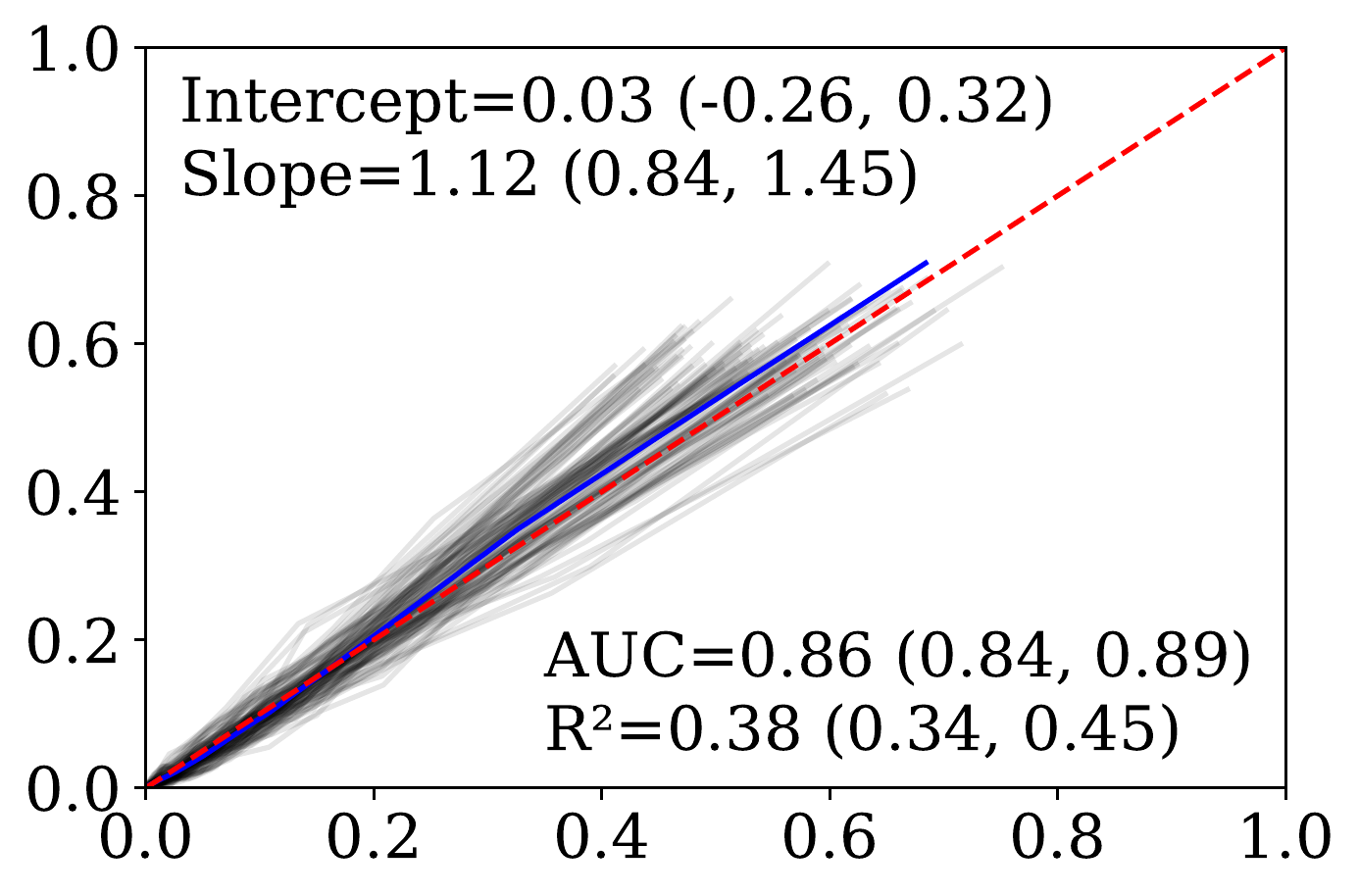} & \includegraphics[width=\plotwidth]{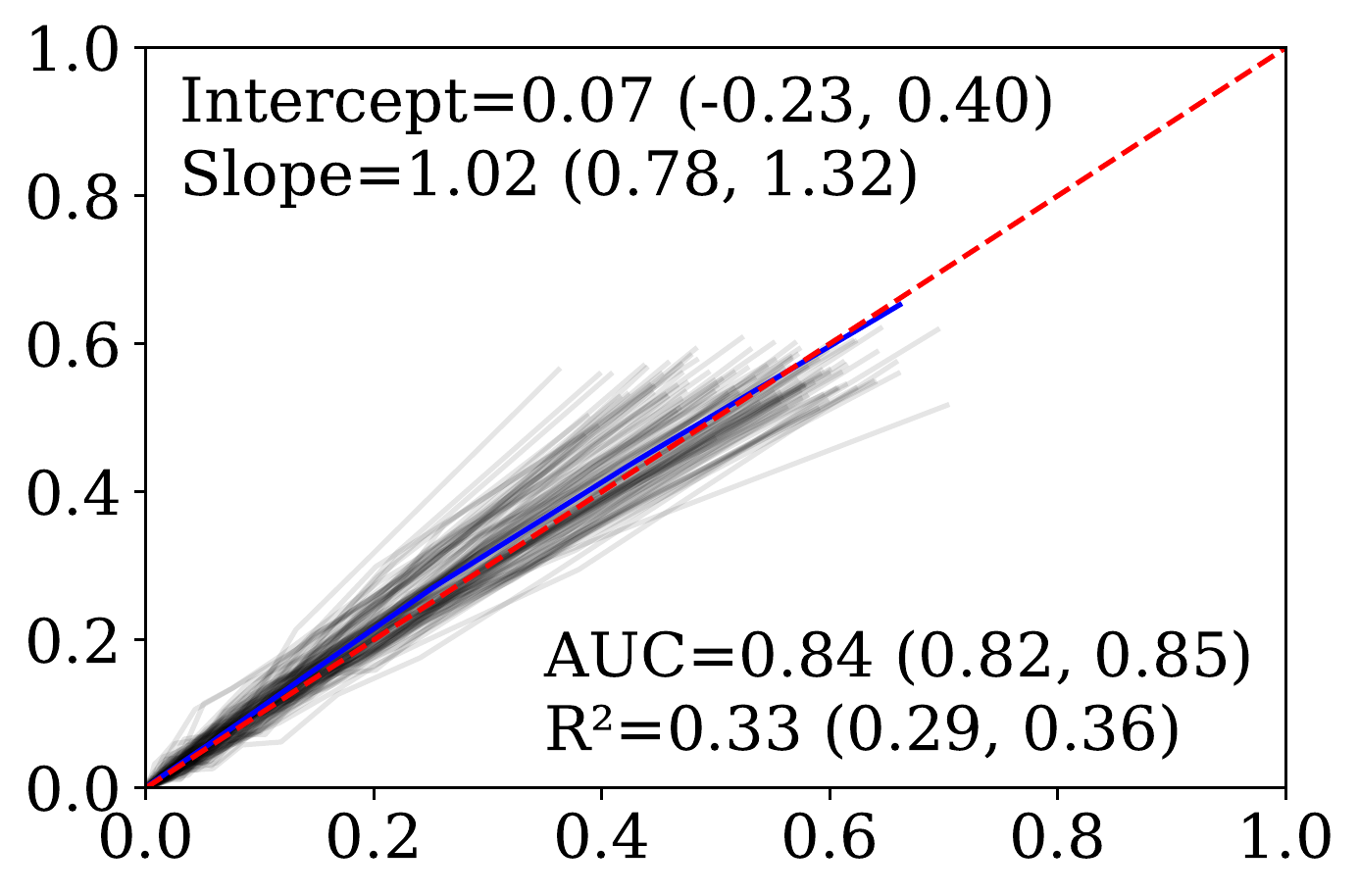} & \includegraphics[width=\plotwidth]{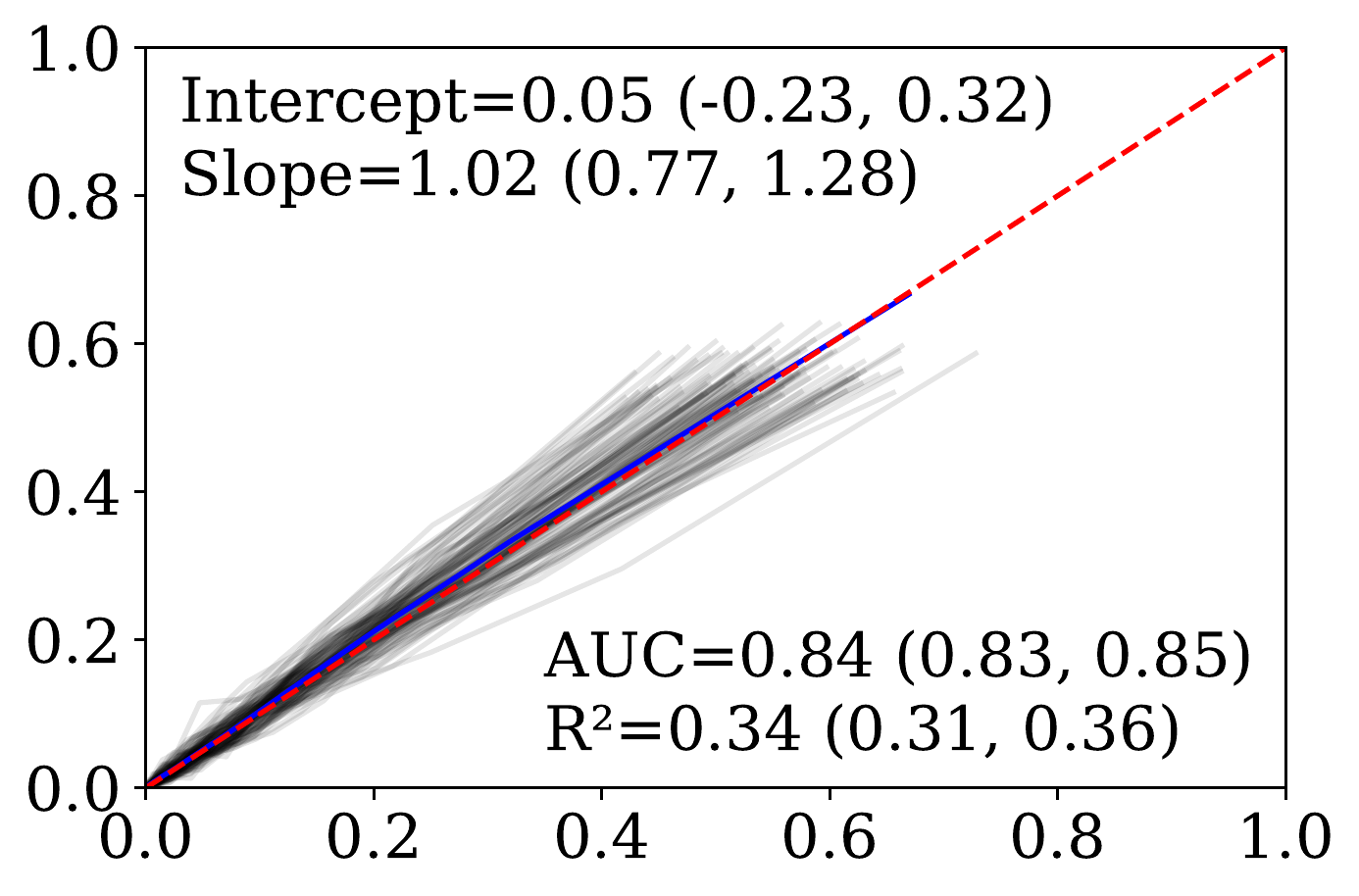}\\
         \centered{\LRnn} & \includegraphics[width=\plotwidth]{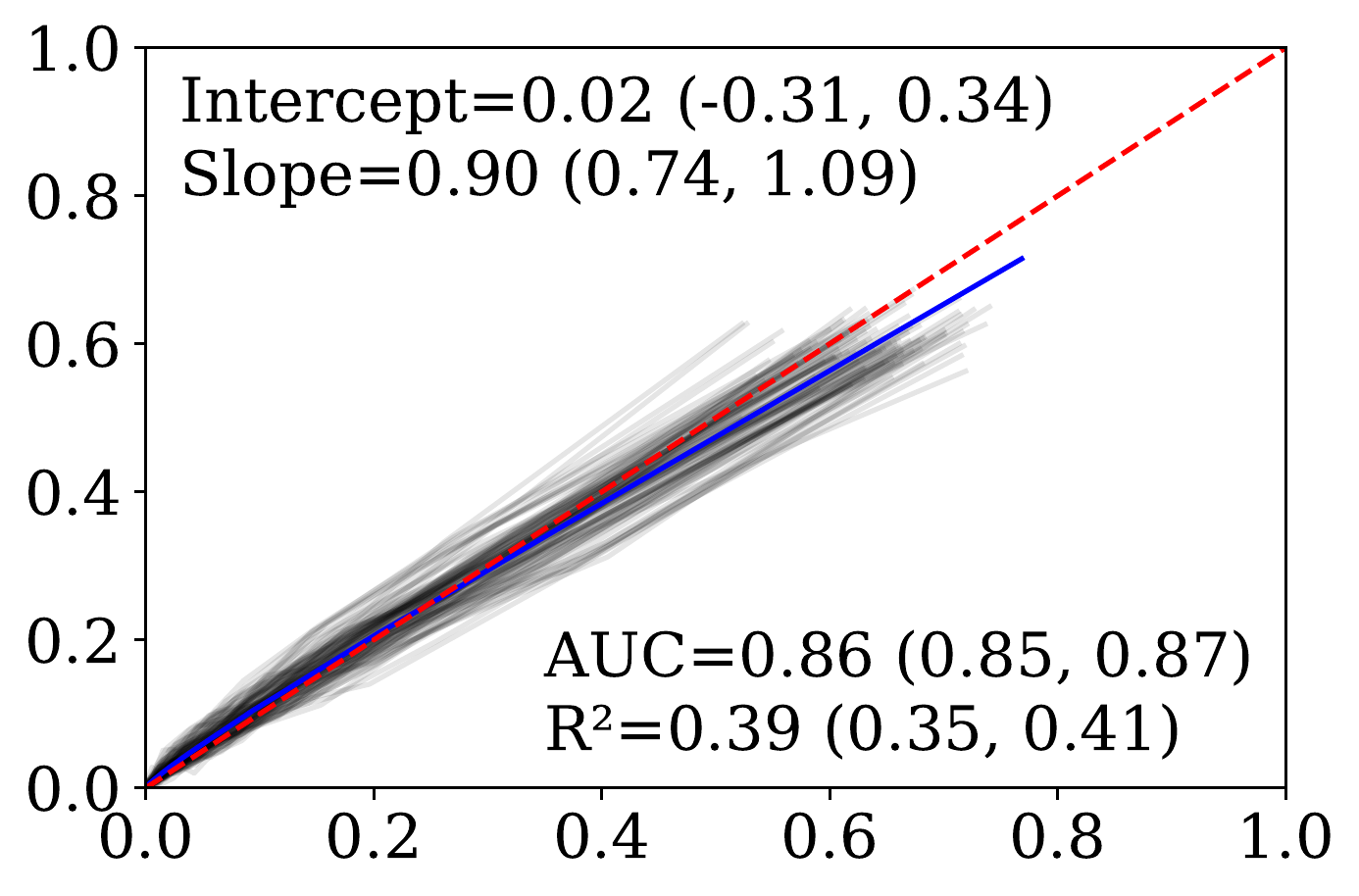} & \includegraphics[width=\plotwidth]{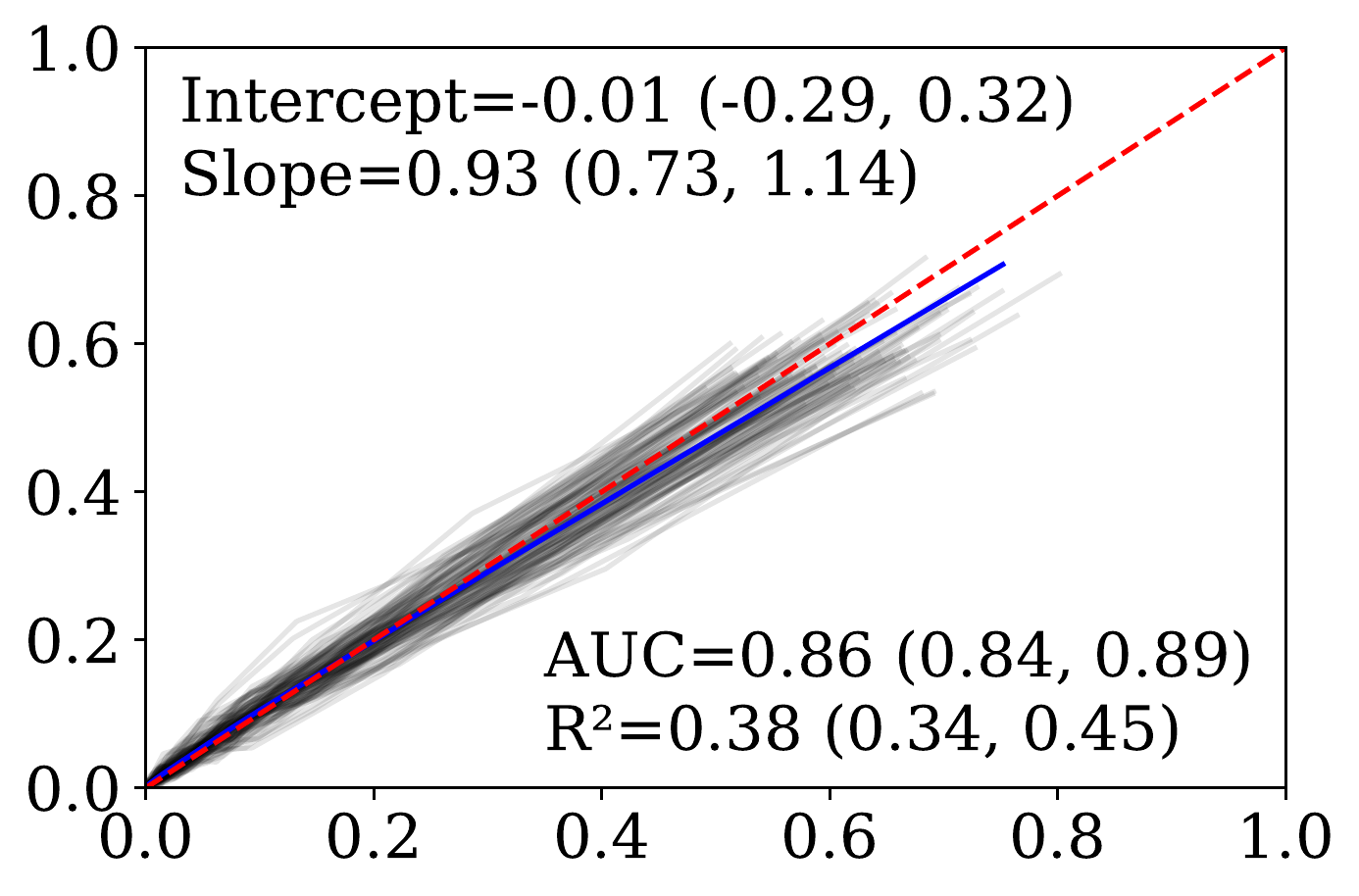} & \includegraphics[width=\plotwidth]{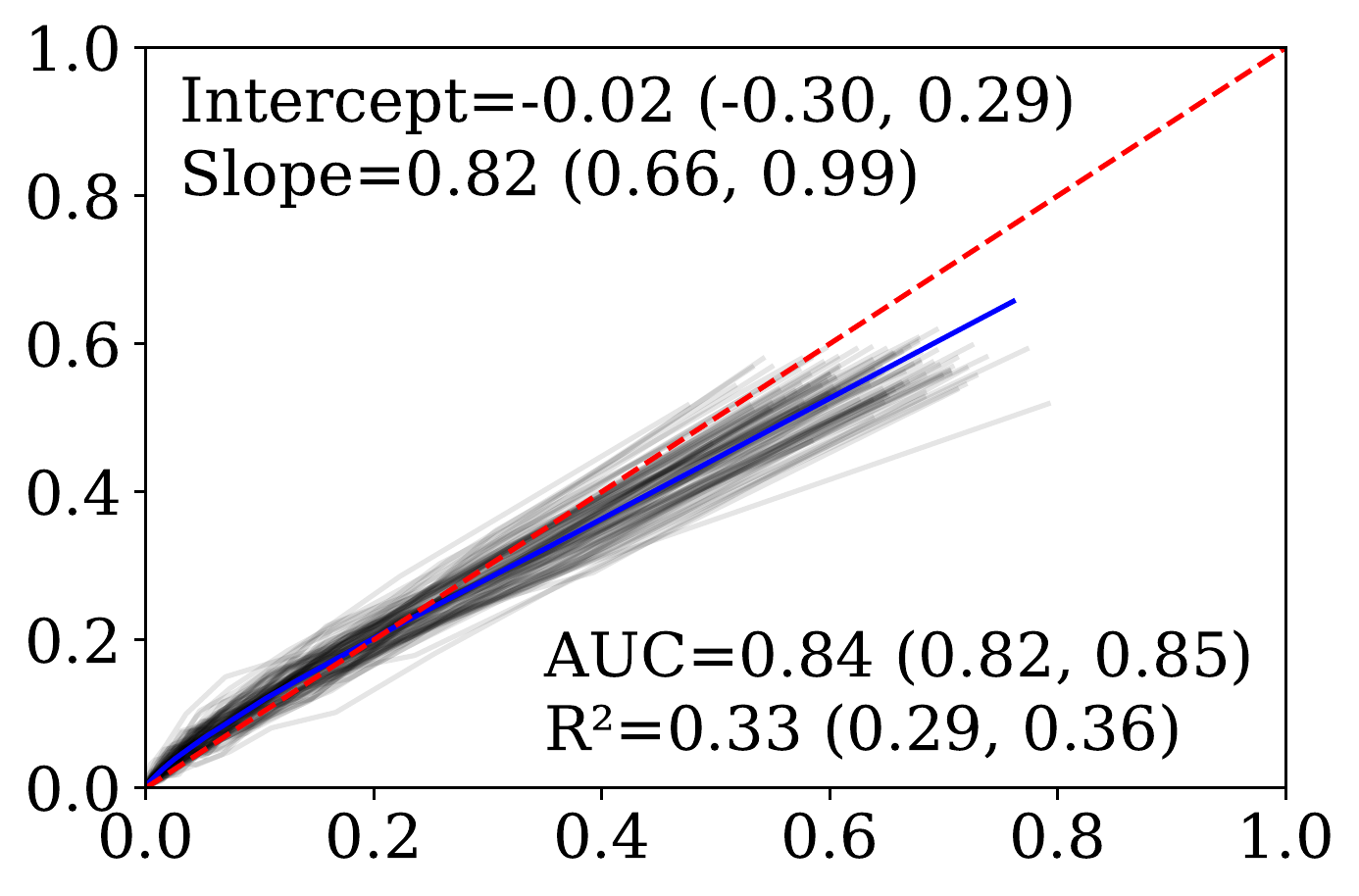} & \includegraphics[width=\plotwidth]{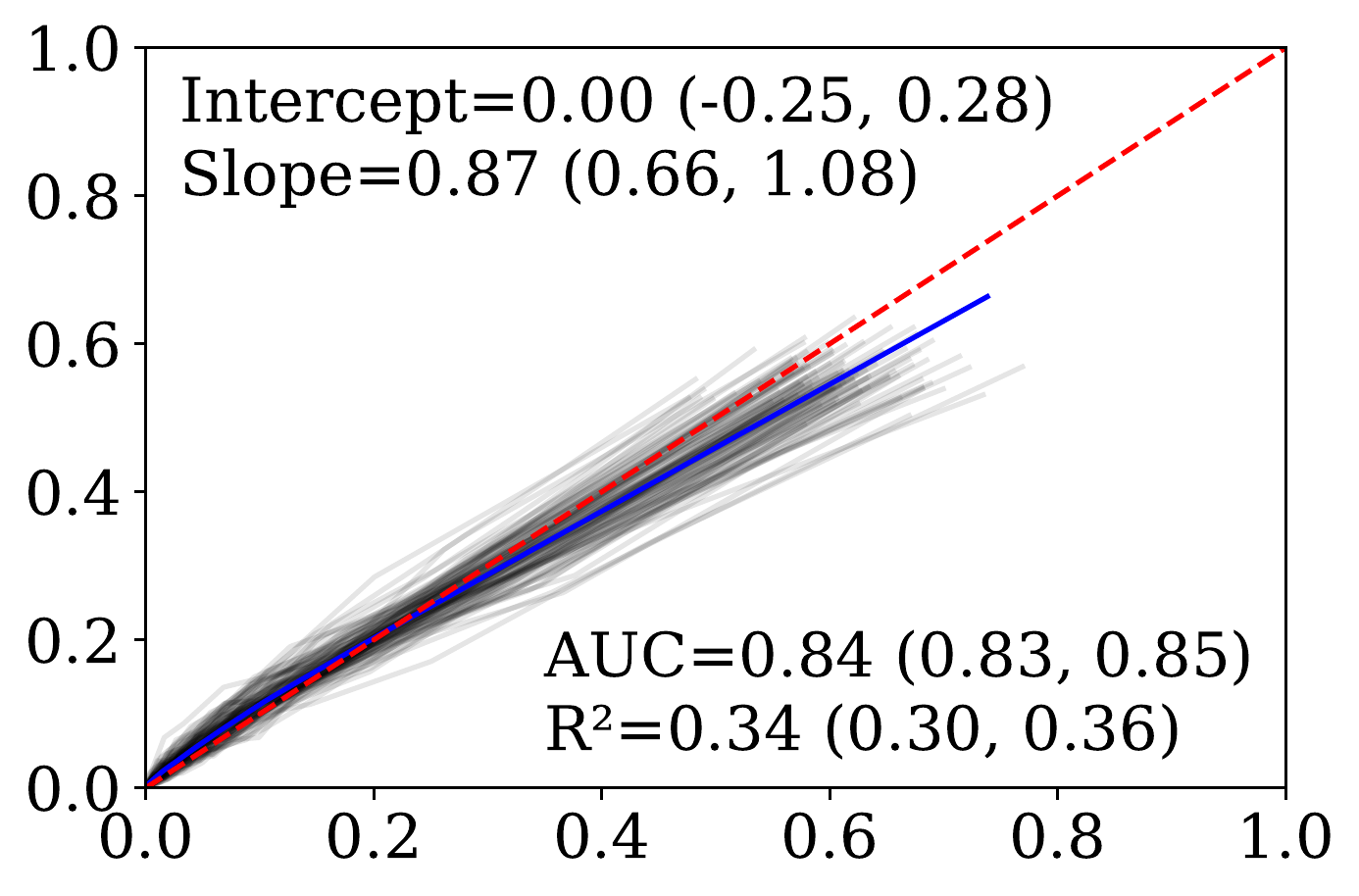}\\
    \end{tabular}}
    \caption{\label{fig:calibration_fans_dysphagia}Predictive performance results for the \textbf{dysphagia simulations}. Lowess-smoothed calibration curves per simulation are plotted in grey. The calibration curve over all repetitions is shown in blue. Perfect calibration, the diagonal, is dashed in red.}
\end{figure}

%% file: sections/BOXPLOTS-CD.tex
\newcommand\boxplotwidthcd{0.46\textwidth}

\begin{figure}
    \centering
    \begin{tabular}{cc}
    \includegraphics[width=\boxplotwidthcd]{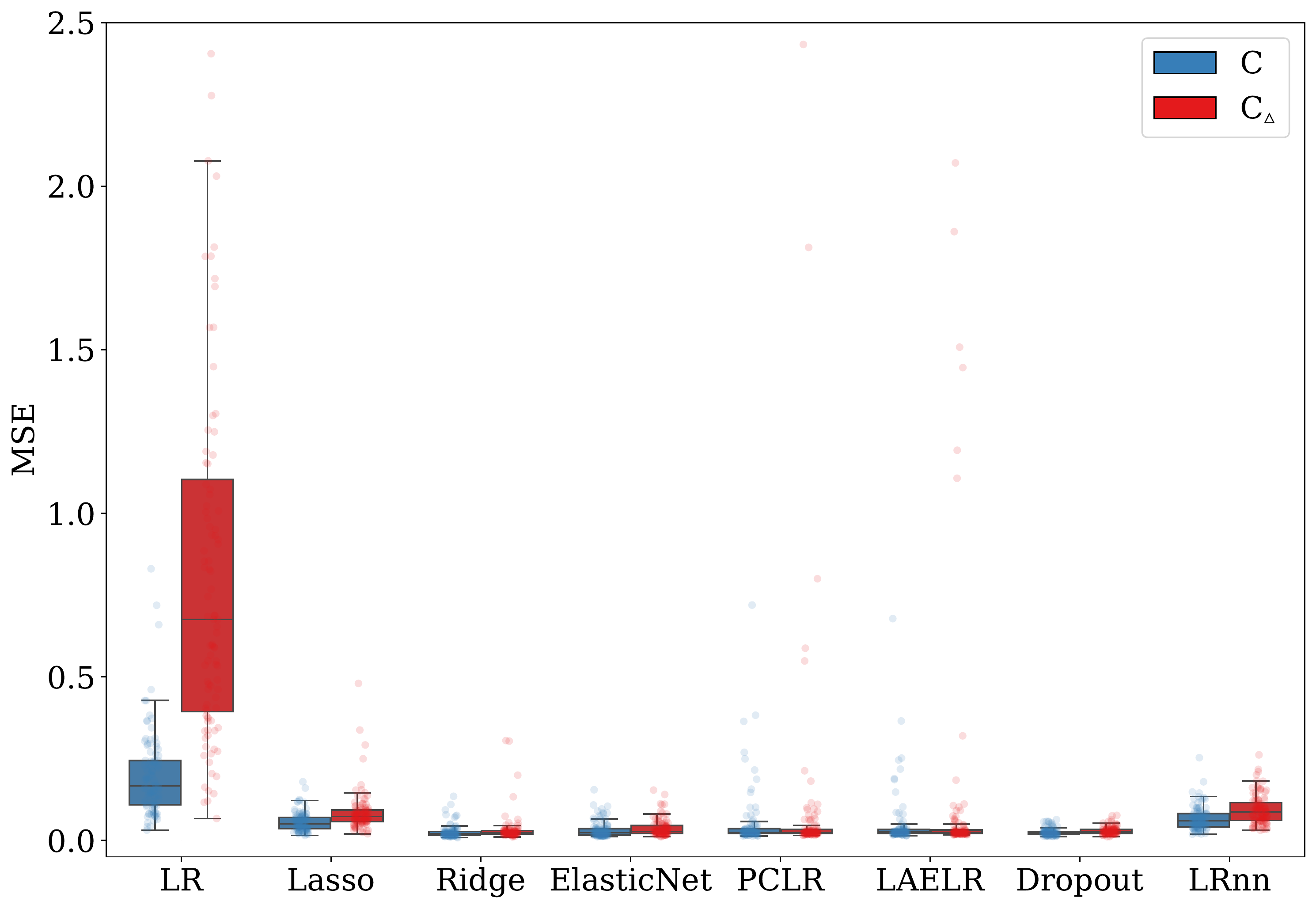}& \includegraphics[width=\boxplotwidthcd]{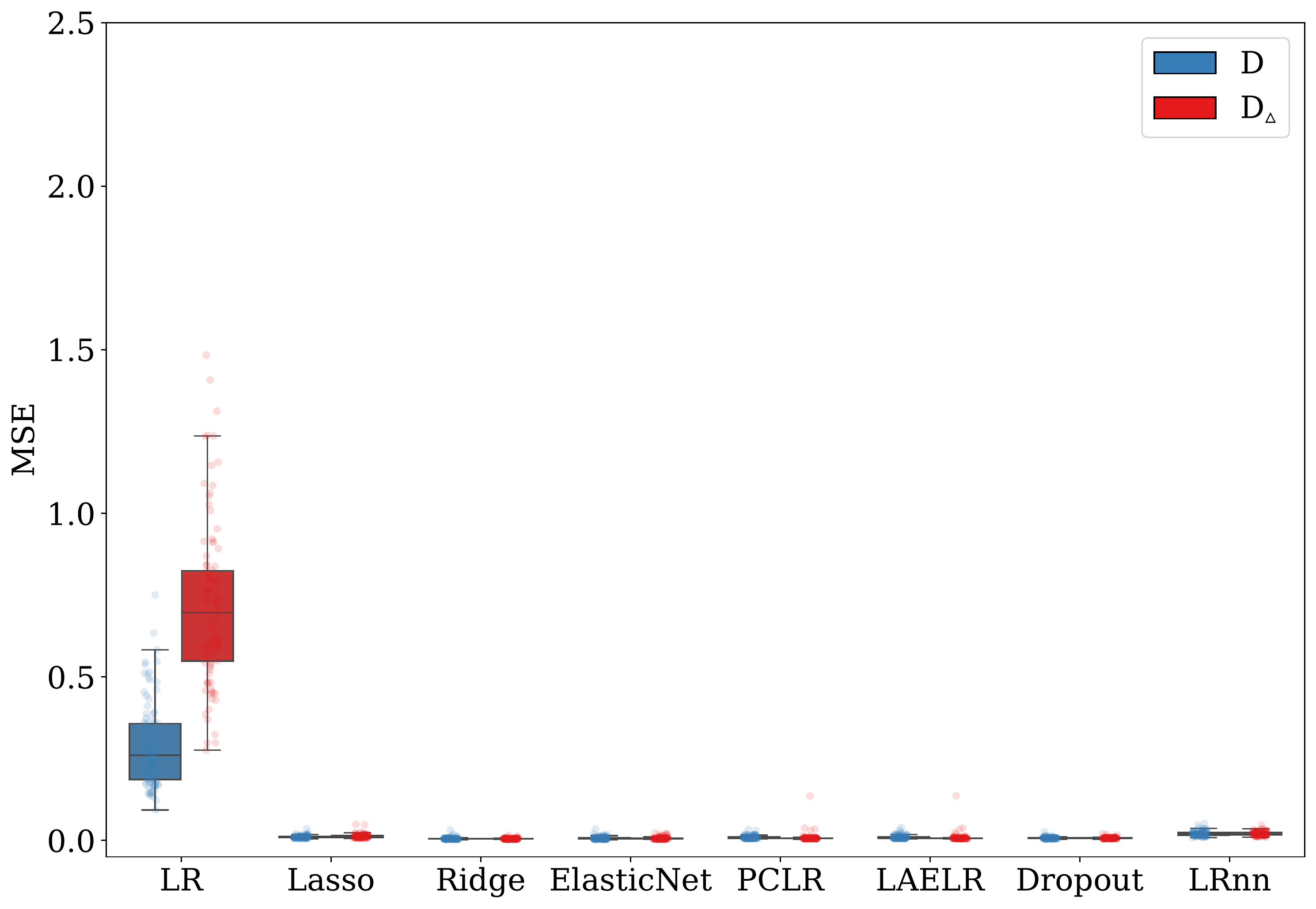}\\
    \end{tabular}
    \caption{\label{fig:mse_results_cd}Per method, the mean squared error between the estimated and the true coefficients for each method, for the \textbf{dysphagia} settings. \textbf{Red} indicates \textbf{high collinearity}, and blue low collinearity.}
\end{figure}

\begin{figure}
    \centering
    \begin{tabular}{cc}
    \includegraphics[width=\boxplotwidthcd]{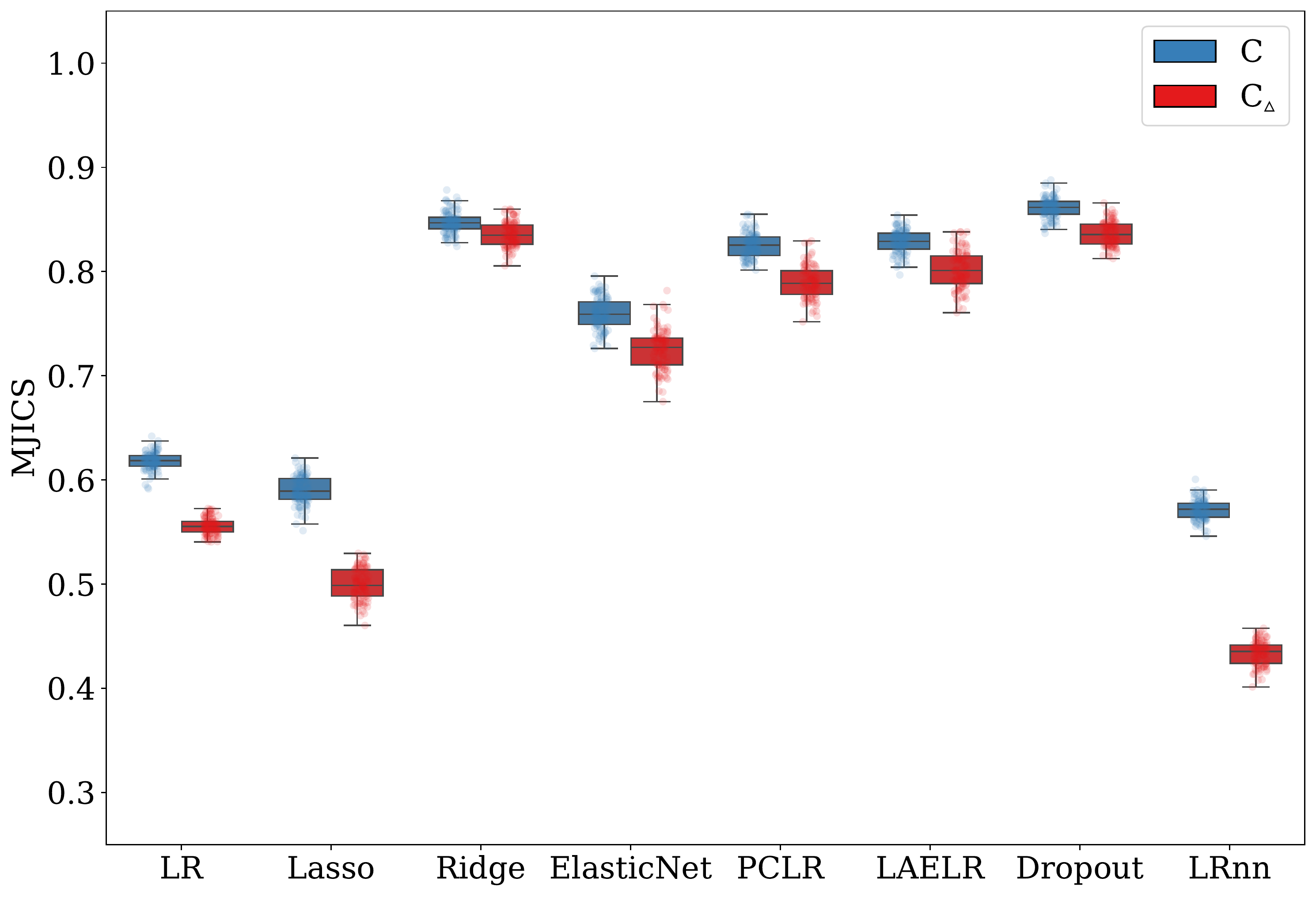}& \includegraphics[width=\boxplotwidthcd]{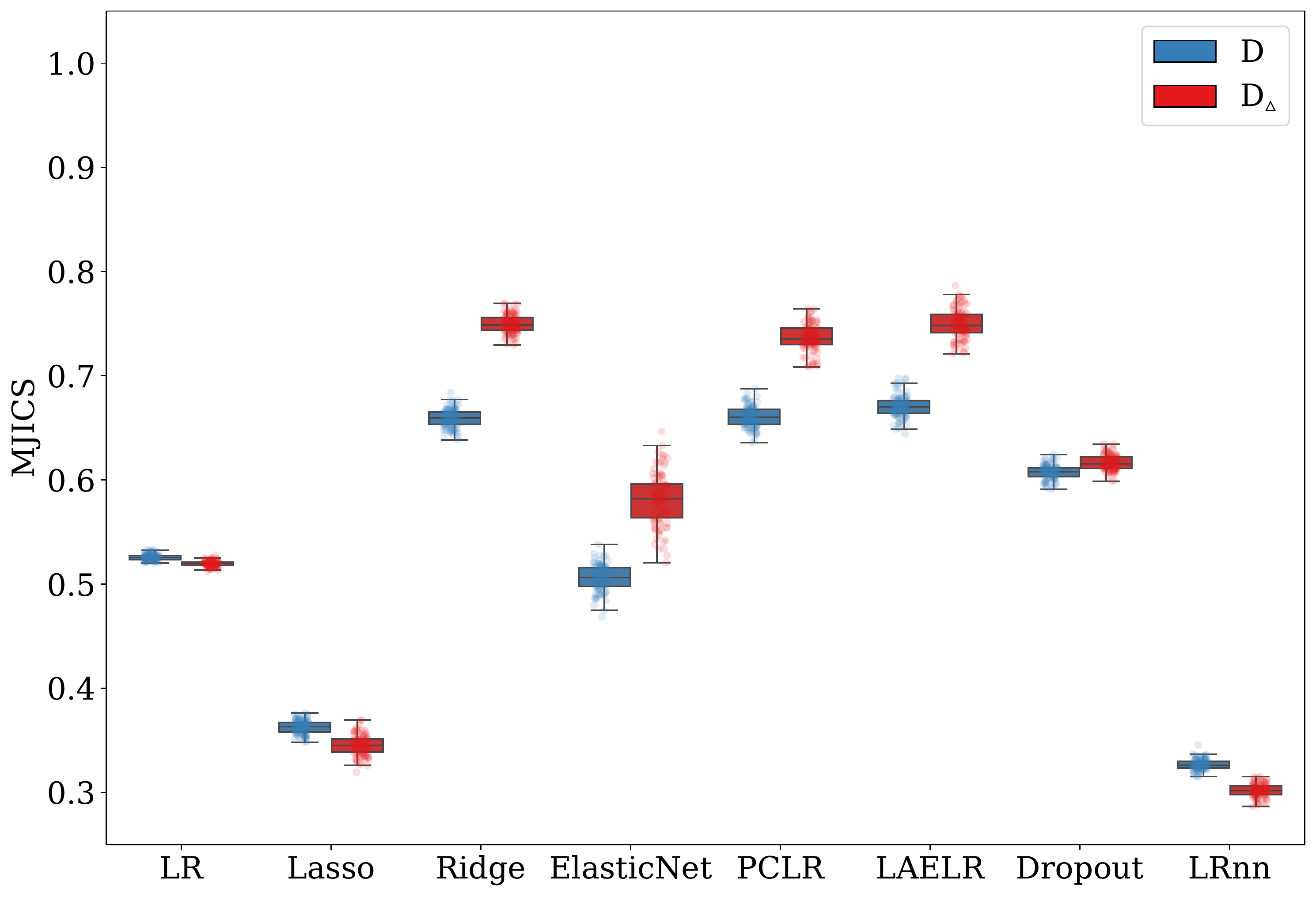}\\
    \end{tabular}
    \caption{\label{fig:mjics_results_cd}Per method, the mean proportion of coefficients with the same direction of effect after repetition for the \textbf{dysphagia} settings. \textbf{Red} indicates \textbf{high collinearity}, and blue low collinearity.}
\end{figure}

%% file: sections/HYPPARAMPLOTS-CD.tex
\newcommand{\hypplotwidthcd}{0.3\textwidth}
\begin{figure}[h!]
    \centering
    \begin{tabular}{cccc}
        \rotatebox{90}{\hspace{1.3cm}\footnotesize Hyperparameter value} &\includegraphics[width=\hypplotwidthcd]{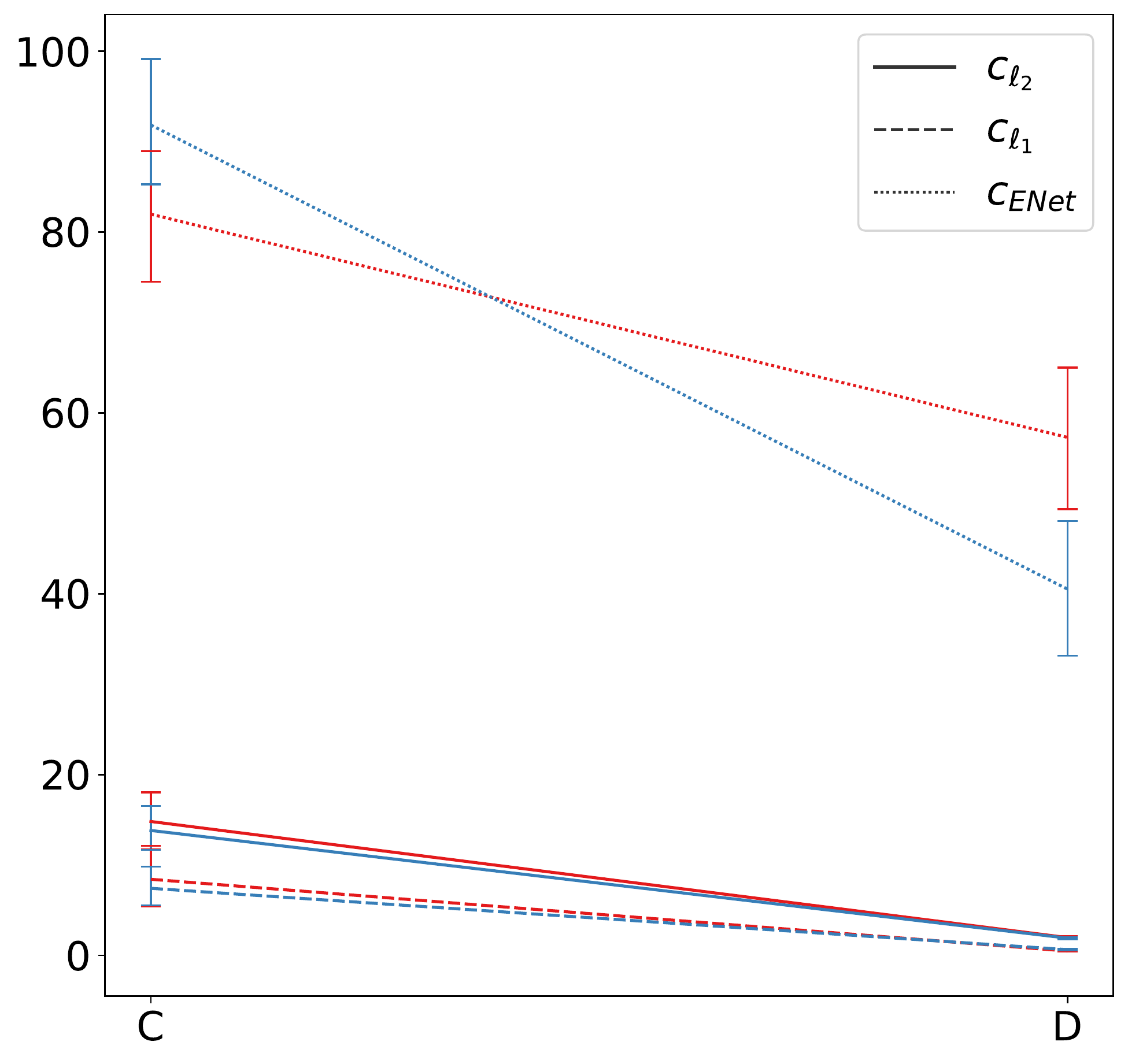} &  \includegraphics[width=\hypplotwidthcd]{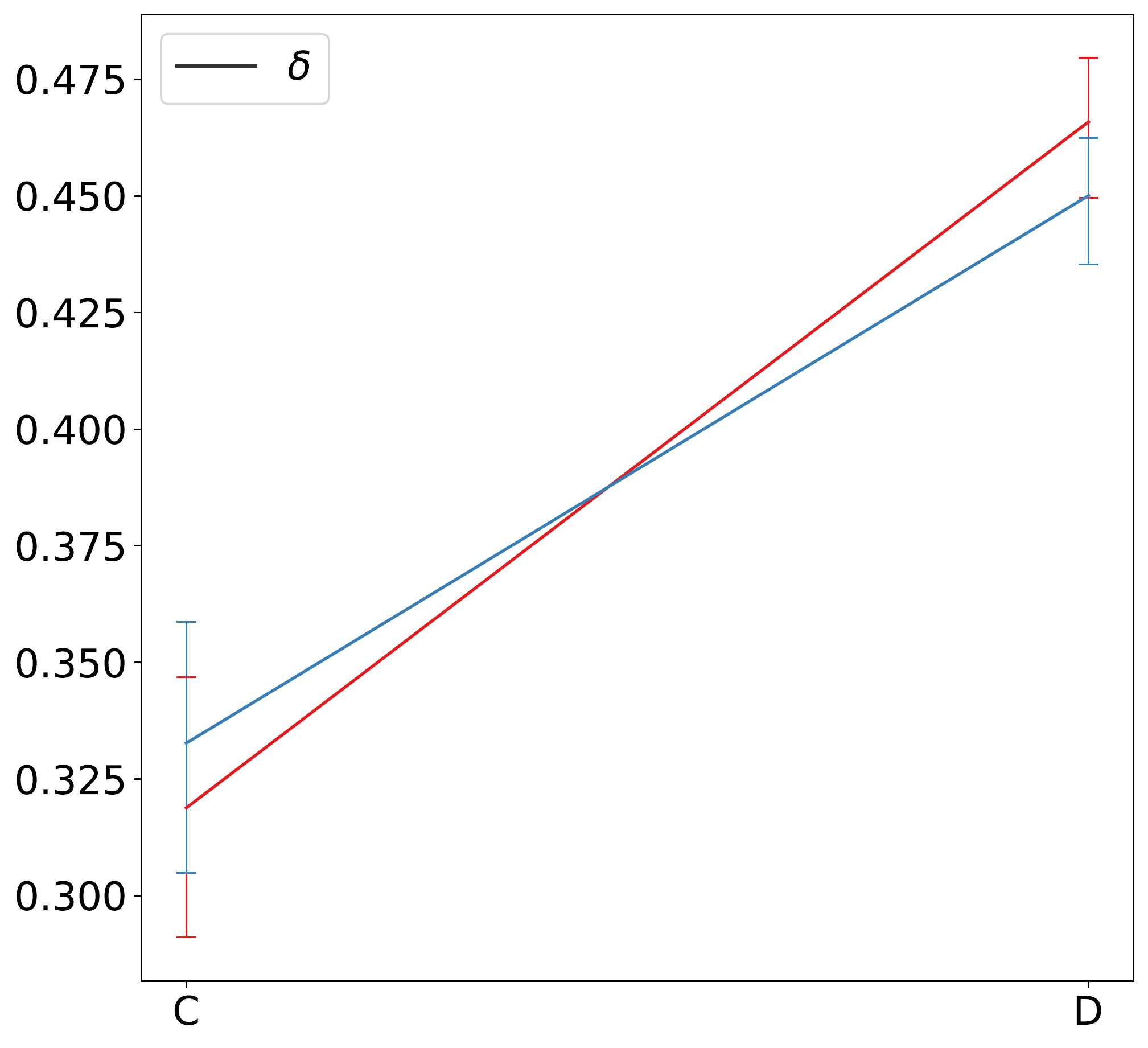} & \includegraphics[width=\hypplotwidthcd]{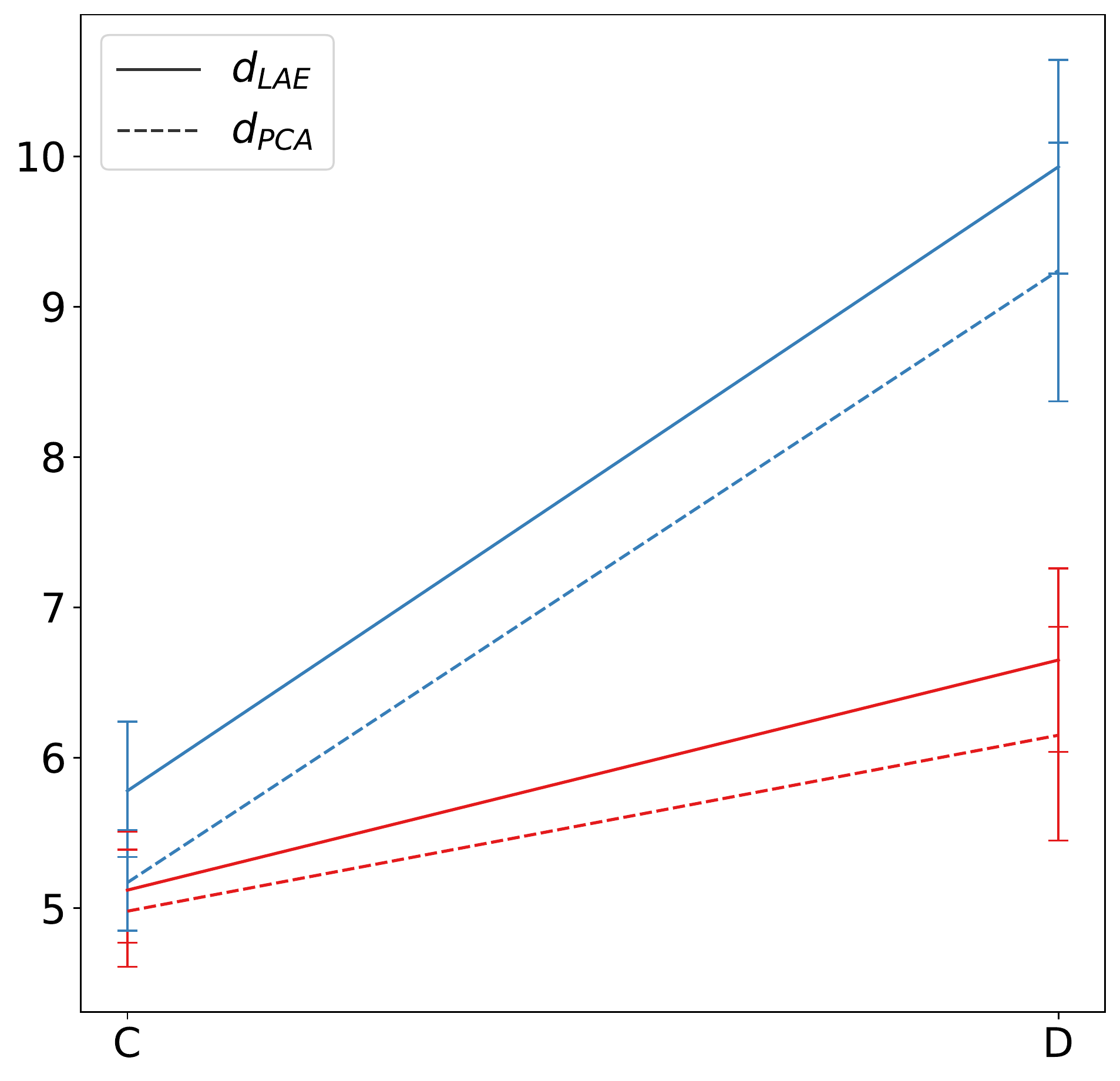}\\
        & (1) & (2) & (3) \\
    \end{tabular}
    \caption{\label{fig:dys_hyperparam_results}Hyperparameter values for \textbf{dysphagia}: per predictor set, C being the small predictor set with relatively high EPV (EPV=6), and D the large predictor set with lower EPV (EPV=2). The \textbf{high collinearity} settings \textbf{in red}, and the \textbf{low collinearity} setting \textbf{in blue}. The methods are distributed across three plots due to their different scales. Hyperparameter notation follows Table \ref{tab:methods}, except for $c_{\textit{ENet}}$, which is the total shrinkage factor for ElasticNet ($c_{\ell_1} + c_{\ell_2}$).}
\end{figure}

%% file: sections/CALIB-GRID-REAL.tex
\begin{figure}[h!]
    \centering
    \resizebox{.95\textwidth}{!}{
    \begin{tabular}{ccccc}
         Method& \xerbasiclowreal{} & \xerexthighreal{} & \dysbasiclowreal{} & \dysexthighreal{}  \\
         \centered{LR}  & \includegraphics[width=\plotwidth]{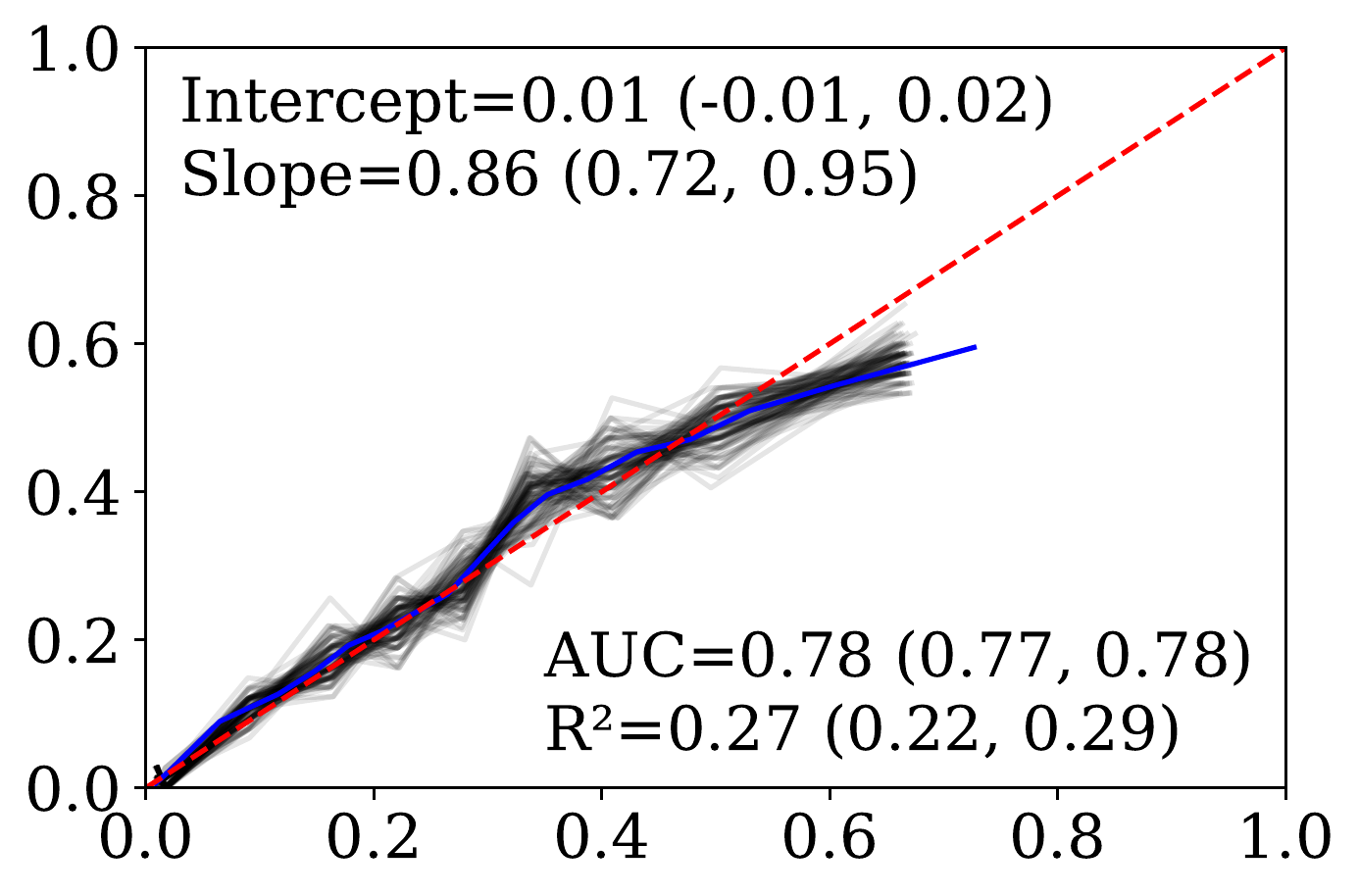} & \includegraphics[width=\plotwidth]{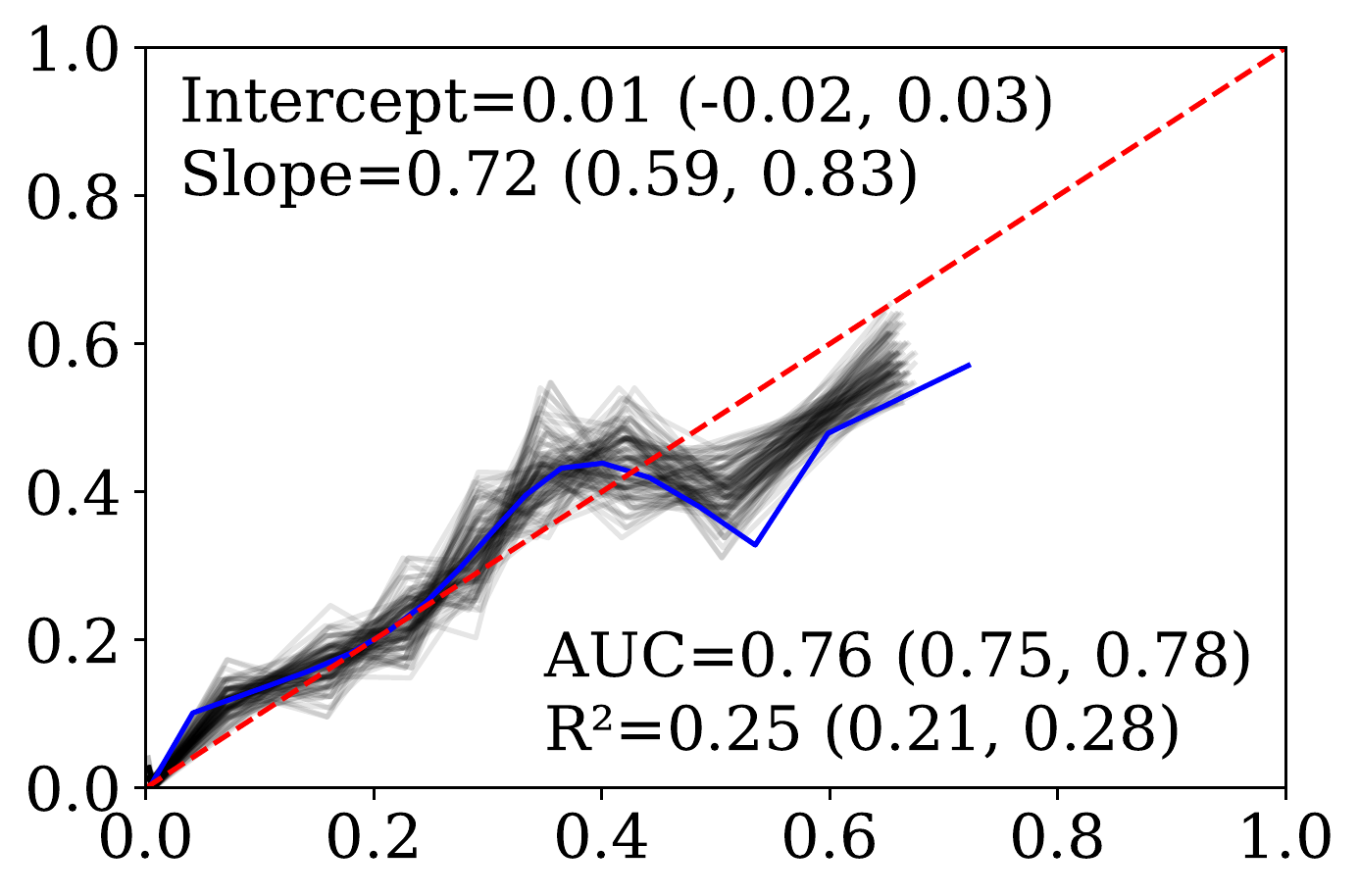} & \includegraphics[width=\plotwidth]{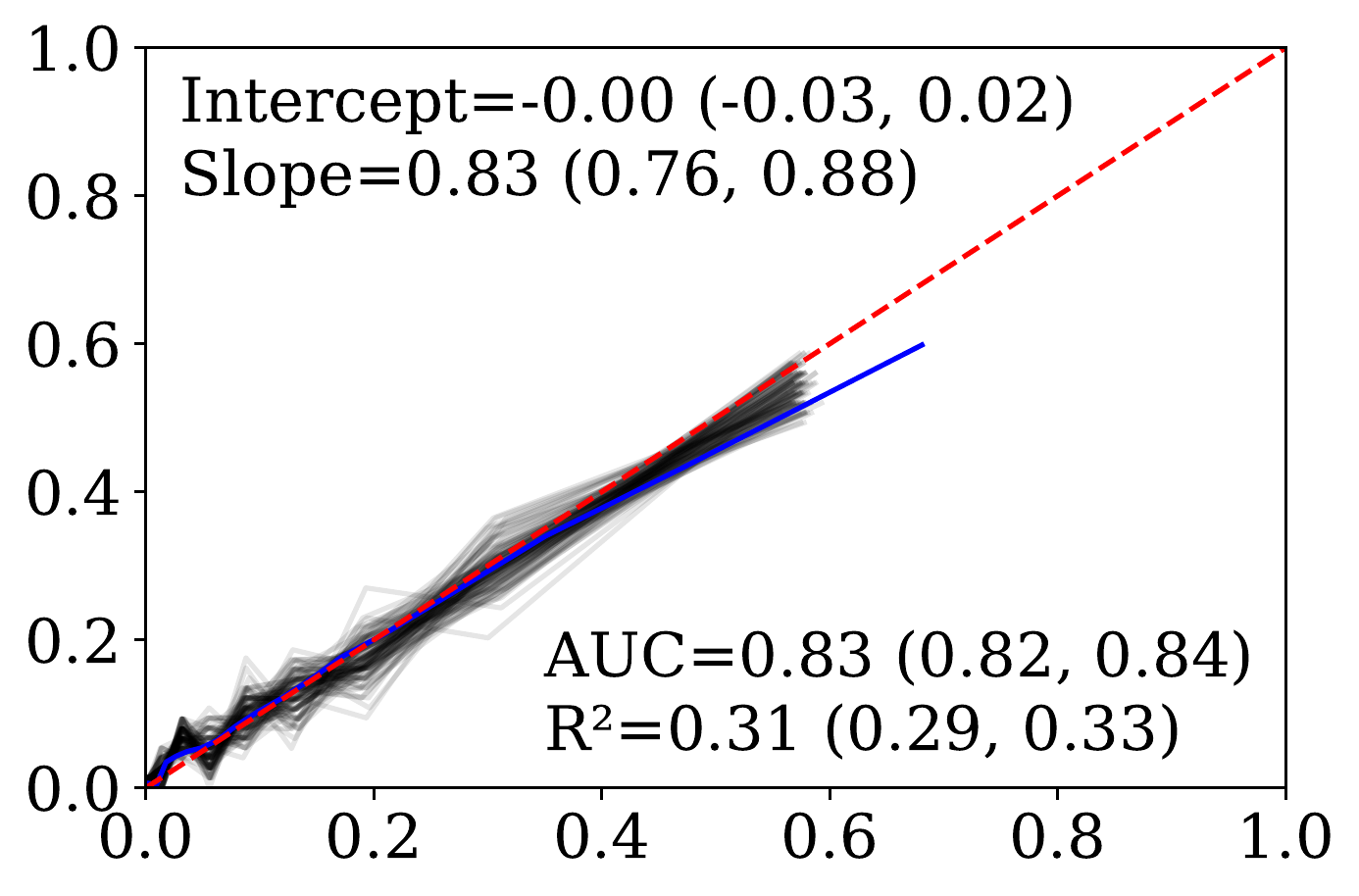} & \includegraphics[width=\plotwidth]{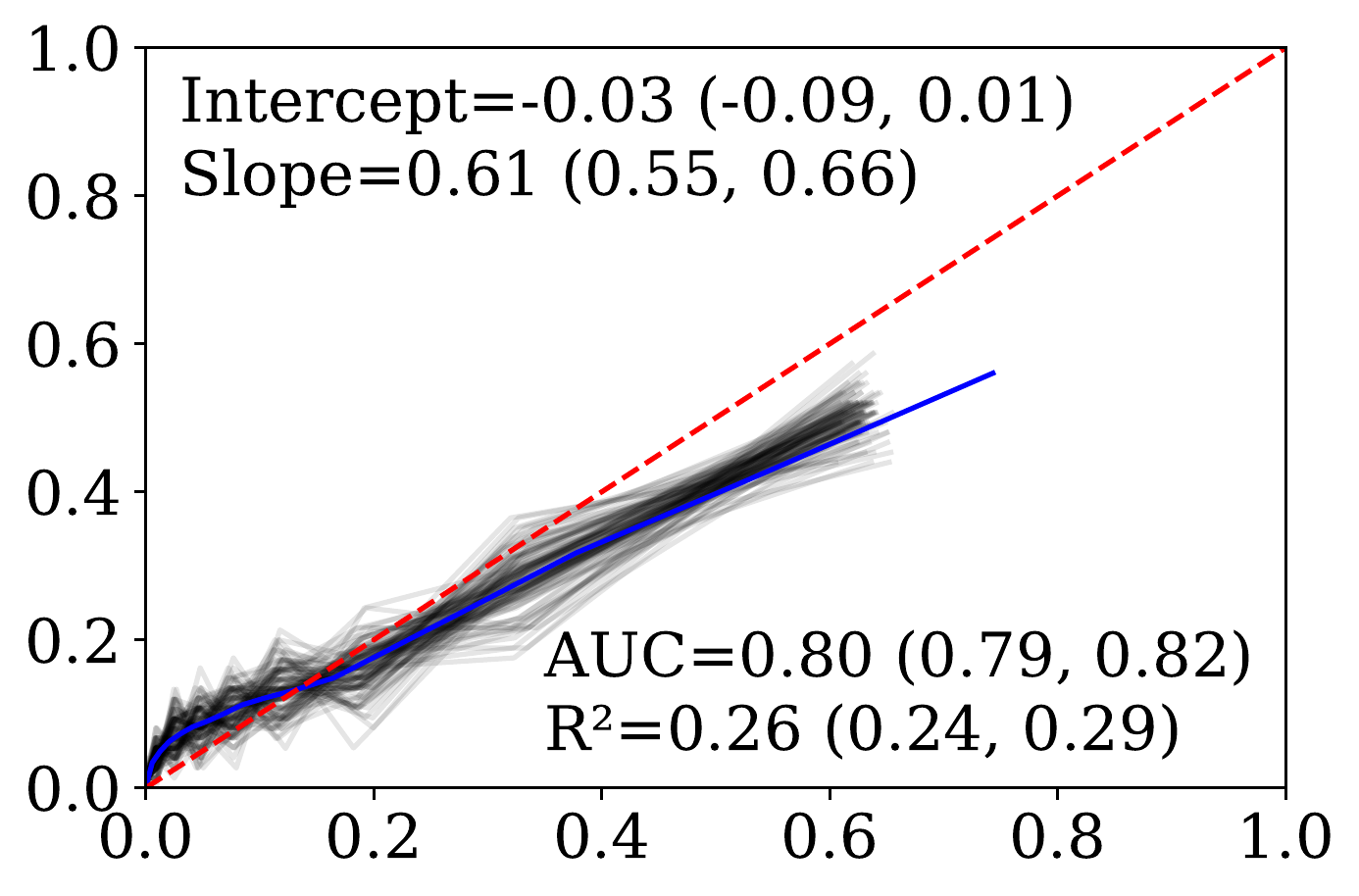}\\
         \centered{Lasso} & \includegraphics[width=\plotwidth]{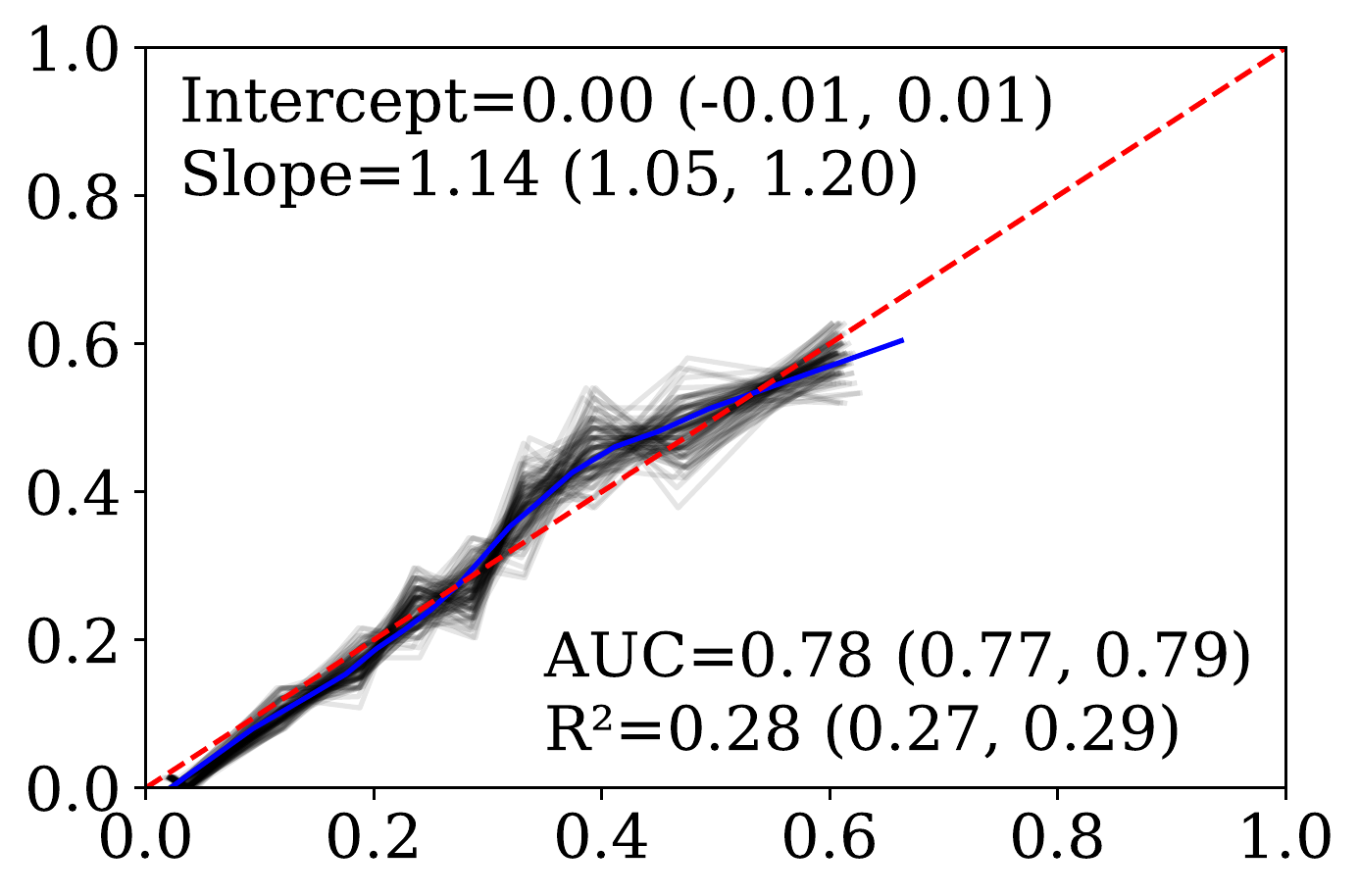} & \includegraphics[width=\plotwidth]{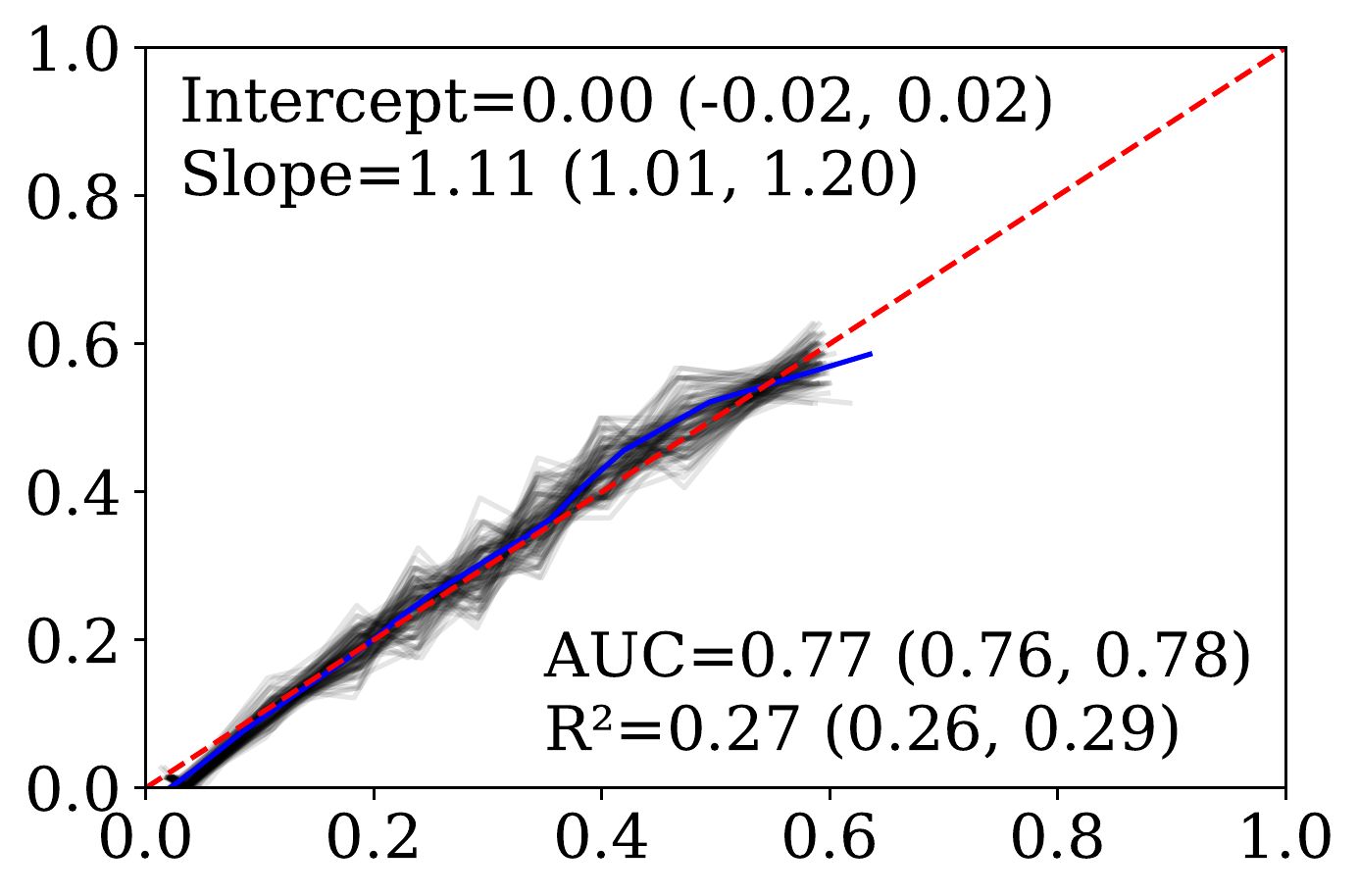} & \includegraphics[width=\plotwidth]{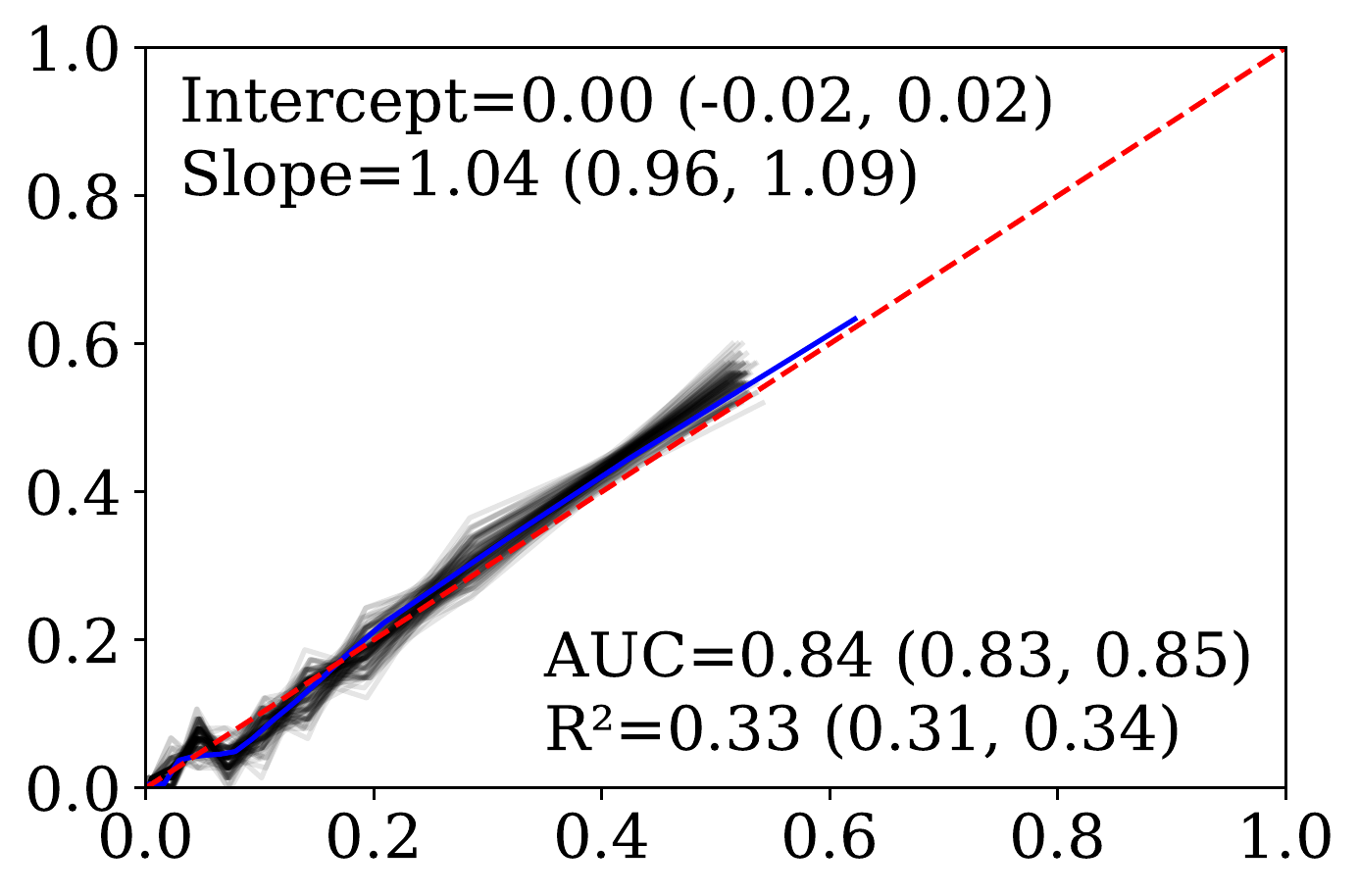} & \includegraphics[width=\plotwidth]{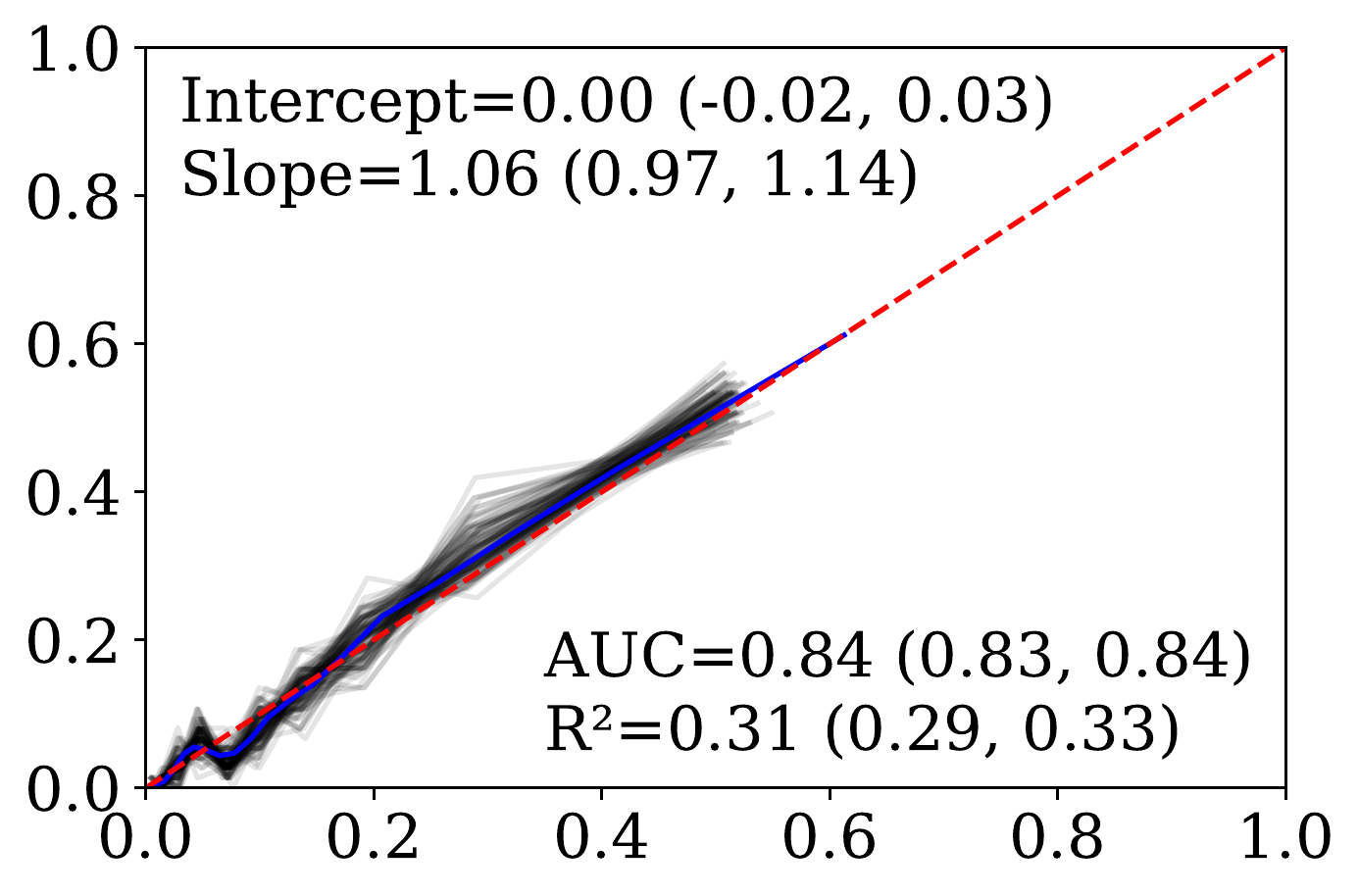}\\
         \centered{Ridge} & \includegraphics[width=\plotwidth]{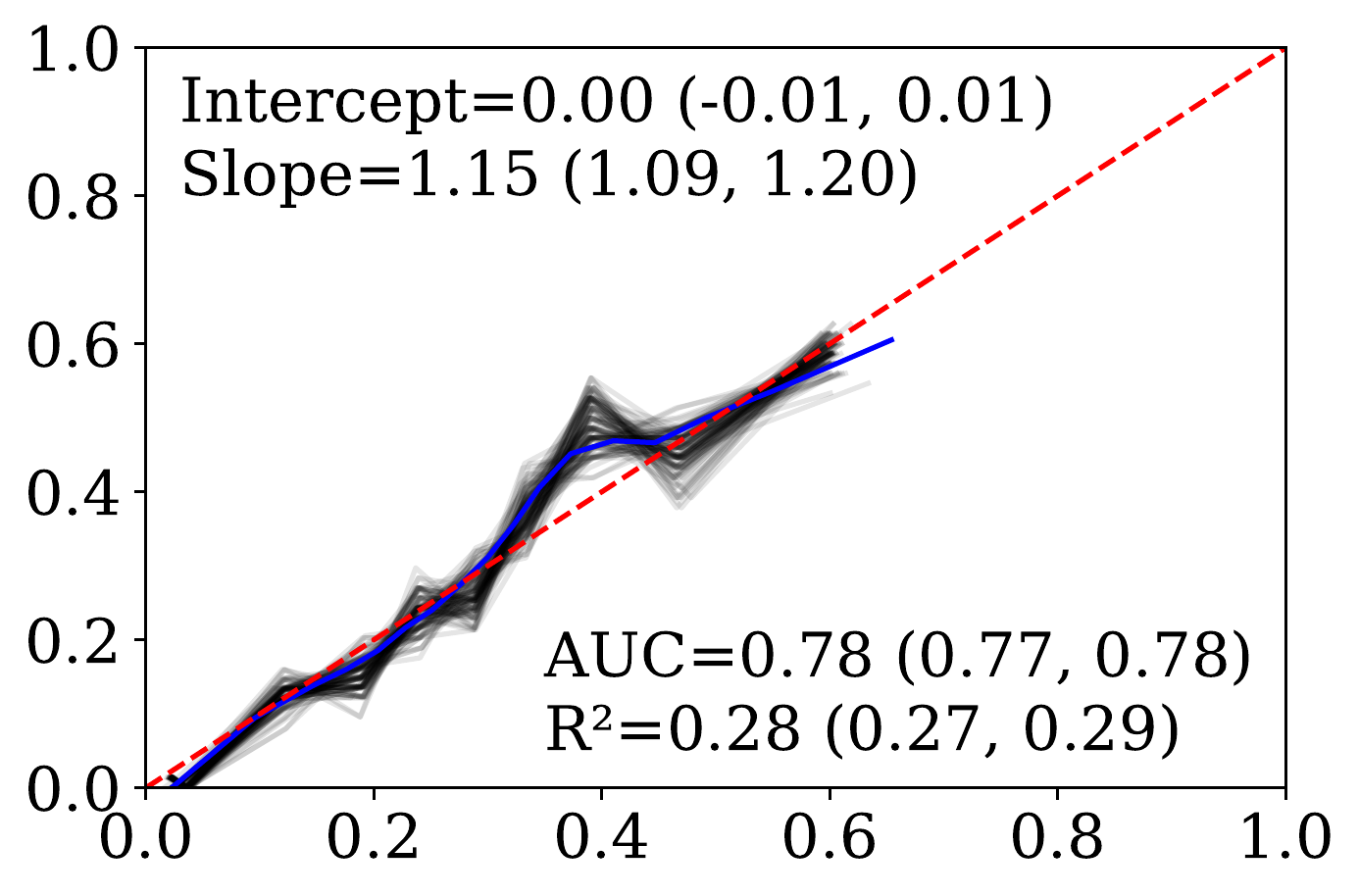} & \includegraphics[width=\plotwidth]{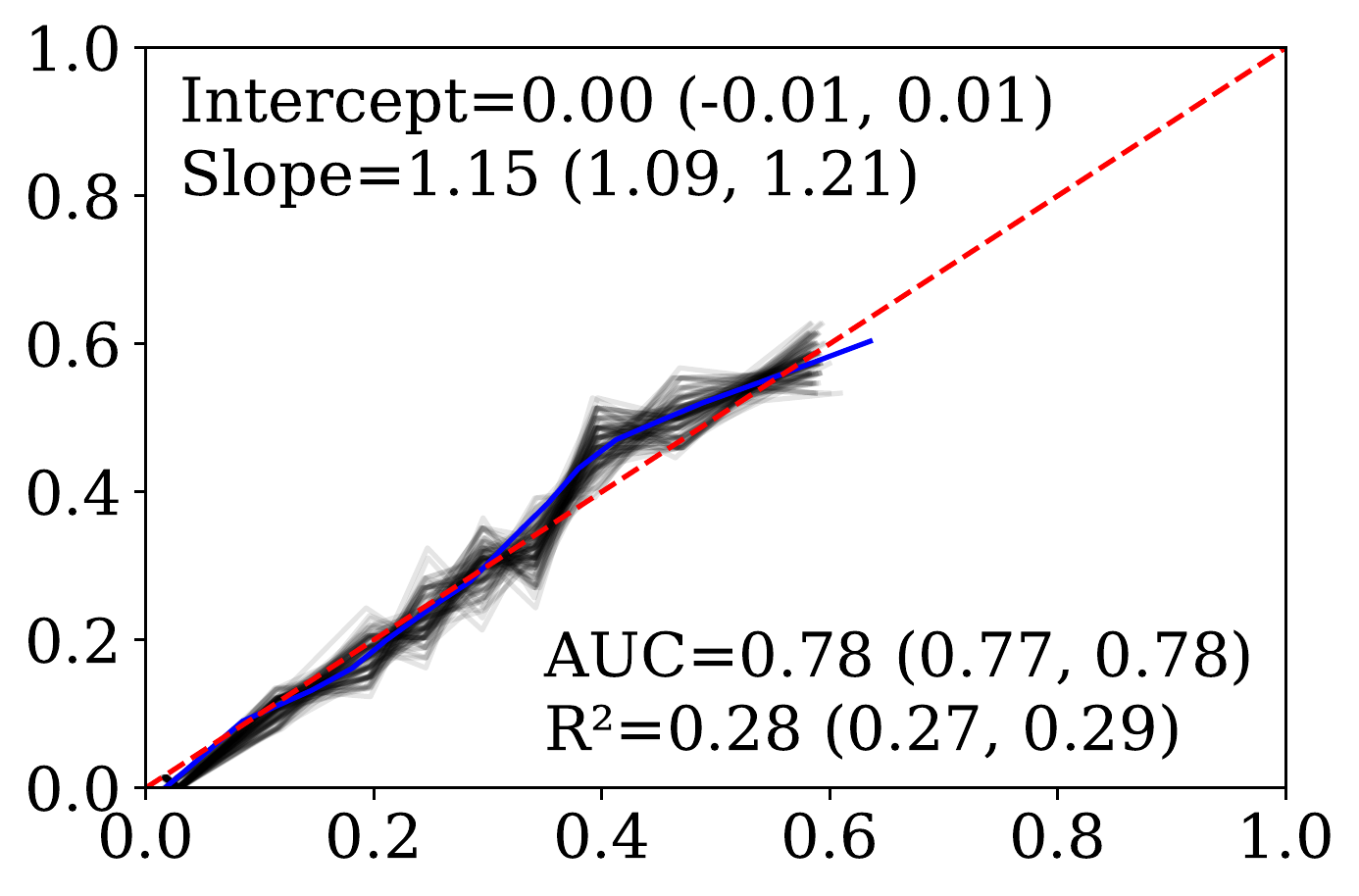} & \includegraphics[width=\plotwidth]{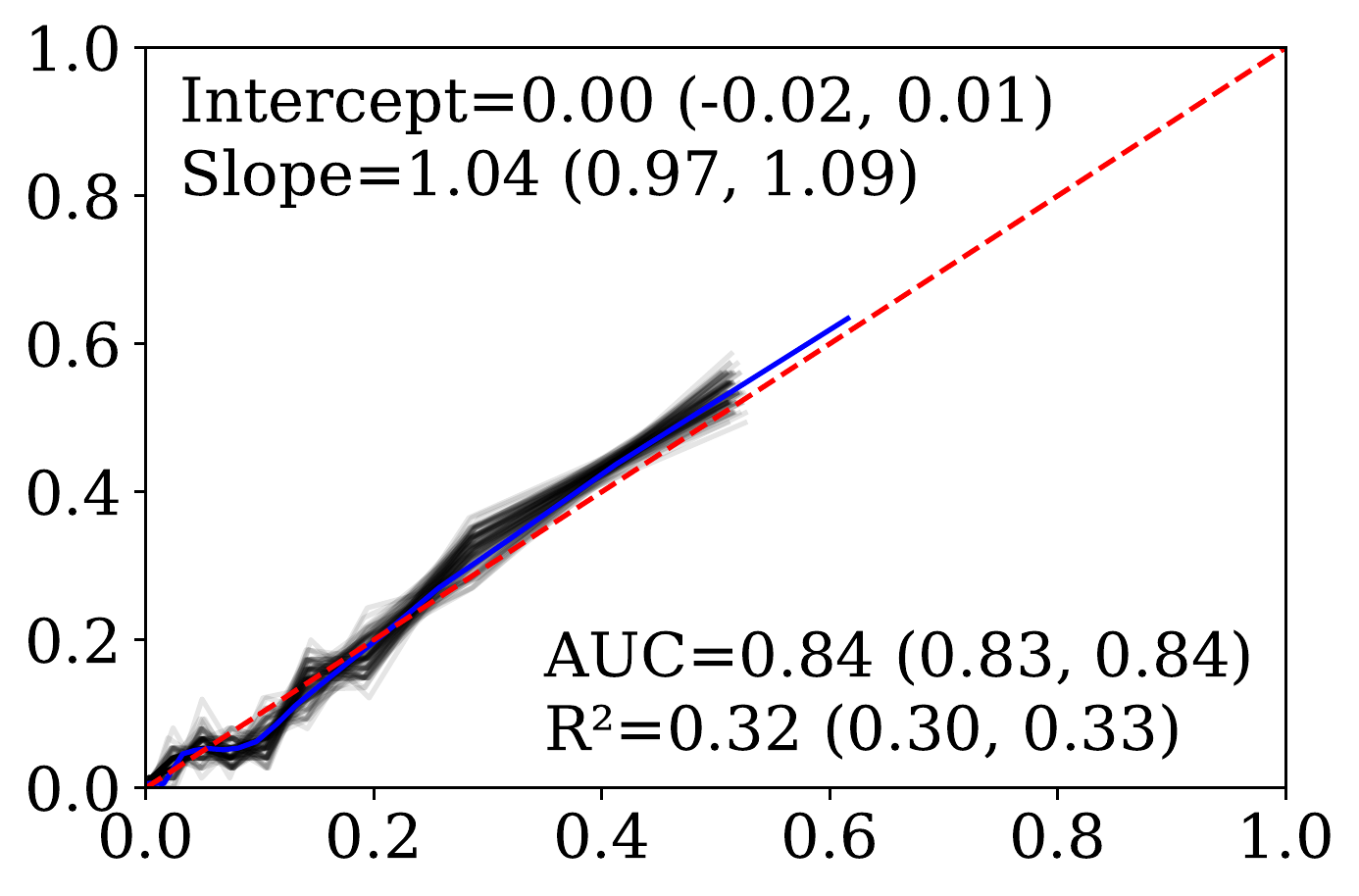} & \includegraphics[width=\plotwidth]{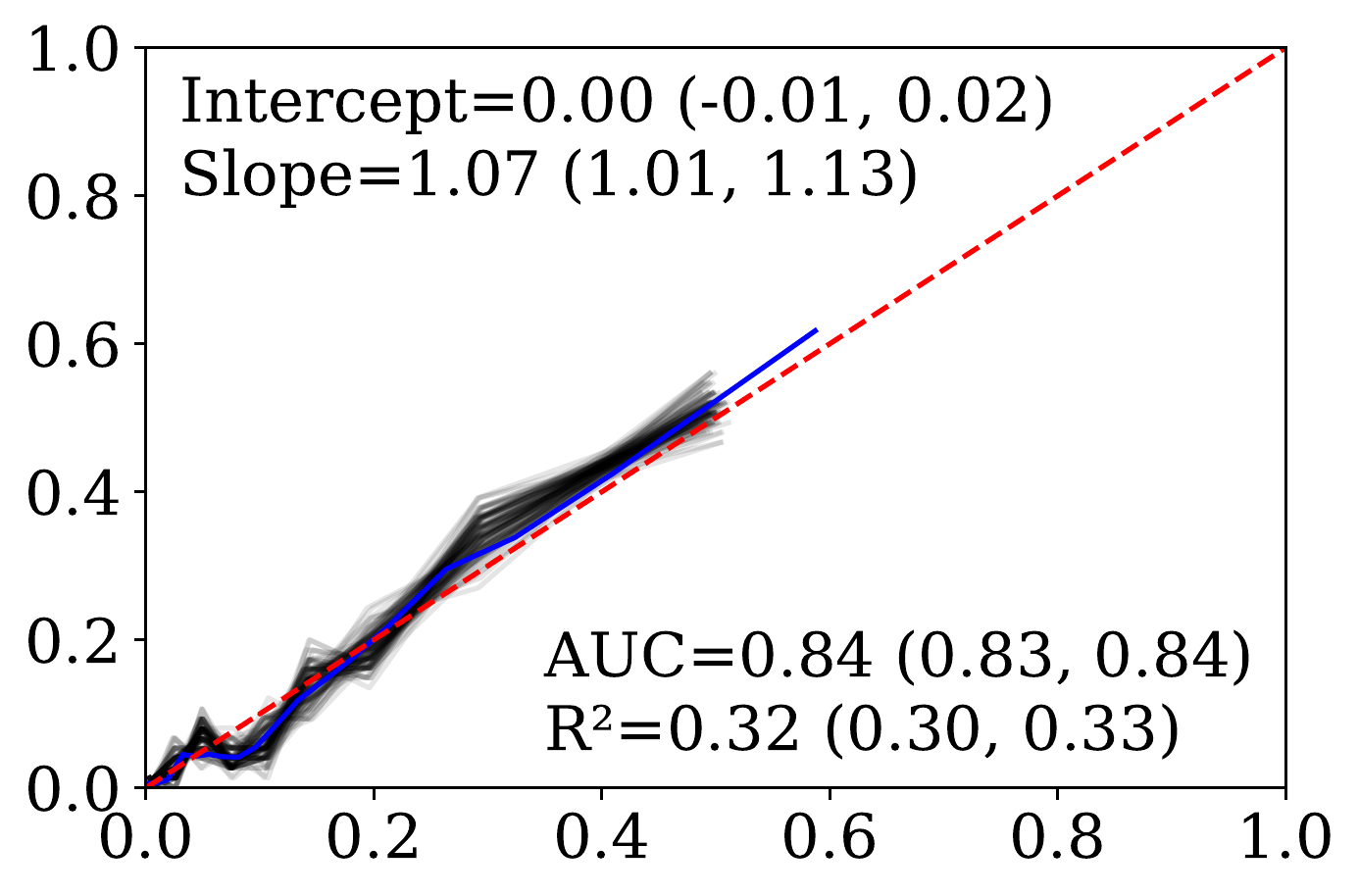}\\
         \centered{ElasticNet} & \includegraphics[width=\plotwidth]{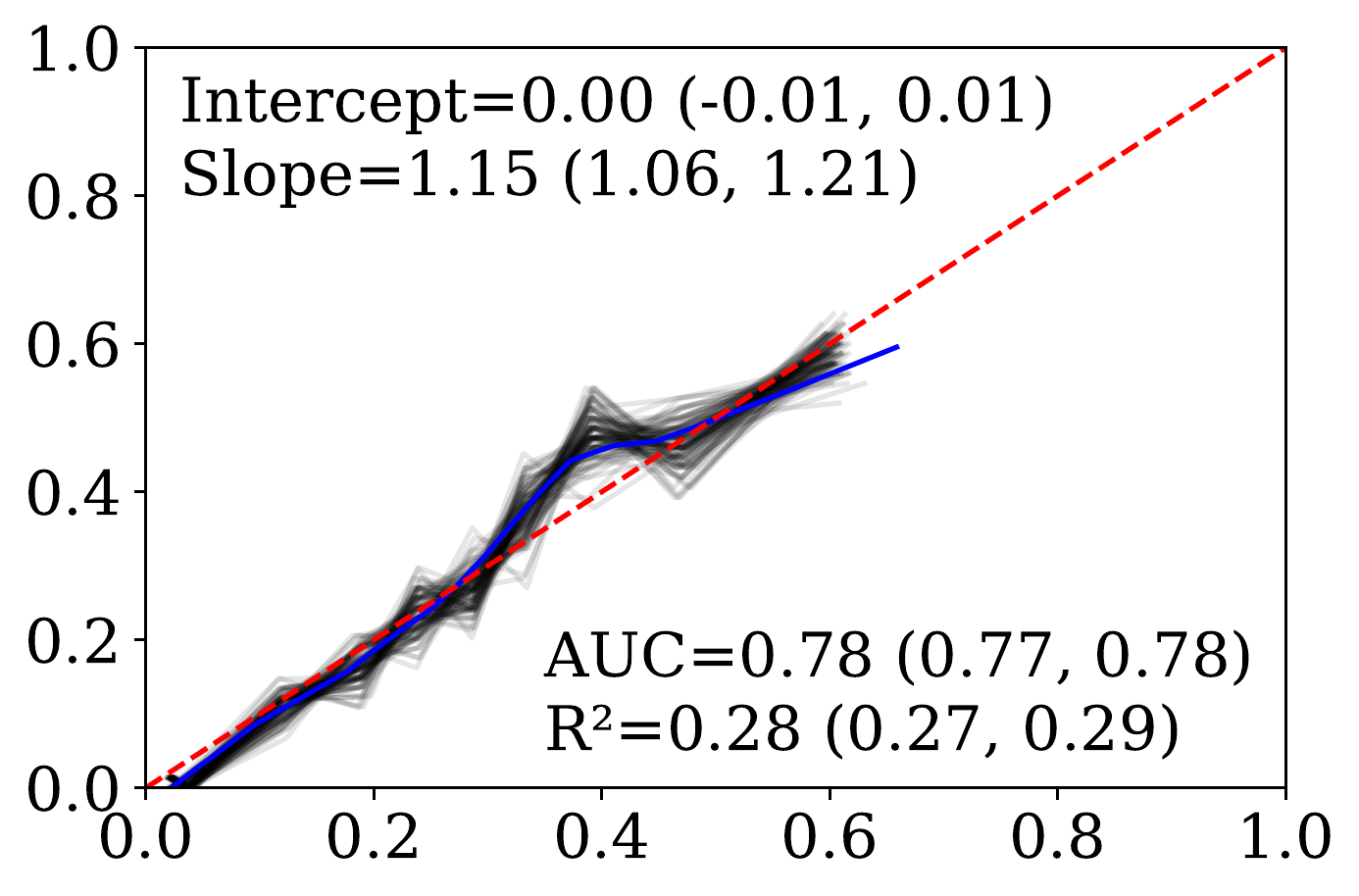} & \includegraphics[width=\plotwidth]{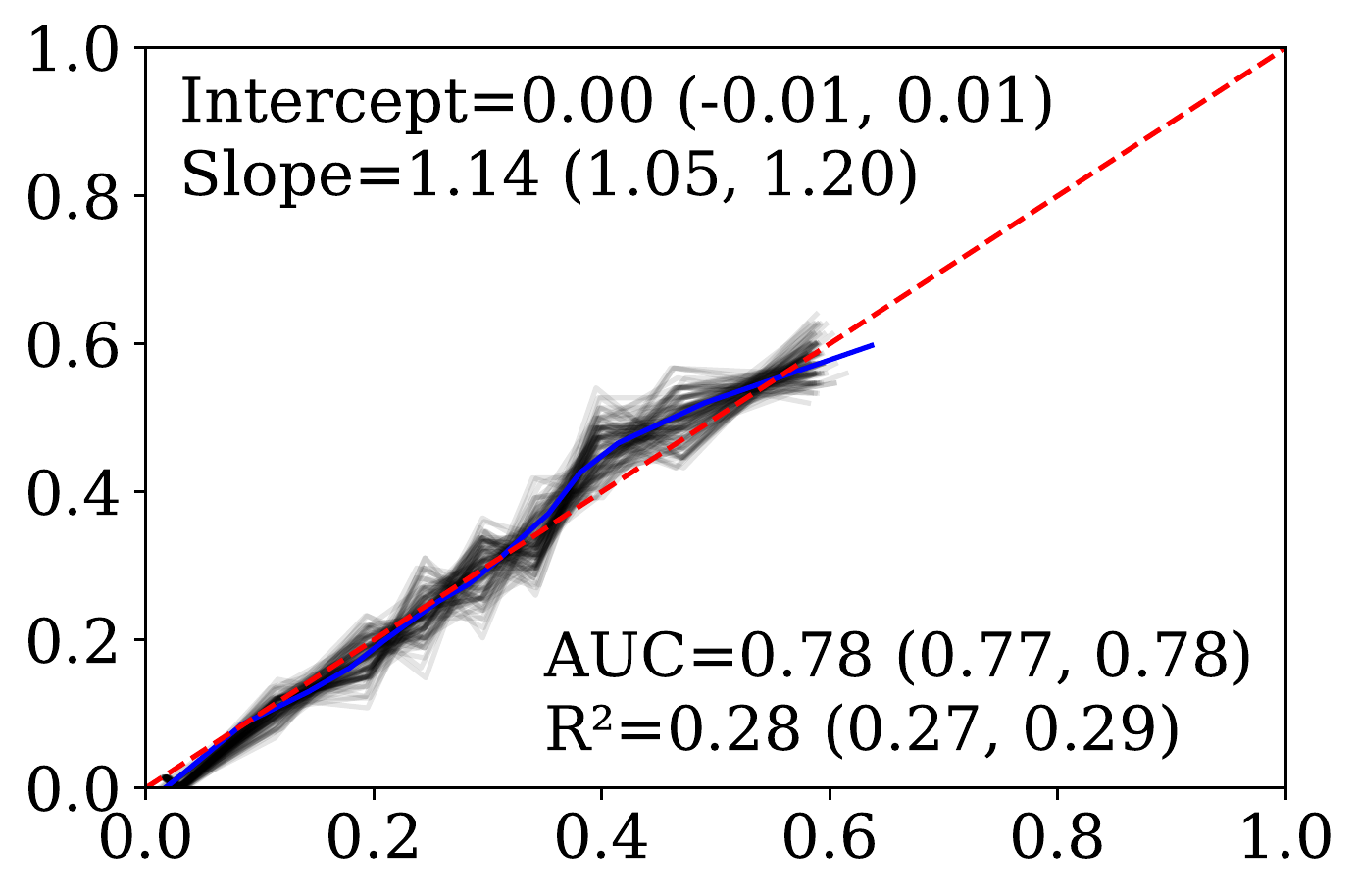} & \includegraphics[width=\plotwidth]{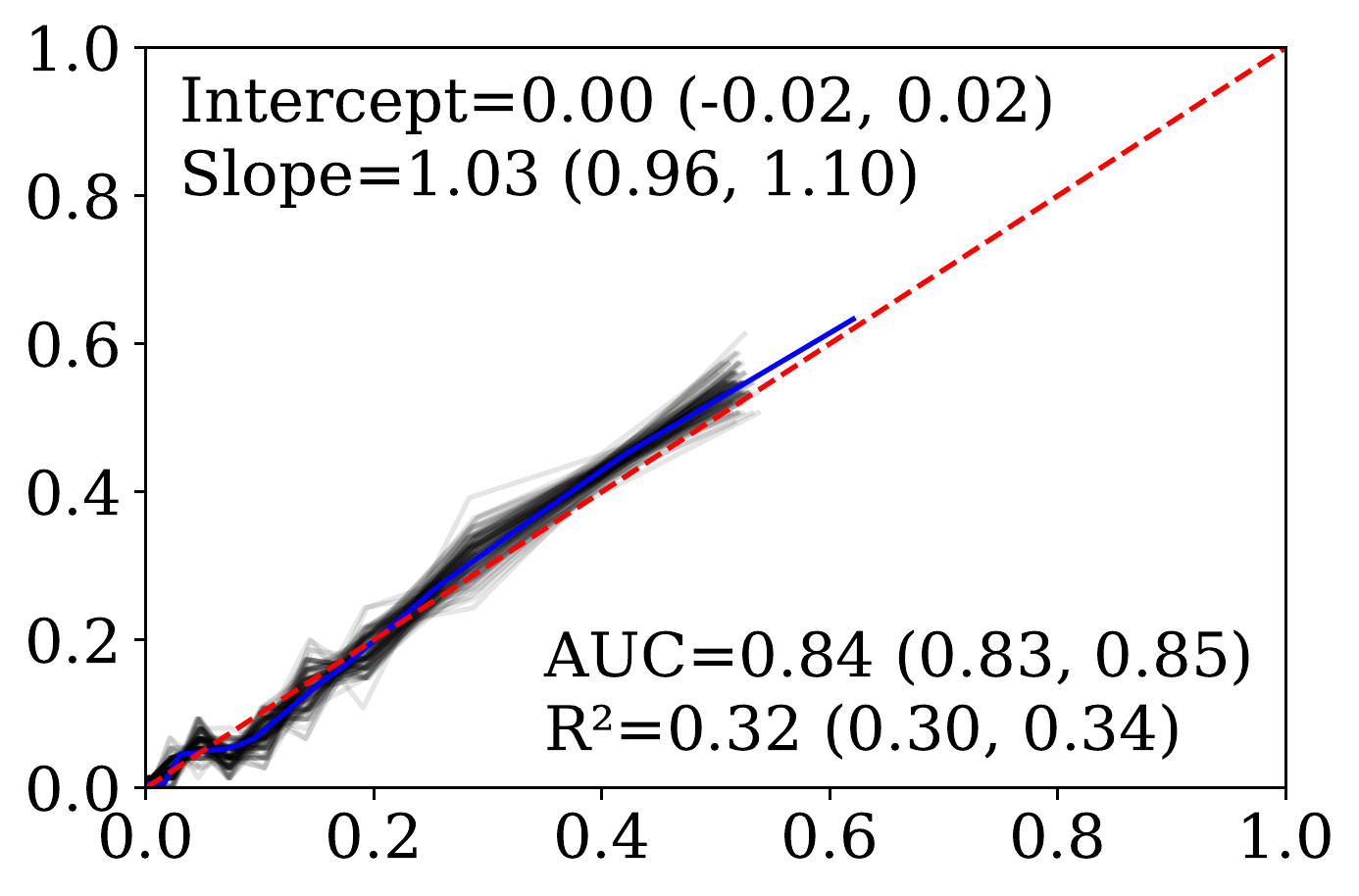} & \includegraphics[width=\plotwidth]{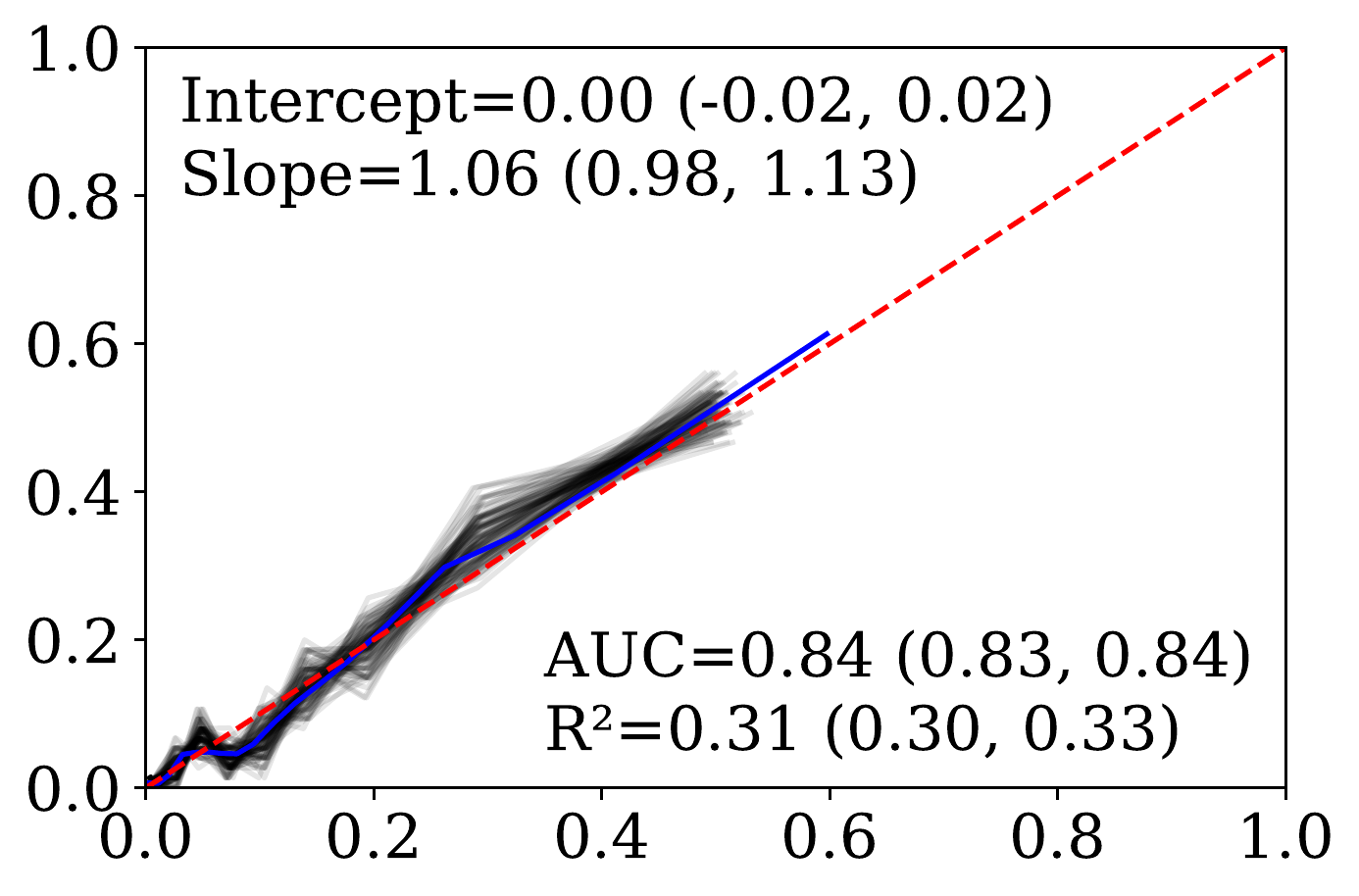}\\
         \centered{PCLR} & \includegraphics[width=\plotwidth]{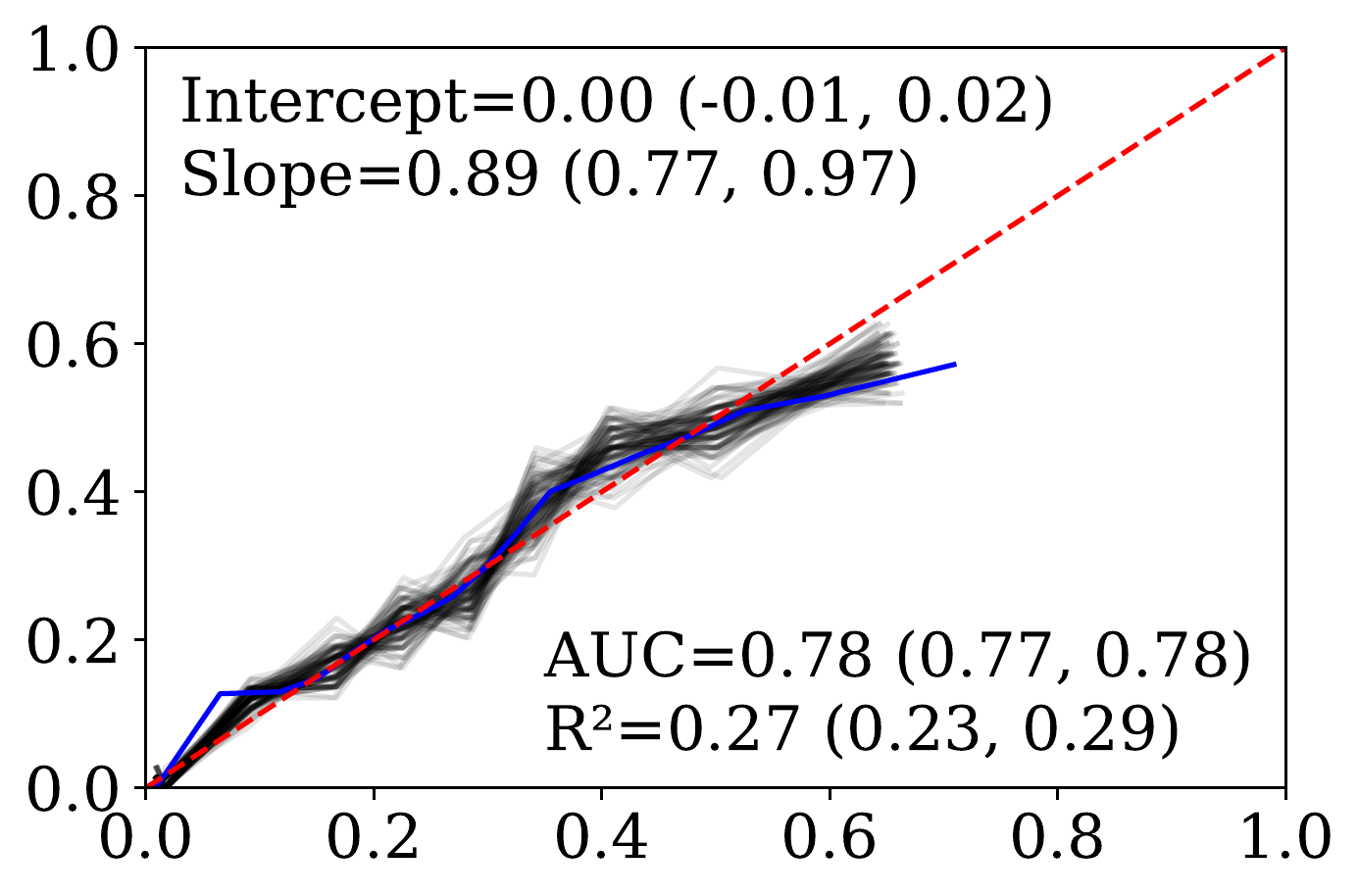} & \includegraphics[width=\plotwidth]{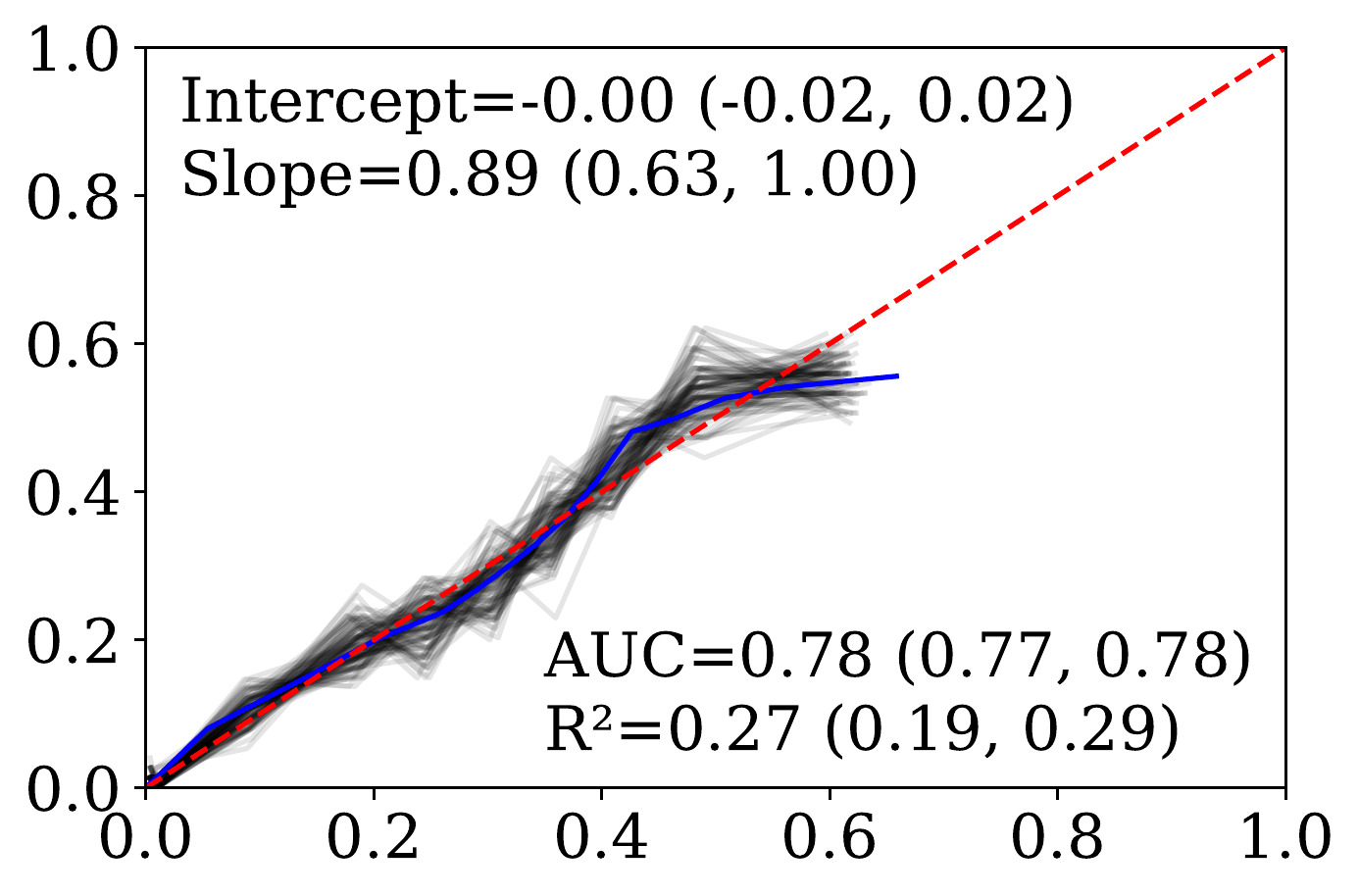} & \includegraphics[width=\plotwidth]{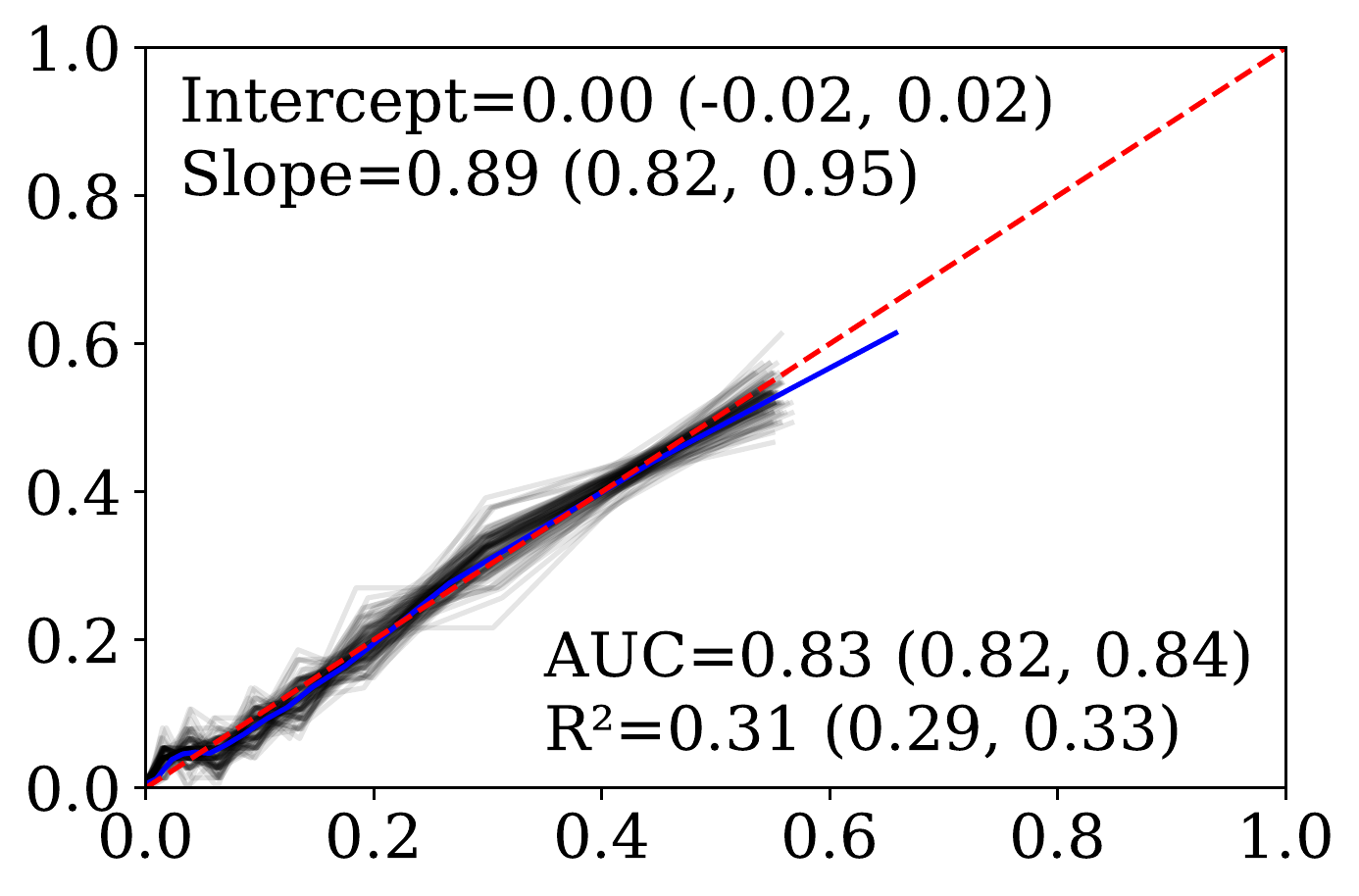} & \includegraphics[width=\plotwidth]{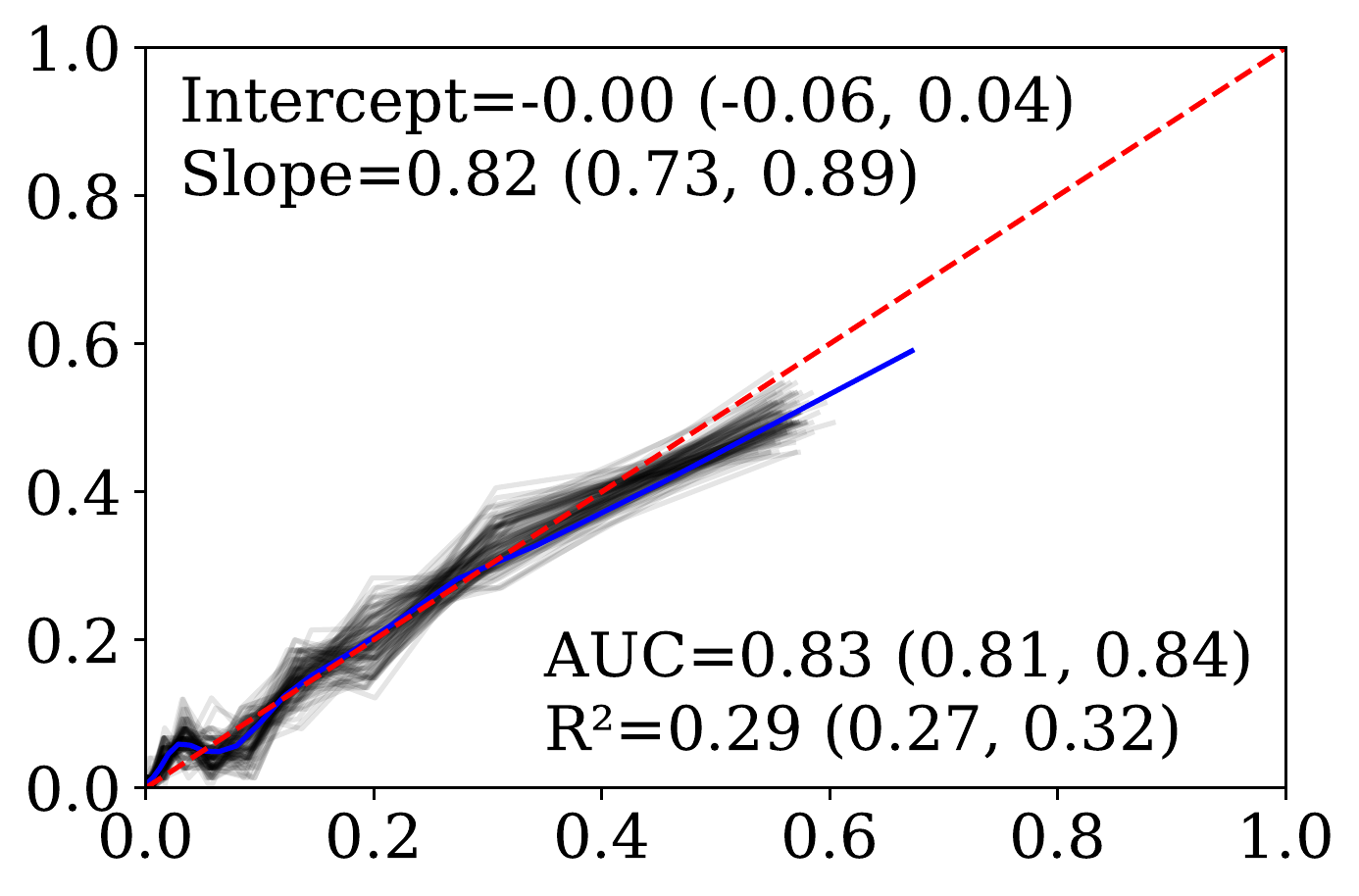}\\
         \centered{LAELR} & \includegraphics[width=\plotwidth]{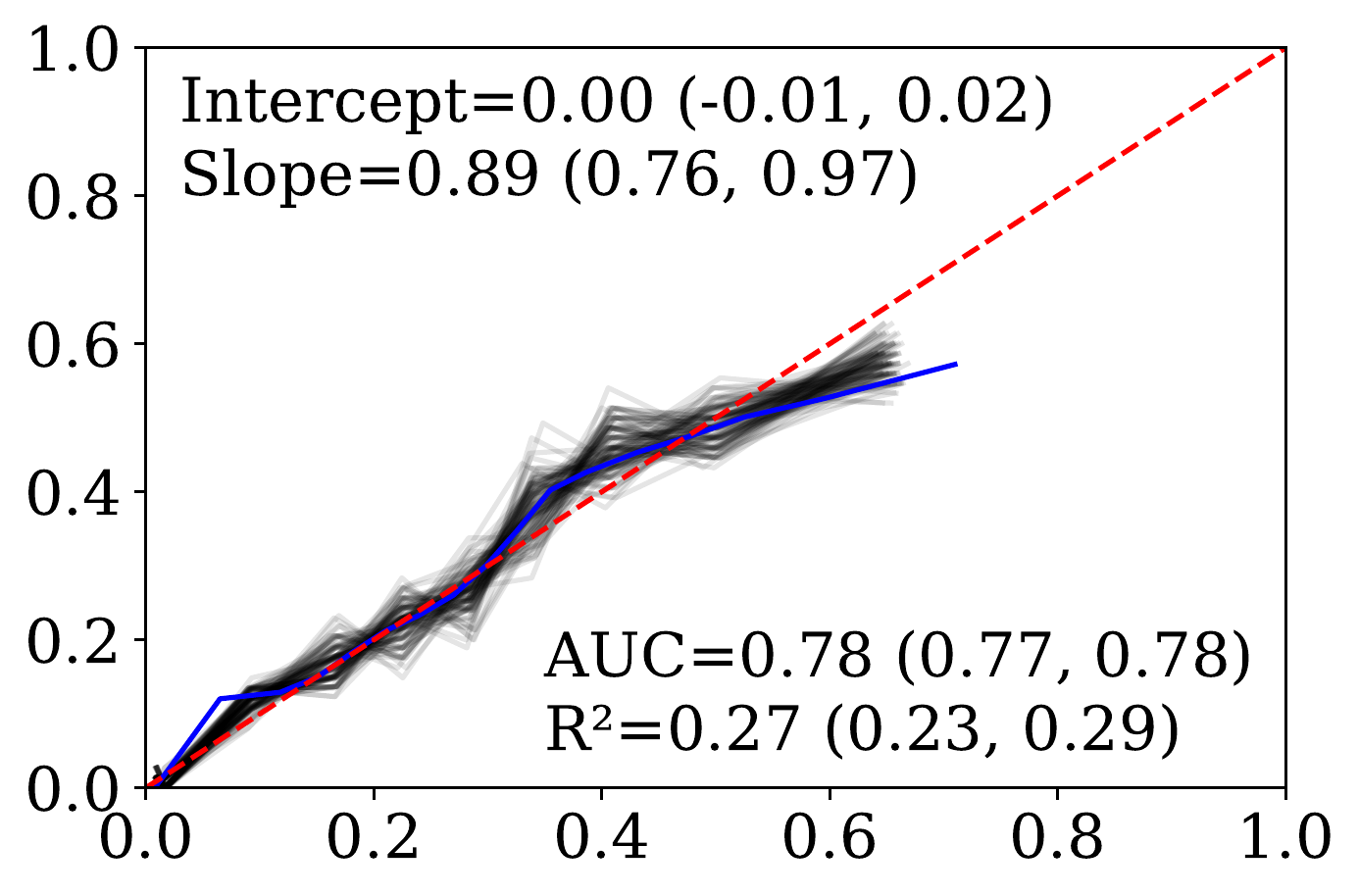} & \includegraphics[width=\plotwidth]{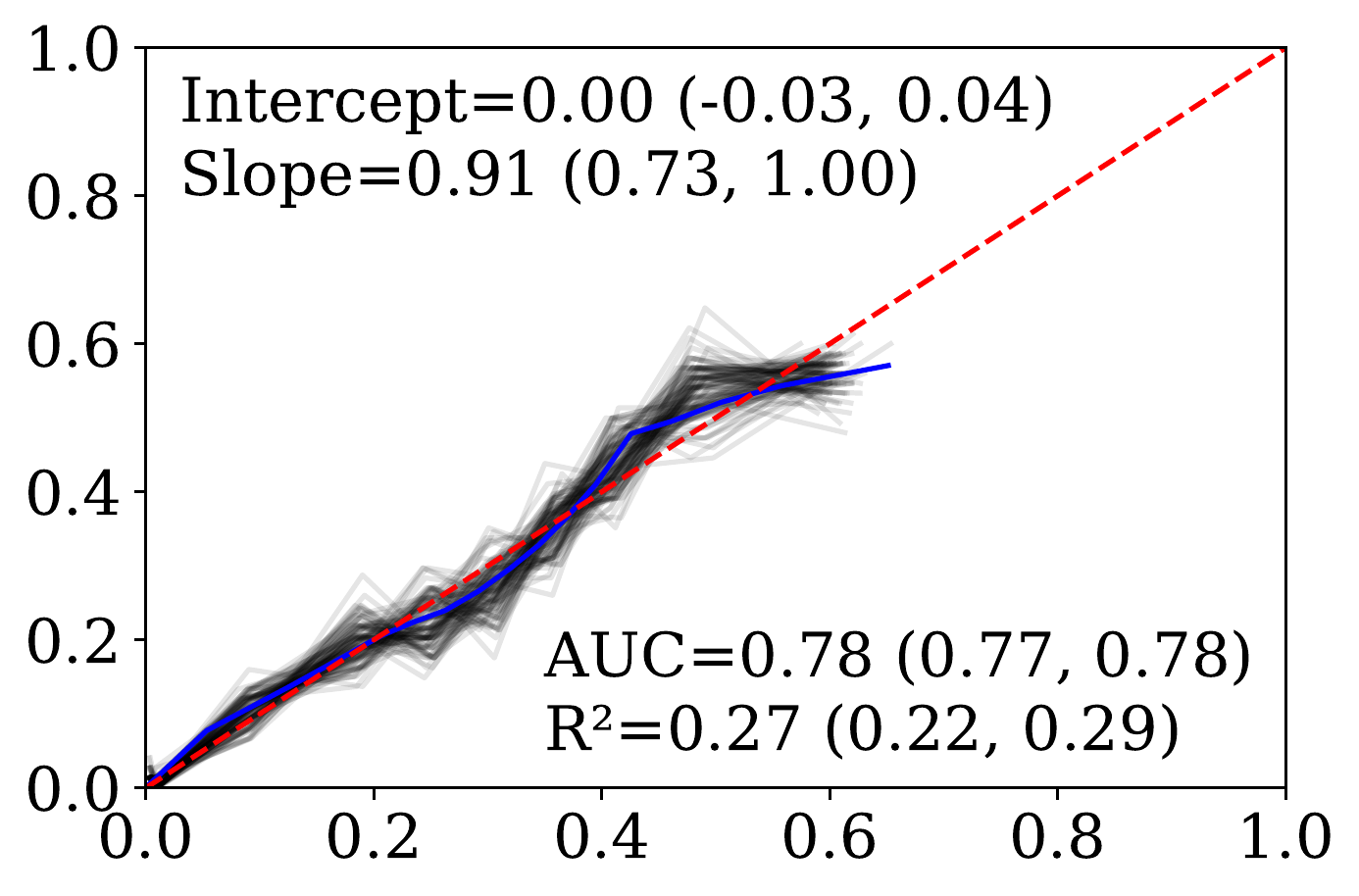} & \includegraphics[width=\plotwidth]{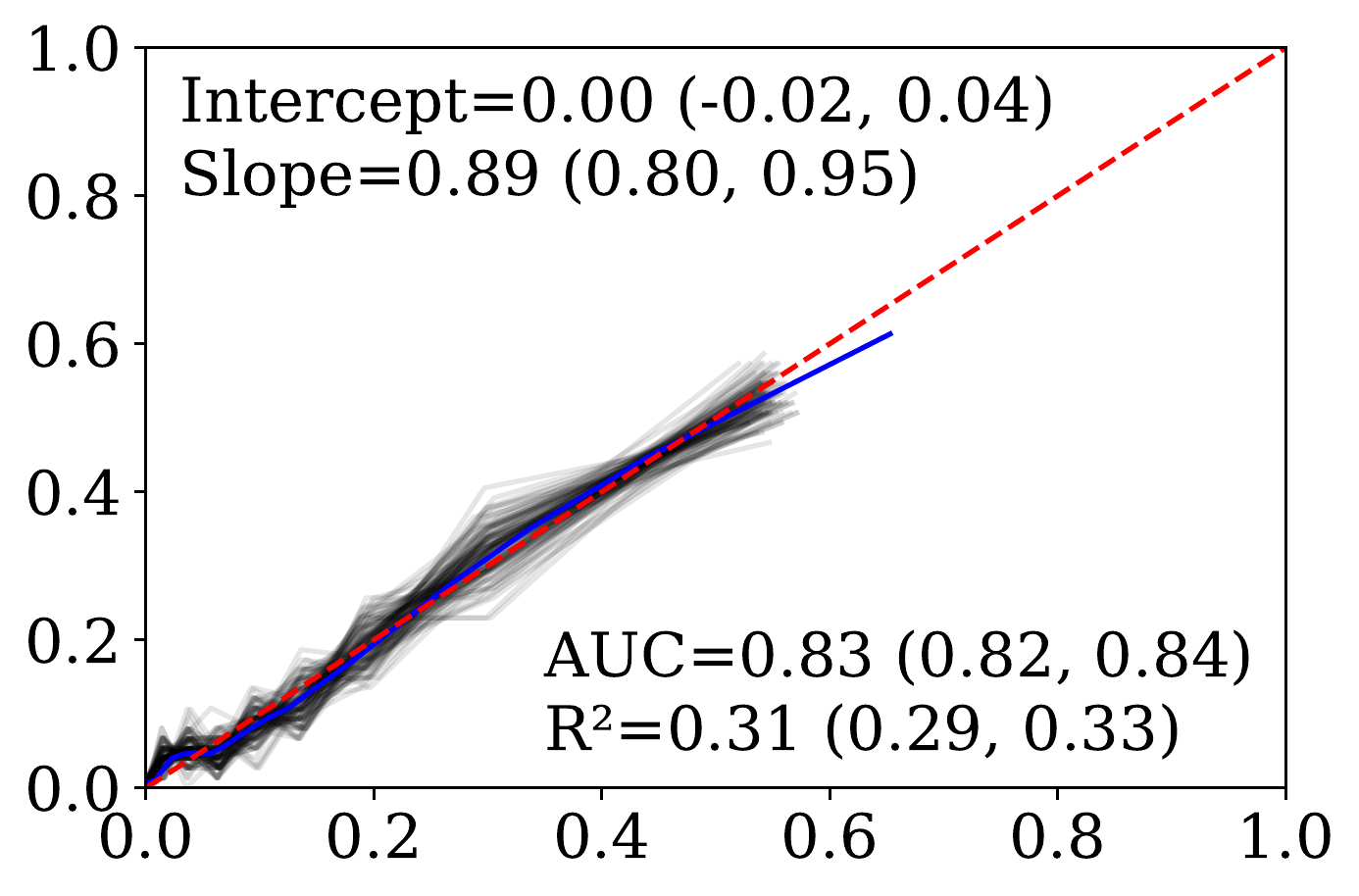} & \includegraphics[width=\plotwidth]{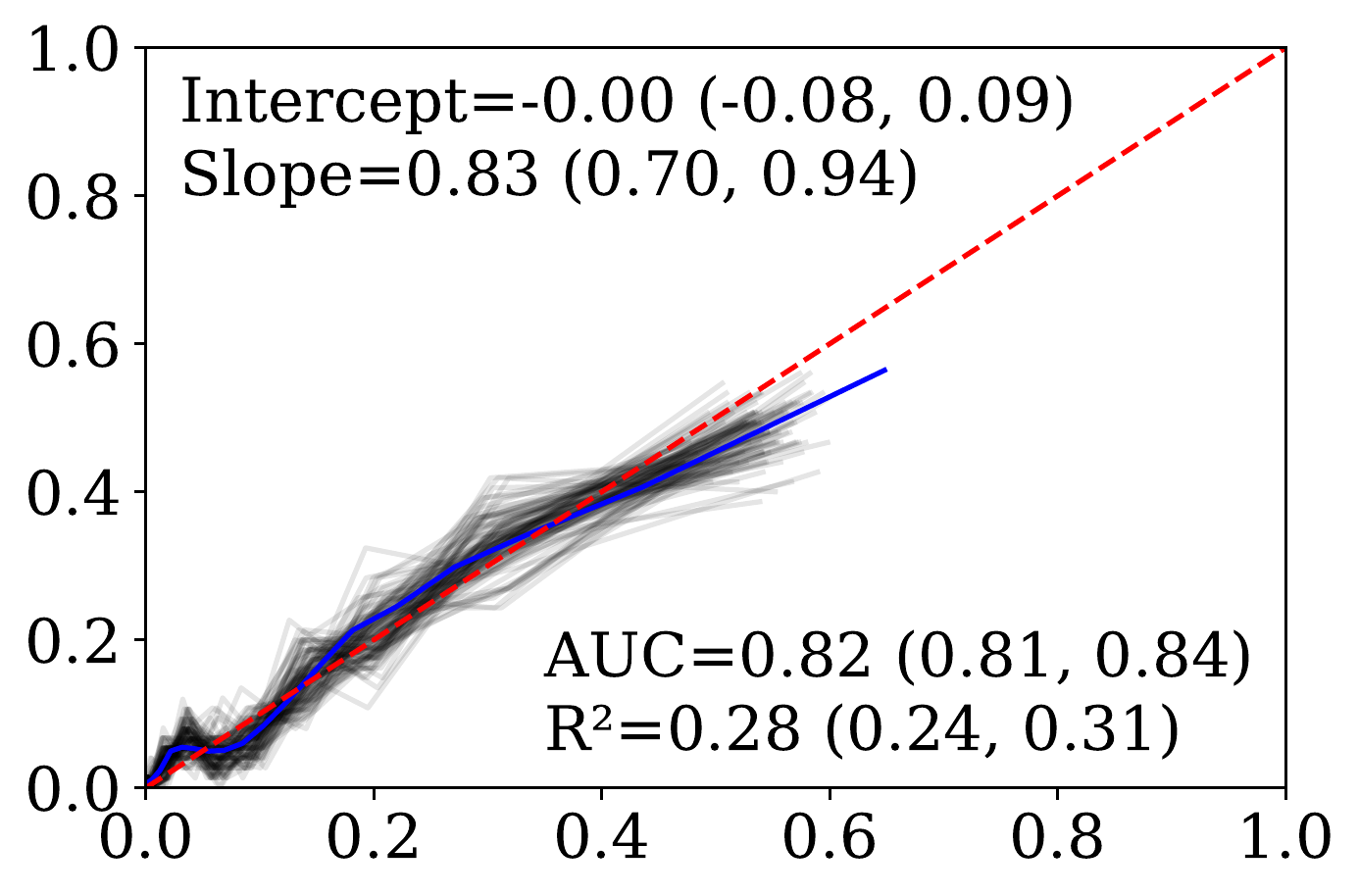}\\
         \centered{Dropout} & \includegraphics[width=\plotwidth]{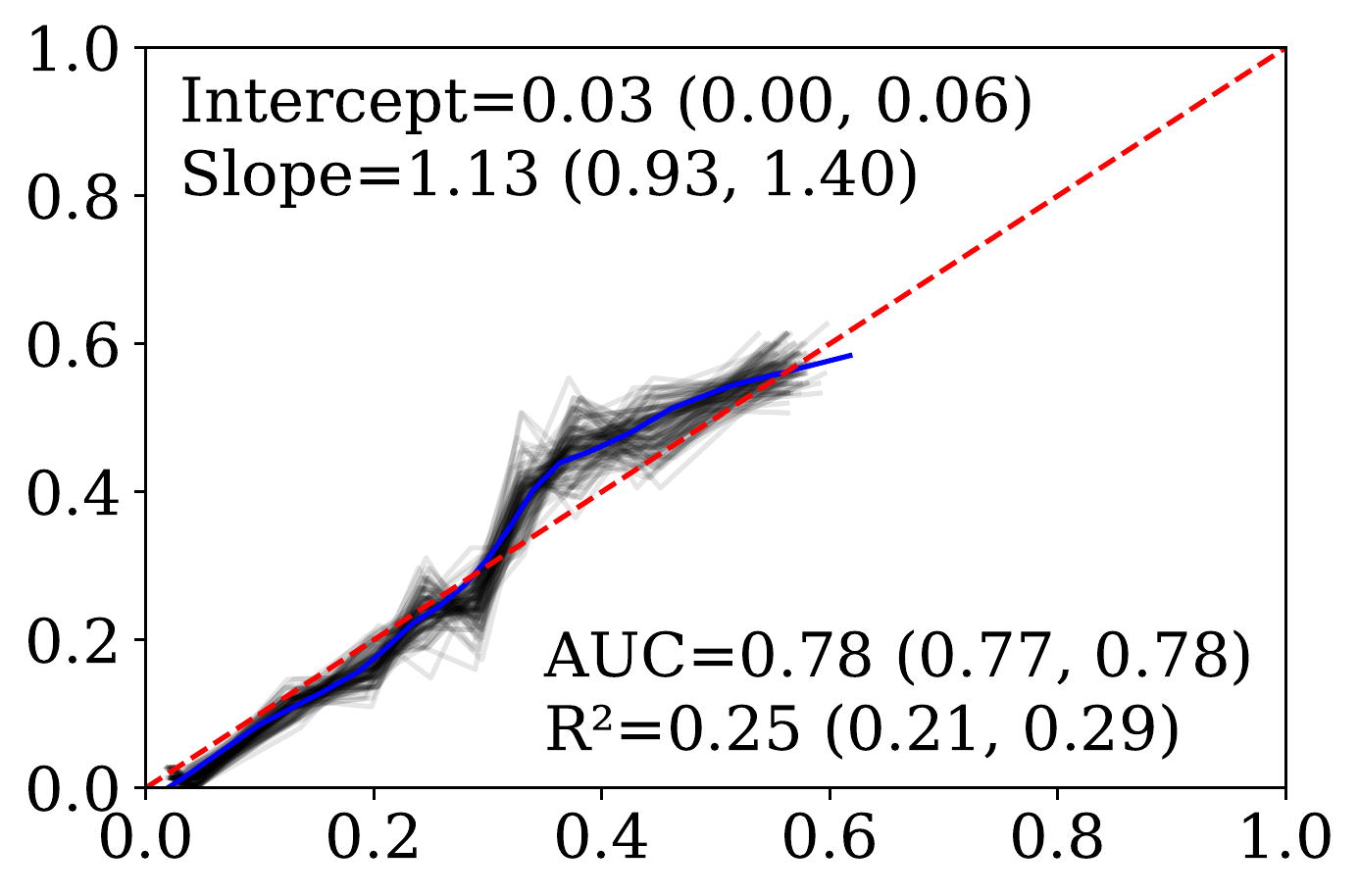} & \includegraphics[width=\plotwidth]{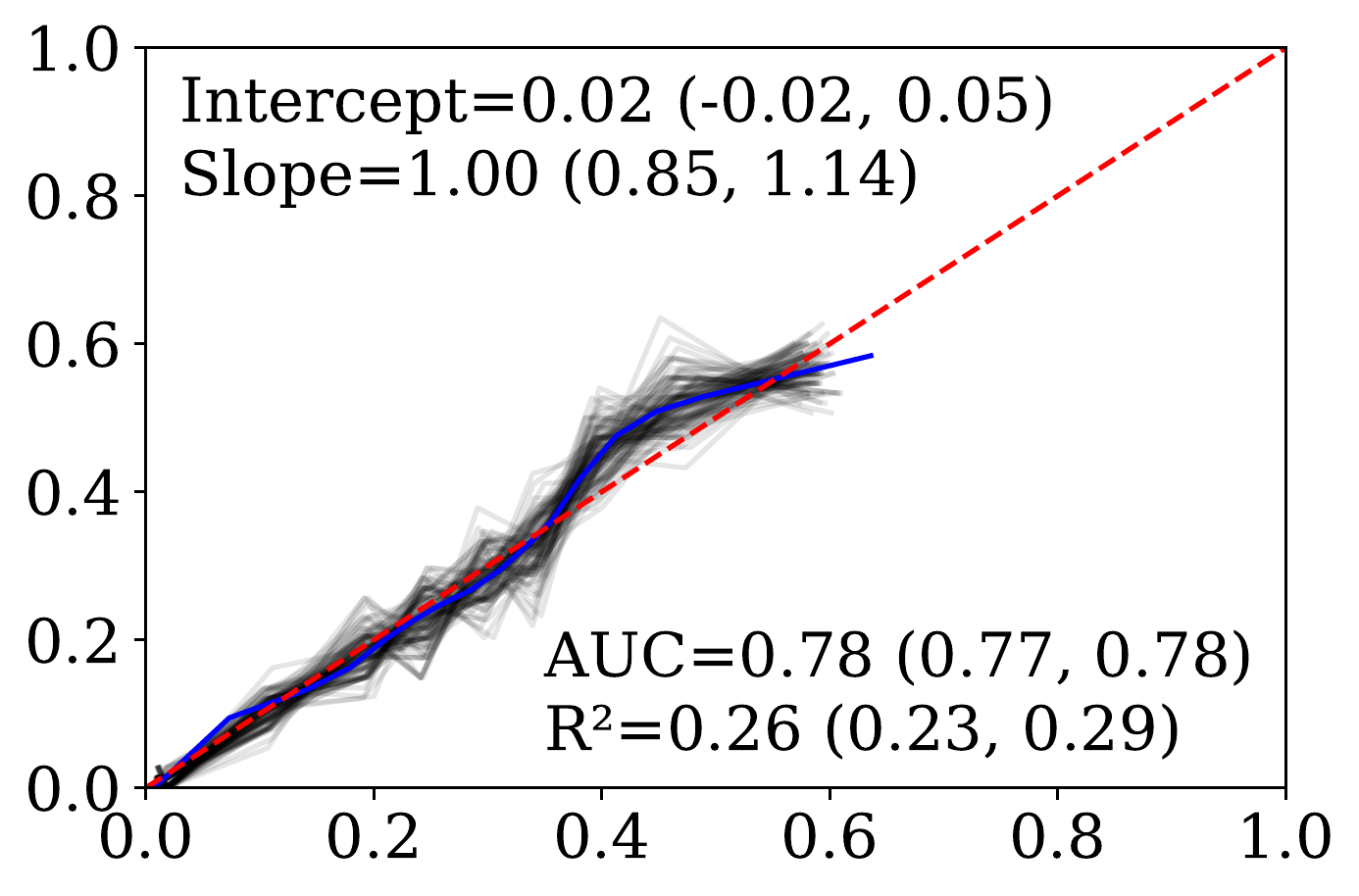} & \includegraphics[width=\plotwidth]{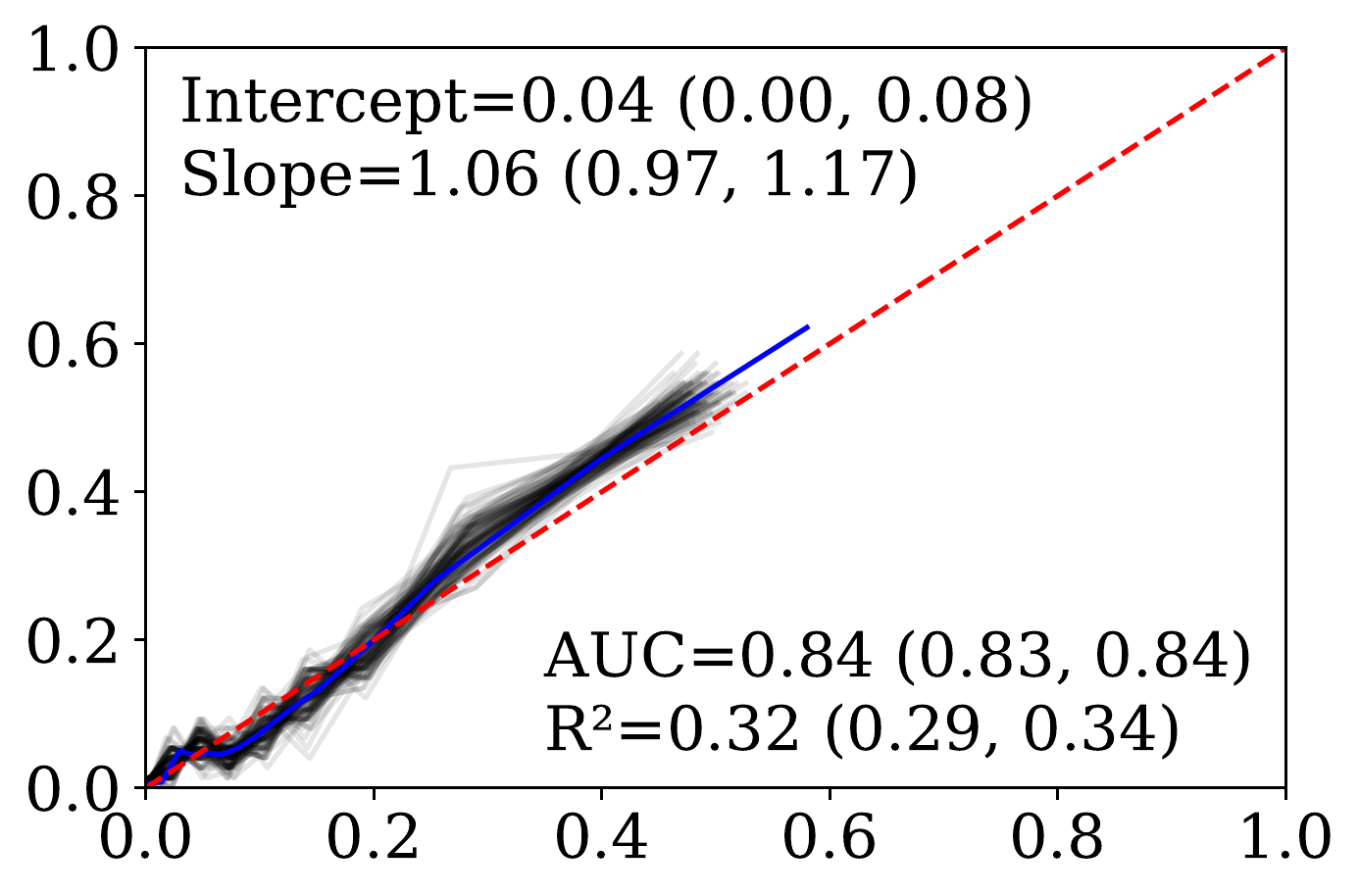} & \includegraphics[width=\plotwidth]{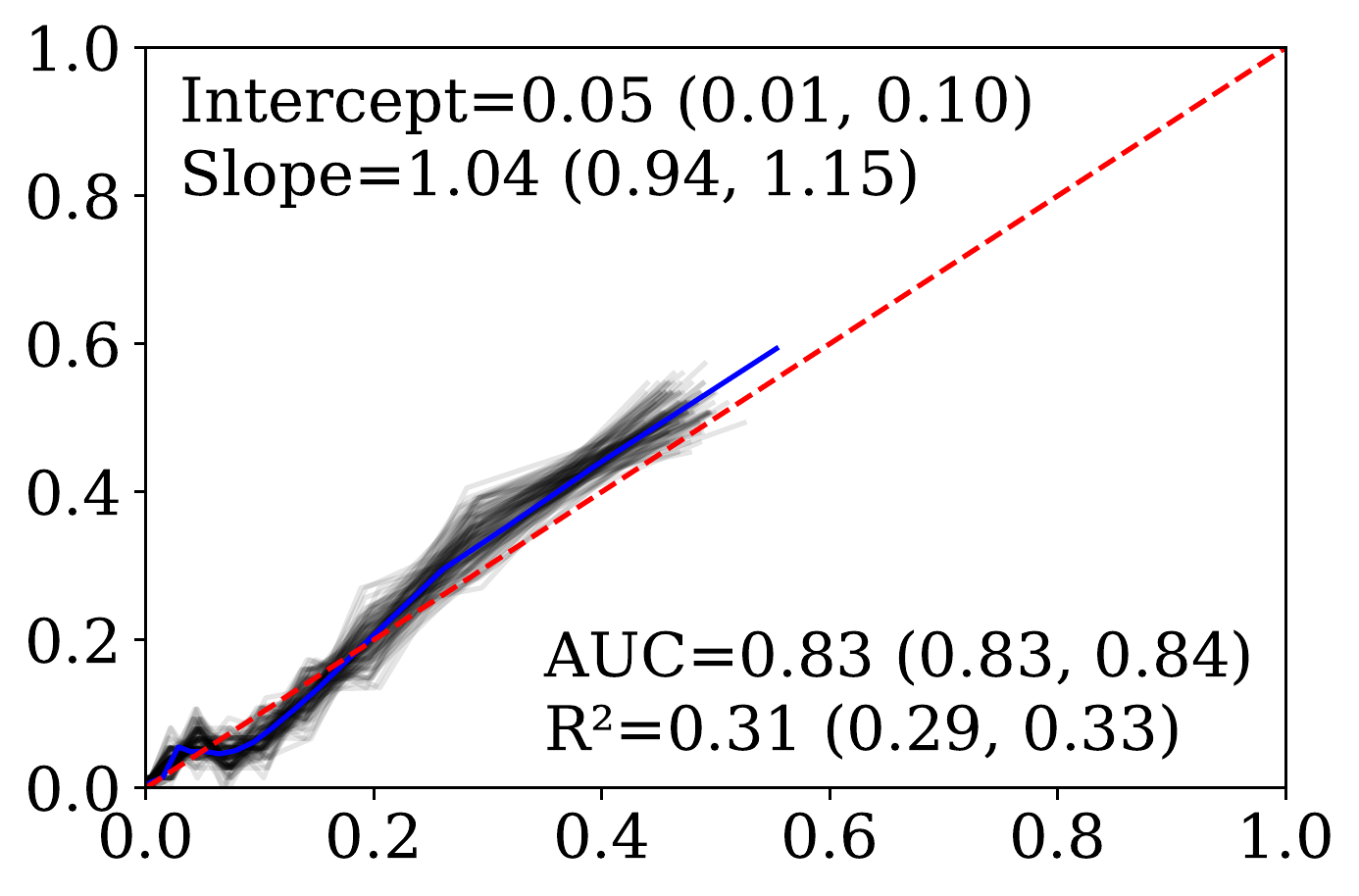}\\
         \centered{\LRnn} & \includegraphics[width=\plotwidth]{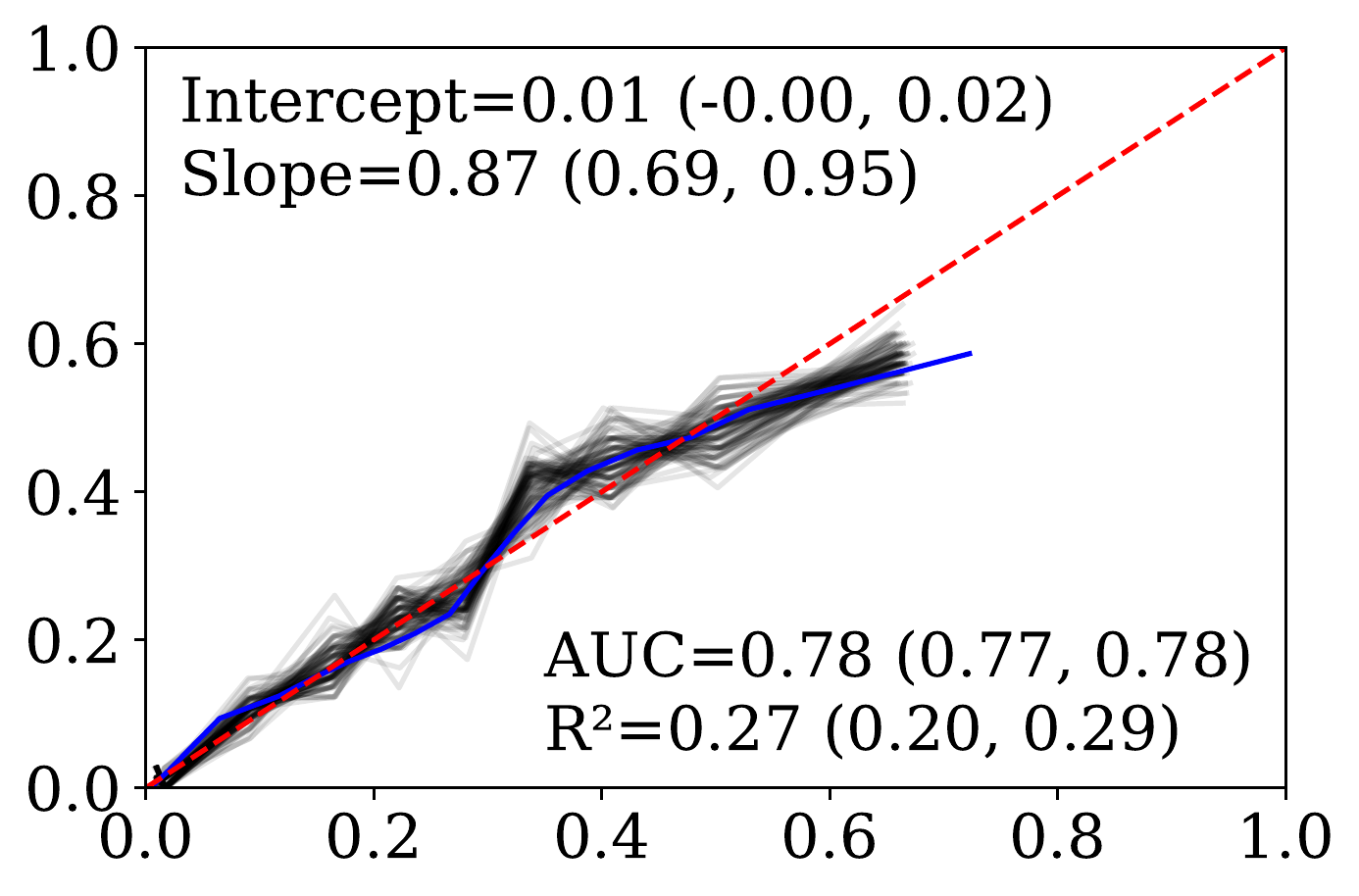} & \includegraphics[width=\plotwidth]{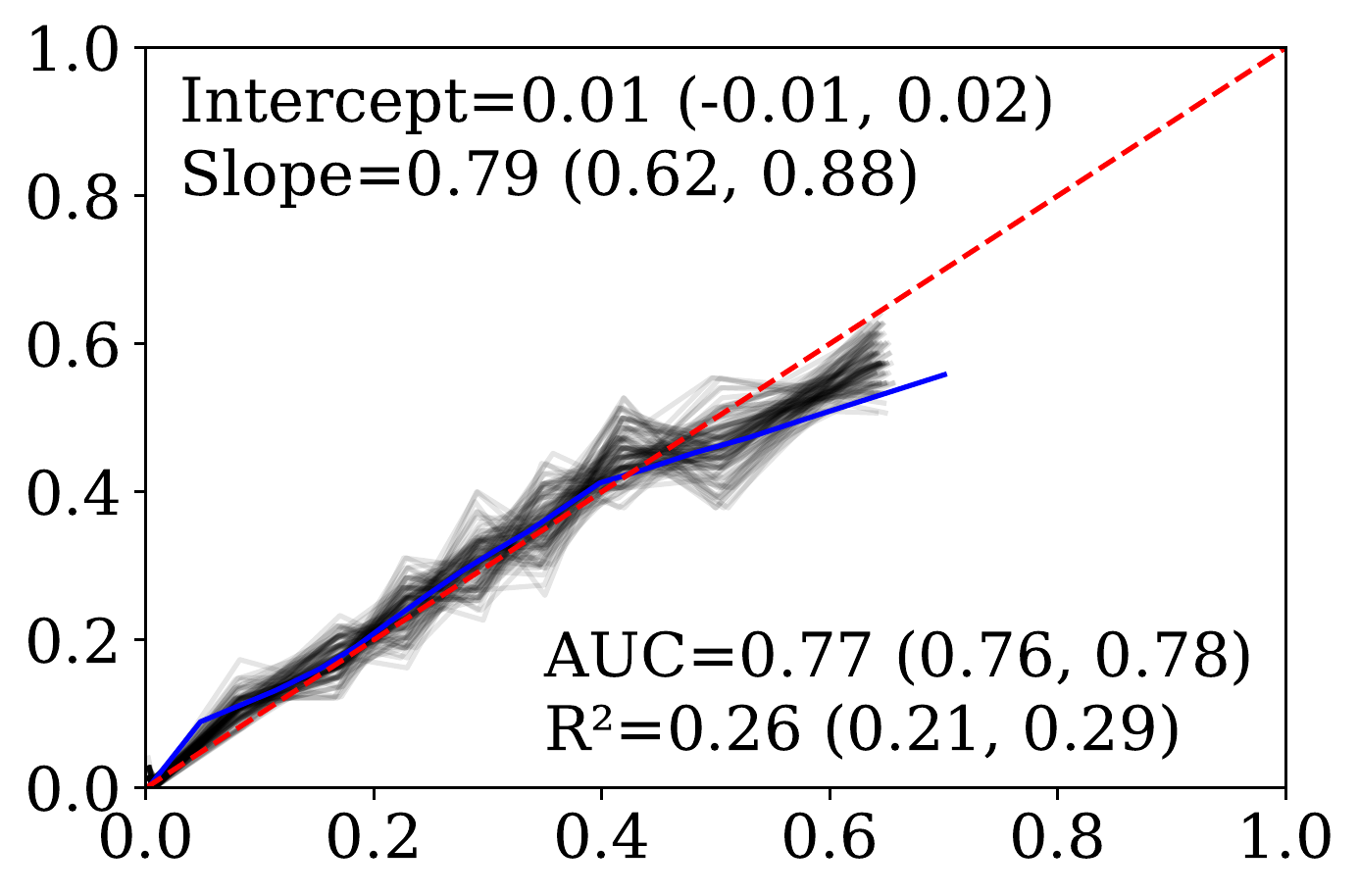} & \includegraphics[width=\plotwidth]{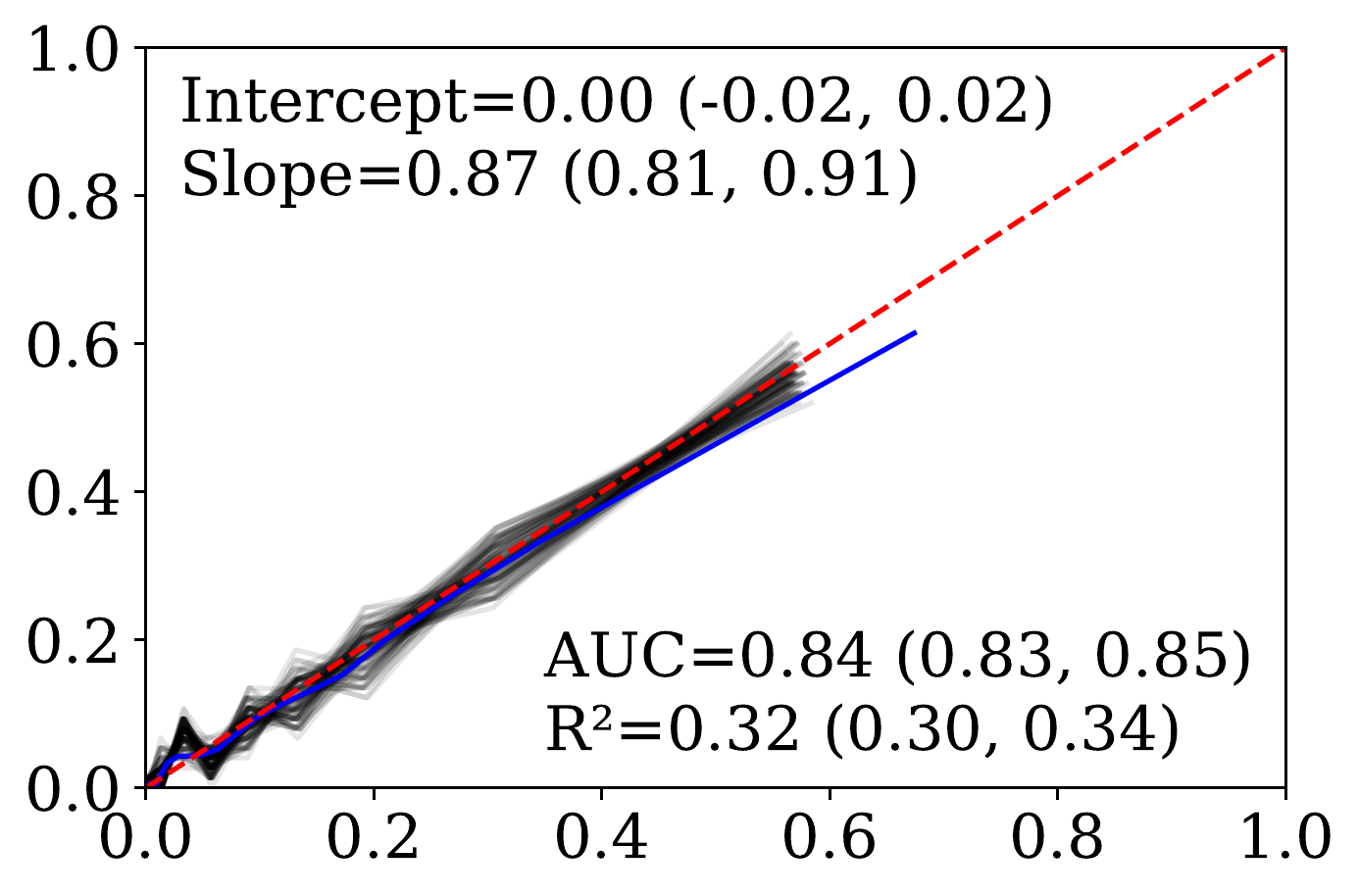} & \includegraphics[width=\plotwidth]{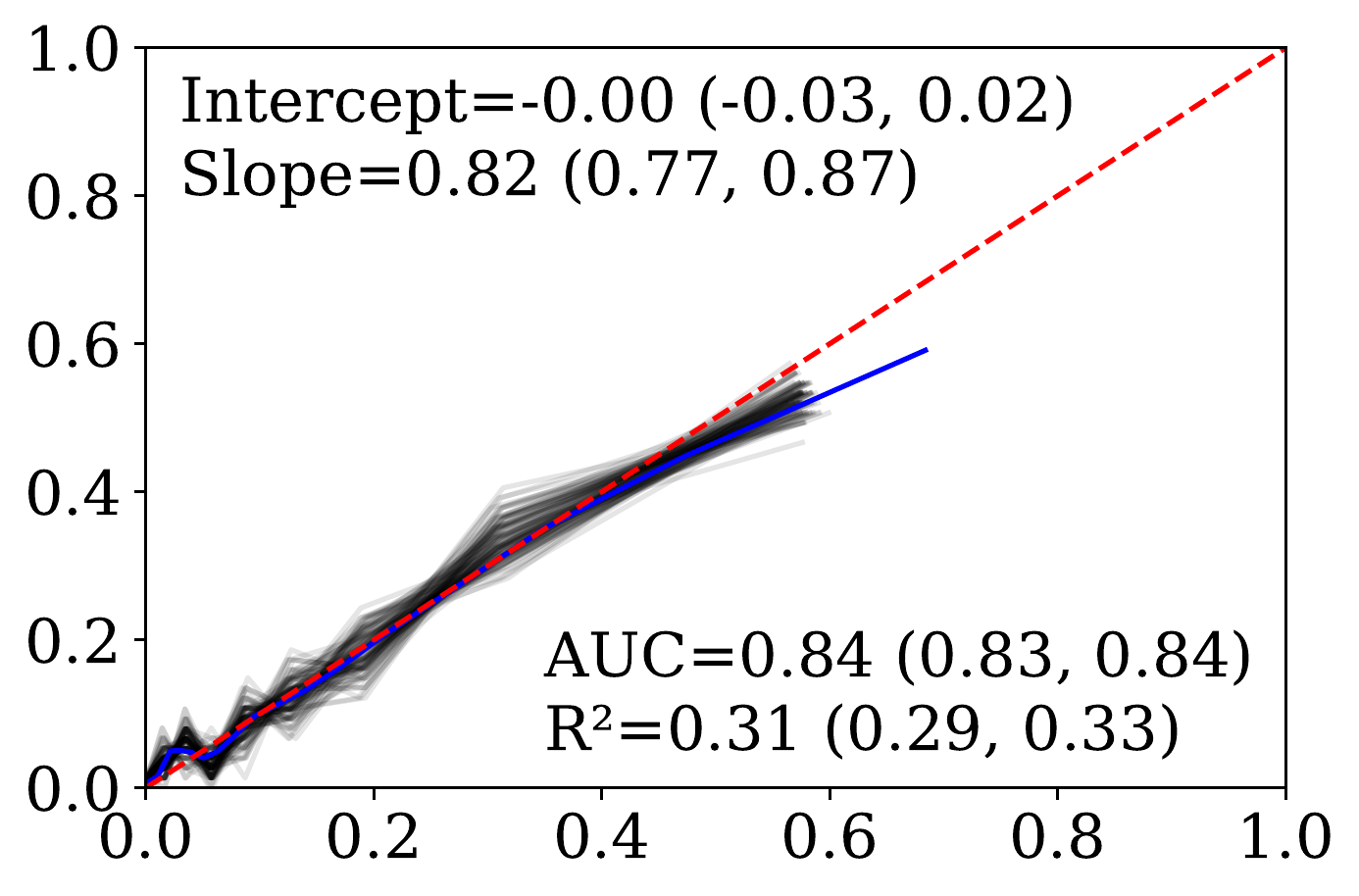}\\
    \end{tabular}}
    \caption{\label{fig:calibration_fans_real}Predictive performance results for the \textbf{real-data xerostomia and dysphagia} settings. Lowess-smoothed calibration curves per simulation are plotted in grey. The calibration curve over all repetitions is shown in blue. Perfect calibration, the diagonal, is dashed in red.}
\end{figure}

%% file: sections/BASELINE-TABS.tex
\newcommand{\tabrowtitle}[1]{#1}
\newcommand{\tabindent}[0]{\hspace{.3cm}}

\begin{table}[h!]
    \centering
     \caption{\label{tab:baseline}General study population characteristics. For age we report the median and standard deviation. For the other variables, we report the number and the percentage of the whole population.}
    \begin{tabular}{llllll}
    \toprule
     & \textbf{UMCG} \\ 
     & (n=740) \\
    \midrule
        \textbf{Patient characteristics (at baseline)}\\
        Age &  63 (10) \\ 
        Gender&\\
        \tabindent{} Men & 552 (75\%) \\ 
        \tabindent{} Women & 188 (25\%) \\ 
        Primary tumor site& \\
        \tabindent{}Pharynx&  356 (48\%) \\ 
        \tabindent{}Larynx & 344 (46\%) \\ 
        \tabindent{}Oral cavity& 40 (5\%) \\ 
        Treatment modality \\
        \tabindent{}Conventional radiotherapy& 159 (21\%)\\ 
        \tabindent{}Accelerated radiotherapy& 287 (39\%)\\ 
        \tabindent{}Chemoradiation &252 (34\%)\\ 
        \tabindent{}Bioradiation & 42 (6\%) \\ 
        T-classification\\
        \tabindent{}Tis-T1 & 136 (18\%) \\ 
        \tabindent{}T2 & 246 (33\%) \\ 
        \tabindent{}T3 & 182 (25\%) \\ 
        \tabindent{}T4 & 176 (24\%) \\ 
        N-classification\\
        \tabindent{}N0 & 356 (48\%) \\ 
        \tabindent{}N1 & 64  (9\%) \\ 
        \tabindent{}N2 & 303 (41\%)\\ 
        \tabindent{}N3 & 17 (2\%)\\ 
        \grayline
        \textbf{Predicted outcomes (at M6)}\\
        Xerostomia $\geq$ Grade 2 &200 (27\%) \\ 
        Dysphagia $\geq$ Grade 2  & 102 (14\%) \\ 
        \bottomrule
    \end{tabular}

\end{table}

\newcommand{\boolcheckmark}{$\{0,1\}$}
\newcommand{\contcheckmark}{$\mathbb{R}$}

\begin{table}[h!]
    \centering
    \caption{\label{tab:variable_sets}Predictors used in each predictor set. \contcheckmark{} refers to continuous predictors, and \boolcheckmark{} refers to binary predictors. PCM refers to the pharyngeal constrictor muscle. Organs at risk were delineated following consensus guidelines.\cite{brouwer2015ct} Dose is expressed in Gray (Gy).} 
\begin{tabular}{lccccc}
   \toprule
   &\xerbasiclowreal{} & \xerexthighreal{} & \dysbasiclowreal{} & \dysexthighreal{}\\
   \textbf{Predictor}&(dim=7) & (dim=19) & (dim=13) & (dim=43) \\\midrule
      \tabrowtitle{Age} & \contcheckmark & \contcheckmark & \contcheckmark & \contcheckmark\\
   \tabrowtitle{Grade 1 complications at baseline} & \boolcheckmark & \boolcheckmark & \boolcheckmark & \boolcheckmark\\
   \tabrowtitle{Grade 2 complications at baseline} & \boolcheckmark & \boolcheckmark & \boolcheckmark & \boolcheckmark\\   
   \tabrowtitle{Submandibular left: mean dose} & \contcheckmark & \contcheckmark & \contcheckmark & \contcheckmark\\
   \tabrowtitle{Submandibular left: V10, V30, V50}& - & \contcheckmark & - & \contcheckmark\\
   \tabrowtitle{Submandibular right: mean dose} & \contcheckmark & \contcheckmark & \contcheckmark & \contcheckmark\\
   \tabrowtitle{Submandibular right: V10, V30, V50}& - & \contcheckmark & - & \contcheckmark\\
   \tabrowtitle{Parotid left: mean dose} & \contcheckmark & \contcheckmark & \contcheckmark & \contcheckmark\\
   \tabrowtitle{Parotid left: V10, V30, V50}& - & \contcheckmark & - & \contcheckmark\\   
   \tabrowtitle{Parotid right: mean dose} & \contcheckmark & \contcheckmark & \contcheckmark & \contcheckmark\\
   \tabrowtitle{Parotid right: V10, V30, V50}& - & \contcheckmark & - & \contcheckmark\\      
   \tabrowtitle{PCM.Superior.Dm} & - & -  & \contcheckmark & \contcheckmark\\
   \tabrowtitle{PCM.Superior V10, 30, 50}& - & -  & - & \contcheckmark\\   
   \tabrowtitle{PCM.Med.Dm} & - & -  & \contcheckmark & \contcheckmark\\
   \tabrowtitle{PCM.Med V10, 30, 50}& - & -  & - & \contcheckmark\\
   \tabrowtitle{PCM.Inferior.Dm} & - & -  & \contcheckmark & \contcheckmark\\
   \tabrowtitle{PCM.Inferior V10, 30, 50}& - & -  & - & \contcheckmark\\
   \tabrowtitle{Supraglottic.Dm} & - & -  & \contcheckmark & \contcheckmark\\
   \tabrowtitle{Supraglottic V10, 30, 50}& - & -  & - & \contcheckmark\\
   \tabrowtitle{OralCavity.Ext.Dm} & - & -  & \contcheckmark & \contcheckmark\\
   \tabrowtitle{OralCavity.Ext. V10, 30, 50}& - & -  & - & \contcheckmark\\
   \tabrowtitle{GlotticArea.Dm} & - & -  & \contcheckmark & \contcheckmark\\
   \tabrowtitle{GlotticArea V10, 30, 50}& - & -  & - & \contcheckmark\\   
 \bottomrule
    \end{tabular}
    
\end{table}

%% file: sections/APP-METHOD-DEFS.tex
In this section we provide a concise but more precise description of each method used in the experiments of this paper, to facilitate easier reproducibility.

\subsubsection{The data}

In this appendix, we use matrix $X = \langle x_1, x_2, ..., x_n \rangle$ of dimension $n \times d$ to describe the $n$ feature vectors representing the $n$ patients. Each feature vector $x_i$ is of dimension $d$. The outcome vector $y$ is used to denote the $n$ outcome labels (one or zero), where $y_i$ is the value at dimension $i$ of $y$, corresponding to the observed ("ground truth") outcome of $x_i$.

\subsubsection{The link function}
With the term \textit{link function}, we refer to the (parameterized) function that maps the predictor values to predicted outcome probability. 
For all methods, the link function used to calculate predicted outcome $\hat{y_i}$ from input $x_i$ is of logistic regression form: 
\begin{align}
\label{eq:pred_func}
    f(x_i, \beta, \beta_0) = \frac{1}{e^{-(\beta x_i  + \beta_0)}}
\end{align}
, where $\beta=\langle \beta_1, \beta_2, ..., \beta_d \rangle$ is the vector of coefficients, and $\beta_0$ the intercept.

\subsubsection{The objective function}
The \textit{objective function} defines what are considered \textit{optimal} coefficients for each method (in some literature referred to as the training criterion). A method's objective function is generally defined to return coefficients $\beta$ and intercept $\beta_0$ that minimize the training error, which is given by a \textit{loss function} (in some literature referred to as the cost function). Since all methods in this study have logistic regression form, the only difference between the compared methods is how coefficients are obtained from the data, i.e., they differ only in their objective function.
The component of the objective function that is shared by all methods is the cross-entropy loss (i.e., the negative log-likelihood):
\newcommand{\mlloss}{L_{\text{ML}}(\beta, \beta_0)}

\begin{align}
    \mlloss{} = - \sum_{i=1}^{n} y \log f(x_i, \beta, \beta_0) + (1-y) \log (1 - f(x_i, \beta, \beta_0))
\end{align}
For our baseline model, regular logistic regression, the corresponding objective consists purely of finding coefficient values $\beta^*$ and intercept value $\beta_0^*$ that minimize the cross-entropy (equivalent to maximum likelihood):
\begin{align}
    \beta^*, \beta_0^* = \argmin_{\beta,  \beta_0} \mlloss{}
\end{align}
In the sections below, we will specify for each method how the corresponding objective differs.

\subsubsection{Loss minimization and hyperparameter tuning}
\label{appendix:loss_minimization}
Unless specified otherwise, each method's loss function is minimized using Adam \cite{kingma2014adam}, for at most 1000 epochs, using early stopping \cite{} with a patience of 500 epochs, and a maximum learning rate of 0.1.
For tuning of important hyperparameters that may come with certain methods, we use nested cross-validation, in a Bayesian optimization setting\cite{snoek2012practical}, using 10 iterations of a Gaussian process with a certain prior and range. The range and prior of each hyperparameter are given in the sections below.

\subsubsection{Lasso}

For Lasso, the only difference with regard to the baseline is an extension of the $\mlloss{}$  with a penalty on the size of the coefficient values, the $\ell_1$-norm of the coefficients, shown in Equation \ref{eq:loss_l1}, where $c_{\ell_1}$ is a hyperparameter determining the inverse importance of the $\ell_1$-penalty.
\begin{multicols}{2}
\begin{align}
\label{eq:loss_l1}
    L_{\ell_1} = \mlloss{} + \frac{1}{c_{\ell_1}} | \beta |
\end{align}\break
\begin{align}
\label{eq:obj_l1}
    \beta^*, \beta_0^* = \argmin_{\beta,  \beta_0} L_{\ell_1}
\end{align}
\end{multicols}
The tuning range for $c_{\ell_1}$ is [$10^{-3},10^2$], and we used a log-linear prior. The final objective is shown in Equation \ref{eq:obj_l1}.

\subsubsection{Ridge}

For Ridge, the modification with regard to the baseline is very similar as for Lasso, except that the penalty on coefficient size is the $\ell_2$-norm of the $\beta$. This results in loss function, and objective functions \ref{eq:loss_l2}, and \ref{eq:obj_l2} respectively.
\begin{multicols}{2}
\begin{align}
\label{eq:loss_l2}
    L_{\ell_2} = \mlloss{} + \frac{1}{c_{\ell_2}} | \beta |^2
\end{align}\break
\begin{align}
\label{eq:obj_l2}
    \beta^*, \beta_0^* = \argmin_{\beta,  \beta_0} L_{\ell_2}
\end{align}
\end{multicols}

For Ridge, the hyperparameter $c_{\ell_2}$ is tuned with the same range and prior as $c_{\ell_1}$ for Lasso.

\subsubsection{Elastic Net}

Elastic Net is the application of both a Lasso and a Ridge penalty to the model. Consequently, the final loss function, and objective function of Elastic Net can be given by Equations \ref{eq:loss_elasticnet}, and \ref{eq:obj_elasticnet} respectively.

\begin{multicols}{2}
\begin{align}
\label{eq:loss_elasticnet}
    L_{\text{ElasticNet}} = \mlloss{} + \frac{1}{c_{\ell_1}} | \beta | +\frac{1}{c_{\ell_2}} | \beta |^2 
\end{align}\break
\begin{align}
\label{eq:obj_elasticnet}
    \beta^*, \beta_0^* = \argmin_{\beta,  \beta_0} L_{\text{ElasticNet}}
\end{align}
\end{multicols}

The tuning procedure of hyperparameters $c_{\ell_1}$ and $c_{\ell_2}$ is the same as for Lasso and Ridge.

\subsubsection{Principal Component Logistic Regression (PLCR)}
In PCLR, the input matrix $X$ is first projected into its principal components, using principal component analysis (PCA), before applying (logistic) regression. PCA aims to find a linear projection $W^*$ that maps $X$ to a (smaller) set of non-correlating latent variables $H := W^* X$ (the principal components) that best capture the variance in $X$. The loss function, and objective to find projection $W^*$ are given by Equations \ref{eq:pca_loss} and \ref{eq:pca_obj} respectively.\cite{udell2015generalized} The number of latent variables (or principal components) is tuned using a linear prior over integer values in [4, $d$].
\begin{multicols}{2}
\begin{align}\label{eq:pca_loss}
    L_{\textsc{pca}}(W) = |X - W^\top W X|^2
\end{align}
\begin{align*}
\text{ }
\end{align*}\break
\begin{align}\label{eq:pca_obj}
    &W^* = \argmin_W L_{\textsc{pca}}(W)
\end{align}
\begin{align*}
\text{subject to } W^\top W = I  
\end{align*}
\end{multicols}
After projecting each input vector $x_i \in X$ to its latent vector $h_i \in H$, a logistic regression $g(h_i, \gamma, \gamma_0)$ is fitted to relate the latent variables $H$ to the outcome $Y$, using coefficients $\gamma$, and intercept $\gamma_0$, following standard maximum likelihood optimization (Equation \ref{eq:ob_g}). 
\begin{align}
\label{eq:ob_g}
    \gamma^*, \gamma_0^* = \argmin_{\gamma,  \gamma_0} L_{\text{ML}}(\gamma, \gamma_0)
\end{align}

In this study, we rewrite PCA projection $W^*$, and logistic regression $g(\cdot)$ to an equivalent link function $f(x_i, \beta, \beta_0)$ as in Equation \ref{eq:pred_func} (this is possible as $W^*$ is a \textit{linear} projection), by setting $\beta := W^\top_{1\leq i \leq h} \gamma^*$, and $\beta_0 := \gamma_0^*$. This way $f(x_i, \beta, \beta_0)$ is equivalent to first projecting $x_i$ to latent vector $h_i$, and afterwards obtaining predicted $\hat{y_i}$ using $g(h_i, \gamma, \gamma_0)$. By doing this, PCLR becomes directly comparable in terms of coefficients to the other methods mentioned in this article.


\subsubsection{Linear Auto-Encoder Logistic Regression (LAELR)}
Linear auto-encoders (LAE) are similar to PCA. The aim of LAE is to find a linear projection $W$ from the input data $X$ to a set of latent variables $H$, that explain the variance in $X$ (in the case of LAE through a linear reconstruction projection $V$). In contrast to PCA, for LAE there is no orthogonality constraint on $H$, and the dimensions of $H$ are not ordered by explained variance. Nevertheless, LAE find projections to the same axis as PCA. \cite{kunin2019loss}
\begin{multicols}{2}
\begin{align}
\label{eq:loss_lae}
    L_{\textsc{lae}}(W,V) = |X - V W X|^2
\end{align}\break
\begin{align}
\label{eq:loss_laelr}
    L_{\textsc{laelr}}(W,V, \gamma, \gamma_0) = L_{\text{ML}}(\gamma, \gamma_0) + \frac{1}{c_{\textsc{lae}}} L_{\textsc{lae}}(W,V)
\end{align}
\end{multicols}

Similar to PCLR, $H$ is related to $Y$ using a logistic regression function  $g(h_i, \gamma, \gamma_0)$, and obtain the final coefficients of $f(x_i, \beta, \beta_0)$ by setting $\beta := W^\top_{1\leq i \leq h} \gamma$, and $\beta_0 := \gamma_0$, and the number of latent variables is tuned using a linear prior over integer values in [4, $d$]. However, the difference between PCLR and LAELR in this study is that instead of optimizing $W$ only on the reconstruction loss (Equation \ref{eq:loss_lae}), we optimize both $W$ and $g(\cdot)$ jointly on the combined loss and corresponding objective, shown in Equations \ref{eq:loss_laelr} and \ref{eq:obj_laelr} respectively. This way, projection $W$ is not purely optimized to explain the variance in $X$, but also, for a part, to facilitate explanation of variance in $Y$. If $c_{\textsc{lae}}$ is very small the objective becomes similar to PCLR, whereas if $c_{\textsc{lae}}$ is very large, the overall objective is similar to standard logistic regression. To empirically balance two loss functions we tune hyperparameter $c_{\textsc{lae}}$ with a log-linear prior in the same range as the penalty of Lasso and Ridge: [$10^{-3},10^2$].

\begin{multicols}{2}
\begin{align}
\label{eq:obj_laelr}
    &W^*, \gamma^*, \gamma_0^* = \argmin_{W, \gamma, \gamma_0} L_{\textsc{laelr}}(W,V, \gamma, \gamma_0)
\end{align}
\end{multicols}

\subsubsection{Dropout regularization}
Dropout training was proposed as a regularization method to prevent co-adaptation of weights in neural networks \cite{hinton2012improving}. Dropout works in iterative gradient-based training procedures, like the one used in the current work (described in Appendix \ref{appendix:loss_minimization}). When using dropout, a sub-model is randomly selected at each training iteration, effectively ``dropping out'' a random percentage $\delta$ of the model's coefficients. This selected sub-model is used to make predictions as part of that training iteration, and the involved coefficients are updated accordingly. Because at each iteration not all coefficients are involved in the model update, co-adaptation of the coefficients is disrupted\tuurfootnote{discouraged}. When training is completed, the coefficients are scaled down by a factor $1-\delta$ to maintain the same expected output of the model during testing as during training (correcting for the fact that during training the full model was never used as a whole). The current work uses the dropout implementation in PyTorch, which is \textit{inverted} dropout (used in most software implementations). In inverted dropout the coefficients are temporarily scaled during training by a factor $\frac{1}{1-\delta}$ instead of after training is completed. This way, no scaling is required when applying the model.
For logistic regression\footnote{For other model architectures than logistic regression, the loss function of dropout is different. The more general formulation is provided in the original article.\cite{hinton2012improving}}, the loss function of dropout can be given by Equation \ref{eq:dropout_loss}, in which we abbreviate $f(x_i, \beta, \beta_0)$ as $f_i$ for clarity.\cite{wager2013dropout} The corresponding objective is given by Equation \ref{eq:dropout_obj}. 

\begin{align}
\label{eq:dropout_loss}
L_{\text{dropout}}(\beta, \beta_0) = \mlloss{} + \frac{1}{2}\frac{\delta}{1-\delta} \sum_{i=1}^{n}\sum_{j=1}^d f_i (1- f_i) x_{ij}^2 \beta_j^2
\end{align}
\begin{align}\label{eq:dropout_obj}
    &W^* = \argmin_W L_{\text{dropout}}(W)
\end{align}


The loss function of dropout for logistic regression models can be summarized in two parts. First, like Ridge, it includes a quadratic penalty on the size of the coefficients, shown on the far right of the equation: $\beta_j^2$. Second, it includes an additional penalty discouraging moderate predictions during training (close to 0.5), shown by $f_i (1-f_i)$. The degree of dropout regularization is determined by hyperparameter $\delta$, which we tune using a linear prior over the interval [0.1, 0.5].

\subsubsection{Non-negative logistic regression (LR$_{\text{NN}}$)}
Sometimes, the coefficient search space can be constrained based on prior knowledge, preventing the model's coefficient estimation procedure from exploring coefficients that are assumed invalid by the modeler. This can help reduce co-adaptation of coefficients, and their inflation, and may improve the model's predictive performance, if the assumption is valid. 

In this study we explore the use of non-negativity constraints on dosage coefficients $\beta_{\textsc{oar}} \subseteq \beta$, as we believe increasing dose to OAR should not result in a decrease in predicted risk of complications. The only difference in this method compared to standard logistic regression is that the feasible coefficient values for all dosage parameters are constrained to the non-negative region during loss minimization, shown in the objective in Eq. \ref{eq:lrnnobj}.

\begin{align}
    \label{eq:lrnnobj}
    \beta^*, \beta_0^* = \argmin_{\beta,  \beta_0} \mlloss{}\text{,   with } \forall_{\beta_i \in \beta_{\text{OAR}}} \beta_i \geq 0
\end{align}

In terms of implementation, we enforce the constraint during our gradient-based minimization through gradient projection \cite{}: setting all negative dosage coefficients to 0 after each coefficient update.

%% file: sections/SAMPLE-SIZE-CALS.tex


In Table \ref{tab:sample_size_calculations}, we report recommended sample size (RSS) calculations. We used the method by Riley et al., (2020)\cite{riley2020calculating}, following their recommendations on the chosen parameters for doing this calculation, reported below in Table \ref{tab:sample_size_calculations}. As expected $R^2_{\text{Nagelkerke}}$ we take the mean of $R^2_{\text{Nagelkerke}}$ values reported in the literature for NTCP models with the same outcomes as the current study: xerostomia\cite{beetz2012ntcp,lee2015patient}, and dysphagia\cite{christianen2016swallowing} six months after radiotherapy.
\begin{table}[h!]
    \centering
    \begin{tabular}{lccccccc}
    \toprule
    Setting & $\Phi$ & $P$ & $\alpha$ & $R^2_{\text{Nagelkerke}}$ & $S$ & $\delta$ & RSS\\\midrule
        \xerbasiclowreal{} &  0.27 & 7 & 0.05 & 0.42 & 0.9 & 0.05 & 315\\
        \xerexthighreal{} & 0.27 & 19 & 0.05 & 0.42 & 0.9 & 0.05 & 794\\
        \dysbasiclowreal{} & 0.14 & 13 & 0.05 & 0.26 & 0.9 & 0.05 & 744\\
        \dysexthighreal{} & 0.14 & 43 & 0.05 & 0.26 & 0.9 & 0.05 & 2460\\
        \bottomrule
    \end{tabular}
    \caption{Parameters used in the sample size calculations for each setting: the anticipated outcome proportion ($\Phi$), the number of predictors ($P$), the absolute margin of error ($\alpha$), Nagelkerke's explained variance ($R^2_{\text{Nagelkerke}}$), the expected uniform shrinkage factor ($S$), and the expected optimism ($\delta$).}
    \label{tab:sample_size_calculations}
\end{table}


%% file: sections/3-TABS-CORR.tex
\newtoggle{inTableHeader}
\toggletrue{inTableHeader}
\newcommand*{\StartTableHeader}{\global\toggletrue{inTableHeader}}%
\newcommand*{\EndTableHeader}{\global\togglefalse{inTableHeader}}%

\let\OldTabular\tabular%
\let\OldEndTabular\endtabular%
\renewenvironment{tabular}{\StartTableHeader\OldTabular}{\OldEndTabular\StartTableHeader}%



\newcommand*{\MinCorrNumber}{-1.0}%
\newcommand*{\MidCorrNumber}{0.0} %
\newcommand*{\MaxCorrNumber}{1}%

\newcommand*{\MinCorrColor}{cyan}%
\newcommand*{\MidCorrColor}{white} %
\newcommand*{\MaxCorrColor}{red}%

\newcommand{\ApplyCorrGradient}[1]{%
  \iftoggle{inTableHeader}{#1}{
    \ifdim #1 pt > \MidCorrNumber pt
        \pgfmathsetmacro{\PercentColor}{max(min(100.0*(#1 - \MidCorrNumber)/(\MaxCorrNumber-\MidCorrNumber),100.0),0.00)} %
        \hspace{-0.33em}\colorbox{\MaxCorrColor!\PercentColor!\MidCorrColor}{\hspace{3pt}#1\hspace{0pt}}
    \else
        \pgfmathsetmacro{\PercentColor}{max(min(100.0*(\MidCorrNumber - #1)/(\MidCorrNumber-\MinCorrNumber),100.0),0.00)} %
        \hspace{-0.33em}\colorbox{\MinCorrColor!\PercentColor!\MidCorrColor}{#1}
    \fi
  }}

\newcolumntype{E}{>{\collectcell\ApplyCorrGradient}l<{\endcollectcell}}
\renewcommand{\arraystretch}{0}
\setlength{\fboxsep}{1mm} 
\setlength{\tabcolsep}{.3pt}

\begin{figure}[h!]
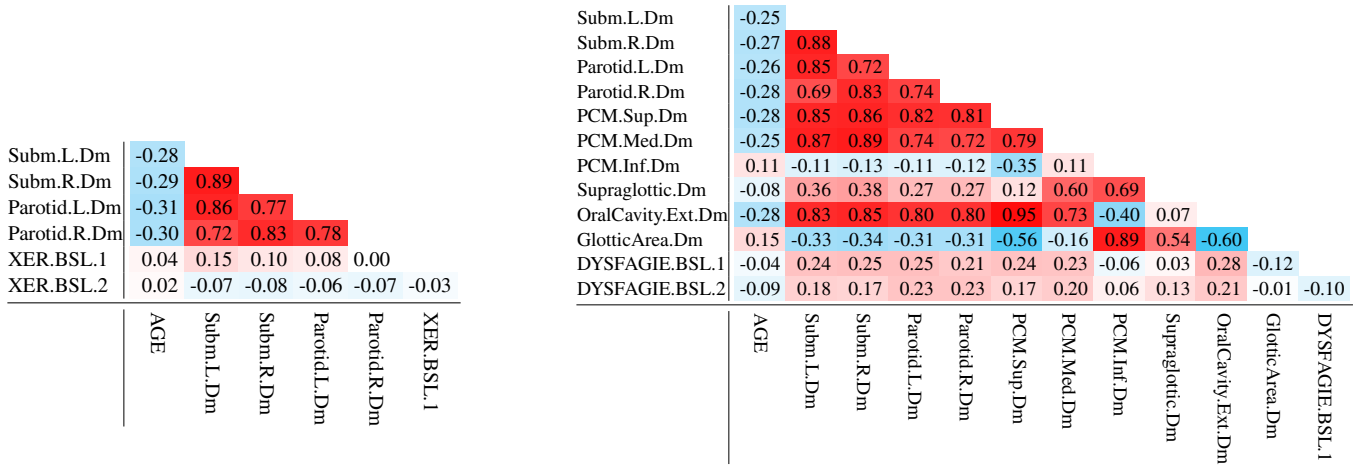

\begin{minipage}{.38\textwidth}
\vspace{1.3cm}
\resizebox{.9\textwidth}{!}{
\begin{tabular}{l|ccccccccc} 
\EndTableHeader \\ 
Subm.L.Dm&&\multicolumn{1}{E}{-0.28}&&&&&&&\\ 
Subm.R.Dm&&\multicolumn{1}{E}{-0.29}&\multicolumn{1}{E}{0.89}&&&&&&\\ 
Parotid.L.Dm&&\multicolumn{1}{E}{-0.31}&\multicolumn{1}{E}{0.86}&\multicolumn{1}{E}{0.77}&&&&&\\ 
Parotid.R.Dm&&\multicolumn{1}{E}{-0.30}&\multicolumn{1}{E}{0.72}&\multicolumn{1}{E}{0.83}&\multicolumn{1}{E}{0.78}&&&&\\ 
XER.BSL.1&&\multicolumn{1}{E}{0.04}&\multicolumn{1}{E}{0.15}&\multicolumn{1}{E}{0.10}&\multicolumn{1}{E}{0.08}&\multicolumn{1}{E}{0.00}&&&\\ 
XER.BSL.2&&\multicolumn{1}{E}{0.02}&\multicolumn{1}{E}{-0.07}&\multicolumn{1}{E}{-0.08}&\multicolumn{1}{E}{-0.06}&\multicolumn{1}{E}{-0.07}&\multicolumn{1}{E}{-0.03}&&\\ 
\midrule&\textcolor{white}{.}&\rotatebox{270}{{AGE}}&\rotatebox{270}{{Subm.L.Dm}}&\rotatebox{270}{{Subm.R.Dm}}&\rotatebox{270}{{Parotid.L.Dm}}&\rotatebox{270}{{Parotid.R.Dm}}&\rotatebox{270}{{XER.BSL.1}}
\end{tabular} }
\end{minipage}
\hfill
\begin{minipage}{.58\textwidth}
\resizebox{\textwidth}{!}{
\begin{tabular}{l|ccccccccccccccc} 
\EndTableHeader \\ 
Subm.L.Dm&&\multicolumn{1}{E}{-0.25}&&&&&&&&&&&&&\\ 
Subm.R.Dm&&\multicolumn{1}{E}{-0.27}&\multicolumn{1}{E}{0.88}&&&&&&&&&&&&\\ 
Parotid.L.Dm&&\multicolumn{1}{E}{-0.26}&\multicolumn{1}{E}{0.85}&\multicolumn{1}{E}{0.72}&&&&&&&&&&&\\ 
Parotid.R.Dm&&\multicolumn{1}{E}{-0.28}&\multicolumn{1}{E}{0.69}&\multicolumn{1}{E}{0.83}&\multicolumn{1}{E}{0.74}&&&&&&&&&&\\ 
PCM.Sup.Dm&&\multicolumn{1}{E}{-0.28}&\multicolumn{1}{E}{0.85}&\multicolumn{1}{E}{0.86}&\multicolumn{1}{E}{0.82}&\multicolumn{1}{E}{0.81}&&&&&&&&&\\ 
PCM.Med.Dm&&\multicolumn{1}{E}{-0.25}&\multicolumn{1}{E}{0.87}&\multicolumn{1}{E}{0.89}&\multicolumn{1}{E}{0.74}&\multicolumn{1}{E}{0.72}&\multicolumn{1}{E}{0.79}&&&&&&&&\\ 
PCM.Inf.Dm&&\multicolumn{1}{E}{0.11}&\multicolumn{1}{E}{-0.11}&\multicolumn{1}{E}{-0.13}&\multicolumn{1}{E}{-0.11}&\multicolumn{1}{E}{-0.12}&\multicolumn{1}{E}{-0.35}&\multicolumn{1}{E}{0.11}&&&&&&&\\
\end{tabular} }
\end{minipage}
    \caption{\label{fig:corr_xer_dys_basic}Correlation matrix for the \xerbasiclow{} setting (left), and the \dysbasiclow{} setting (right).}
\end{figure}

\begin{figure}[h!]
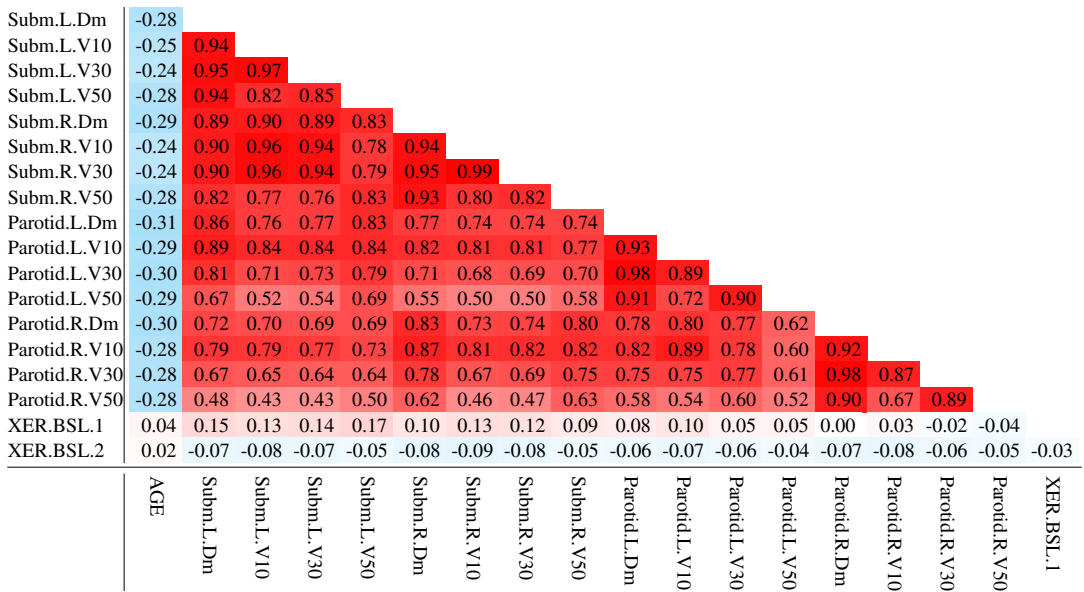

\resizebox{.8\textwidth}{!}{
\begin{tabular}{l|ccccccccccccccccccccc} 
\EndTableHeader \\ 
Subm.L.Dm&&\multicolumn{1}{E}{-0.28}&&&&&&&&&&&&&&&&&&&\\ 
Subm.L.V10&&\multicolumn{1}{E}{-0.25}&\multicolumn{1}{E}{0.94}&&&&&&&&&&&&&&&&&&\\ 
Subm.L.V30&&\multicolumn{1}{E}{-0.24}&\multicolumn{1}{E}{0.95}&\multicolumn{1}{E}{0.97}&&&&&&&&&&&&&&&&&\\ 
Subm.L.V50&&\multicolumn{1}{E}{-0.28}&\multicolumn{1}{E}{0.94}&\multicolumn{1}{E}{0.82}&\multicolumn{1}{E}{0.85}&&&&&&&&&&&&&&&&\\ 
Subm.R.Dm&&\multicolumn{1}{E}{-0.29}&\multicolumn{1}{E}{0.89}&\multicolumn{1}{E}{0.90}&\multicolumn{1}{E}{0.89}&\multicolumn{1}{E}{0.83}&&&&&&&&&&&&&&&\\ 
Subm.R.V10&&\multicolumn{1}{E}{-0.24}&\multicolumn{1}{E}{0.90}&\multicolumn{1}{E}{0.96}&\multicolumn{1}{E}{0.94}&\multicolumn{1}{E}{0.78}&\multicolumn{1}{E}{0.94}&&&&&&&&&&&&&&\\ 
Subm.R.V30&&\multicolumn{1}{E}{-0.24}&\multicolumn{1}{E}{0.90}&\multicolumn{1}{E}{0.96}&\multicolumn{1}{E}{0.94}&\multicolumn{1}{E}{0.79}&\multicolumn{1}{E}{0.95}&\multicolumn{1}{E}{0.99}&&&&&&&&&&&&&\\
\end{tabular} }
    \caption{Correlation matrix for the \xerexthigh{} setting.}
    \label{fig:corr_xer_extended}
\end{figure}



\newpage\clearpage

\begin{landscape}
\begin{figure}[h!]
\resizebox{1.3\textwidth}{!}{
\begin{tabular}{l|ccccccccccccccccccccccccccccccccccccccccccccc} 
\EndTableHeader \\ 
Subm.L.Dm&&\multicolumn{1}{E}{-0.25}&&&&&&&&&&&&&&&&&&&&&&&&&&&&&&&&&&&&&&&&&&&\\ 
Subm.L.V10&&\multicolumn{1}{E}{-0.22}&\multicolumn{1}{E}{0.93}&&&&&&&&&&&&&&&&&&&&&&&&&&&&&&&&&&&&&&&&&&\\ 
Subm.L.V30&&\multicolumn{1}{E}{-0.23}&\multicolumn{1}{E}{0.95}&\multicolumn{1}{E}{0.96}&&&&&&&&&&&&&&&&&&&&&&&&&&&&&&&&&&&&&&&&&\\ 
Subm.L.V50&&\multicolumn{1}{E}{-0.23}&\multicolumn{1}{E}{0.94}&\multicolumn{1}{E}{0.80}&\multicolumn{1}{E}{0.85}&&&&&&&&&&&&&&&&&&&&&&&&&&&&&&&&&&&&&&&&\\ 
Subm.R.Dm&&\multicolumn{1}{E}{-0.27}&\multicolumn{1}{E}{0.88}&\multicolumn{1}{E}{0.91}&\multicolumn{1}{E}{0.88}&\multicolumn{1}{E}{0.81}&&&&&&&&&&&&&&&&&&&&&&&&&&&&&&&&&&&&&&&\\ 
Subm.R.V10&&\multicolumn{1}{E}{-0.22}&\multicolumn{1}{E}{0.90}&\multicolumn{1}{E}{0.97}&\multicolumn{1}{E}{0.93}&\multicolumn{1}{E}{0.77}&\multicolumn{1}{E}{0.93}&&&&&&&&&&&&&&&&&&&&&&&&&&&&&&&&&&&&&&\\ 
Subm.R.V30&&\multicolumn{1}{E}{-0.22}&\multicolumn{1}{E}{0.89}&\multicolumn{1}{E}{0.96}&\multicolumn{1}{E}{0.93}&\multicolumn{1}{E}{0.78}&\multicolumn{1}{E}{0.95}&\multicolumn{1}{E}{0.98}&&&&&&&&&&&&&&&&&&&&&&&&&&&&&&&&&&&&&\\
\end{tabular} }
    \caption{Correlation matrix for the \dysexthigh{} setting.}
    \label{fig:corr_dys_ext}
\end{figure}
\end{landscape}

%% file: sections/COEFFS.tex
\newcommand*{\MinCoefNumber}{-2.0}%
\newcommand*{\MidCoefNumber}{0.0} %
\newcommand*{\MaxCoefNumber}{1}%

\newcommand*{\MinCoefColor}{cyan}%
\newcommand*{\MidCoefColor}{white} %
\newcommand*{\MaxCoefColor}{red}%

\newcommand{\ApplyCoefGradient}[1]{%
  \iftoggle{inTableHeader}{#1}{
    \ifdim #1 pt > \MidCoefNumber pt
        \pgfmathsetmacro{\PercentColor}{max(min(100.0*(#1 - \MidCoefNumber)/(\MaxCoefNumber-\MidCoefNumber),100.0),0.00)} %
        \hspace{-0.33em}\colorbox{\MaxCoefColor!\PercentColor!\MidCoefColor}{\hspace{3pt}#1\hspace{0pt}}
    \else
        \pgfmathsetmacro{\PercentColor}{max(min(100.0*(\MidCoefNumber - #1)/(\MidCoefNumber-\MinCoefNumber),100.0),0.00)} %
        \hspace{-0.33em}\colorbox{\MinCoefColor!\PercentColor!\MidCoefColor}{#1}
    \fi
  }}

\newcolumntype{R}{>{\collectcell\ApplyCoefGradient}l<{\endcollectcell}}
\renewcommand{\arraystretch}{0}
\setlength{\tabcolsep}{2pt}

\input{sections/COEFFS-SIM}

\input{sections/COEFFS-REAL}

%% file: sections/COEFFS-SIM.tex

\begin{table}[h!]
    \caption{\label{tab:coef:xerbasiclow}Mean model coefficients for \xerbasiclow{}.}
\resizebox{\textwidth}{!}{
\input{img/coeff_tables/CompareMethods-sim-A}}
\end{table}

\begin{table}[h!]
    \caption{\label{tab:coef:xerbasichigh}Mean model coefficients for \xerbasichigh{}.}
\resizebox{\textwidth}{!}{
\input{img/coeff_tables/CompareMethods-sim-A+}}
\end{table}

\begin{table}[h!]
    \caption{\label{tab:coef:xerextlow}Mean model coefficients for \xerextlow{}.}
\resizebox{\textwidth}{!}{
\input{img/coeff_tables/CompareMethods-sim-B}}
\end{table}

\begin{table}[h!]
    \caption{\label{tab:coef:xerexthigh}Mean model coefficients for \xerexthigh{}.}
\resizebox{\textwidth}{!}{
\input{img/coeff_tables/CompareMethods-sim-B+}}
\end{table}

\begin{table}[h!]
    \caption{\label{tab:coef:dysbasiclow}Mean model coefficients for \dysbasiclow{}.}
\resizebox{\textwidth}{!}{
\input{img/coeff_tables/CompareMethods-sim-C}}
\end{table}

\begin{table}[h!]
    \caption{\label{tab:coef:dysbasichigh}Mean model coefficients for \dysbasichigh{}.}
\resizebox{\textwidth}{!}{
\input{img/coeff_tables/CompareMethods-sim-C+}}
\end{table}

\begin{table}[h!]
    \caption{\label{tab:coef:dysextlow}Mean model coefficients for \dysextlow{}.}
\resizebox{\textwidth}{!}{
\input{img/coeff_tables/CompareMethods-sim-D}}
\end{table}

\begin{table}[h!]
    \caption{\label{tab:coef:dysexthigh}Mean model coefficients for \dysexthigh{}.}
\resizebox{\textwidth}{!}{
\input{img/coeff_tables/CompareMethods-sim-D+}}
\end{table}

%% file: img/coeff_tables/CompareMethods-sim-A.tex
\begin{tabular}{lRlRlRlRlRlRlRlRlRlRlRlRlRlRlRlRlR}\toprule
 &\multicolumn{2}{l}{LR}&\multicolumn{2}{l}{Lasso}&\multicolumn{2}{l}{Ridge}&\multicolumn{2}{l}{ElasticNet}&\multicolumn{2}{l}{PCLR}&\multicolumn{2}{l}{LAELR}&\multicolumn{2}{l}{Dropout}&\multicolumn{2}{l}{LRnn}\EndTableHeader\\
\midrule \negsum&-0.13&$^{(-0.71, 0.00)}$&-0.01&$^{(0.00, 0.00)}$&-0.01&$^{(-0.09, 0.00)}$&0.00&$^{(0.00, 0.00)}$&-0.04&$^{(-0.66, 0.00)}$&-0.03&$^{(-0.42, 0.00)}$&-0.00&$^{(0.00, 0.00)}$&0.00&$^{(0.00, 0.00)}$\\
\possum&1.48&$^{(1.10, 2.04)}$&1.24&$^{(0.94, 1.52)}$&1.20&$^{(0.93, 1.56)}$&1.20&$^{(0.94, 1.50)}$&1.38&$^{(1.06, 2.04)}$&1.37&$^{(1.06, 1.94)}$&1.20&$^{(0.83, 1.55)}$&1.36&$^{(1.09, 1.63)}$\\
\negprop&0.16&$^{(0.00, 0.50)}$&0.01&$^{(0.00, 0.00)}$&0.01&$^{(0.00, 0.25)}$&0.00&$^{(0.00, 0.00)}$&0.04&$^{(0.00, 0.50)}$&0.03&$^{(0.00, 0.25)}$&0.00&$^{(0.00, 0.00)}$&0.00&$^{(0.00, 0.00)}$\\
\posprop&0.83&$^{(0.50, 1.00)}$&0.86&$^{(0.50, 1.00)}$&0.99&$^{(0.75, 1.00)}$&0.96&$^{(0.75, 1.00)}$&0.96&$^{(0.50, 1.00)}$&0.96&$^{(0.75, 1.00)}$&1.00&$^{(1.00, 1.00)}$&0.84&$^{(0.50, 1.00)}$\\
\midrule Intercept&-1.32&$^{(-1.58, -1.12)}$&-1.27&$^{(-1.53, -1.07)}$&-1.26&$^{(-1.52, -1.08)}$&-1.26&$^{(-1.50, -1.09)}$&-1.32&$^{(-1.58, -1.12)}$&-1.32&$^{(-1.57, -1.12)}$&-1.27&$^{(-1.52, -1.06)}$&-1.32&$^{(-1.58, -1.12)}$\\
AGE&-0.05&$^{(-0.25, 0.15)}$&-0.04&$^{(-0.23, 0.10)}$&-0.05&$^{(-0.22, 0.11)}$&-0.04&$^{(-0.21, 0.10)}$&-0.05&$^{(-0.26, 0.14)}$&-0.05&$^{(-0.26, 0.15)}$&-0.06&$^{(-0.23, 0.12)}$&-0.05&$^{(-0.25, 0.14)}$\\
Subm.L.Dm&0.19&$^{(-0.51, 0.82)}$&0.21&$^{(-0.00, 0.72)}$&0.24&$^{(-0.00, 0.39)}$&0.22&$^{(0.00, 0.55)}$&0.28&$^{(-0.36, 0.63)}$&0.28&$^{(-0.36, 0.44)}$&0.26&$^{(0.09, 0.40)}$&0.23&$^{(0.00, 0.81)}$\\
\bottomrule
\end{tabular}

%% file: img/coeff_tables/CompareMethods-sim-A+.tex
\begin{tabular}{lRlRlRlRlRlRlRlRlRlRlRlRlRlRlRlRlR}\toprule
 &\multicolumn{2}{l}{LR}&\multicolumn{2}{l}{Lasso}&\multicolumn{2}{l}{Ridge}&\multicolumn{2}{l}{ElasticNet}&\multicolumn{2}{l}{PCLR}&\multicolumn{2}{l}{LAELR}&\multicolumn{2}{l}{Dropout}&\multicolumn{2}{l}{LRnn}\EndTableHeader\\
\midrule \negsum&-0.65&$^{(-3.03, 0.00)}$&0.00&$^{(0.00, 0.00)}$&-0.01&$^{(-0.15, 0.00)}$&0.00&$^{(0.00, 0.00)}$&-0.30&$^{(-3.76, 0.00)}$&-0.34&$^{(-3.66, 0.00)}$&0.00&$^{(0.00, 0.00)}$&0.00&$^{(0.00, 0.00)}$\\
\possum&2.08&$^{(1.17, 3.74)}$&1.27&$^{(0.97, 1.65)}$&1.27&$^{(1.00, 1.64)}$&1.26&$^{(1.00, 1.63)}$&1.74&$^{(1.15, 5.87)}$&1.78&$^{(1.15, 5.81)}$&1.28&$^{(0.99, 1.68)}$&1.43&$^{(1.13, 1.84)}$\\
\negprop&0.28&$^{(0.00, 0.50)}$&0.00&$^{(0.00, 0.00)}$&0.01&$^{(0.00, 0.25)}$&0.00&$^{(0.00, 0.00)}$&0.06&$^{(0.00, 0.50)}$&0.08&$^{(0.00, 0.50)}$&0.00&$^{(0.00, 0.00)}$&0.00&$^{(0.00, 0.00)}$\\
\posprop&0.71&$^{(0.50, 1.00)}$&0.85&$^{(0.50, 1.00)}$&0.99&$^{(0.75, 1.00)}$&0.97&$^{(0.75, 1.00)}$&0.93&$^{(0.50, 1.00)}$&0.92&$^{(0.50, 1.00)}$&1.00&$^{(1.00, 1.00)}$&0.70&$^{(0.50, 1.00)}$\\
\midrule Intercept&-1.35&$^{(-1.56, -1.04)}$&-1.28&$^{(-1.49, -1.02)}$&-1.28&$^{(-1.48, -1.01)}$&-1.28&$^{(-1.48, -1.01)}$&-1.34&$^{(-1.56, -1.04)}$&-1.34&$^{(-1.56, -1.04)}$&-1.29&$^{(-1.53, -1.01)}$&-1.34&$^{(-1.56, -1.04)}$\\
AGE&0.02&$^{(-0.77, 0.85)}$&-0.01&$^{(-0.16, 0.09)}$&-0.01&$^{(-0.20, 0.14)}$&-0.01&$^{(-0.18, 0.14)}$&0.04&$^{(-0.67, 1.22)}$&0.04&$^{(-0.73, 1.19)}$&-0.00&$^{(-0.20, 0.17)}$&0.02&$^{(-0.24, 0.35)}$\\
Subm.L.Dm&0.26&$^{(-2.04, 2.82)}$&0.15&$^{(-0.00, 0.73)}$&0.24&$^{(0.13, 0.37)}$&0.20&$^{(0.00, 0.38)}$&0.32&$^{(-2.22, 3.98)}$&0.31&$^{(-2.09, 3.95)}$&0.26&$^{(0.12, 0.39)}$&0.25&$^{(0.00, 0.97)}$\\
\bottomrule
\end{tabular}

%% file: img/coeff_tables/CompareMethods-sim-B.tex
\begin{tabular}{lRlRlRlRlRlRlRlRlRlRlRlRlRlRlRlRlR}\toprule
 &\multicolumn{2}{l}{LR}&\multicolumn{2}{l}{Lasso}&\multicolumn{2}{l}{Ridge}&\multicolumn{2}{l}{ElasticNet}&\multicolumn{2}{l}{PCLR}&\multicolumn{2}{l}{LAELR}&\multicolumn{2}{l}{Dropout}&\multicolumn{2}{l}{LRnn}\EndTableHeader\\
\midrule \negsum&-1.18&$^{(-2.25, -0.34)}$&-0.01&$^{(-0.20, 0.00)}$&-0.04&$^{(-0.22, 0.00)}$&-0.02&$^{(-0.17, 0.00)}$&-0.06&$^{(-0.56, 0.00)}$&-0.07&$^{(-0.46, 0.00)}$&-0.05&$^{(-0.22, 0.00)}$&0.00&$^{(0.00, 0.00)}$\\
\possum&2.81&$^{(1.98, 3.92)}$&1.39&$^{(1.00, 1.82)}$&1.44&$^{(1.01, 1.81)}$&1.42&$^{(1.04, 1.78)}$&1.65&$^{(1.26, 2.51)}$&1.65&$^{(1.21, 2.25)}$&1.51&$^{(1.16, 1.98)}$&1.65&$^{(1.26, 2.02)}$\\
\negprop&0.34&$^{(0.19, 0.50)}$&0.01&$^{(0.00, 0.06)}$&0.05&$^{(0.00, 0.19)}$&0.03&$^{(0.00, 0.19)}$&0.04&$^{(0.00, 0.25)}$&0.05&$^{(0.00, 0.25)}$&0.06&$^{(0.00, 0.19)}$&0.00&$^{(0.00, 0.00)}$\\
\posprop&0.63&$^{(0.44, 0.81)}$&0.65&$^{(0.44, 0.88)}$&0.92&$^{(0.75, 1.00)}$&0.83&$^{(0.44, 1.00)}$&0.94&$^{(0.69, 1.00)}$&0.93&$^{(0.62, 1.00)}$&0.89&$^{(0.69, 1.00)}$&0.56&$^{(0.44, 0.75)}$\\
\midrule Intercept&-1.37&$^{(-1.71, -1.10)}$&-1.27&$^{(-1.55, -1.02)}$&-1.25&$^{(-1.53, -1.01)}$&-1.26&$^{(-1.53, -1.03)}$&-1.33&$^{(-1.64, -1.08)}$&-1.33&$^{(-1.64, -1.08)}$&-1.29&$^{(-1.61, -1.04)}$&-1.35&$^{(-1.67, -1.09)}$\\
AGE&-0.03&$^{(-0.31, 0.28)}$&-0.02&$^{(-0.20, 0.17)}$&-0.03&$^{(-0.18, 0.13)}$&-0.03&$^{(-0.17, 0.12)}$&-0.04&$^{(-0.23, 0.20)}$&-0.03&$^{(-0.27, 0.21)}$&-0.03&$^{(-0.19, 0.12)}$&-0.02&$^{(-0.25, 0.27)}$\\
Subm.L.Dm&0.05&$^{(-1.12, 1.05)}$&0.08&$^{(-0.00, 0.47)}$&0.08&$^{(-0.02, 0.20)}$&0.09&$^{(-0.02, 0.24)}$&0.10&$^{(0.01, 0.18)}$&0.10&$^{(-0.02, 0.24)}$&0.09&$^{(-0.02, 0.24)}$&0.06&$^{(0.00, 0.53)}$\\
\bottomrule
\end{tabular}

%% file: img/coeff_tables/CompareMethods-sim-B+.tex
\begin{tabular}{lRlRlRlRlRlRlRlRlRlRlRlRlRlRlRlRlR}\toprule
 &\multicolumn{2}{l}{LR}&\multicolumn{2}{l}{Lasso}&\multicolumn{2}{l}{Ridge}&\multicolumn{2}{l}{ElasticNet}&\multicolumn{2}{l}{PCLR}&\multicolumn{2}{l}{LAELR}&\multicolumn{2}{l}{Dropout}&\multicolumn{2}{l}{LRnn}\EndTableHeader\\
\midrule \negsum&-3.99&$^{(-7.41, -1.42)}$&-0.10&$^{(-1.37, 0.00)}$&-0.09&$^{(-0.93, 0.00)}$&-0.03&$^{(-0.50, 0.00)}$&-0.14&$^{(-1.33, 0.00)}$&-0.30&$^{(-4.09, 0.00)}$&-0.06&$^{(-0.58, 0.00)}$&0.00&$^{(0.00, 0.00)}$\\
\possum&5.52&$^{(2.92, 8.67)}$&1.43&$^{(0.97, 2.89)}$&1.46&$^{(1.04, 2.51)}$&1.39&$^{(1.01, 1.96)}$&1.68&$^{(1.26, 3.17)}$&1.83&$^{(1.23, 5.58)}$&1.50&$^{(1.09, 2.09)}$&1.57&$^{(1.23, 2.00)}$\\
\negprop&0.45&$^{(0.31, 0.62)}$&0.02&$^{(0.00, 0.25)}$&0.04&$^{(0.00, 0.38)}$&0.01&$^{(0.00, 0.19)}$&0.06&$^{(0.00, 0.44)}$&0.06&$^{(0.00, 0.50)}$&0.06&$^{(0.00, 0.31)}$&0.00&$^{(0.00, 0.00)}$\\
\posprop&0.54&$^{(0.38, 0.69)}$&0.60&$^{(0.31, 0.88)}$&0.95&$^{(0.56, 1.00)}$&0.87&$^{(0.31, 1.00)}$&0.93&$^{(0.50, 1.00)}$&0.93&$^{(0.50, 1.00)}$&0.90&$^{(0.69, 1.00)}$&0.39&$^{(0.25, 0.56)}$\\
\midrule Intercept&-1.39&$^{(-1.66, -1.09)}$&-1.29&$^{(-1.52, -1.02)}$&-1.29&$^{(-1.53, -1.01)}$&-1.29&$^{(-1.55, -1.01)}$&-1.35&$^{(-1.61, -1.06)}$&-1.35&$^{(-1.61, -1.06)}$&-1.32&$^{(-1.57, -1.04)}$&-1.37&$^{(-1.63, -1.07)}$\\
AGE&-0.04&$^{(-0.29, 0.15)}$&-0.03&$^{(-0.21, 0.06)}$&-0.03&$^{(-0.21, 0.07)}$&-0.03&$^{(-0.20, 0.08)}$&-0.02&$^{(-0.25, 0.16)}$&-0.02&$^{(-0.27, 0.10)}$&-0.03&$^{(-0.20, 0.11)}$&-0.04&$^{(-0.26, 0.17)}$\\
Subm.L.Dm&0.06&$^{(-1.60, 1.40)}$&0.08&$^{(-0.00, 0.62)}$&0.08&$^{(-0.03, 0.19)}$&0.08&$^{(-0.00, 0.25)}$&0.10&$^{(-0.08, 0.26)}$&0.09&$^{(-0.14, 0.26)}$&0.08&$^{(-0.08, 0.20)}$&0.07&$^{(0.00, 0.51)}$\\
\bottomrule
\end{tabular}

%% file: img/coeff_tables/CompareMethods-sim-C.tex
\begin{tabular}{lRlRlRlRlRlRlRlRlRlRlRlRlRlRlRlRlR}\toprule
 &\multicolumn{2}{l}{LR}&\multicolumn{2}{l}{Lasso}&\multicolumn{2}{l}{Ridge}&\multicolumn{2}{l}{ElasticNet}&\multicolumn{2}{l}{PCLR}&\multicolumn{2}{l}{LAELR}&\multicolumn{2}{l}{Dropout}&\multicolumn{2}{l}{LRnn}\EndTableHeader\\
\midrule \negsum&-1.02&$^{(-2.38, -0.21)}$&-0.05&$^{(-0.50, 0.00)}$&-0.08&$^{(-0.55, 0.00)}$&-0.06&$^{(-0.54, 0.00)}$&-0.18&$^{(-1.69, 0.00)}$&-0.15&$^{(-1.65, 0.00)}$&-0.06&$^{(-0.25, 0.00)}$&0.00&$^{(0.00, 0.00)}$\\
\possum&3.48&$^{(2.44, 5.28)}$&2.07&$^{(1.50, 3.11)}$&2.06&$^{(1.58, 3.00)}$&2.06&$^{(1.49, 2.92)}$&2.49&$^{(1.83, 4.17)}$&2.46&$^{(1.81, 4.16)}$&2.01&$^{(1.50, 2.73)}$&2.52&$^{(1.93, 3.62)}$\\
\negprop&0.32&$^{(0.20, 0.50)}$&0.02&$^{(0.00, 0.20)}$&0.08&$^{(0.00, 0.30)}$&0.05&$^{(0.00, 0.20)}$&0.08&$^{(0.00, 0.30)}$&0.07&$^{(0.00, 0.30)}$&0.07&$^{(0.00, 0.20)}$&0.00&$^{(0.00, 0.00)}$\\
\posprop&0.67&$^{(0.50, 0.80)}$&0.69&$^{(0.50, 0.90)}$&0.90&$^{(0.70, 1.00)}$&0.85&$^{(0.50, 1.00)}$&0.91&$^{(0.60, 1.00)}$&0.91&$^{(0.70, 1.00)}$&0.92&$^{(0.80, 1.00)}$&0.64&$^{(0.40, 0.90)}$\\
\midrule Intercept&-2.96&$^{(-3.58, -2.49)}$&-2.69&$^{(-3.16, -2.28)}$&-2.66&$^{(-3.12, -2.27)}$&-2.67&$^{(-3.14, -2.28)}$&-2.89&$^{(-3.48, -2.44)}$&-2.88&$^{(-3.52, -2.44)}$&-2.69&$^{(-3.19, -2.33)}$&-2.93&$^{(-3.56, -2.47)}$\\
AGE&0.01&$^{(-0.28, 0.25)}$&0.00&$^{(-0.17, 0.18)}$&-0.01&$^{(-0.19, 0.17)}$&-0.01&$^{(-0.18, 0.18)}$&0.03&$^{(-0.30, 0.27)}$&0.04&$^{(-0.31, 0.27)}$&-0.01&$^{(-0.22, 0.17)}$&0.02&$^{(-0.25, 0.26)}$\\
Subm.L.Dm&0.22&$^{(-0.75, 1.07)}$&0.19&$^{(-0.00, 0.66)}$&0.22&$^{(-0.12, 0.53)}$&0.22&$^{(-0.07, 0.54)}$&0.28&$^{(-0.17, 0.88)}$&0.28&$^{(-0.16, 0.80)}$&0.23&$^{(0.05, 0.44)}$&0.21&$^{(0.00, 0.64)}$\\
\bottomrule
\end{tabular}

%% file: img/coeff_tables/CompareMethods-sim-C+.tex
\begin{tabular}{lRlRlRlRlRlRlRlRlRlRlRlRlRlRlRlRlR}\toprule
 &\multicolumn{2}{l}{LR}&\multicolumn{2}{l}{Lasso}&\multicolumn{2}{l}{Ridge}&\multicolumn{2}{l}{ElasticNet}&\multicolumn{2}{l}{PCLR}&\multicolumn{2}{l}{LAELR}&\multicolumn{2}{l}{Dropout}&\multicolumn{2}{l}{LRnn}\EndTableHeader\\
\midrule \negsum&-3.14&$^{(-7.37, -0.74)}$&-0.08&$^{(-1.10, 0.00)}$&-0.11&$^{(-1.31, 0.00)}$&-0.06&$^{(-0.61, 0.00)}$&-0.31&$^{(-3.40, 0.00)}$&-0.39&$^{(-4.87, 0.00)}$&-0.08&$^{(-0.50, 0.00)}$&0.00&$^{(0.00, 0.00)}$\\
\possum&5.40&$^{(2.87, 10.86)}$&1.88&$^{(1.24, 3.37)}$&1.88&$^{(1.30, 3.21)}$&1.82&$^{(1.33, 2.82)}$&2.36&$^{(1.63, 6.30)}$&2.44&$^{(1.59, 6.79)}$&1.84&$^{(1.23, 2.72)}$&2.33&$^{(1.62, 3.33)}$\\
\negprop&0.40&$^{(0.20, 0.60)}$&0.03&$^{(0.00, 0.20)}$&0.08&$^{(0.00, 0.40)}$&0.04&$^{(0.00, 0.20)}$&0.10&$^{(0.00, 0.50)}$&0.10&$^{(0.00, 0.50)}$&0.09&$^{(0.00, 0.30)}$&0.00&$^{(0.00, 0.00)}$\\
\posprop&0.60&$^{(0.40, 0.80)}$&0.60&$^{(0.30, 0.90)}$&0.90&$^{(0.60, 1.00)}$&0.82&$^{(0.40, 1.00)}$&0.88&$^{(0.50, 1.00)}$&0.88&$^{(0.50, 1.00)}$&0.89&$^{(0.70, 1.00)}$&0.48&$^{(0.30, 0.70)}$\\
\midrule Intercept&-2.89&$^{(-3.43, -2.41)}$&-2.65&$^{(-3.22, -2.22)}$&-2.62&$^{(-3.10, -2.24)}$&-2.62&$^{(-3.06, -2.21)}$&-2.82&$^{(-3.40, -2.37)}$&-2.82&$^{(-3.39, -2.37)}$&-2.66&$^{(-3.13, -2.26)}$&-2.86&$^{(-3.41, -2.40)}$\\
AGE&0.07&$^{(-0.70, 0.79)}$&0.02&$^{(-0.18, 0.34)}$&0.02&$^{(-0.22, 0.29)}$&0.02&$^{(-0.18, 0.28)}$&0.07&$^{(-0.25, 0.36)}$&0.09&$^{(-0.26, 0.59)}$&0.02&$^{(-0.20, 0.25)}$&0.06&$^{(-0.26, 0.42)}$\\
Subm.L.Dm&0.22&$^{(-2.42, 3.20)}$&0.10&$^{(-0.00, 0.54)}$&0.15&$^{(-0.15, 0.38)}$&0.14&$^{(-0.01, 0.39)}$&0.16&$^{(-0.39, 0.44)}$&0.24&$^{(-0.34, 1.86)}$&0.14&$^{(-0.17, 0.33)}$&0.17&$^{(0.00, 0.97)}$\\
\bottomrule
\end{tabular}

%% file: img/coeff_tables/CompareMethods-sim-D.tex
\begin{tabular}{lRlRlRlRlRlRlRlRlRlRlRlRlRlRlRlRlR}\toprule
 &\multicolumn{2}{l}{LR}&\multicolumn{2}{l}{Lasso}&\multicolumn{2}{l}{Ridge}&\multicolumn{2}{l}{ElasticNet}&\multicolumn{2}{l}{PCLR}&\multicolumn{2}{l}{LAELR}&\multicolumn{2}{l}{Dropout}&\multicolumn{2}{l}{LRnn}\EndTableHeader\\
\midrule \negsum&-7.10&$^{(-10.95, -4.18)}$&-0.05&$^{(-0.29, 0.00)}$&-0.32&$^{(-1.22, -0.01)}$&-0.19&$^{(-1.10, 0.00)}$&-0.41&$^{(-1.25, 0.00)}$&-0.40&$^{(-1.57, 0.00)}$&-0.48&$^{(-0.98, -0.17)}$&0.00&$^{(0.00, 0.00)}$\\
\possum&9.90&$^{(6.87, 13.90)}$&1.91&$^{(1.23, 2.92)}$&2.35&$^{(1.64, 3.61)}$&2.17&$^{(1.40, 3.25)}$&2.77&$^{(1.86, 4.09)}$&2.75&$^{(1.75, 4.25)}$&2.63&$^{(1.98, 3.58)}$&2.85&$^{(2.05, 3.63)}$\\
\negprop&0.44&$^{(0.35, 0.55)}$&0.03&$^{(0.00, 0.10)}$&0.15&$^{(0.03, 0.30)}$&0.08&$^{(0.00, 0.25)}$&0.16&$^{(0.00, 0.35)}$&0.16&$^{(0.00, 0.40)}$&0.21&$^{(0.10, 0.33)}$&0.00&$^{(0.00, 0.00)}$\\
\posprop&0.54&$^{(0.42, 0.62)}$&0.46&$^{(0.33, 0.62)}$&0.76&$^{(0.62, 0.90)}$&0.64&$^{(0.33, 0.88)}$&0.76&$^{(0.55, 0.97)}$&0.77&$^{(0.55, 0.97)}$&0.72&$^{(0.57, 0.85)}$&0.34&$^{(0.20, 0.45)}$\\
\midrule Intercept&-3.08&$^{(-3.81, -2.62)}$&-2.42&$^{(-2.82, -2.09)}$&-2.43&$^{(-2.92, -2.09)}$&-2.43&$^{(-2.84, -2.10)}$&-2.67&$^{(-3.20, -2.31)}$&-2.65&$^{(-3.20, -2.26)}$&-2.58&$^{(-3.02, -2.25)}$&-2.79&$^{(-3.33, -2.42)}$\\
AGE&0.02&$^{(-0.48, 0.43)}$&-0.00&$^{(-0.13, 0.12)}$&0.00&$^{(-0.11, 0.13)}$&-0.00&$^{(-0.13, 0.13)}$&0.01&$^{(-0.26, 0.32)}$&0.01&$^{(-0.21, 0.34)}$&0.01&$^{(-0.16, 0.18)}$&0.06&$^{(-0.27, 0.35)}$\\
Subm.L.Dm&0.14&$^{(-1.82, 1.98)}$&0.06&$^{(-0.00, 0.38)}$&0.07&$^{(-0.03, 0.17)}$&0.07&$^{(-0.00, 0.26)}$&0.07&$^{(-0.05, 0.19)}$&0.06&$^{(-0.07, 0.16)}$&0.08&$^{(-0.04, 0.21)}$&0.04&$^{(0.00, 0.31)}$\\
\bottomrule
\end{tabular}

%% file: img/coeff_tables/CompareMethods-sim-D+.tex
\begin{tabular}{lRlRlRlRlRlRlRlRlRlRlRlRlRlRlRlRlR}\toprule
 &\multicolumn{2}{l}{LR}&\multicolumn{2}{l}{Lasso}&\multicolumn{2}{l}{Ridge}&\multicolumn{2}{l}{ElasticNet}&\multicolumn{2}{l}{PCLR}&\multicolumn{2}{l}{LAELR}&\multicolumn{2}{l}{Dropout}&\multicolumn{2}{l}{LRnn}\EndTableHeader\\
\midrule \negsum&-12.37&$^{(-17.94, -7.79)}$&-0.05&$^{(-0.62, 0.00)}$&-0.15&$^{(-0.90, 0.00)}$&-0.12&$^{(-0.75, 0.00)}$&-0.26&$^{(-1.67, 0.00)}$&-0.27&$^{(-1.70, 0.00)}$&-0.38&$^{(-1.11, -0.09)}$&0.00&$^{(0.00, 0.00)}$\\
\possum&14.95&$^{(10.05, 19.91)}$&1.75&$^{(1.16, 2.67)}$&2.04&$^{(1.38, 3.15)}$&1.97&$^{(1.33, 3.10)}$&2.46&$^{(1.75, 4.65)}$&2.50&$^{(1.70, 4.57)}$&2.43&$^{(1.86, 3.48)}$&2.57&$^{(1.85, 3.63)}$\\
\negprop&0.46&$^{(0.38, 0.55)}$&0.02&$^{(0.00, 0.10)}$&0.08&$^{(0.00, 0.25)}$&0.06&$^{(0.00, 0.23)}$&0.10&$^{(0.00, 0.30)}$&0.10&$^{(0.00, 0.38)}$&0.18&$^{(0.05, 0.33)}$&0.00&$^{(0.00, 0.00)}$\\
\posprop&0.53&$^{(0.45, 0.60)}$&0.46&$^{(0.28, 0.68)}$&0.84&$^{(0.68, 0.97)}$&0.71&$^{(0.28, 0.95)}$&0.82&$^{(0.57, 1.00)}$&0.83&$^{(0.57, 1.00)}$&0.74&$^{(0.60, 0.88)}$&0.27&$^{(0.15, 0.38)}$\\
\midrule Intercept&-3.02&$^{(-3.63, -2.57)}$&-2.44&$^{(-2.89, -2.09)}$&-2.46&$^{(-2.86, -2.12)}$&-2.47&$^{(-2.84, -2.14)}$&-2.66&$^{(-3.16, -2.29)}$&-2.65&$^{(-3.13, -2.26)}$&-2.61&$^{(-3.01, -2.27)}$&-2.77&$^{(-3.33, -2.40)}$\\
AGE&0.01&$^{(-0.32, 0.37)}$&-0.01&$^{(-0.16, 0.11)}$&-0.02&$^{(-0.15, 0.13)}$&-0.02&$^{(-0.15, 0.12)}$&-0.02&$^{(-0.29, 0.27)}$&-0.01&$^{(-0.21, 0.24)}$&-0.02&$^{(-0.22, 0.15)}$&0.01&$^{(-0.31, 0.35)}$\\
Subm.L.Dm&0.00&$^{(-2.43, 1.84)}$&0.04&$^{(-0.00, 0.36)}$&0.05&$^{(0.01, 0.12)}$&0.05&$^{(-0.00, 0.14)}$&0.05&$^{(-0.02, 0.12)}$&0.05&$^{(-0.01, 0.11)}$&0.06&$^{(-0.04, 0.17)}$&0.03&$^{(0.00, 0.32)}$\\
\bottomrule
\end{tabular}

%% file: sections/COEFFS-REAL.tex
\newpage
\clearpage

\begin{table}[h!]
    \caption{\label{tab:coef:xerbasiclowreal}Mean model coefficients for \xerbasiclowreal{}.}
\resizebox{\textwidth}{!}{
\input{img/coeff_tables/CompareMethods-real-A}}
\end{table}

\begin{table}[h!]
    \caption{\label{tab:coef:xerexthighreal}Mean model coefficients for \xerexthighreal{}.}
\resizebox{\textwidth}{!}{
\input{img/coeff_tables/CompareMethods-real-B+}}
\end{table}

\newpage
\clearpage
\begin{table}[h!]
    \caption{\label{tab:coef:dysbasiclowreal}Mean model coefficients for \dysbasiclowreal{}.}
\resizebox{\textwidth}{!}{
\input{img/coeff_tables/CompareMethods-real-C}}
\end{table}

\newpage
\clearpage
\begin{table}[h!]
    \caption{\label{tab:coef:dysexthighreal}Mean model coefficients for \dysexthighreal{}.}
\resizebox{\textwidth}{!}{
\input{img/coeff_tables/CompareMethods-real-D+}}
\end{table}

%% file: img/coeff_tables/CompareMethods-real-A.tex
\begin{tabular}{lRlRlRlRlRlRlRlRlRlRlRlRlRlRlRlRlR}\toprule
 &\multicolumn{2}{l}{LR}&\multicolumn{2}{l}{Lasso}&\multicolumn{2}{l}{Ridge}&\multicolumn{2}{l}{ElasticNet}&\multicolumn{2}{l}{PCLR}&\multicolumn{2}{l}{LAELR}&\multicolumn{2}{l}{Dropout}&\multicolumn{2}{l}{LRnn}\EndTableHeader\\
\midrule \negsum&-0.08&$^{(-0.09, -0.06)}$&0.00&$^{(0.00, 0.00)}$&0.00&$^{(0.00, 0.00)}$&0.00&$^{(0.00, 0.00)}$&0.00&$^{(0.00, 0.00)}$&0.00&$^{(0.00, 0.00)}$&0.00&$^{(0.00, 0.00)}$&0.00&$^{(0.00, 0.00)}$\\
\possum&1.78&$^{(1.76, 1.79)}$&1.36&$^{(1.31, 1.41)}$&1.34&$^{(1.31, 1.38)}$&1.35&$^{(1.29, 1.40)}$&1.72&$^{(1.69, 1.74)}$&1.72&$^{(1.69, 1.74)}$&1.26&$^{(1.17, 1.37)}$&1.72&$^{(1.70, 1.73)}$\\
\negprop&0.25&$^{(0.25, 0.25)}$&0.00&$^{(0.00, 0.00)}$&0.00&$^{(0.00, 0.00)}$&0.00&$^{(0.00, 0.00)}$&0.00&$^{(0.00, 0.00)}$&0.00&$^{(0.00, 0.00)}$&0.00&$^{(0.00, 0.00)}$&0.00&$^{(0.00, 0.00)}$\\
\posprop&0.75&$^{(0.75, 0.75)}$&0.92&$^{(0.75, 1.00)}$&1.00&$^{(1.00, 1.00)}$&1.00&$^{(1.00, 1.00)}$&1.00&$^{(1.00, 1.00)}$&1.00&$^{(1.00, 1.00)}$&1.00&$^{(1.00, 1.00)}$&0.94&$^{(0.75, 1.00)}$\\
\midrule Intercept&-1.51&$^{(-1.53, -1.49)}$&-1.35&$^{(-1.37, -1.33)}$&-1.34&$^{(-1.36, -1.33)}$&-1.34&$^{(-1.37, -1.31)}$&-1.52&$^{(-1.54, -1.51)}$&-1.52&$^{(-1.54, -1.44)}$&-1.34&$^{(-1.38, -1.31)}$&-1.52&$^{(-1.53, -1.50)}$\\
AGE&-0.02&$^{(-0.02, -0.02)}$&-0.01&$^{(-0.03, -0.00)}$&-0.04&$^{(-0.05, -0.04)}$&-0.02&$^{(-0.04, -0.01)}$&-0.02&$^{(-0.03, -0.02)}$&-0.02&$^{(-0.03, -0.02)}$&-0.05&$^{(-0.08, -0.02)}$&-0.02&$^{(-0.02, -0.01)}$\\
Subm.L.Dm&-0.08&$^{(-0.09, -0.06)}$&0.02&$^{(0.01, 0.05)}$&0.20&$^{(0.16, 0.21)}$&0.12&$^{(0.06, 0.17)}$&0.37&$^{(0.25, 0.46)}$&0.34&$^{(0.17, 0.46)}$&0.25&$^{(0.22, 0.29)}$&0.03&$^{(0.00, 0.08)}$\\
\bottomrule
\end{tabular}

%% file: img/coeff_tables/CompareMethods-real-B+.tex
\begin{tabular}{lRlRlRlRlRlRlRlRlRlRlRlRlRlRlRlRlR}\toprule
 &\multicolumn{2}{l}{LR}&\multicolumn{2}{l}{Lasso}&\multicolumn{2}{l}{Ridge}&\multicolumn{2}{l}{ElasticNet}&\multicolumn{2}{l}{PCLR}&\multicolumn{2}{l}{LAELR}&\multicolumn{2}{l}{Dropout}&\multicolumn{2}{l}{LRnn}\EndTableHeader\\
\midrule \negsum&-3.08&$^{(-3.44, -2.77)}$&-0.00&$^{(0.00, 0.00)}$&0.00&$^{(0.00, 0.00)}$&0.00&$^{(0.00, 0.00)}$&0.00&$^{(0.00, 0.00)}$&-0.02&$^{(-0.22, 0.00)}$&-0.00&$^{(-0.03, 0.00)}$&0.00&$^{(0.00, 0.00)}$\\
\possum&5.06&$^{(4.74, 5.53)}$&1.37&$^{(1.29, 1.47)}$&1.50&$^{(1.47, 1.54)}$&1.49&$^{(1.41, 1.55)}$&2.10&$^{(2.05, 2.16)}$&2.13&$^{(2.03, 2.43)}$&1.69&$^{(1.63, 1.76)}$&2.23&$^{(2.14, 2.65)}$\\
\negprop&0.47&$^{(0.44, 0.50)}$&0.00&$^{(0.00, 0.00)}$&0.00&$^{(0.00, 0.00)}$&0.00&$^{(0.00, 0.00)}$&0.00&$^{(0.00, 0.00)}$&0.02&$^{(0.00, 0.25)}$&0.00&$^{(0.00, 0.06)}$&0.00&$^{(0.00, 0.00)}$\\
\posprop&0.53&$^{(0.50, 0.56)}$&0.79&$^{(0.69, 0.88)}$&1.00&$^{(1.00, 1.00)}$&0.99&$^{(0.88, 1.00)}$&1.00&$^{(1.00, 1.00)}$&0.97&$^{(0.69, 1.00)}$&0.98&$^{(0.94, 1.00)}$&0.77&$^{(0.69, 0.88)}$\\
\midrule Intercept&-1.80&$^{(-1.97, -1.74)}$&-1.36&$^{(-1.42, -1.33)}$&-1.39&$^{(-1.41, -1.37)}$&-1.39&$^{(-1.42, -1.35)}$&-1.67&$^{(-1.72, -1.65)}$&-1.69&$^{(-1.75, -1.65)}$&-1.50&$^{(-1.52, -1.47)}$&-1.77&$^{(-1.99, -1.72)}$\\
AGE&-0.05&$^{(-0.06, -0.05)}$&-0.01&$^{(-0.03, -0.00)}$&-0.03&$^{(-0.03, -0.02)}$&-0.02&$^{(-0.04, -0.01)}$&0.04&$^{(-0.01, 0.07)}$&0.04&$^{(0.00, 0.06)}$&-0.03&$^{(-0.06, 0.01)}$&-0.04&$^{(-0.04, -0.03)}$\\
Subm.L.Dm&-0.29&$^{(-0.33, -0.26)}$&0.01&$^{(0.00, 0.03)}$&0.06&$^{(0.05, 0.06)}$&0.04&$^{(0.03, 0.06)}$&0.14&$^{(0.11, 0.16)}$&0.14&$^{(-0.03, 0.17)}$&0.07&$^{(0.03, 0.11)}$&0.00&$^{(0.00, 0.03)}$\\
\bottomrule
\end{tabular}

%% file: img/coeff_tables/CompareMethods-real-C.tex
\begin{tabular}{lRlRlRlRlRlRlRlRlRlRlRlRlRlRlRlRlR}\toprule
 &\multicolumn{2}{l}{LR}&\multicolumn{2}{l}{Lasso}&\multicolumn{2}{l}{Ridge}&\multicolumn{2}{l}{ElasticNet}&\multicolumn{2}{l}{PCLR}&\multicolumn{2}{l}{LAELR}&\multicolumn{2}{l}{Dropout}&\multicolumn{2}{l}{LRnn}\EndTableHeader\\
\midrule \negsum&-0.53&$^{(-0.58, -0.48)}$&0.00&$^{(0.00, 0.00)}$&0.00&$^{(0.00, 0.00)}$&0.00&$^{(0.00, 0.00)}$&-0.00&$^{(-0.03, 0.00)}$&-0.01&$^{(-0.06, 0.00)}$&-0.00&$^{(0.00, 0.00)}$&0.00&$^{(0.00, 0.00)}$\\
\possum&3.38&$^{(3.32, 3.44)}$&2.30&$^{(2.22, 2.40)}$&2.26&$^{(2.20, 2.32)}$&2.28&$^{(2.17, 2.40)}$&2.70&$^{(2.54, 2.86)}$&2.76&$^{(2.57, 2.96)}$&2.40&$^{(2.25, 2.59)}$&2.95&$^{(2.90, 3.01)}$\\
\negprop&0.33&$^{(0.30, 0.40)}$&0.00&$^{(0.00, 0.00)}$&0.00&$^{(0.00, 0.00)}$&0.00&$^{(0.00, 0.00)}$&0.01&$^{(0.00, 0.10)}$&0.02&$^{(0.00, 0.10)}$&0.00&$^{(0.00, 0.00)}$&0.00&$^{(0.00, 0.00)}$\\
\posprop&0.55&$^{(0.50, 0.70)}$&0.79&$^{(0.60, 1.00)}$&1.00&$^{(1.00, 1.00)}$&0.97&$^{(0.90, 1.00)}$&0.97&$^{(0.90, 1.00)}$&0.97&$^{(0.90, 1.00)}$&0.99&$^{(0.90, 1.00)}$&0.94&$^{(0.80, 1.00)}$\\
\midrule Intercept&-3.02&$^{(-3.05, -2.99)}$&-2.69&$^{(-2.75, -2.66)}$&-2.72&$^{(-2.75, -2.68)}$&-2.71&$^{(-2.76, -2.64)}$&-3.02&$^{(-3.10, -2.95)}$&-3.05&$^{(-3.14, -2.97)}$&-2.80&$^{(-2.89, -2.73)}$&-3.04&$^{(-3.08, -3.02)}$\\
AGE&0.02&$^{(0.01, 0.03)}$&0.00&$^{(-0.01, 0.02)}$&0.01&$^{(0.00, 0.01)}$&0.00&$^{(-0.01, 0.02)}$&0.05&$^{(0.04, 0.07)}$&0.06&$^{(0.04, 0.09)}$&0.01&$^{(-0.02, 0.04)}$&0.03&$^{(0.02, 0.04)}$\\
Subm.L.Dm&0.12&$^{(0.08, 0.15)}$&0.03&$^{(0.00, 0.10)}$&0.17&$^{(0.16, 0.19)}$&0.11&$^{(0.05, 0.16)}$&0.29&$^{(0.16, 0.39)}$&0.31&$^{(0.19, 0.41)}$&0.21&$^{(0.16, 0.28)}$&0.11&$^{(0.05, 0.18)}$\\
\bottomrule
\end{tabular}

%% file: img/coeff_tables/CompareMethods-real-D+.tex
\begin{tabular}{lRlRlRlRlRlRlRlRlRlRlRlRlRlRlRlRlR}\toprule
 &\multicolumn{2}{l}{LR}&\multicolumn{2}{l}{Lasso}&\multicolumn{2}{l}{Ridge}&\multicolumn{2}{l}{ElasticNet}&\multicolumn{2}{l}{PCLR}&\multicolumn{2}{l}{LAELR}&\multicolumn{2}{l}{Dropout}&\multicolumn{2}{l}{LRnn}\EndTableHeader\\
\midrule \negsum&-9.10&$^{(-9.73, -8.70)}$&-0.04&$^{(-0.12, 0.00)}$&-0.15&$^{(-0.18, -0.12)}$&-0.12&$^{(-0.17, -0.07)}$&-0.46&$^{(-1.20, -0.08)}$&-0.30&$^{(-0.93, -0.04)}$&-0.26&$^{(-0.51, -0.09)}$&0.00&$^{(0.00, 0.00)}$\\
\possum&13.39&$^{(12.86, 14.06)}$&1.98&$^{(1.81, 2.21)}$&2.43&$^{(2.38, 2.52)}$&2.29&$^{(2.15, 2.45)}$&4.38&$^{(3.07, 5.75)}$&3.76&$^{(2.61, 5.08)}$&3.23&$^{(2.95, 3.62)}$&4.82&$^{(4.65, 4.98)}$\\
\negprop&0.46&$^{(0.42, 0.50)}$&0.03&$^{(0.00, 0.07)}$&0.08&$^{(0.07, 0.10)}$&0.08&$^{(0.05, 0.10)}$&0.20&$^{(0.07, 0.30)}$&0.14&$^{(0.03, 0.30)}$&0.14&$^{(0.10, 0.20)}$&0.00&$^{(0.00, 0.00)}$\\
\posprop&0.53&$^{(0.50, 0.57)}$&0.42&$^{(0.35, 0.53)}$&0.87&$^{(0.85, 0.88)}$&0.81&$^{(0.65, 0.88)}$&0.71&$^{(0.62, 0.85)}$&0.80&$^{(0.65, 0.95)}$&0.80&$^{(0.70, 0.88)}$&0.50&$^{(0.42, 0.57)}$\\
\midrule Intercept&-3.57&$^{(-3.61, -3.54)}$&-2.49&$^{(-2.57, -2.42)}$&-2.61&$^{(-2.63, -2.59)}$&-2.58&$^{(-2.62, -2.53)}$&-3.13&$^{(-3.36, -2.93)}$&-3.04&$^{(-3.32, -2.82)}$&-2.84&$^{(-2.91, -2.80)}$&-3.42&$^{(-3.48, -3.38)}$\\
AGE&0.08&$^{(0.07, 0.09)}$&0.00&$^{(-0.01, 0.01)}$&0.01&$^{(0.00, 0.01)}$&0.01&$^{(-0.00, 0.02)}$&0.03&$^{(-0.03, 0.06)}$&0.01&$^{(-0.08, 0.05)}$&0.01&$^{(-0.03, 0.04)}$&0.05&$^{(0.04, 0.06)}$\\
Subm.L.Dm&0.54&$^{(0.49, 0.61)}$&0.03&$^{(0.00, 0.11)}$&0.08&$^{(0.08, 0.09)}$&0.07&$^{(0.05, 0.09)}$&0.05&$^{(0.00, 0.10)}$&0.07&$^{(0.03, 0.14)}$&0.12&$^{(0.07, 0.16)}$&0.06&$^{(0.00, 0.16)}$\\
\bottomrule
\end{tabular}

%% file: main.bbl
\begin{thebibliography}{10}
\providecommand \doibase [0]{http://dx.doi.org/}%

\bibitem{teipel2015relative}
Teipel SJ, Kurth J, Krause B, Grothe MJ, Initiative ADN, others . The relative
  importance of imaging markers for the prediction of Alzheimer's disease
  dementia in mild cognitive impairment—beyond classical regression. {\it
  NeuroImage: Clinical} 2015\string; 8\string: 583--593.

\bibitem{westerhuis2012prediction}
Westerhuis ME, Schuit E, Kwee A, et al. Prediction of neonatal metabolic
  acidosis in women with a singleton term pregnancy in cephalic presentation.
  {\it American Journal of Perinatology} 2012\string; 29(03)\string: 167--174.

\bibitem{narchi2018prediction}
Narchi H, AlBlooshi A. Prediction equations of forced oscillation technique:
  the insidious role of collinearity. {\it Respiratory research} 2018\string;
  19(1)\string: 48.

\bibitem{van2020key}
Bosch V.~dL, Schuit E, Laan v.~dHP, et al. Key challenges in normal tissue
  complication probability model development and validation: towards a
  comprehensive strategy. {\it Radiotherapy and Oncology} 2020.

\bibitem{van2019method}
Van Der~Schaaf A, Bosch V.~dL, Both S, Schuit E, Langendijk J. EP-1914 A method
  to deal with highly correlated explanatory variables in the development of
  NTCP models. {\it Radiotherapy and Oncology} 2019\string; 133\string: S1040.

\bibitem{schisterman2017collinearity}
Schisterman EF, Perkins NJ, Mumford SL, Ahrens KA, Mitchell EM. Collinearity
  and causal diagrams--a lesson on the importance of model specification. {\it
  Epidemiology (Cambridge, Mass.)} 2017\string; 28(1)\string: 47.

\bibitem{farrar1967multicollinearity}
Farrar DE, Glauber RR. Multicollinearity in regression analysis: the problem
  revisited. {\it The Review of Economic and Statistics} 1967\string: 92--107.

\bibitem{harrell2015regression}
Harrell~Jr FE. {\it Regression modeling strategies: with applications to linear
  models, logistic and ordinal regression, and survival analysis}.
\newblock Springer .
\newblock 2015.

\bibitem{schuit2013unexpected}
Schuit E, Groenwold RH, Harrell FE, et al. Unexpected predictor--outcome
  associations in clinical prediction research: causes and solutions. {\it
  CMAJ} 2013\string; 185(10)\string: E499--E505.

\bibitem{moons2009prognosis}
Moons KG, Altman DG, Vergouwe Y, Royston P. Prognosis and prognostic research:
  application and impact of prognostic models in clinical practice. {\it Bmj}
  2009\string; 338\string: b606.

\bibitem{tibshirani1996regression}
Tibshirani R. Regression shrinkage and selection via the lasso. {\it Journal of
  the Royal Statistical Society: Series B (Methodological)} 1996\string;
  58(1)\string: 267--288.

\bibitem{hoerl1970ridge}
Hoerl AE, Kennard RW. Ridge regression: Biased estimation for nonorthogonal
  problems. {\it Technometrics} 1970\string; 12(1)\string: 55--67.

\bibitem{zou2005regularization}
Zou H, Hastie T. Regularization and variable selection via the elastic net.
  {\it Journal of the Royal Statistical Society: Series B (Statistical
  Methodology)} 2005\string; 67(2)\string: 301--320.

\bibitem{riley2020penalisation}
Riley RD, Snell KI, Martin GP, et al. Penalisation and shrinkage methods
  produced unreliable clinical prediction models especially when sample size
  was small. {\it Journal of Clinical Epidemiology} 2020.

\bibitem{hinton2012improving}
Hinton GE, Srivastava N, Krizhevsky A, Sutskever I, Salakhutdinov RR. Improving
  neural networks by preventing co-adaptation of feature detectors. {\it arXiv
  preprint arXiv:1207.0580} 2012.

\bibitem{wager2013dropout}
Wager S, Wang S, Liang PS. Dropout training as adaptive regularization. {\it
  Advances in neural information processing systems} 2013\string; 26\string:
  351--359.

\bibitem{kendall1965course}
Kendall MG, others . {\it Course in multivariate analysis}.
\newblock Charles Griffin \& Co. .
\newblock 1965.

\bibitem{aguilera2006using}
Aguilera AM, Escabias M, Valderrama MJ. Using principal components for
  estimating logistic regression with high-dimensional multicollinear data.
  {\it Computational Statistics \& Data Analysis} 2006\string; 50(8)\string:
  1905--1924.

\bibitem{suarthana2010diagnostic}
Suarthana E, Vergouwe Y, Moons KG, et al. A diagnostic model for the detection
  of sensitization to wheat allergens was developed and validated in bakery
  workers. {\it Journal of clinical epidemiology} 2010\string; 63(9)\string:
  1011--1019.

\bibitem{kunin2019loss}
Kunin D, Bloom J, Goeva A, Seed C. Loss Landscapes of Regularized Linear
  Autoencoders. {\it International Conference on Machine Learning} 2019\string:
  3560--3569.

\bibitem{hull1997xerox}
Hull D, Grefenstette G, Schulze B, Gaussier E, Sch{\"u}tze H. Xerox TREC-5 site
  report: Routing, filtering, NLP, and Spanish tracks. {\it NIST special
  publication} 1997(500238)\string: 167--180.

\bibitem{calamai1987projected}
Calamai PH, Mor{\'e} JJ. Projected gradient methods for linearly constrained
  problems. {\it Mathematical Programming} 1987\string; 39(1)\string: 93--116.

\bibitem{dritschilo1978complication}
Dritschilo A, Chaffey J, Bloomer W, Marck A. The complication probability
  factor: A method for selection of radiation treatment plans. {\it The British
  journal of radiology} 1978\string; 51(605)\string: 370--374.

\bibitem{langendijk2013selection}
Langendijk JA, Lambin P, De~Ruysscher D, Widder J, Bos M, Verheij M. Selection
  of patients for radiotherapy with protons aiming at reduction of side
  effects: the model-based approach. {\it Radiotherapy and Oncology}
  2013\string; 107(3)\string: 267--273.

\bibitem{wolbarst1982optimization}
Wolbarst AB, Chin LM, Svensson GK. Optimization of radiation therapy:
  integral-response of a model biological system. {\it International Journal of
  Radiation Oncology* Biology* Physics} 1982\string; 8(10)\string: 1761--1769.

\bibitem{kierkels2016multivariable}
Kierkels RG, Wopken K, Visser R, et al. Multivariable normal tissue
  complication probability model-based treatment plan optimization for grade
  2--4 dysphagia and tube feeding dependence in head and neck radiotherapy.
  {\it Radiotherapy and Oncology} 2016\string; 121(3)\string: 374--380.

\bibitem{morris2019using}
Morris TP, White IR, Crowther MJ. Using simulation studies to evaluate
  statistical methods. {\it Statistics in Medicine} 2019\string; 38(11)\string:
  2074--2102.

\bibitem{neter1989applied}
Neter J, Wasserman W, Kutner MH. {\it Applied linear regression models}.
\newblock Irwin Homewood, IL .
\newblock 1989.

\bibitem{snoek2012practical}
Snoek J, Larochelle H, Adams RP. Practical bayesian optimization of machine
  learning algorithms. {\it Advances in Neural Information Processing Systems}
  2012\string: 2951--2959.

\bibitem{kim2009estimating}
Kim JH. Estimating classification error rate: Repeated cross-validation,
  repeated hold-out and bootstrap. {\it Computational Statistics \& Data
  Analysis} 2009\string; 53(11)\string: 3735--3745.

\bibitem{pedregosa2011scikit}
Pedregosa F, Varoquaux G, Gramfort A, et al. Scikit-learn: Machine learning in
  Python. {\it Journal of Machine Learning Research} 2011\string; 12\string:
  2825--2830.

\bibitem{paszke2019pytorch}
Paszke A, Gross S, Massa F, et al. Pytorch: An imperative style,
  high-performance deep learning library. {\it Advances in Neural Information
  Processing Systems} 2019\string: 8026--8037.

\bibitem{van2016calibration}
Van~Calster B, Nieboer D, Vergouwe Y, De~Cock B, Pencina MJ, Steyerberg EW. A
  calibration hierarchy for risk models was defined: from utopia to empirical
  data. {\it Journal of Clinical Epidemiology} 2016\string; 74\string:
  167--176.

\bibitem{cohen2013applied}
Cohen J, Cohen P, West SG, Aiken LS. {\it Applied multiple
  regression/correlation analysis for the behavioral sciences}.
\newblock Routledge .
\newblock 2013.

\bibitem{dormann2013collinearity}
Dormann CF, Elith J, Bacher S, et al. Collinearity: a review of methods to deal
  with it and a simulation study evaluating their performance. {\it Ecography}
  2013\string; 36(1)\string: 27--46.

\bibitem{pavlou2016review}
Pavlou M, Ambler G, Seaman S, De~Iorio M, Omar RZ. Review and evaluation of
  penalised regression methods for risk prediction in low-dimensional data with
  few events. {\it Statistics in Medicine} 2016\string; 35(7)\string:
  1159--1177.

\bibitem{brouwer2015ct}
Brouwer CL, Steenbakkers RJ, Bourhis J, et al. CT-based delineation of organs
  at risk in the head and neck region: DAHANCA, EORTC, GORTEC, HKNPCSG, NCIC
  CTG, NCRI, NRG Oncology and TROG consensus guidelines. {\it Radiotherapy and
  Oncology} 2015\string; 117(1)\string: 83--90.

\bibitem{kingma2014adam}
Kingma DP, Ba J. Adam: A method for stochastic optimization. {\it arXiv
  preprint arXiv:1412.6980} 2014.

\bibitem{udell2015generalized}
Udell M. {\it Generalized Low Rank Models}.
\newblock Stanford University .
\newblock 2015.

\bibitem{riley2020calculating}
Riley RD, Ensor J, Snell KI, et al. Calculating the sample size required for
  developing a clinical prediction model. {\it Bmj} 2020\string; 368.

\bibitem{beetz2012ntcp}
Beetz I, Schilstra C, Schaaf v.~dA, et al. NTCP models for patient-rated
  xerostomia and sticky saliva after treatment with intensity modulated
  radiotherapy for head and neck cancer: the role of dosimetric and clinical
  factors. {\it Radiotherapy and Oncology} 2012\string; 105(1)\string:
  101--106.

\bibitem{lee2015patient}
Lee TF, Liou MH, Ting HM, et al. Patient-and therapy-related factors associated
  with the incidence of xerostomia in nasopharyngeal carcinoma patients
  receiving parotid-sparing helical tomotherapy. {\it Scientific Reports}
  2015\string; 5(1)\string: 1--13.

\bibitem{christianen2016swallowing}
Christianen ME, Schaaf v.~dA, Laan v.~dHP, et al. Swallowing sparing intensity
  modulated radiotherapy (SW-IMRT) in head and neck cancer: clinical validation
  according to the model-based approach. {\it Radiotherapy and Oncology}
  2016\string; 118(2)\string: 298--303.

\end{thebibliography}
